\documentclass[a4paper]{book}

\usepackage{tocbibind}
\usepackage{amsmath}
\usepackage{amssymb}
\usepackage{graphicx}
\usepackage[english]{babel}
\usepackage[round, sort, numbers, authoryear]{natbib}
\usepackage[breaklinks = true,linktocpage,pagebackref,colorlinks =
true,linkcolor = black,urlcolor = black, citecolor = black, anchorcolor =
black, hyperindex = true, hyperfigures] {hyperref}
\usepackage[twoside,margin=1in
]{geometry}
\usepackage[onehalfspacing]{setspace}
\usepackage[grey]{quotchap}
\usepackage{enumerate}

\usepackage{amsmath}
\usepackage{amssymb}

\newcommand{\Msun}{\hbox{$\rm\thinspace \text{M}_{\odot}$}}

\newcommand{\ms}{$\,{\rm M}_\mathrm{\odot}$}

\newcommand{\diff}[2]{\frac{\partial (#1)}{\partial #2}} 
\newcommand{\diffb}[2]{\frac{\partial #1}{\partial #2}} 
\newcommand{\difft}[2]{\frac{{\rm d} #1}{{\rm d} #2}}
\newcommand{\bi}[1]{\textbf{\textit{#1}}}

\begin{document}
\pagenumbering{roman}



\pagestyle{empty}
\cleardoublepage
\begin{center}
{\vspace*{10mm} {{\includegraphics[width=30mm]{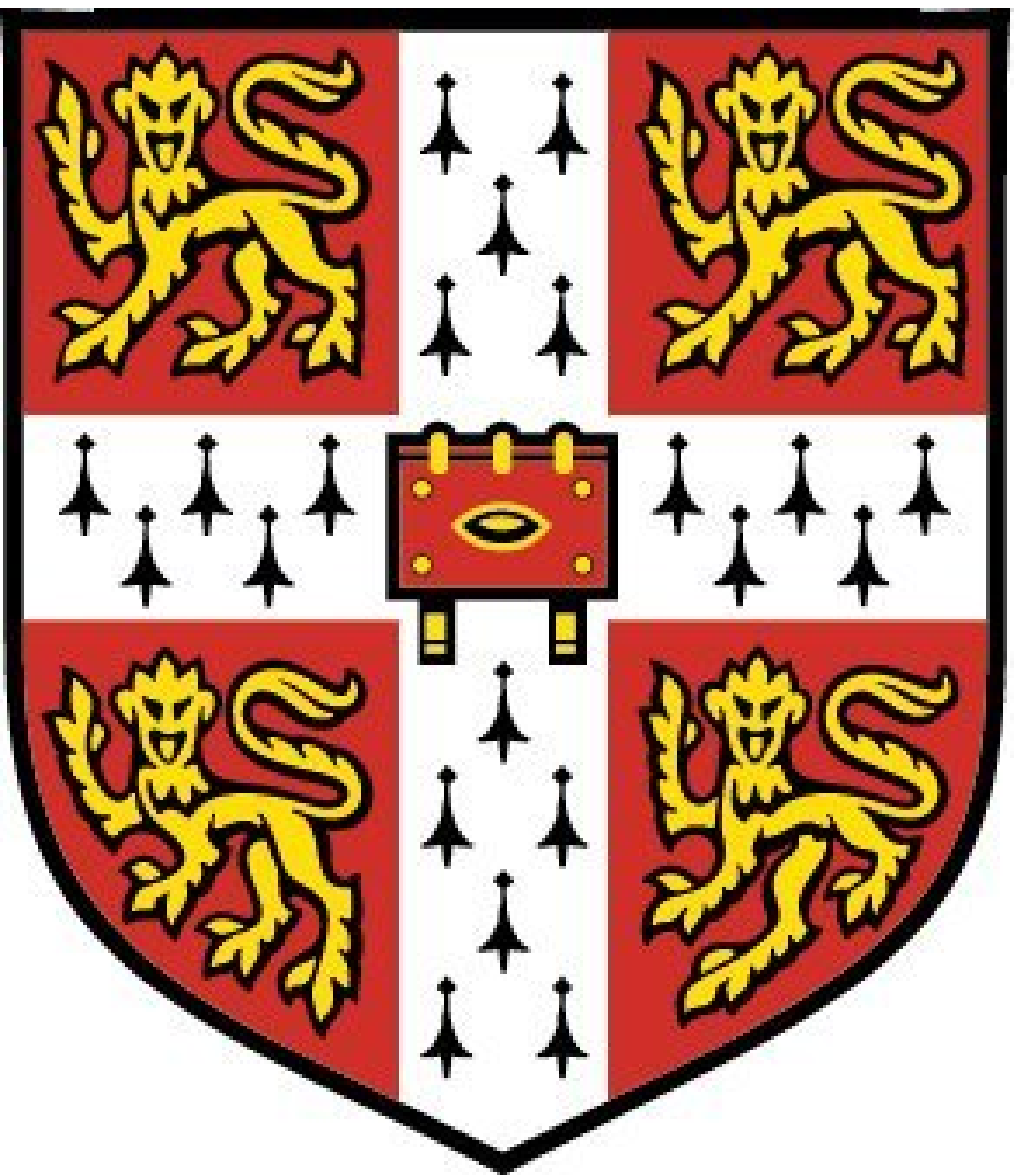}} \par} 
\vspace*{10mm}}
{{\Large{\sc University of Cambridge\\Institute of Astronomy}} \par}
{\vspace*{10mm}}
{{\Large A dissertation submitted for the degree of\\Doctor of Philosophy} \par}
{\vspace*{10mm}}
{{\Huge{\bf  Rotation and magnetism\\in massive stars}}\par}
{\vspace*{10mm}}
{{\Huge{Adrian Thomas Potter}} \par}
{\large \vspace*{10mm} {{\includegraphics[width=30mm]{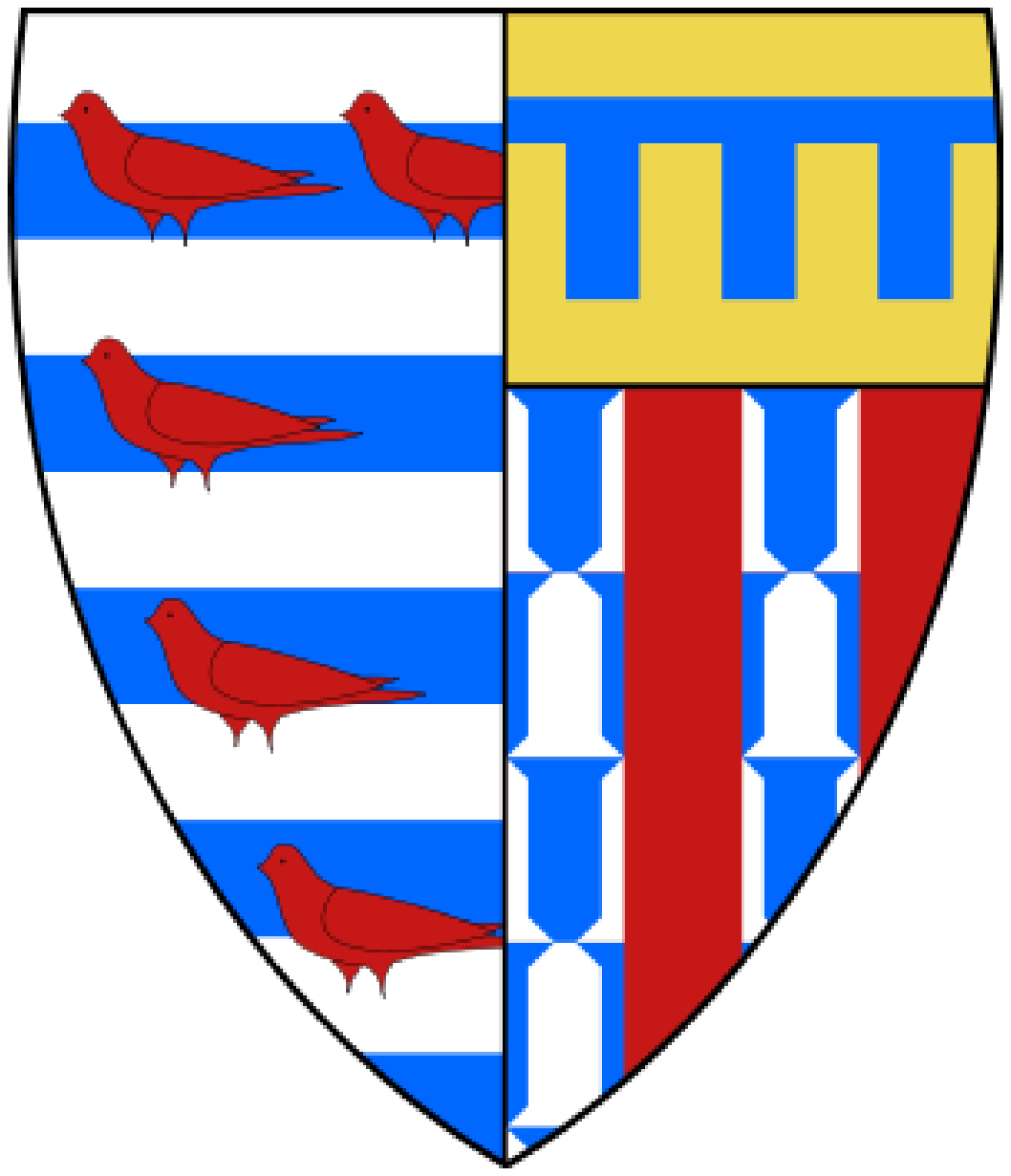}} \par} }
{\vspace*{5mm}}
{{\Large{\sc Pembroke College}}\par}
{\vspace*{10mm}}
{{\it Under the supervision of Dr. Christopher A. Tout}\par}
{\vspace*{5mm}}
{{\it Submitted to the Board of Graduate Studies\\ 10 May, 2012}\par}
  \end{center}
\null\vfill



\frontmatter

\large
\chapter*{}

\begin{center}

For everyone who helped to get me here

\end{center}

\tableofcontents
\chapter{Declaration}

\singlespacing

\begin{quote}

I hereby declare that my dissertation entitled {\it Rotation and magnetism in massive stars} is not substantially the same as any that I have submitted for a degree or diploma or other qualification at any other university. I further state that no part of my thesis has already been or is being concurrently submitted for any such degree, diploma or other qualification. This dissertation is the result of my own work and includes nothing which is the outcome of work done in collaboration except where specifically indicated in the text. Those parts which have been published or accepted for publication are:

\begin{itemize}
\item Material from chapter \ref{ch1} is largely intended as a literature review and so draws heavily on the references contained therein. Material from section \ref{ch1.sec.ce} was submitted for the Certificate of Postgraduate Study to the University of Cambridge.
\item Material from chapters \ref{ch2} and \ref{ch3} has been published as: Potter~A.~T., Tout~C.~A. and Eldridge~J.~J., 2012, ``Towards a unified model of stellar rotation'', Monthly Notices of the Royal Astronomical Society, 419, 788-759 and was completed in collaboration with these authors.
\item Material from chapters \ref{ch2} and \ref{ch4} has been accepted for publication as: Potter~A.~T., Brott~I. and Tout~C.~A., ``Towards a unified model of stellar rotation II: Model-dependent characteristics of stellar populations'', Monthly Notices of the Royal Astronomical Society and was completed in collaboration with these authors.
\item Material from chapters \ref{ch2} and \ref{ch5} has been accepted for publication as: Potter~A.~T., Chitre~S.~M. and Tout~C.~A., ``Stellar evolution of massive stars with a radiative alpha--omega dynamo'', Monthly Notices of the Royal Astronomical Society and was completed in collaboration with these authors.
\item Material from chapter \ref{ch6} has been published as: Potter~A.~T. and Tout~C.~A., 2010, ``Magnetic field evolution of white dwarfs in strongly interacting binary star systems'', Monthly Notices of the Royal Astronomical Society, 402, 1072-1080 and was completed in collaboration with that author.
\end{itemize}
\noindent This dissertation contains fewer than $60,000$ words.


\vspace*{2.0cm}

\noindent Adrian Potter\\

\noindent \today

\end{quote}

\onehalfspacing

\begin{savequote}[60mm]
Edmund: It's taken me seven years, and it's perfect... My magnum opus, Baldrick. Everybody has one novel in them, and this is mine.
 
Baldrick: And this is mine. My magnificent octopus.

 (Blackadder the Third, Ink and Incapability)
\end{savequote}

\chapter{Acknowledgements}

\begin{quote}

For this piece of work I am indebted to the invaluable help and support of countless people over the past four years. Without their time and effort none of this work would have been possible.

First and foremost I would like to give the greatest of thanks to my supervisor, Christopher Tout. His hard work, knowledge and guidance have been an absolutely essential for driving this work forwards and I am thoroughly grateful. I am also particularly thankful for the support of John Eldridge who stepped in to fill the void whilst Christopher was on sabbatical. Not only that but his firm knowledge of the Cambridge stellar evolution code has allowed me to overcome countless hurdles. I would also like to thank Ines Brott and Shashikumar Chitre for their ongoing support.

The success of a PhD is dependent on all of those people behind the scenes who make things possible. Firmly at the top of that list is Andrea Kuesters who has always been there when I needed her the most. I have always been able count on her regardless of what life sent my way and for that I am eternally grateful. More than that we've had some fantastic times and even when she tried to break my hand one Christmas, I treasure every moment we've spent together. I also couldn't have asked for a better year group at the Institute of Astronomy. Amy, Becky, Ryan, Warrick, Steph, Alex, Jon, James, other James, Dom and Yin-Zhe, you've been fantastic, thank you. I'd also like to thank Chrissie, Amy, Natasha, Barny, Sam, Mark C, James, Samantha, Mark W, Simon, Bahar, Sophie and Lucy for all the great times during the past four years. In addition I'd like to thank Sian Owen, Margaret Harding and Becky Coombs for their tireless administrative efforts. My family have also been a constant source of support. They shouldered the financial burden of my degrees and have always been just a phone call away whenever I needed them.  

Finally I'd like to give my thanks to Christine whose love, support and encouragement has been unwavering. Her time and understanding has been so important in helping me to cope with pressure of finishing my eight years in Cambridge and I'm very thankful for everything she's given me.

\end{quote}
\begin{savequote}[60mm]
...so smart it's got a PhD from Cambridge.
(Blackadder Goes Forth, Private Plane)
\end{savequote}

\chapter{Summary}

Rotation has a number of important effects on the evolution of stars. Apart from structural changes because of the centrifugal force, turbulent mixing and meridional circulation can dramatically affect a star's chemical evolution. This leads to changes in the surface temperature and luminosity as well as modifying its lifetime. Rotation decreases the surface gravity, causes enhanced mass loss and leads to surface abundance anomalies of various chemical isotopes all of which have been observed. The replication of these physical effects with simple stellar evolution models is very difficult and has resulted in the use of numerous different formulations to describe the physics. We have adapted the Cambridge stellar evolution code to incorporate a number of different physical models for rotation, including several treatments of angular momentum transport in convection zones. We compare detailed grids of stellar evolution models along with simulated stellar populations to identify the key differences between them. We then consider how these models relate to observed data.

Models of rotationally-driven dynamos in stellar radiative zones have suggested that magnetohydrodynamic transport of angular momentum and chemical composition can dominate over the otherwise purely hydrodynamic processes. If this is the case then a proper consideration of the interaction between rotation and magnetic fields is essential. We have adapted our purely hydrodynamic model to include the evolution of the magnetic field with a pair of time-dependent advection--diffusion equations coupled with the equations for the evolution of the angular momentum distribution and stellar structure. This produces a much more complete, though still reasonably simple, model for the magnetic field evolution. We consider how the surface field strength varies during the main-sequence evolution and compare the surface enrichment of nitrogen for a simulated stellar population with observations.

Strong magnetic fields are also observed at the end of the stellar lifetime.  The surface magnetic field strength of white dwarfs is observed to vary from very little up to $10^9$G. As well as considering the main-sequence evolution of magnetic fields we also look at how the strongest magnetic fields in white dwarfs may be generated by dynamo action during the common envelope phase of strongly interacting binary stars. The resulting magnetic field depends strongly on the electrical conductivity of the white dwarf, the lifetime of the convective envelope and the variability of the magnetic dynamo. We assess the various energy sources available and estimate necessary lifetimes of the common envelope.

\listoftables
\listoffigures


\mainmatter
\pagenumbering{arabic}
\begin{savequote}[60mm]
Focus on the journey, not the destination. Joy is found not in finishing an activity but in doing it. (Greg Anderson)
\end{savequote}

\chapter{Introduction}
\label{ch1}

The study of the effects of rotation on stars is notoriously difficult because of the challenge to introduce them in a consistent yet sufficiently simple way. Rotation's strong connection with the evolution of magnetic fields through dynamo mechanisms means that a great deal of interesting behaviour arises from their introduction into stellar models. Rotation and magnetic fields are present in almost all areas of astrophysics at all scales and are so very deserving of attention. Massive stars are of particular interest because they are largely responsible for driving the heavy element chemical evolution of the Universe. They are far hotter and more luminous than our own Sun and burn through their nuclear fuel much faster. During the late stages of their evolution, many of the heavier elements that are so important to us on Earth, are formed and at the end of their lives they explode in huge supernovae, scattering their ashes over huge distances. The remnants of these explosions eventually begin to collapse again to form new stars and planets. Rotation and magnetic fields not only affect the structure of stars but also the way in which they evolve. Understanding how rotation and magnetic fields affect stars is therefore of great importance for understanding how the Universe evolves as a whole.

\section{Stellar evolution}

At its simplest level, stellar evolution is the study of why stars exist and how they change over time. The subject has a long history and beyond the fairly straightforward basic principles, a huge number of subtle and interesting effects have been identified that cause a plethora of fascinating behaviours. In this work we focus on two of these, rotation and magnetism.

\subsection{The mechanical equilibrium of stars}

Stars are incredible objects. They power the Universe through nuclear fusion and make life possible through the formation of heavy elements. Yet basic models of stars can be constructed from extremely simple principles. At each point in a star, the stellar material is being acted on by two major forces.

\begin{figure}
\centering
\includegraphics[width=0.99\textwidth]{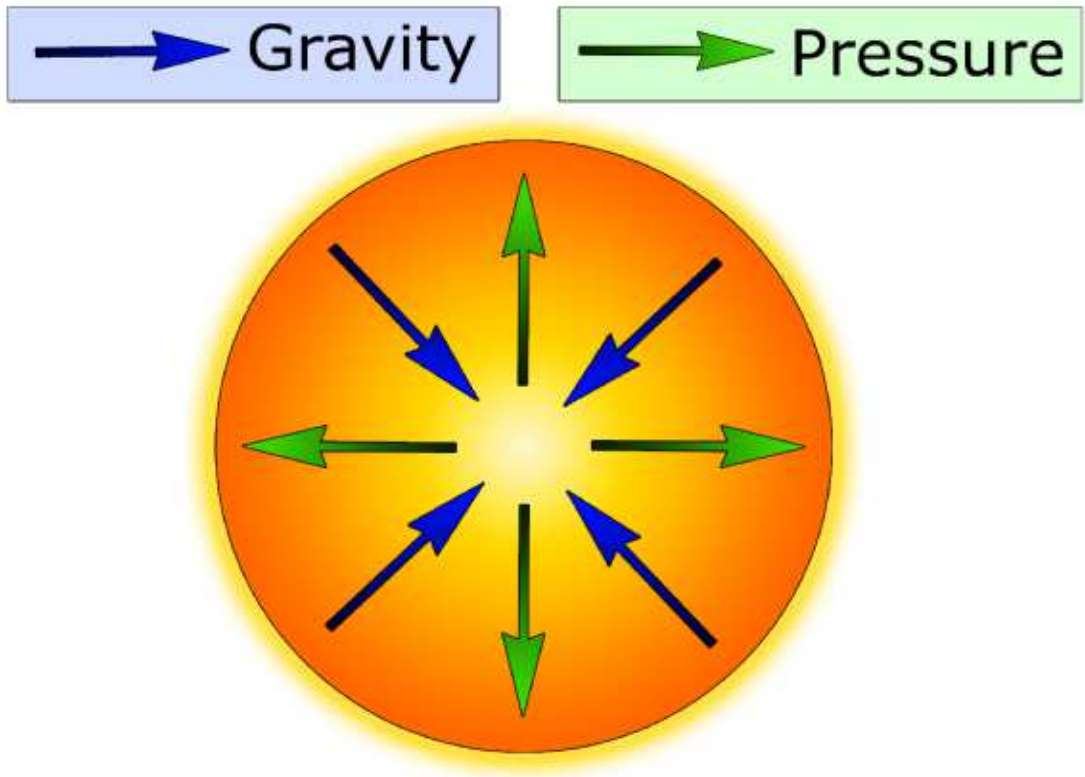}
\caption[Schematic of the two principal forces acting in a hydrostatic star]{Schematic diagram showing the forces of gravity and pressure acting in a star in hydrostatic equilibrium.}
\label{ch1.fig.hydrostatic}
\end{figure}

\begin{itemize}
\item Gravity: Stars are extremely massive objects. The mass of the Sun is $2\times 10^{30}$\,kg. By comparison, the mass of the Earth is $6 \times 10^{24}$\,kg, roughly a million times smaller. Because the gravitational pull of an object is proportional to its mass, the surface gravity of the Sun is roughly 28 times stronger than on Earth\footnote{The surface gravity of the Sun is not a million times larger than on Earth because the Sun has a much larger radius.}. In a more extreme case, white dwarfs, which are stars that have expended their nuclear fuel, have a diameter comparable to the Earth and mass comparable to the Sun. In these stars, the surface gravity can be millions of times greater than on Earth. Gravity is extremely important in stars and is constantly pulling all material towards the centre. 
\item Pressure: This is the combination of the outward force due to the gas and radiation within a star. Pressure is a measure of the overall outward force exerted by ions, electrons and even photons in a material. A gas (or alternatively a liquid or plasma) is made up of atoms undergoing rapid, random motions. For any enclosed gas, the surface of the enclosure must exert a force on any atoms that collide with it to prevent them from passing through. The sum of the force from all of the collisions is called pressure. In a star material isn't enclosed in a container but atoms in the gas do collide with each other. Imagine a completely permeable boundary in a star. Atoms cross in both directions. If more atomic collisions occur below the boundary (i.e. the pressure is higher)  than above it then more atoms will be ejected in the upwards direction than are deflected downwards. There is therefore a tendency for a net upward transfer of momentum, or in other words the material below the boundary exerts an upwards force on the material above the boundary. Hence we are not so much interested in the total pressure in a star but how the pressure changes at different levels.
\end{itemize}

\noindent Because gravity and pressure both act isotropically (i.e. equally in all directions) stars tend towards spherical symmetry\footnote{I recall one departmental meeting where we spent over an hour discussing the fundamental reason why stars are spherical.}.

The balance of these two forces, pressure acting outwards and gravity acting towards the centre, keeps stars in perfect equilibrium. We call this situation hydrostatic equilibrium, illustrated in Fig.~\ref{ch1.fig.hydrostatic}. In stars this equilibrium is, thankfully, extremely stable. The change caused by the introduction of the centrifugal force\footnote{A force which does exist! We refer the reader to \url{http://xkcd.com/123/}.}, which arises because of rotation, is one of the main focuses of this work.

\subsection{The main sequence}

Stars proceed through a number of important stages of evolution during their lifetimes. Stars are formed from protostellar clouds which collapse under their own gravity, this stage of evolution is commonly referred to as the pre--main sequence. Eventually the internal pressure and temperature become large enough to ignite the fusion of hydrogen to helium. This is the start of the main sequence. Beyond this point, nuclear fusion becomes the primary energy source for the star. The fusion of hydrogen into helium halts the collapse of the star which then remains in a stable, quiescent state over a time period varying between a few million and many billions of years depending on the mass of the star. Eventually a star burns through all of the hydrogen in the core. When the central hydrogen abundance reaches zero, the star starts to rapidly expand. This marks the end of the main sequence and the star transitions into the various phases of giant evolution.

Hydrogen is converted into helium through two primary mechanisms, the pp chain and the CNO cycle. Stars begin their lives composed of around one quarter helium and three quarters hydrogen by mass. There are also small quantities of heavier elements often present (in particular carbon, nitrogen and oxygen, which are produced in the late stages of the evolution of massive stars). The rate of the CNO cycle depends strongly on how abundant these elements are. 

The pp chain dominates the nuclear reactions at temperatures lower than around $2 \times 10^7$\,K and is split into three different branches. The first few reactions of each chain are the same. They start with the formation of a deuterium nucleus, $^2$H, from two hydrogen nuclei

\begin{equation}
{\rm ^1H+^1H \to ^2H+e^++\nu}
\end{equation}

\noindent and release a positron, e$^+$, which annihilates quickly with an electron, e, and a neutrino, $\nu$. Another proton then fuses with the deuterium nucleus so that

\begin{equation}
{\rm ^1H+^2H \to ^3He + \gamma},
\end{equation}

\noindent releasing an energetic photon, $\gamma$. From this point on the reaction chains are different. 

\begin{itemize}
\item ppI: This reaction dominates between around $10^7$\,K and $1.4 \times 10^7$\,K. In this process, two $^3$He nuclei fuse to produce $^4$He,

\begin{equation}
{\rm ^3He+^3He \to ^4He + 2^1H + \gamma}.
\end{equation}

\noindent The ppI chain produces 26.2 MeV per $^4$He nucleus produced.

\item ppII: This reaction dominates between around $1.4 \times 10^7$\,K and $2.3\times 10^7$\,K. The reactions of the ppII chain are

\begin{equation}
{\rm ^3He+^4He\to ^7Be+\gamma},
\end{equation}
\begin{equation}
{\rm ^7Be+e^-\to ^7Li+\nu},
\end{equation}
\noindent and
\begin{equation}
{\rm ^7Li+^1H\to 2^4He}.
\end{equation}

\noindent The ppII chain produces 25.7 MeV of energy per $^4$He nucleus.

\item ppIII: In this branch of the chain, $^7$Be is produced as in the ppII chain but the reaction progresses as

\begin{equation}
{\rm ^7Be + ^1H \to ^8Be^* +\gamma},
\end{equation}
\begin{equation}
{\rm ^8Be^* \to ^8Be + e^+ + \nu +\gamma},
\end{equation}
\noindent and
\begin{equation}
{\rm ^8Be\to 2^4He},
\end{equation}

\noindent where $^8$Be$^*$ is an unstable isotope of beryllium. This branch of the reaction is important only when the temperature is greater than $2.3 \times 10^7$\,K and generates 19.3 MeV of energy.
\end{itemize}

If the temperature exceeds $2\times 10^7$\,K then nuclear reactions are dominated by the CNO cycle which generates 23.8 MeV of energy per $^4$He nucleus. The reactions of these cycles are

\begin{equation}
{\rm ^{12}C + ^1H \to ^{13}N + \gamma},
\end{equation}
\begin{equation}
{\rm ^{13}N \to ^{13}C + e^+ + \nu},
\end{equation}
\begin{equation}
{\rm ^{13}C + ^1H \to ^{14}N + \gamma},
\end{equation}
\begin{equation}
{\rm ^{14}N + ^1H \to ^{15}O + \gamma},
\end{equation}
\begin{equation}
{\rm ^{15}O \to ^{15}N + e^+ + \nu},
\end{equation}
\noindent and
\begin{equation}
{\rm ^{15}N + ^1H \to ^{12}C +^4He + \gamma}.
\end{equation}

\begin{figure}
\centering
\includegraphics[width=0.99\textwidth]{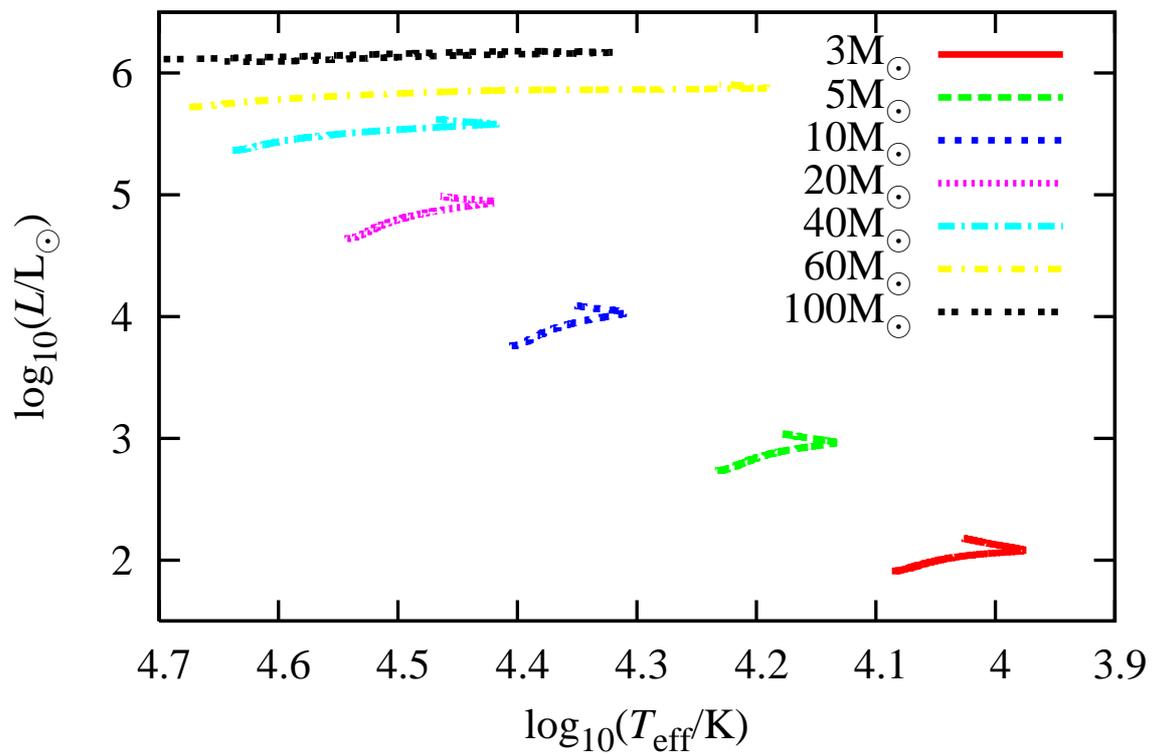}
\caption[HR diagram for the main--sequence evolution for a range of masses of non--rotating stars]{The Hertzsprung--Russell diagram for the main--sequence evolution of non--rotating solar--metallicity stars for a range of initial masses between $3$\Msun and $100$\Msun. The horizontal axis is for the effective surface temperature, $T_{\rm eff}$, and the vertical axis is for the stellar luminosity, $L$.}
\label{ch1.fig.ms}
\end{figure}

Evidently the nuclear processes that cause the evolution of a star across the main sequence are very dependent on the temperature of the stellar material, this is mainly influenced by the mass of the star. Stars of different masses evolve quite differently. In this piece of work we focus solely on massive stars. The exact definition of what massive means varies between authors but we consider such stars to have convective cores and radiative outer envelopes. This applies to all stars more massive than approximately $1.2$\Msun. Typically we refer to intermediate--mass stars as stars with masses between $1.2$\Msun\ and $10$\Msun. High--mass stars are those stars more massive than $10$\Msun. We show the main--sequence evolution in the Hertzsprung--Russell diagram for stars with a range of masses in Fig.~\ref{ch1.fig.ms}. The Hertzsprung--Russell diagram relates the temperature and luminosity evolution of stars.

Of particular importance to us in this work is the main--sequence lifetime of a star. A star that lives for longer has more time for rotational mixing to transport material between the core and the surface. In addition, stars that live for longer, have more time to be spun down by magnetic braking and there is more time for the magnetic field to decay. Any difference in the stellar lifetime therefore has serious consequences for the evolution of the angular momentum distribution and the magnetic field. As shown in Fig.~\ref{ch1.fig.ms}, the luminosity of a main--sequence star increases rapidly with mass. This is because of the higher temperature and pressure in the cores of massive stars and the lower opacity of their outer envelopes. In fact, the stellar luminosity increases roughly as a power law such that $L\propto M^{3.5}$ where $L$ is the stellar luminosity and $M$ is the stellar mass. This power law breaks down above around $M\approx 50$\Msun\ because the opacity in this range is dominated by electron scattering which is not strongly dependent on temperature. The lifetime of a star depends on how rapidly it burns through its fuel (i.e. the luminosity) and how much fuel there is to burn. The latter increases in proportion to the stellar mass. The main--sequence lifetime therefore varies as $\tau_{\rm ms}\propto M^{-2.5}$. Therefore, the main--sequence lifetime of a star is far shorter for more massive stars.

In this work we focus on the main--sequence evolution of stars. Whilst models of rotation and magnetic fields have often been extended on to the giant branch, it is much more difficult to get a convergent model owing to the emergence of convective shells. The dramatic change in the mechanism for angular momentum transport across these shells means that models tend to be far less stable than their main--sequence counterparts. A consequence is that, whilst models might progress to further stages of evolution, the progress of a model is likely to be very dependent on the stellar mass and the particular stages of evolution the model has to traverse.

\section{Rotation in massive stars}

Stars rotate because it is actually quite hard for them not to. Every object in our own Solar System is rotating; the Earth, the Sun, the planets, the asteroids and the moons. So it is reasonable to expect that objects beyond the Solar System also rotate which is indeed what we observe. Stellar rotation arises from two simple ideas, turbulence in the interstellar medium and conservation of angular momentum. Stars form from giant clouds of gas that collapse under their own gravity. This material has large--scale turbulence and so different fluid parcels are moving in different directions. When a section of the cloud starts to collapse, it is therefore highly probable that the material has non--zero total angular momentum, an average rotation in one particular direction around the centre. The amount of angular momentum the cloud has varies greatly because of the random nature of turbulence. As the cloud collapses, it retains its total angular momentum and, much as an ice skater does when she\footnote{This choice of pronoun is because it specifically refers to Christine Yallup.} draws in her arms, rotates at an increasingly rapid rate the further it contracts. The gas almost certainly sheds some of its angular momentum through mass loss from the system during the various stages of star formation but almost all of the angular momentum would have to be lost from the system for the star to have minimal rotation. For a review of the star formation process we direct the reader to \citet{McKee07}. Rotation in stars is therefore a rather normal feature of their structure. One of the key questions in this work is how fast does a star need to be rotating before it has a significant effect on the structure and evolution and, when it does, what exactly are those effects.

\subsection{Changes in stellar structure}

\begin{figure}
\centering
\includegraphics[width=0.99\textwidth]{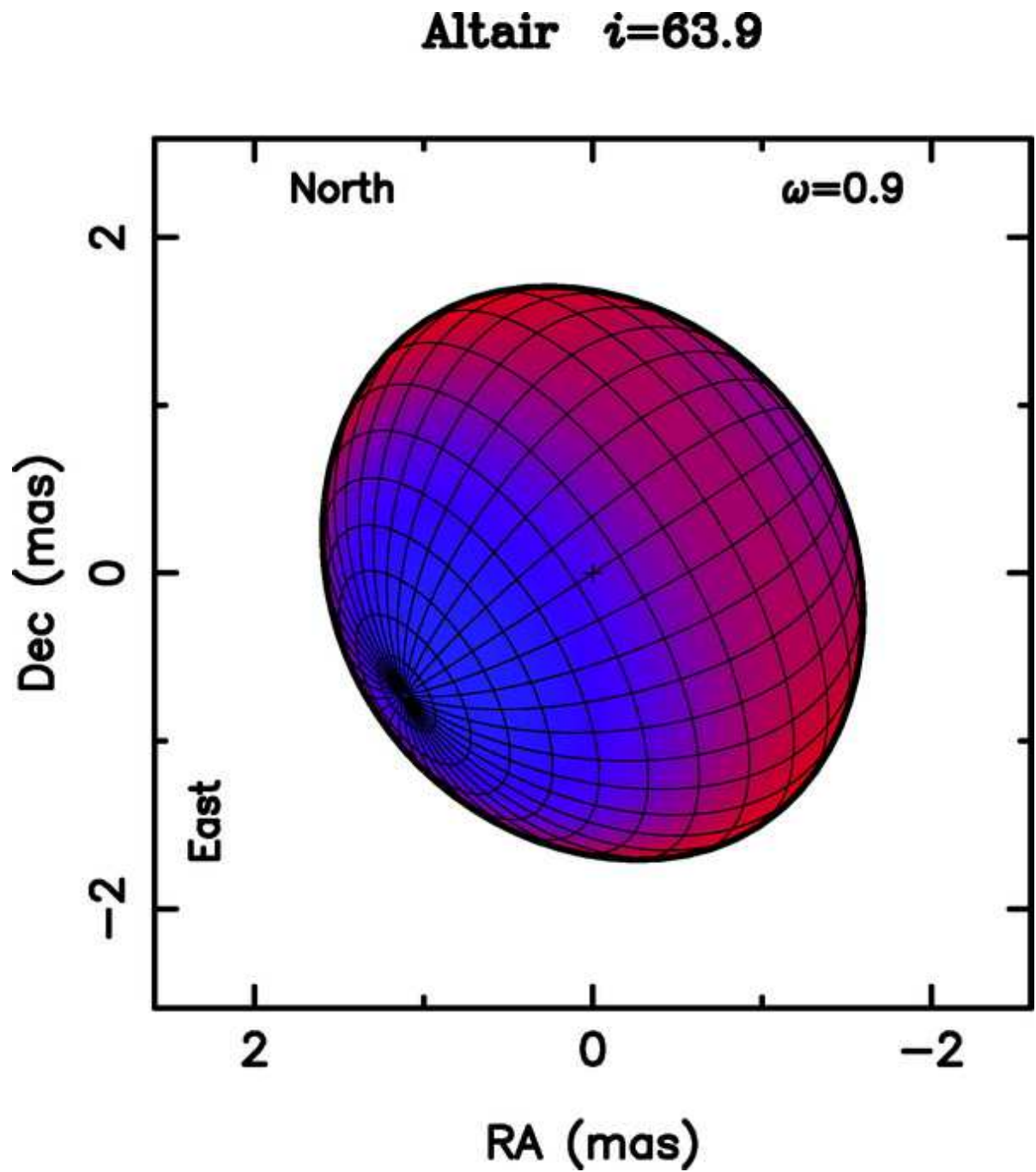}
\caption[High--resolution image of the visible surface of the rapidly rotating star Altair]{A false--colour rendering of the visible surface of the rapidly rotating star, Altair, from \citet{Peterson06}. Red indicates lower luminosity and blue indicates higher luminosity. It is also indicative of the temperature which ranges from 8,740\,K at the poles to 6,890\,K at the equator.}
\label{ch1.fig.altair}
\end{figure}

Owing to the Earth's rotation, it is over 20\,km wider at the equator than it is from pole to pole. The same effect happens in stars except, in a high proportion of cases, the rotation is sufficiently rapid that the effect is far more pronounced. In fact the equatorial radius can approach 1.5 times the polar radius. This figure not only comes from theoretical models but also from long--baseline interferometric observations of the rapidly rotating stars Achernar \citep{Carciofi08} and Altair \citep{Peterson06}. A map of the visible surface of Altair by \citet{Peterson06} is shown in Fig.~\ref{ch1.fig.altair}. In this figure, the distortion to the shape of the star is clearly visible. It is also notable that the star is hotter at the poles than at the equator. We discuss this further in section~\ref{ch1.sec.vonzeipel}.

We can approximate the distortion of stars owing to rotation with a number of simple models. One of the most common models is that of McLaurin Spheroids which assumes that the star is a body of constant density with uniform rotation velocity. The other is the Roche model which assumes the gravitational potential is that of the mass of the star concentrated at the centre. Typically the second approximation is a better fit to simulations and observations owing to the strong density gradient in stars. For an introduction to McLaurin Spheroids we direct the reader to section~41.1 of \citet{Kippenhahn94}.

In the Roche model we assume a spherically symmetric gravitational potential

\begin{equation}
\Phi_{\rm grav}=-\frac{GM}{r}
\end{equation}

\noindent where $G$ is the gravitational constant, $M$ is the mass of the star and $r$ is the distance from the centre. This is the gravitational potential of mass $M$ concentrated at the origin. If the star rotates as a solid body then we can represent the potential of the centrifugal force by

\begin{equation}
\Phi_{\rm rot}=-\frac{1}{2}s^2\Omega^2
\end{equation}

\noindent where $s$ is the perpendicular distance from the rotation axis and $\Omega$ is the angular velocity. Let $z$ be the perpendicular distance from the equatorial plane so that $r^2=s^2+z^2$. The total potential is then

\begin{equation}
\Phi=\Phi_{\rm grav}+\Phi_{\rm rot}=-\frac{GM}{\sqrt{s^2+z^2}}-\frac{s^2\Omega^2}{2}.
\end{equation}

\noindent As in section \ref{ch1.sec.ce} the stellar surface is expected to lie along lines of constant potential, $\Phi={\rm constant}$. For Roche models, the critical rotation rate, above which the star becomes unbound, is given by

\begin{equation}
\Omega_{\rm crit}=\sqrt{\frac{8GM}{27R_{\rm p}^3}}
\end{equation}

\begin{figure}
\centering
\includegraphics[width=0.99\textwidth]{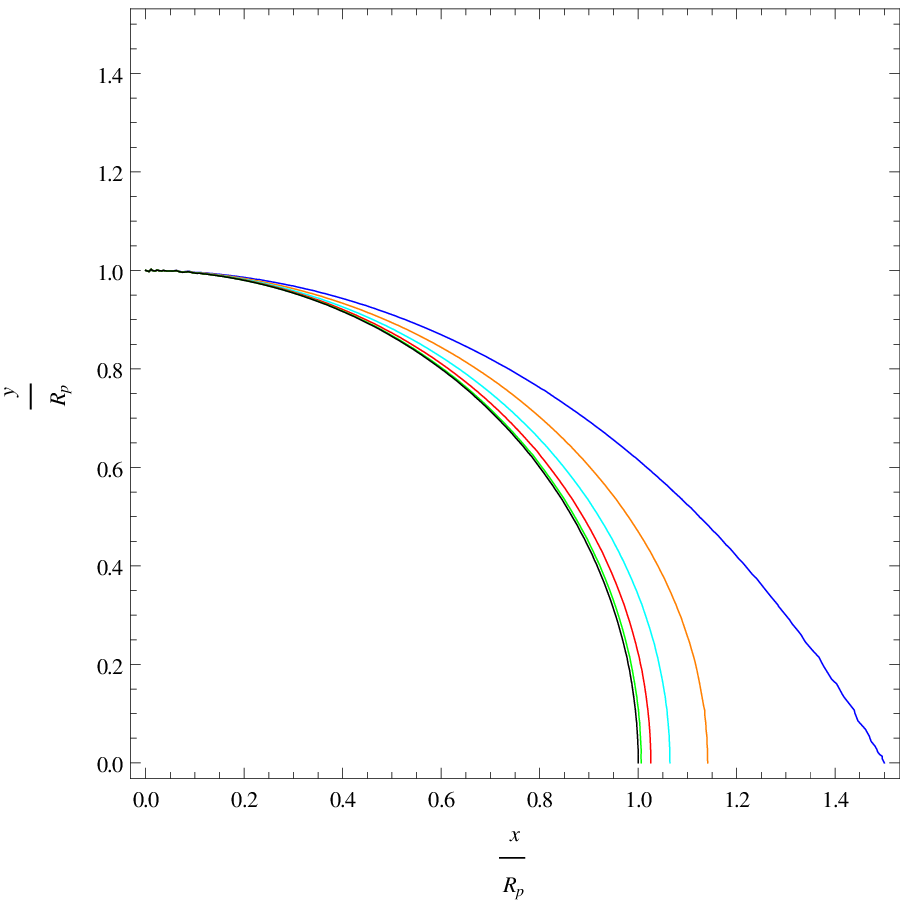}
\caption[The distortion of stars with different rotation rates calculated using the Roche Model]{The distortion of stars rotating at different rates calculated with the Roche Model. The contours are for different values of $\zeta=\frac{\Omega}{\Omega_{\rm crit}}$; $\zeta=0$ (black), $\zeta=0.2$ (green), $\zeta=0.4$ (red), $\zeta=0.6$ (cyan), $\zeta=0.8$ (orange), $\zeta=1.0$ (blue). The degree of distortion is reasonably small even for rotation rates up to 80\% of critical rotation. For very high rotation rates the ratio of the equatorial radius to the polar radius tends to 1.5.}
\label{ch1.fig.rot}
\end{figure}

\noindent where $R_{\rm p}$ is the polar radius. Fig.~\ref{ch1.fig.rot} shows how the shape of stars in the Roche Model changes with rotation rate. We note that stars rotating at $80\%$ of their critical rotation rate still only have an equatorial radius which is approximately $14\%$ larger than their polar radius.

\subsection{The Von Zeipel paradox}
\label{ch1.sec.vonzeipel}

The condition for radiative equilibrium outside burning regions in a hydrostatic star is such that the divergence of the radiative flux, $\bi{F}$, is $0$ (i.e. $\nabla\cdot\bi{F}=0$). This condition ensures that the amount of heat flux is conserved as it passes through an arbitrary fluid parcel, or rather that no energy is either created or destroyed as it is transported through the star. \citet{VonZeipel24} considered how radiative equilibrium would be affected by rotation and concluded that a rotating, hydrostatic star could not simultaneously be in radiative equilibrium. This is commonly known as the {\it Von Zeipel paradox}. For a complete description we direct the reader to \citet{Tassoul78}. We go through a basic derivation here. The thermal flux at some point in a star is given by

\begin{equation}
\bi{F}=-\frac{4acT^3}{3\kappa\rho}\nabla T,
\end{equation}

\noindent where $a=7.5646\times 10^{-15}{\rm erg\,cm^{-3}\,K^{-4}}$ is the radiation--density constant, $c$ is the speed of light, $T$ is the temperature, $\kappa$ is the opacity and $\rho$ is the density. All of these quantities can be written in terms of the total potential, $\Phi$, which includes the force of gravity and rotation such that

\begin{equation}
\label{ch1.eq.flux}
\bi{F}=\left(-\frac{4a3T^3}{3\kappa\rho}\difft{T}{\Phi}\right)\nabla\Phi=f(\Phi)\nabla\Phi,
\end{equation}

\noindent where $f(\Phi)$ is some unknown function of the potential. The radiative flux is therefore perpendicular to contours of constant $\Phi$. If we calculate the divergence of $\bi{F}$ from equation (\ref{ch1.eq.flux}) we find

\begin{equation}
\label{ch1.eq.flux2}
\nabla\cdot\bi{F}=f'(\Phi)(\nabla\Phi)^2+f(\Phi)\nabla^2\Phi.
\end{equation}

\begin{figure}
\centering
\includegraphics[width=0.99\textwidth]{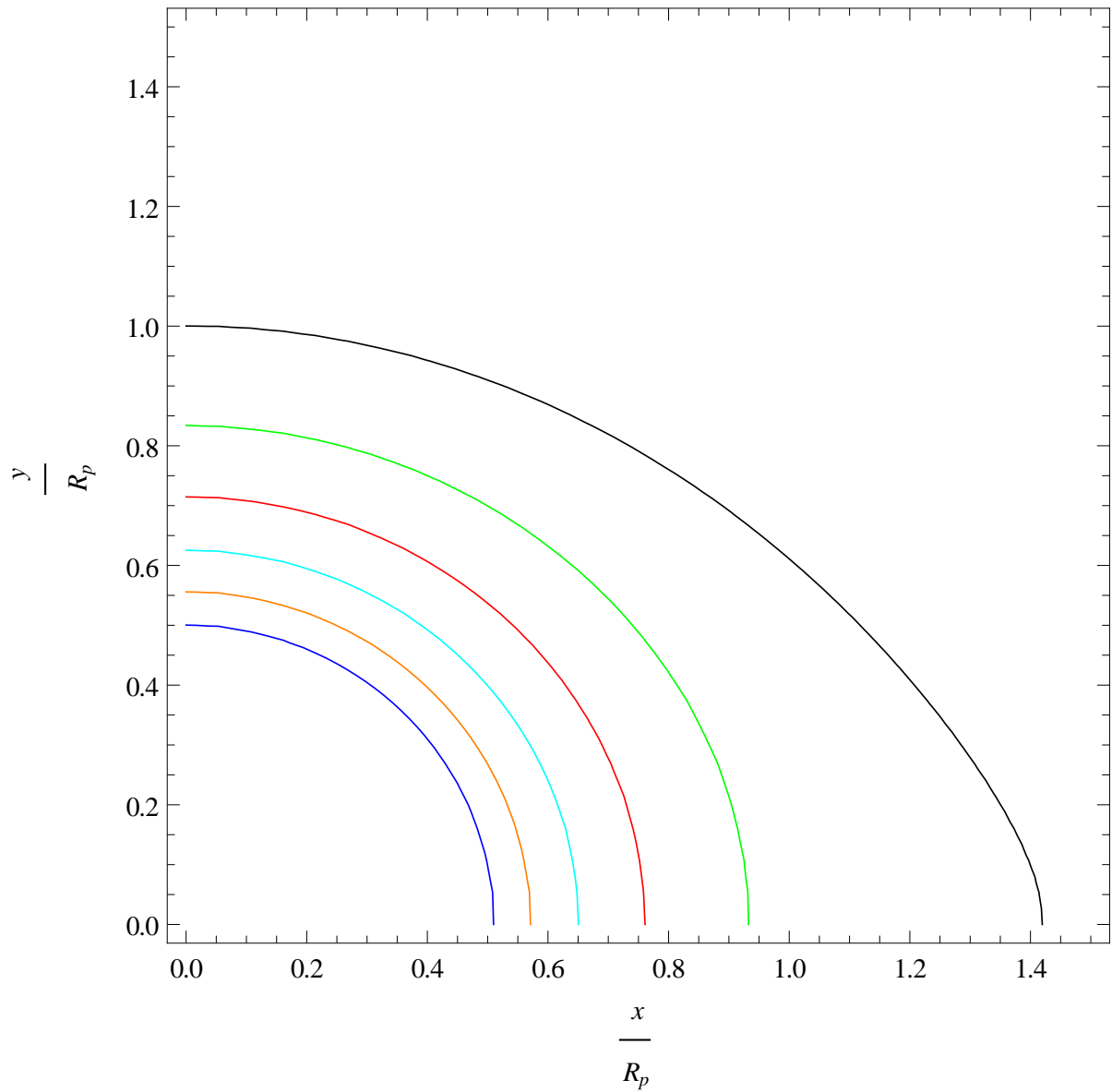}
\caption[Contours of constant potential in a rotating star calculated using the Roche Model]{The shape of contours of constant potential in a star rotating at $99\%$ of its critical rotation rate calculated using the Roche Model. The contours are for different values of the potentail, $\Phi$. Given that the potential at the surface of the star is $\Phi_{\rm surf}$ the contours as for, $\Phi=\Phi_{\rm surf}$ (black), $\Phi=1.2\Phi_{\rm surf}$ (green), $\Phi=1.4\Phi_{\rm surf}$ (red), $\Phi=1.6\Phi_{\rm surf}$ (cyan), $\Phi=1.8\Phi_{\rm surf}$ (orange), $\Phi=2\Phi_{\rm surf}$ (blue). Note that the contours are closer together at the pole than at the equator.}
\label{ch1.fig.rot2}
\end{figure}

\noindent From Poisson's equation we know that $\nabla^2\Phi=4\pi G\rho - 2\Omega^2$ where $G$ is the gravitational constant and $\Omega$ is the angular velocity which is also constant along lines of constant $\Phi$. However, $\nabla\Phi$ is not constant at different co--latitudes, the contours of constant $\Phi$ in a rotating star are closer together at the poles than they are at the equator. This is illustrated in Fig.~\ref{ch1.fig.rot2}. Therefore the right--hand side of equation~(\ref{ch1.eq.flux2}) is non--zero and radiative equilibrium cannot be established.

\citet{VonZeipel24} took this further and established a relation between the effective gravity and effective temperatures; $T_{\rm eff}(\Omega,\theta)\propto g_{\rm eff}(\Omega,\theta)^{\frac{1}{4}}$. This is Von Zeipel's theorem, derivations of which can be found in \citet{Tassoul78} and \citet{Maeder09}. A brief derivation is given in appendix~\ref{ap1.vonzeipel}. On the stellar surface the effective gravity is stronger at the poles than at the equator owing to the action of the centrifugal force so we similarly expect the effective surface temperature to be higher at the poles than at the equator. This is exactly the observation of rapidly rotating star Altair as shown in Fig.~\ref{ch1.fig.altair}.

As rotation breaks the radiative equilibrium within the star, additional processes must occur in order to bring the star back into equilibrium. As a result of the Von Zeipel paradox, we know that in the absence of these other processes, there will be different degrees of heating and cooling of material across surfaces with constant potential. This gives rise to buoyant forces and results in a bulk motion of material which manifests as a circulation current within the star. Transport of thermal energy by these currents re--establishes radiative equilibrium. Strictly speaking we can no longer claim that rotating stars are in hydrostatic equilibrium but these circulation currents are sufficiently weak that they do not greatly affect the hydrostatic balance between pressure, gravity and rotation. Various forms of the meridional circulation have been used over the years. The most common is that of \citet{Sweet50} but other theories have been used more recently by \citet{Zahn92} and \citet{Maeder00}.

\subsection{The Kelvin--Helmholtz instability}

\begin{figure}
\centering
\includegraphics[width=0.99\textwidth]{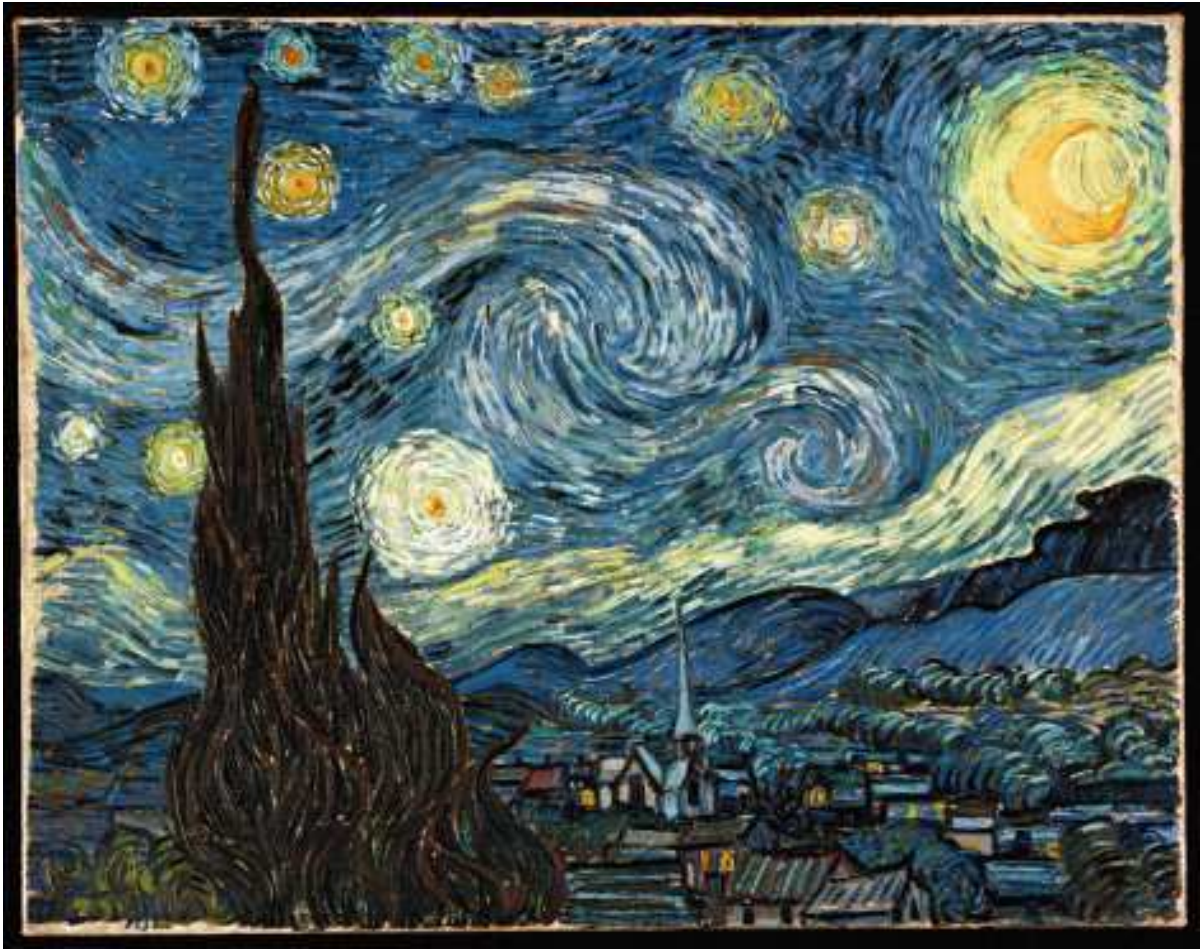}
\caption[Van Gogh's ``La Nuit \'{E}toil\'{e}e'']{Van Gogh's ``La Nuit \'{E}toil\'{e}e''. The white clouds in the centre are widely regarded as undergoing Kelvin--Helmholtz instabilities.}
\label{ch1.fig.gogh}
\end{figure}
\begin{figure}
\centering
\includegraphics[width=0.45\textwidth]{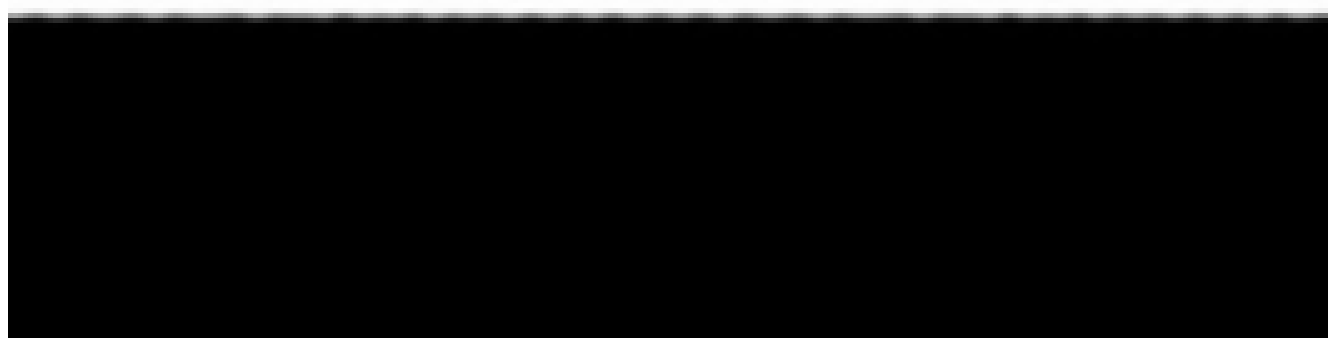}\quad\includegraphics[width=0.45\textwidth]{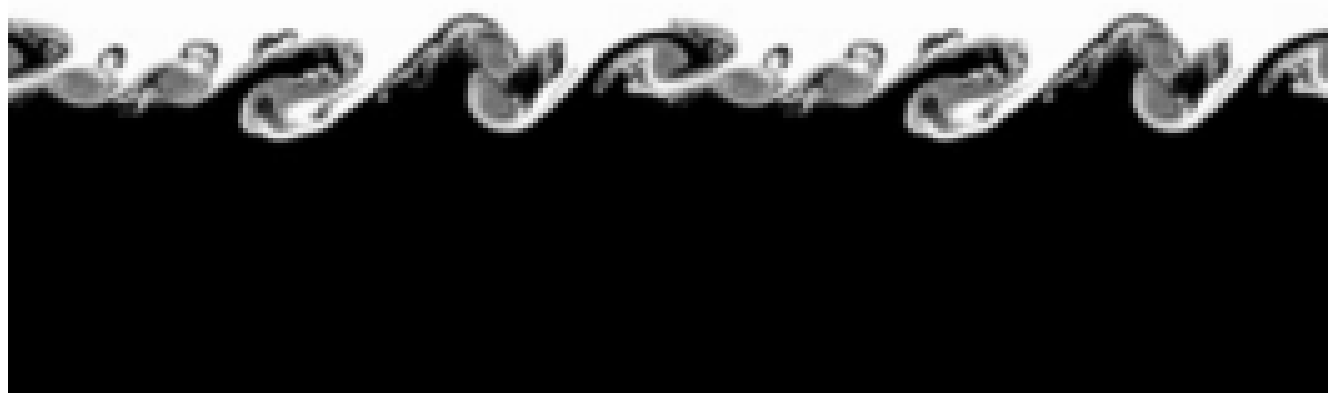}\\\includegraphics[width=0.45\textwidth]{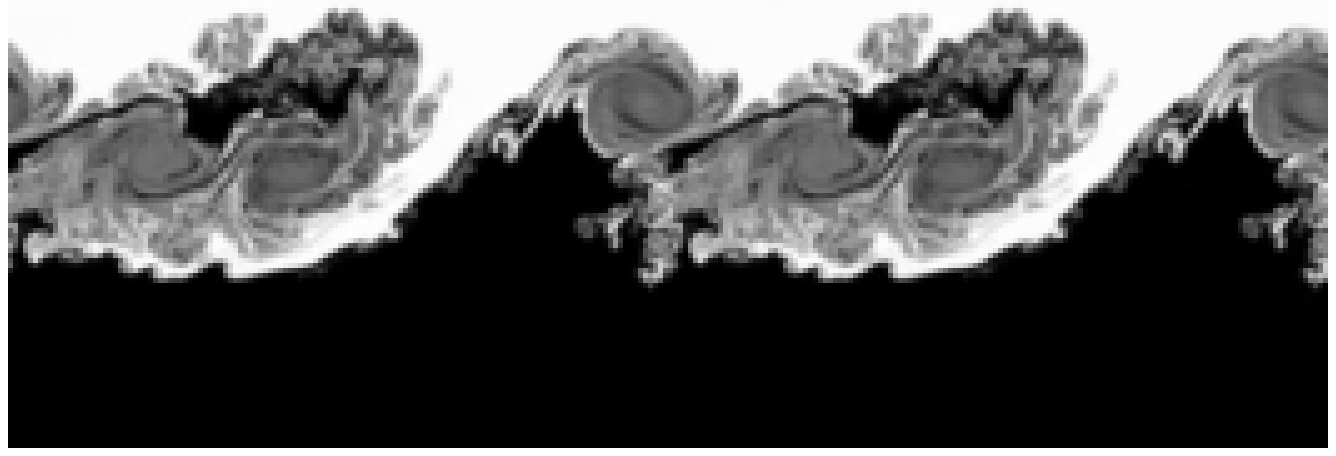}\quad\includegraphics[width=0.45\textwidth]{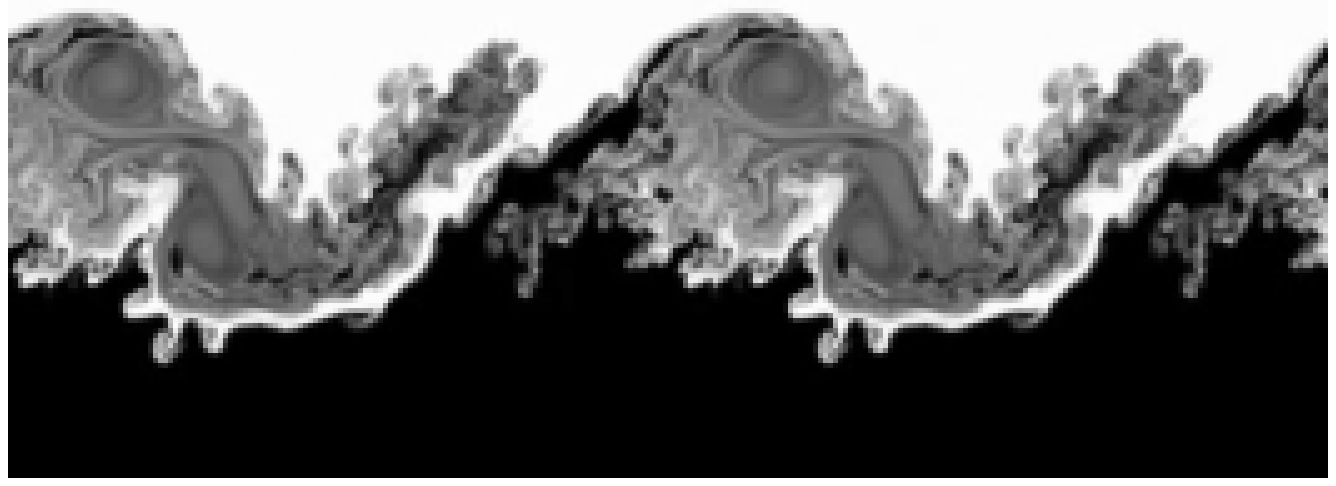}\\\includegraphics[width=0.45\textwidth]{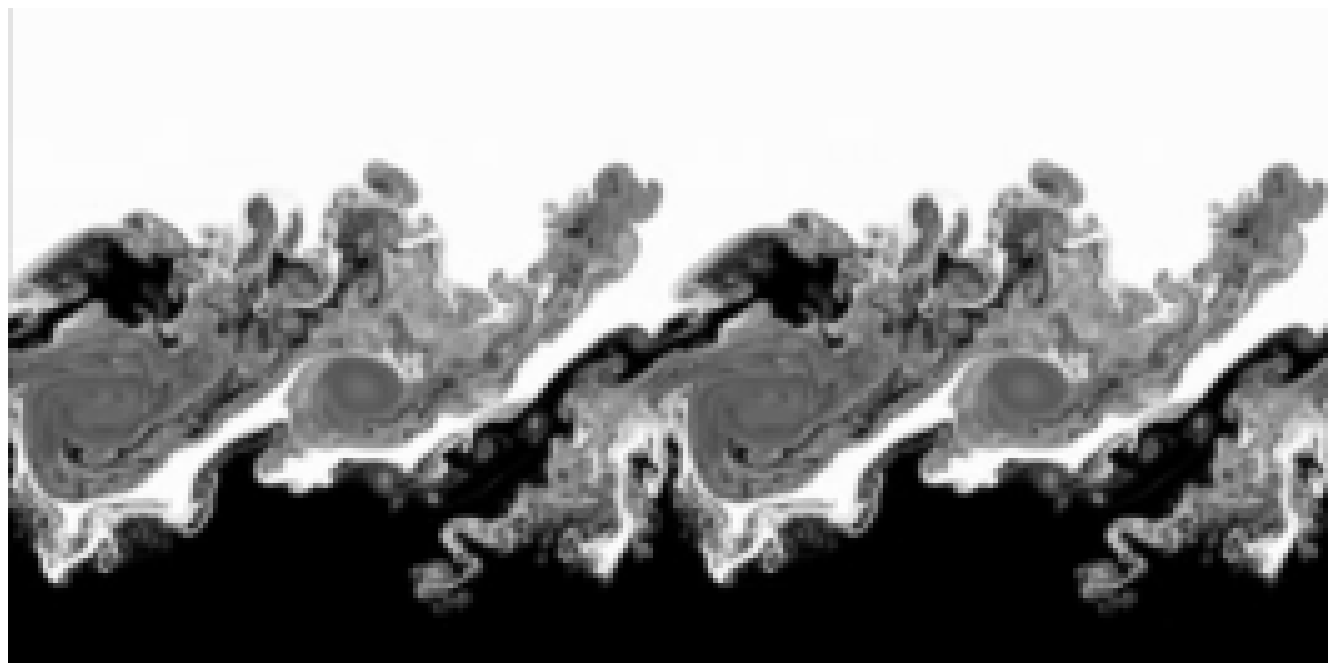}\quad\includegraphics[width=0.45\textwidth]{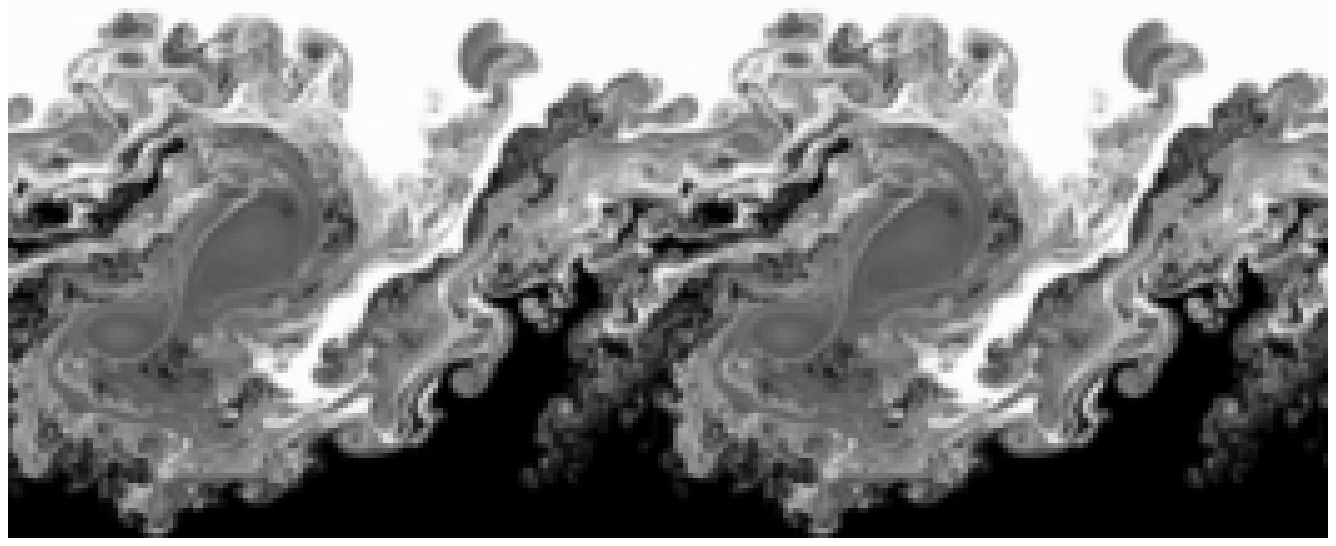}
\caption[Numerical simulation of the Kelvin--Helmholtz instability]{Numerical simulation of the Kelvin--Helmholtz instability. The white fluid is moving to the right and the black fluid is moving to the left. Eddies form at the interface between the two regions and the subsequent turbulence leads to mixing of the two fluids. The simulation is shown at $t=0\,{\rm s}$ (top left), $t=1\,{\rm s}$ (top right), $t=3\,{\rm s}$ (middle left), $t=5\,{\rm s}$ (middle right), $t=6\,{\rm s}$ (bottom left) and $t=7\,{\rm s}$ (bottom right).}
\label{ch1.fig.kh}
\end{figure}

Certain types of fluid flow are susceptible to a range of instabilities. When an instability occurs, an otherwise simple flow develops complicated secondary motions and eventually may descend into turbulence. 
The particular instability which is of most interest to us is the Kelvin--Helmholtz instability and occurs at the interface between two fluids moving at different velocities. This is an extremely common instability and occasionally can be observed in certain cloud formations. In fact, it is widely considered \citep[e.g][]{Forste96} that Van Gogh's ``La Nuit \'{E}toil\'{e}e'' (Fig.~\ref{ch1.fig.gogh}) depicts this phenomenon. A numerical simulation of the phenomenon is shown in Fig.~\ref{ch1.fig.kh}\footnote{The slides in this figure were taken from an open--source movie of a numerical simulation of the instability at \url{http://en.wikipedia.org/wiki/File:Kelvin-Helmholtz_Instability.ogv}. It's my personal favourite animation of the instability and I've yet to find a more illustrative simulation in the literature. Unfortunately many details of the model such as the initial conditions and code used are unavailable.}. 


It is quite easy to demonstrate the existence of this instability from linear analysis of the equations of fluid dynamics. We restrict ourselves to the simple case of two fluids separated by a discrete, horizontal interface. In stellar environments the change in density is continuous and so the analysis is somewhat more complicated and the result is known as the Taylor--Goldstein equation. Below the interface the fluid velocity is $\bi{u}_1=u_1\bi{e}_x$, the density is $\rho_1$ and the pressure is $p_1=p_0-\rho_1 g z$ where $z$ is the vertical coordinate. Similarly, above the interface the fluid velocity is $\bi{u}_2=u_2\bi{e}_x$, the density is $\rho_2$ and the pressure is $p_2=p_0-\rho_2gz$. We assume the background flow above and below the boundary are both moving in the $x$--direction. We could generalize this for arbitrary flow directions above and below the interface but for simplicity we look at the example where the flows above and below the interface are in the same direction. The equilibrium position of the interface is at $z=0$ but we perturb it so that the interface is at

\begin{equation}
z=\xi=\xi_0\exp\left(i\left(kx+ly\right)+st\right).
\end{equation}

\noindent We assume that the background flow is irrotational ($\nabla\times\bi{u}=0$) and incompressible ($\nabla\cdot\bi{u}=0$) so that we can represent the fluid velocity by the gradient of a potential field, $\bi{u}=\nabla\phi$ where

\begin{equation}
\phi=
\begin{cases}
\phi_1,& z>\xi, \\\phi_2,& z<\xi
\end{cases}
\end{equation}

\noindent and

\begin{equation}
\nabla^2\phi_i=0.
\end{equation}

\noindent In the limit of very large $z$

\begin{equation}
\nabla\phi_1\to \bi{u}_1 \quad {\rm as} \quad z \to -\infty
\end{equation}

\noindent and

\begin{equation}
\nabla\phi_2\to \bi{u}_2 \quad {\rm as} \quad z \to +\infty.
\end{equation}

\noindent The kinematic boundary condition states that fluid on the surface, $\xi$, must remain on the surface. This implies

\begin{equation}
\diffb{\phi_i}{z}=\frac{{\rm D}\xi}{{\rm D}t}=\diffb{\xi}{t}+\diffb{\phi_i}{x}\diffb{\xi}{x}+\diffb{\phi_i}{y}\diffb{\xi}{y}\quad {\rm on} \quad z=\xi \quad {\rm for } \quad {\rm i=1,2}.
\end{equation}

\noindent The dynamic boundary condition is derived from the Bernoulli principle and ensures that pressure is balanced across the boundary

\begin{equation} 
\rho_1\left(\frac{1}{2}\left(\nabla\phi_1\right)^2-\frac{1}{2}u_1^2+\diffb{\phi_1}{t}-gz\right)=\rho_2\left(\frac{1}{2}\left(\nabla\phi_2\right)^2-\frac{1}{2}u_2^2+\diffb{\phi_2}{t}-gz\right) \quad {\rm on} \quad z=\xi.
\end{equation}

\noindent To investigate linear stability we set

\begin{equation}
\phi_2=u_2 x + \tilde{\phi}_2 \quad {\rm for} \quad \ z>\xi
\end{equation}

\noindent and

\begin{equation}
\phi_1=u_1 x + \tilde{\phi}_1 \quad {\rm for} \quad z<\xi.
\end{equation}

\noindent We substitute this solution into the boundary conditions and neglect second order terms. This gives

\begin{align}
\nabla^2\tilde{\phi}_1&=0&z<0,\label{ch1.eq.kh1}\\
\nabla^2\tilde{\phi}_2&=0&z>0,\label{ch1.eq.kh2}\\
\nabla\tilde{\phi}_1&\to 0&z\to -\infty,\label{ch1.eq.kh3}\\
\nabla\tilde{\phi}_2&\to 0&z\to +\infty,\label{ch1.eq.kh4}\\
\diffb{\tilde{\phi}_i}{z}&=\diffb{\xi}{t}+u_i\diffb{\xi}{x}&z=0,\ i=1,2\label{ch1.eq.kh5}\\
\end{align}
\noindent and
\begin{align}
\rho_1\left(u_1\diffb{\tilde{\phi}_1}{x}+\diffb{\tilde{\phi}_1}{t}+g\xi\right)=\rho_2\left(u_2\diffb{\tilde{\phi}_2}{x}+\diffb{\tilde{\phi}_2}{t}+g\xi\right)&&z=0.\label{ch1.eq.kh6}\\
\end{align}

\noindent The solution of equations (\ref{ch1.eq.kh1}) to (\ref{ch1.eq.kh4}) is

\begin{equation}
\tilde{\phi}_1=A_1{\rm e}^{qz}{\rm e}^{i(kx+ly)+st}
\end{equation}

\noindent and

\begin{equation}
\tilde{\phi}_2=A_2{\rm e}^{-qz}{\rm e}^{i(kx+ly)+st},
\end{equation}

\noindent where $q^2=k^2+l^2$ and $A_i$ are constants to be determined. Substituting these solutions into equation (\ref{ch1.eq.kh5}) gives

\begin{equation}
A_1=-\frac{s+iku_1}{q}\xi_0
\end{equation}

\noindent and

\begin{equation}
A_2=\frac{s+iku_2}{q}\xi_0.
\end{equation}

\noindent Further substituting these solutions into equation (\ref{ch1.eq.kh6}) gives

\begin{equation}
\rho_1\left(qg+(s+iku_1)^2\right)=\rho_2\left(qg-(s+iku_2)^2\right)
\end{equation}

\noindent which simplifies to

\begin{equation}
\label{ch1.eq.s}
s=-ik\left(\frac{\rho_1u_1+\rho_2u_2}{\rho_1+\rho_2}\right)\pm\left(\frac{k^2\rho_1\rho_2(u_1-u_2)^2}{(\rho_1+\rho_2)^2}-\frac{qg(\rho_1-\rho_2)}{\rho_1+\rho_2}\right)^{\frac{1}{2}}.
\end{equation}

\noindent We get instability in the system if the real part of $s$ is positive. This can only occur when

\begin{equation}
\frac{\sqrt{k^2+l^2}}{k^2}g<\frac{\rho_1\rho_2(u_1-u_2)^2}{(\rho_1-\rho_2)^2}.
\end{equation}

\noindent This criterion is met whenever there is shear (i.e. $u_1-u_2\neq 0$) for sufficiently large $k$. However, for very large frequencies, surface tension effects can stabilise the boundary. The main conclusion we draw from this criterion is that the system becomes more unstable for higher shear and more stable if the density difference between the two layers is large.

\subsection{Observations of rotational velocities}
\label{ch1.sec.rotobs}

\begin{figure}
\centering
\includegraphics[width=0.7\textwidth]{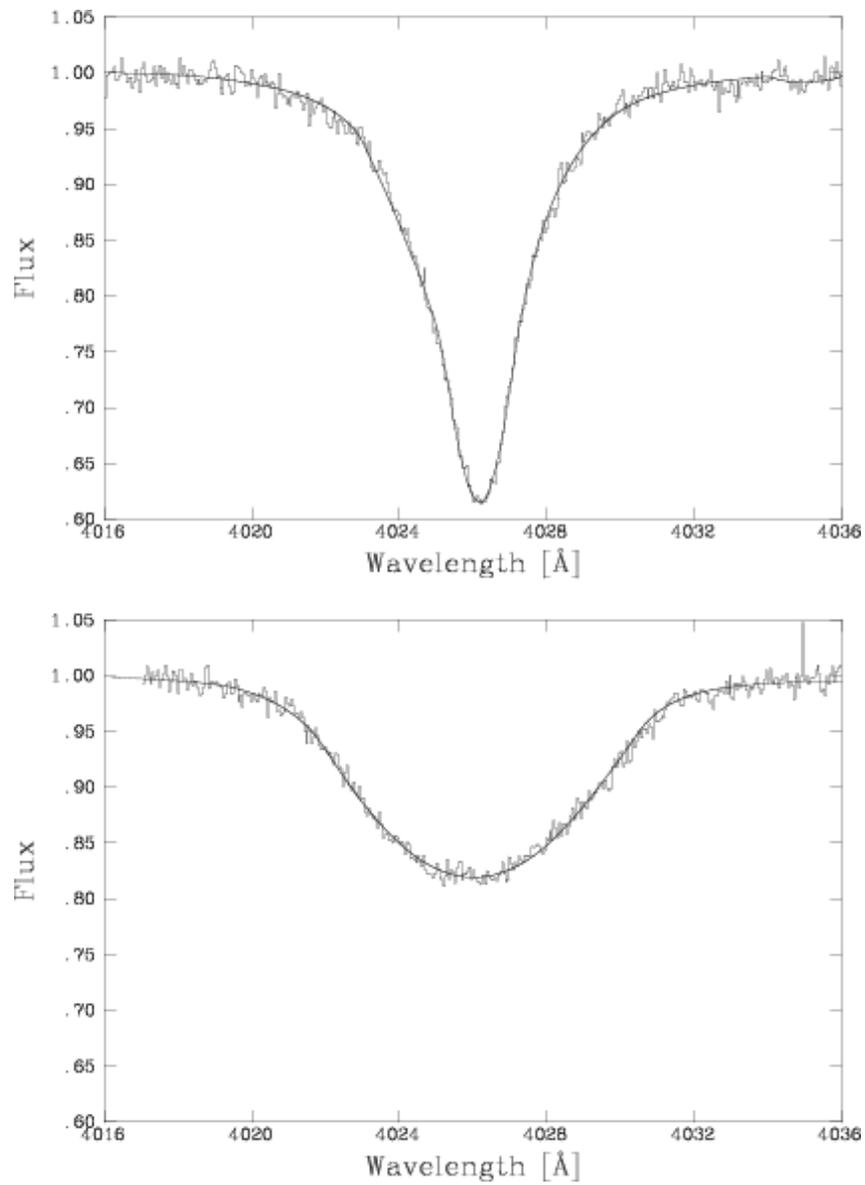}
\caption[Rotational broadening of absorption lines in stellar spectra]{The rotational broadening of absorption lines in stellar spectra. The top panel is for a star with estimated rotation velocity of $75{\rm \,km\,s^{-1}}$ and the bottom panel is for a star with estimated rotation velocity of $330{\rm \,km\,s^{-1}}$. The figure is from \citet{Dufton06}.}
\label{ch1.fig.spectra}
\end{figure}

\begin{figure}
\centering
\includegraphics[width=0.99\textwidth]{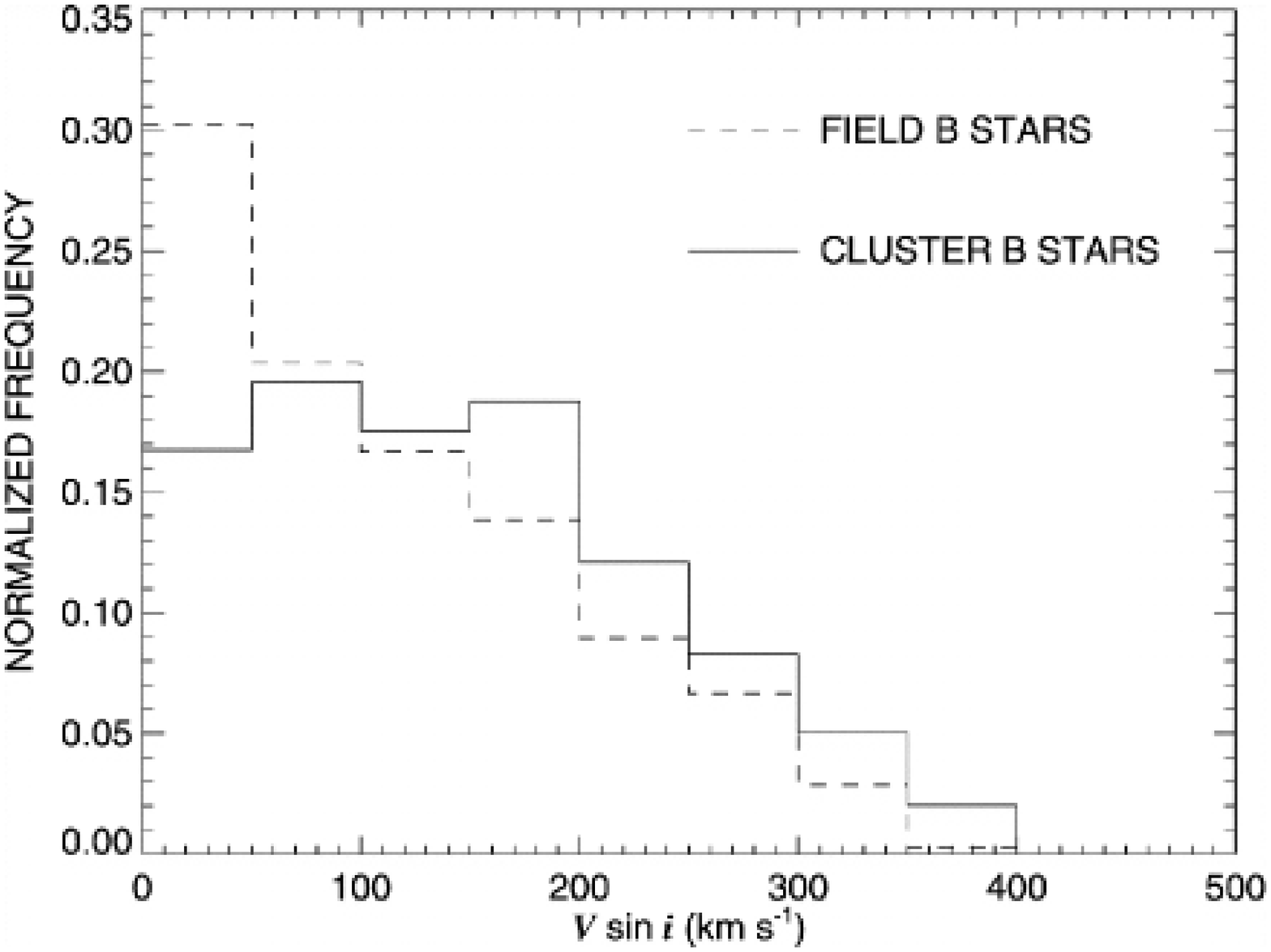}
\caption[Rotational velocity distribution of B--stars]{Histograms of the projected surface rotational velocity, $V \sin i$, for field and cluster B stars. The figure is from \citet{Huang06} who used the data of \citet{Abt02}.}
\label{ch1.fig.vsini}
\end{figure}

The observations of rotation rates of massive stars are nothing new. Much of the analysis over the past three decades has come from the data of the Bright Stars Catalogue \citep{Hoffleit82} but there have been a number of significant updates \citep[e.g.][]{Abt02, Strom05, Huang06}. 

There are a number of ways in which the surface rotation rate can be measured. One of these is to measure the difference in the redshift between the light emitted from different sides. If the star is rotating then one side moves towards the observer and so the light is shifted towards higher frequencies (blue--shift). On the other side of the star, the emission surface recedes from the observer and so the light is shifted towards lower frequencies (red--shift). This is a good technique but largely impractical for all but the closest stars because extremely high resolution is needed to distinguish between light emitted from two different sides of the star. The more common way to measure the surface rotation is to use the broadening of absorption lines in stellar spectra. Stars emit light across a wide range of frequencies. Different isotopes in stellar atmospheres absorb light very strongly at very specific frequencies. The change in the amount of light being emitted at a specific frequency indicates the abundance of a particular isotope and the shape of the absorption feature can be modelled with simulations of stellar atmospheres. One particular observed effect is that, in rotating stars, the overall amount of light absorbed is the same but the width of the line is significantly broader. The line width is typically measured by looking at the FHWM (full width half maximum) which is the width of the absorption feature at half its depth. The larger the FHWM is, the faster the star is rotating. Fig.~\ref{ch1.fig.spectra} shows an example absorption feature from \citet{Dufton06}. This shows the same feature in two stars with different rotation rates. Note that the feature is significantly broader in the more rapidly rotating star.

Stellar populations of massive stars have been observed to exhibit a range of different rotation rates. Fig.~\ref{ch1.fig.vsini} shows the distribution of surface rotation rates for a sample of Galactic B stars from \citet{Huang06} who used the data of \citet{Abt02}. The field stars have mean surface rotation velocity of $113\,{\rm km\,s^{-1}}$ and for cluster stars the mean velocity is $148{\rm \,km\,s^{-1}}$. \citet{Strom05} similarly concluded that populations of massive stars in dense clusters had average rotation velocities that were significantly higher than field stars. It is not known exactly why this happens. \citet{Strom05} suggest it could either be that in dense clusters it is harder to form wide binaries. Wide binaries would reduce the average angular momentum of the stars because much of the angular momentum of the stellar material would instead be absorbed as orbital angular momentum of the binary systems. An alternative explanation is that dense star clusters affect the way material is accreted on to stars during the formation phase. This affects the angular momentum of the population when it reaches the main sequence. It has also been suggested by \citet{Dufton11} and \citet{deMink11} that the most rapidly rotating stars arise because of angular momentum transfer between stars in binary systems. We do not examine these effects in this work but simply stress that rotation is a common feature of massive stars and hence it is important to continue developing our understanding of how it affects physical processes in stars.

Much of the current work regarding stellar rotation uses the data of \citet{Dufton06} and \citet{Hunter09} from the VLT--FLAMES survey of massive stars \citep{Evans05}. Using the rotational velocities derived from the survey, \citet{Hunter09} estimated the chemical compositions at the surface of many of the massive stars in the survey. As described in section~\ref{ch1.sec.mixing}, we expect rotation to cause abundance anomalies in rotating stars by transporting material from the core to the surface. The data of \citet{Hunter09} allows us to see exactly how rotation and chemical abundance anomalies are related. A sample of this data is shown in Fig.~\ref{ch5.fig.dufton}. It is primarily this data that we use to examine how closely our models for stellar rotation fit with real stars.

\subsection{Chemical mixing in massive stars}
\label{ch1.sec.mixing}

Chemical anomalies in O and B stars have been studied for over 40 years. \citet{Walborn70} identified a number of stars in this luminosity range whose measured nitrogen abundances were significantly different to other similar stars in the same population and which could not be classified into any other existing stellar types. These observations were expanded by \citet{Dufton72} who looked in detail at two B--type stars and found abundance anomalies in the nitrogen, neon, silicon and magnesium. However, these were attributed to chemical inhomogeneities in the interstellar medium rather than to rotation. This phenomenon was later also observed in the Magellanic Clouds \citep{Trundle04}. The chemical abundances for a large number of stars from the Milky Way, Small Magellanic Cloud and Large Magellanic Cloud have recently been analysed by \citet{Hunter09} who found many stars with peculiar chemical abundances, particularly nitrogen. For most stars, there is a strong correlation between nitrogen enrichment and rotation. However, there are two notable classes of stars which do not obey this rule.

\begin{enumerate}
\item Chemically peculiar stars are observed that have slow surface rotation but an unusually high nitrogen abundance compared to the rest of the population. \citet{Hunter09} suggest that these stars might be the result of magnetic fields but asserts that the magnetic fields in these stars are of fossil origin (c.f. section~\ref{ch1.sec.fossil}) because dynamo--driven magnetic fields based on current models \citep[e.g.][]{Spruit02} would be present in all the observed stars. This is not necessarily true as discussed in chapter~\ref{ch5}.
\item A number of stars were observed with rapid surface rotation but low surface nitrogen enrichment. Whilst this would be the case for very young stars, such stars would fall outside of the observational limits of the VLT--FLAMES survey \citep{Brott11b}. The stars in this category also have relatively low surface gravity and so are almost certainly not young enough to fit this explanation. It has been speculated that these stars could be the result of binary evolution but this has thus far remained untested.
\end{enumerate}

\section{Stellar magnetism}

Along with rotation, stellar magnetism is a property of main--sequence stars that is often regarded as of secondary importance. In many stars this may be the case but it is increasingly becoming recognised that magnetic fields may play a major role in the evolution of some stars, particularly in populations of massive stars. As with rotation, magnetic fields are not an alien phenomenon. The Earth has a magnetic field strength of approximately $0.1$\,G and the Sun has a surface field strength of, on average, around $1$\,G. However, the complex behaviour of magnetic fields means that their generation and evolution is difficult to model and our knowledge of how this occurs within our own Solar System, where our observations are somewhat more detailed than other systems, is still a very active area of research. However, we do know that extremely strong fields are not an uncommon feature of other stellar systems. The most striking example of this is the case of magnetars, neutron stars with extremely strong fields, which can have field strengths of order $10^{15}$\,G.

\subsection{Observations of magnetism in massive stars}

The first stellar magnetic field reported outside of our Solar System was by \citet{Babcock47} for a chemically peculiar A star, 78 Vir. Chemically peculiar stars have distinctly different surface compositions across a number of elements when compared to their surrounding populations and are commonly referred to as Ap and Bp stars for chemically peculiar A and B stars respectively. Around $10$\% of A stars are estimated to belong in the Ap classification \citep{Moss01}. It has been suggested that all chemically peculiar Ap stars result from the action of magnetic fields \citep{Auriere07}. In fact, \citet{Auriere07} suggested further that there may exist a minimal magnetic field strength, below which no stable field can be sustained given that so few Ap stars are observed with field strengths less than around $300$\,G.

Studies have shown that these chemically peculiar stars contain extremely slow rotators and in general have slower rotation than normal stars of similar temperature \citep{Mathys04}. This suggests that a mechanism for magnetic braking is likely to be operating in these stars (c.f. section~\ref{ch2.sec.braking}). It has been suggested that chemical peculiarity arises because of slow rotation \citep{Abt00} but an alternative scenario, and the one we consider in chapter~\ref{ch5}, is that slow rotation and chemical peculiarity are both results of the magnetic field evolution and are otherwise not causally related.

Observations of the structure of magnetic fields in magnetic stars typically find that they have significant large--scale structure and in most cases are well approximated by a simple dipolar field \citep{Mathys09}. This contrasts with the more complex field geometries found in less massive stars \citep{Wade03}. Measured field strengths for these stars can exceed $20$\,kG \citep{Borra78} and though this is rare, field strengths of around $7.5$\,kG are not uncommon \citep[e.g.][]{Bagnulo04, Hubrig05, Kudryavtsev06}. The determination of the magnetic fields in more massive stars (i.e. O stars) is far more difficult because they have few spectral lines and these are often too broad for use with standard techniques. That said, a number of magnetic O stars have been identified. The star $\theta^1$ OriC has an observed field of $550$\,G \citep{Donati02} and HD 191612 has observed field strength of 220\,G \citep{Donati06}.

Observations of the magnetic fields of massive stars come from a number of different methods. The most common is high--resolution spectropolarimetry. Magnetic fields cause a shift in the spectra in circularly polarised light. Therefore, by looking at the difference in the spectra produced from both right and left polarisations we can obtain an estimate for the magnetic field strength. If the field is sufficiently strong and there are sharp absorption features in the spectra it is sometimes possible to see the splitting of absorption lines in normal spectra without having to look at different polarisations. These methods are less useful at higher rotation rates because of the rotational broadening discussed in section~\ref{ch1.sec.rotobs}. In these situations, the largest absorption features become too broad to get accurate measurements. For fast rotators, hydrogen Balmer lines are often used to determine the magnetic field strength because rotational broadening has a much smaller effect on them owing to their larger intrinsic line width. This method has notably been used by the {\sc fors}--1 instrument at the Very Large Telescope \citep[e.g.][]{Bagnulo02, Hubrig08}. For a more detailed review of the observational techniques used to determine the magnetic field configuration in massive stars we direct the reader to \citet{Mathys09}.

More recently, a great deal of work to expand our knowledge about massive magnetic stars has been conducted by the MiMeS (Magnetism in Massive Stars) collaboration\footnote{\url{http://www.physics.queensu.ca/~wade/mimes}} \citep{Wade09}. The survey involves over 1,000 hours using the high--resolution spectropolarimeters ESPaDOnS and Narval. The results of this survey are only now beginning to reach maturity and so we expect a great deal of future work will examine how this new data relates to our current theoretical understanding of the magnetic field evolution of massive stars. Interestingly, in their preliminary observations they identified a star with field strength around $2$\,kG rotating at around $290{\rm \,km\,s^{-1}}$ \citep{Grunhut12}. This is unusually quick for a very magnetic star but its low mass ($M=5.5$\Msun) is consistent with the model we shall present in chapter~\ref{ch5}. \citet{Grunhut11} also report the discovery of a number of new magnetic O and B stars. Notably the incidence of magnetic fields is around $8\%$ although it is significantly higher in B stars than O stars. However, whether this is a genuine reflection of actual stellar populations or a result of small number statistics will only be resolved as more data becomes available.

\subsection{Stellar dynamos}
\label{ch1.sec.dynamo}

A dynamo is any process that converts mechanical or kinetic energy into magnetic or electrical energy. Mechanical dynamos are very common and are used for generating all of our household electricity. However, these dynamos use solid magnets where as stars are made up of plasma which has a lot more freedom of motion. This makes the problem of sustaining a dynamo in stars significantly more challenging. In a stellar dynamo, energy is transferred between the kinetic motion of the gas and its associated magnetic field. As the fluid moves, the magnetic field lines are deformed, broken and reconnected. This complex interplay drives the generation of new field but also results in significant Ohmic decay. A dynamo--driven field is only possible if the rate of magnetic energy generation exceeds its overall dissipation. The proposition that dynamos might be responsible for sustaining stellar magnetic fields dates back as far as \citet{Larmor19} although the existence of self--sustaining dynamo action was not proved until several decades later \citep{Backus58, Herzenberg58}.

The study of stellar dynamos is important because simple magnetic fields are subject to a number of instabilities. These instabilities cause the rapid dissipation of any large--scale field and so unless a stable configuration exists (section~\ref{ch1.sec.fossil}) then a dynamo is required to regenerate the large--scale field. We stress that it is the large--scale field we are interested in. Small--scale fields are relatively straightforward to generate but the transformation of those fields into a cohesive large--scale field is somewhat more difficult. It was \citet{Cowling33} who showed that axisymmetric magnetic fields were intrinsically unstable. Further instabilities have also been demonstrated to affect simple fields \citep[e.g.][]{Parker58,Tayler73}. For a history of stellar dynamos we direct the reader to \citet{Weiss05}.

Stellar magnetic fields are most often considered to derive from a specific dynamo, known as the $\alpha$--$\Omega$ dynamo\footnote{Sometimes $\omega$ is used instead of $\Omega$ depending on the author.}. We shall discuss other possible dynamos later in this section. In the $\alpha$--$\Omega$ dynamo, toroidal flux\footnote{Poloidal and toroidal refer to the two different components of the magnetic field. Toroidal refers to the component that encircles the rotation axis, poloidal refers to the component that points along the rotation axis and radially outwards. The divergence--free nature of magnetic fields means that only two components are necessary to describe them.} is generated from the poloidal flux by the action of shear (i.e. differential rotation). This is the $\Omega$--effect \citep{Cowling45}. The regeneration of the poloidal field comes from correlations of the small--scale turbulent motions and is known as the $\alpha$--effect \citep{Parker55}. We shall see where this comes from later in this section.

The study of dynamo mechanisms often comes from the use of mean field magnetohydrodynamics (MHD), developed by \citet{Steenbeck66} and later by \citet{Moffat70}. Here we present a derivation for a simple $\alpha$--$\Omega$ dynamo following the method of \citet{Roberts72}. We assume a background fluid flow $\bi{U}$ and magnetic field $\bi{B}$ contained within the volume $V$. By Maxwell's equations, the magnetic field must be divergence free, $\nabla\cdot\bi{B}=0$, and evolves according to the induction equation,

\begin{equation}
\label{ch1.eq.induction}
\diffb{\bi{B}}{t}=\nabla\left(\bi{U} \times \bi{B}\right) + \nabla\times\left(\eta_0\nabla\times\bi{B}\right),
\end{equation}

\noindent where $\eta_0$ is the magnetic diffusivity. The principle of mean--field MHD is that the fluid velocity and magnetic field may be split into two components; one that varies on large scales and one that varies on small scales. This is not necessarily valid in stars, particularly because turbulent energy cascades operate over a continuous range of scales. However, it does provide a reasonable starting point. We write

\begin{equation}
\bi{U} = \overline{\bi{U}}+ \bi{U}',\quad \bi{B}=\overline{\bi{B}}+\bi{B}',
\end{equation}

\noindent where an over bar denotes some suitable average across small scales and a primed quantity only varies over small scales (i.e. $\overline{\bi{B}'}=0$). If we substitute these quantities into equation equation~(\ref{ch1.eq.induction}) we get

\begin{equation}
\label{ch1.eq.induction2}
\diffb{\overline{\bi{B}}+\bi{B}'}{t}=\nabla\left(\overline{\bi{U}} \times \overline{\bi{B}} + \overline{\bi{U}} \times \bi{B}' + \bi{U}'\times\overline{\bi{B}} + \bi{U}'\times \bi{B}'\right) + \nabla\times\left(\eta_0\nabla\times\left(\overline{\bi{B}}+\bi{B'}\right)\right).
\end{equation}

\noindent By taking our small--scale average of this equation we get

\begin{equation}
\label{ch1.eq.induction3}
\diffb{\overline{\bi{B}}}{t}=\nabla\left(\overline{\bi{U}} \times \overline{\bi{B}} + \mathcal{E}\right) +  \nabla\times\left(\eta_0\nabla\times\overline{\bi{B}}\right)
\end{equation}

\noindent where $\mathcal{E}=\overline{\bi{U}'\times \bi{B}'}$. Finally, by subtracting equation (\ref{ch1.eq.induction3}) from equation (\ref{ch1.eq.induction2}) we get

\begin{equation}
\label{ch1.eq.induction4}
\diffb{\bi{B}'}{t}=\nabla\left(\overline{\bi{U}} \times \bi{B}' + \bi{U}'\times\overline{\bi{B}} + \mathcal{E}'\right) +  \nabla\times\left(\eta_0\nabla\times\bi{B}'\right)
\end{equation}

\noindent where $\mathcal{E}'=\bi{U}'\times \bi{B}'-\overline{\bi{U}'\times \bi{B}'}$. More complicated parameterizations exist \citep[see the reviews of][]{Brandenburg05, Brandenburg09} but for a simple dynamo model we take the second order correlation approximation \citep[SOCA;][]{Steenbeck66} which simplifies to $\mathcal{E}=\alpha \bi{B} -\beta \nabla \times \bi{B}$\footnote{SOCA actually gives $\mathcal{E}_{\rm i}=\alpha_{\rm ij}B_{\rm j}+\beta_{\rm ijk}\diffb{B_{\rm j}}{x_{\rm k}}$ but we take constant, isotropic values for the $\alpha$ and $\beta$ tensors for simplicity.}. Substituting into equation~(\ref{ch1.eq.induction3}) and dropping the over bars we get

\begin{equation}
\label{ch1.eq.induction5}
\diffb{\bi{B}}{t}=\nabla\times\left(\alpha \bi{B}\right)+\nabla\left(\bi{U} \times \bi{B}\right) +  \nabla\times\left(\eta\nabla\times\bi{B}\right)
\end{equation}

\noindent where $\eta=\eta_0+\beta$. This is the origin of the $\alpha$--effect. We expect the microscopic diffusion coefficient $\eta_0\ll\beta$ because it acts across a much smaller length scale. We therefore take $\eta=\beta$ for the remainder of this dissertation. If we now look at the simple case in spherical polar coordinates, $\bi{U}=\Omega(r)\bi{e}_{\phi}$ and $\bi{B}=\left(B_r,B_{\theta},B_{\phi}\right)=B_{\phi}\bi{e}_{\phi}+\nabla\times\left(A(r)\bi{e}_{\phi}\right)$ then equation~\ref{ch1.eq.induction5} becomes

\begin{equation}
\label{ch1.eq.toroidal}
\diffb{B_{\phi}}{t}=rB_r\sin\theta\diffb{\Omega}{r}-(\nabla\times(\eta\nabla\times\bi{B}))_{\phi}
\end{equation}
\noindent and
\begin{equation}
\label{ch1.eq.poloidal}
\diffb{A}{t}=\alpha B_{\phi}-\nabla\times(\eta\nabla\times A\bi{e}_{\phi}).
\end{equation}

\noindent Strictly speaking, an $\alpha$ term should appear in equation (\ref{ch1.eq.toroidal}) but we assume that it is dominated by the shear term. If there were no shear and we included just the $\alpha$ terms we would get an $\alpha^2$--dynamo. Similarly, an $\alpha^2$--$\Omega$ dynamo is one where an $\alpha$--term is included in both equations as well as the shear term in the poloidal equation. The various types of dynamo model relate to the complexity of the terms included from the mean field parameter $\mathcal{E}$ and how the small--scale motions are translated into a large scale effect. For our purposes, the $\alpha$--$\Omega$ dynamo model is currently the most suitable theoretical set up. From equations~(\ref{ch1.eq.toroidal}) and~(\ref{ch1.eq.poloidal}) we see the action of the shear in generating toroidal flux from the radial magnetic field and the $\alpha$--effect generating poloidal field from the toroidal component. We explore more details of this model in section~\ref{ch2.sec.magnetism} and look at the subsequent results in chapter~\ref{ch5}.

\subsection{Fossil fields}
\label{ch1.sec.fossil}

In chapter~\ref{ch5} we examine a model for a radiative dynamo operating in the envelope of rotating massive stars. However, another popular theory for the existence of massive magnetic stars is that the fields are primordial in origin, this is the fossil fields argument. The theory, proposed by \citet{Cowling45}\footnote{Cowling's paper actually suggests how the Sun's magnetic field formed. We now know that the Sun's magnetic field actually has a dynamo origin.}, is that as a star is formed from the inter--stellar medium, weak fields within that material can become enhanced through flux conservation as the material collapses. If enough of the magnetic field can survive the pre--mainsequence evolution, particularly when the star is largely convective, then a star arrives on the main sequence with a substantial magnetic field. 

One of the main issues with this theory in the past has been that simple magnetic field configurations are prone to instability \citep[e.g.][]{Cowling33,Parker58,Tayler73}. If this is the case then we should only see strong magnetic fields at the start of the main sequence if magnetic fields could survive long enough to reach the main sequence at all. However, recent work by \citet{Braithwaite06} has uncovered certain magnetic field configurations that are stable for time scales similar to the main sequence lifetimes and so it seems possible that fossil fields might survive through the main sequence. Furthermore it seems that arbitrary field configurations may relax to these stable states \citep{Mathis11}. Even if dynamos do operate in stellar interiors this is an extremely important discovery. However, at the moment it is somewhat unclear how these field geometries are affected by turbulence driven by rotation and magnetic fields. 

Another issue is that stars are expected to go through a stage of being almost entirely convective during the pre--main sequence. This is expected to cause severe disruption of any existing fossil field. \citet{Moss03} examined the effect of pre--mainsequence evolution on fossil fields and found that significant fields might survive through to the main sequence. The proportion of flux which survives is strongly dependent on the magnetic diffusion coefficient. If it is too high then almost no flux is expected to survive.

Finally, the fossil field argument cannot yet explain the proportion of massive stars that support magnetic fields and why some stars reach the main sequence with strong fields whilst others do not. This is particularly puzzling when magnetic fields are considered to be an essential part of the star formation process. The proposition that the variation in magnetic fields arises because of variations in the interstellar medium is quite reasonable but it does shift the problem to an earlier stage of evolution rather than solving it.

In their study, \citet{Alecian08} found that from a study of 55 A and B stars on the pre--main sequence. Around $7\%$ were found to support significant magnetic fields. A number of these were determined to be almost entirely radiative and so the possibility that magnetic fields originate through a convective dynamo was considered unlikely. However, \citet{Alecian08} did not consider the alternative action of a radiative dynamo.

\section{Common envelope evolution}
\label{ch1.sec.ce}

Binary star systems, where two stars orbit each other, play a crucial role in astronomy. In most situations, the evolution of the two stars in a binary system can be modelled independently. However this simplification breaks down if the two components get close enough for strong interactions to occur. If there is sufficient mass in the system and the separation is small enough then tidal and gravitational effects become important. By assuming zero eccentricity and that the two stars behave as point masses we can model binary systems by transforming to a frame co--rotating with the system. The subsequent force potential, $\Phi$, called the Roche potential, is

\begin{equation}
\Phi(x,y,z)=-\frac{\omega^2}{2}\tilde{\Phi}\left(\frac{x}{a},\frac{y}{a},\frac{z}{a}\right),
\end{equation}
\begin{equation}
\tilde{\Phi}\left(x,y,z\right)=\frac{2q}{1+q}\frac{1}{r_1}+\frac{2}{1+q}\frac{1}{r_2}+\left(x-\frac{1}{1+q}\right)^2+y^2,
\end{equation}
where the mass of the two stars are $m_1$ at the origin and $m_2$ at $(a,0,0)$, $a$ is the orbital separation,
\[r_1^2=x^2+y^2+z^2,\]\[ r_2^2=(x-a)^2+y^2+z^2,\]\[ q=\frac{m_1}{m_2}\]
and
\[ \omega^2=\frac{G(m_1+m_2)}{a^3}.\]

\begin{figure}
\centering
\includegraphics[width=0.99\textwidth]{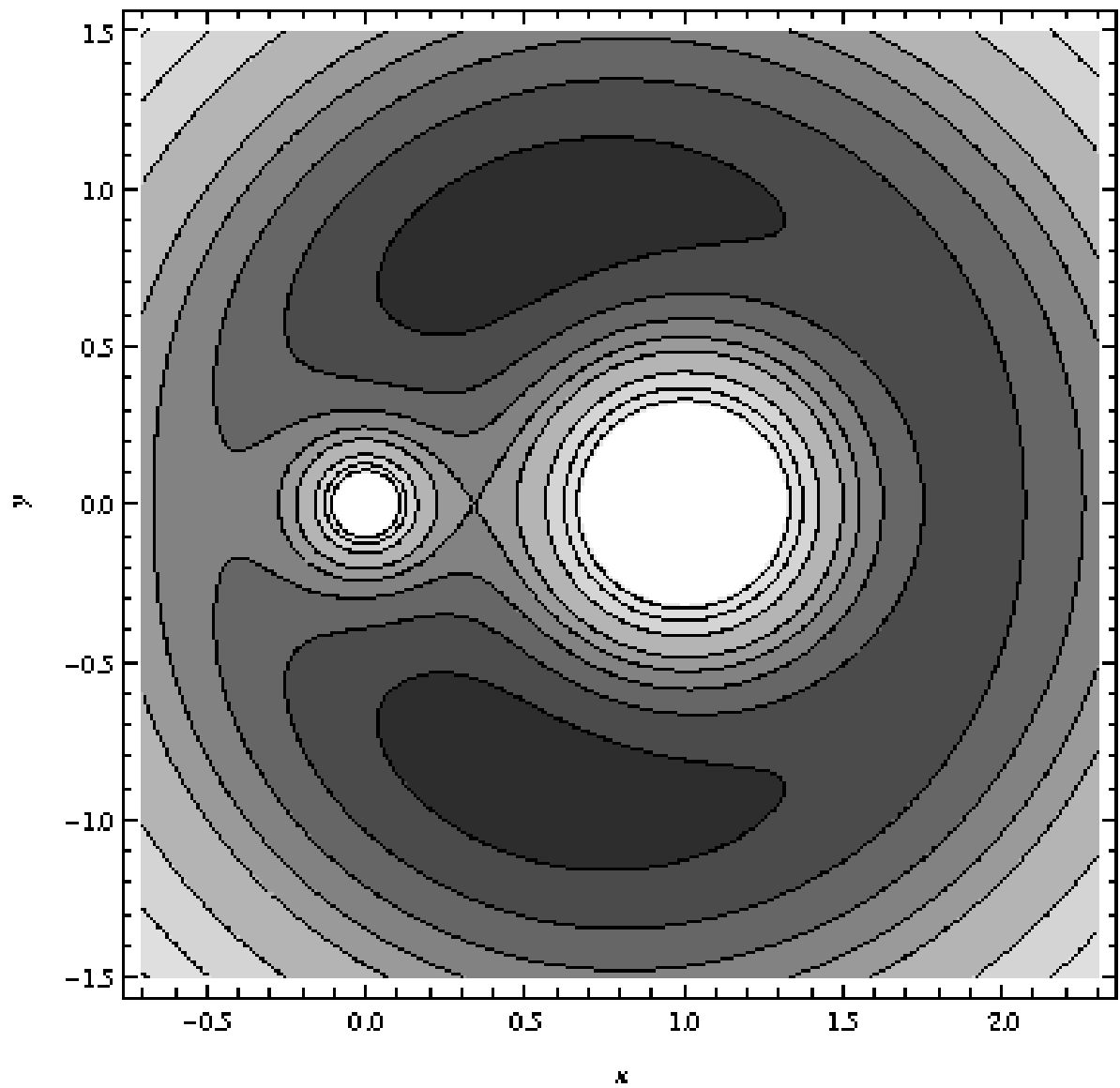}
\caption[An example of the Roche potential for two orbiting point masses]{An example of the Roche potential, $\tilde{\Phi}$, for two orbiting point masses with dimensionless parameters $q=0.2$ and $a=1$. The figure shows several equipotentials including the critical Roche Lobe.}
\label{ch1.fig.roche}
\end{figure}

\noindent An example of the equipotentials of this configuration is shown in Fig. \ref{ch1.fig.roche}. In hydrostatic equilibrium the surfaces of the stars lie along these equipotentials. This is evident from the static Navier--Stokes equation

\begin{equation}
\nabla P=\nabla \Phi,
\end{equation}

\noindent where $P$ is the pressure of the material. The stars are therefore left as almost undisturbed spheres if they are sufficiently far within the critical potential which crosses at the point where the gravitational forces of each star and the centrifugal force balance. The closer they get to this critical potential, the more distorted they become. If one of the stars becomes sufficiently large so that it fills over the critical potential surface, referred to as its Roche Lobe, then mass overflows on to the companion star. The Roche Lobe radius is the radius of a sphere with the same volume and is usually estimated by \citeauthor{Eggleton83}'s \citeyearpar{Eggleton83} formula which is accurate to within $1\%$ for all $q$,

\begin{equation}
\frac{R_L}{a}=\frac{0.49 q^{2/3}}{0.69 q^{2/3}+\ln(1+q^{1/3})}.
\label{ch1.eq.RLobe}
\end{equation}

\noindent If mass transfer is stable and the material has sufficient angular momentum the transferred mass forms an accretion disc around the companion. However in some circumstances the transfer of mass is rapid and the mass cannot be accreted fast enough. In these cases the transferred material forms a thick layer around the companion. This matter is not co--rotating nor does it reach hydrostatic equilibrium and if mass transfer continues it overflows the Roche Lobe of the companion and engulfs the whole system. This is what we refer to as a common envelope (hereinafter CE). The basic principle of CE evolution was proposed by \citet{Paczynski76}

During a CE phase the envelope does not co--rotate with the orbit of the remnant stars and so they feel a drag force. This leads to energy and angular momentum transfer between the orbit and the CE, potentially expelling the CE, at least in part, and a reduction of the orbital separation of the stars. CE evolution is essential to explain the existence of short period binary stars which contain at least one compact object (a white dwarf, neutron star or black hole). The reduction in orbit is important because the compact object must have been formed from an asymptotic giant branch (AGB) or red giant (RG) star which originally had an envelope much larger than the final binary separation. There must therefore exist a mechanism to remove a large part of the energy and angular momentum from the progenitor system. For typical systems, up to $90\%$ of the orbital energy must be removed. Examples of systems that have undergone a CE phase are cataclysmic variables, binary puslsars, low--mass X--ray binaries and planetary nebulae with close binary nuclei.

There are a number of possible mechanisms that can lead to Roche Lobe overflow and a subsequent CE phase. The most common is the expansion of the outer stellar envelope once a star exhausts the hydrogen at its centre. Once this happens, the star expands and may fill its Roche Lobe, depending on the mass ratio and orbital separation. For a common envelope phase to ensue, mass transfer must be unstable. This requires the mass--losing (donor) star to have developed a deep enough convective envelope and a large enough mass ratio. Provided these conditions are met, mass is transferred on the time scale $\tau_{\rm{dyn}}\approx GM^2/RL$ where $M,R\,{\rm and}\,L$ are the mass, radius and luminosity of the donor star respectively \citep{Paczynski71}.

In order for unstable mass transfer to proceed, the mass ratio must be sufficiently high, typically $q>0.8$. This can be simply demonstrated. The Roche Lobe radius, $R_L$, may be approximated by \citep{Iben84}

\begin{equation}
\frac{R_L}{a}=0.52\left(\frac{M_1}{M_1+M_2}\right)^{0.44}.
\end{equation}

\noindent We refer to the {\it primary} as the mass transferring star that initiated the common envelope phase and the {\it secondary} as its companion. The masses of the primary and secondary are represented by $M_1$ and $M_2$ respectively. Where appropriate, the additional subscripts `c' and `e' are used to describe the core and envelope of each star respectively (i.e. $M_1=M_{1{\rm c}}+M_{1{\rm e}}$). Other variables are labelled with the same convention. The mass ratio is defined as $q=M_{1}/M_{2}$ unless otherwise stated.
We take ${\rm d}M_2=-f\,{\rm d}M_1$ where $f\leq 1$ and ${\rm d}M_i$ is the change in mass of star $i$. This may be rewritten as

\begin{equation}
{\rm d}\ln R_L=2{\rm d}\ln J_{\rm orb}-2{\rm d}\ln M_1\left(0.78-fq-0.28\left(\frac{1-f}{1+q}\right)\right),
\end{equation}

\noindent where $J_{\rm{orb}}=a^2\omega\frac{q}{1+q}M_2$ is the orbital angular momentum which we assume is conserved, $\omega^2=\frac{G(M_1+M_2)}{a^3}$ is the orbital frequency and $a$ is the orbital separation. Thus, for $f=1$, the Roche Lobe shrinks as a result of mass transfer when $q>0.78$. Even if the primary doesn't have a deep convective envelope then, provided the mass ratio is high, unstable mass transfer still occurs.

The CE phase continues as long as the luminosity is high enough to keep driving material outwards. Eventually this is no longer the case and any material that has not been ejected can no longer be supported. If the giant star contains nuclear burning shells then once the thickness of the layer outside the shell drops below a critical value the shell is extinguished and the remains of the giant envelope contract. Eventually the giant shrinks within its Roche Lobe and the CE phase ends.

The original formulation of CE evolution assumed transfer of orbital energy of the binary to gravitational binding energy of the envelope with some constant transfer efficiency, $\alpha_{\rm{CE}}$ \citep{Webbink76, Livio88}, where

\begin{equation}
\alpha_{CE}=\frac{\Delta E_{\rm{bind}}}{\Delta E_{\rm{orb}}},
\label{ch1.eq.alpha1}
\end{equation}

 \noindent $\Delta E_{\rm{bind}}$ is the change in the binding energy and $\Delta E_{\rm{orb}}$ is the change in orbital energy. This equation may be written explicitly \citep{Webbink84, Nelemans00} as

\begin{equation}
\frac{M_{\rm{1e}}M_{1}}{\lambda R_{1}}=\alpha_{\rm{CE}}\left[\frac{M_{2}M_{1{\rm c}}}{2a_{\rm{f}}}-\frac{M_{1}M_{2}}{2a_{\rm{i}}}\right],
\label{ch1.eq.alpha2}
\end{equation}

\noindent where $a_{\rm{i}}$ and $a_{\rm{f}}$ are the initial and final orbital separations respectively. The exact form of equation (\ref{ch1.eq.alpha1}) differs in some work \citep[e.g.][]{deKool90}. We also note that the use of $\alpha$ here differs from \citet{Tutukov79} who take $\alpha$ simply as the ratio of the initial and final orbital energies. The parameter $\lambda$ depends on the structure of the envelope and for giant stars may be approximated by the relation \citep{Hjellming87}

\begin{equation}
\lambda^{-1}\approx3.000-3.816m_{\rm{e}}+1.041m^2_{\rm{e}}+0.067m^3_{\rm{e}}+0.136m_{\rm{e}}^4,
\end{equation}

\begin{figure}
\centering
\includegraphics[width=0.99\textwidth]{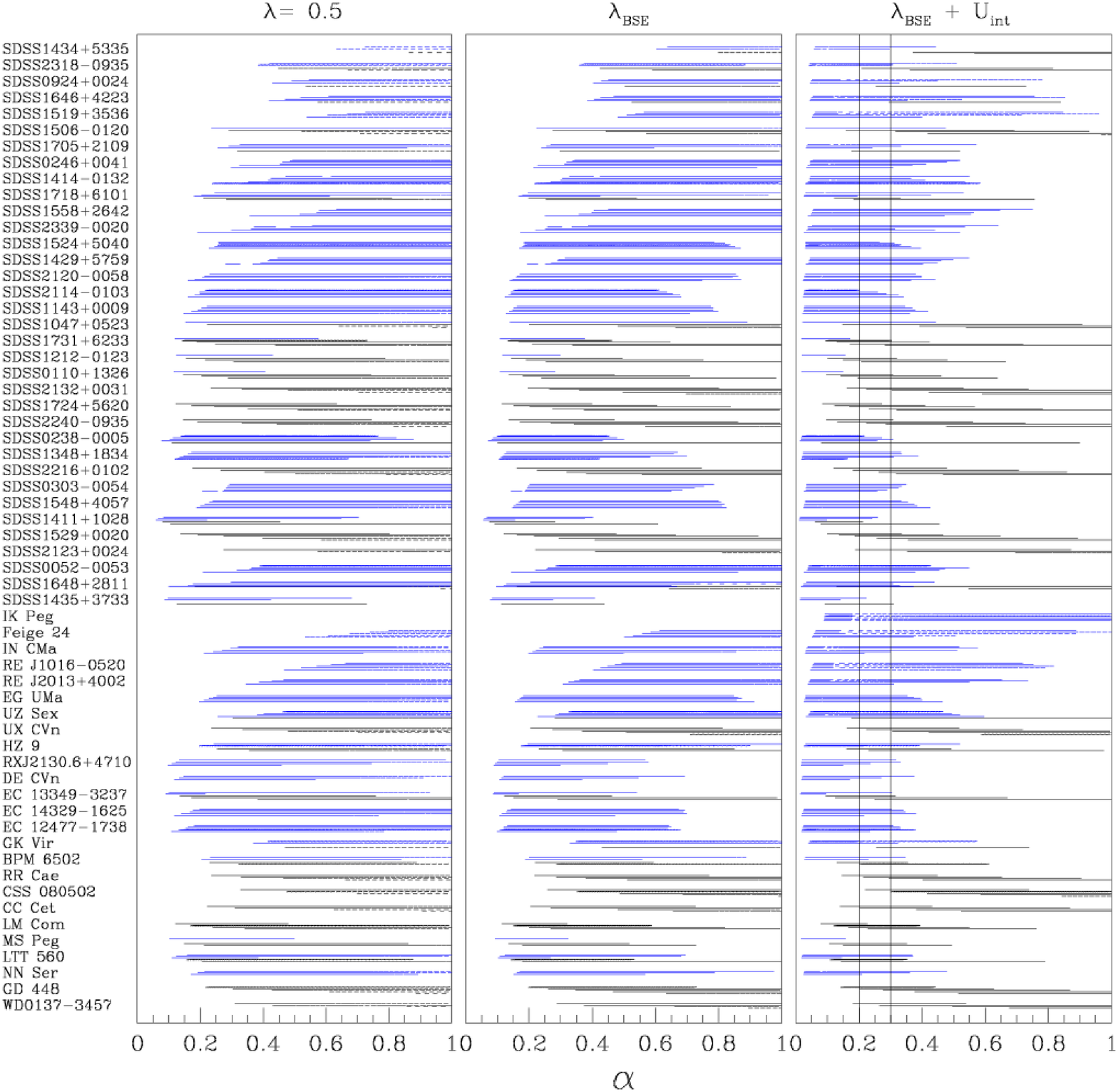}
\caption[Reconstructed values of the common envelope efficiency parameter for a number of systems]{Reconstructed values of $\alpha_{\rm CE}$ taken from \citet{Zorotovic10}. The figures shows the reconstructed values for several values of $\lambda$. The term $\lambda_{\rm BSE}$ refers to the case when the value is reconstructed using the {\sc bse} code of \citet{Hurley02}. The rightmost panel shows the effect of including the internal energy of the star in the calculation of $\alpha_{\rm CE}$.}
\label{ch1.fig.alpha}
\end{figure}

\noindent where $m_{\rm{e}}=M_{\rm{1e}}/M_1$. Several authors \citep[e.g.][]{Podsiadlowski03, Dewi00} take $\lambda=0.5$ but, owing to the phenomenological nature of the theory there is a great deal of uncertainty regarding the values taken by real systems. Alternatively it may be absorbed into the uncertainty factor $\alpha_{\rm{CE}}$ \citep[e.g.][]{Nelemans05}. Given this process of energy transfer, the ratio of initial to final orbital separations is

\begin{equation}
\frac{a_f}{a_i}=\frac{M_{1c}}{M_1}\left[1+\left(\frac{2 a_i}{\alpha_{\rm{CE}}\lambda R_1}\right)\left(\frac{M_{\rm{1e}}}{M_2}\right)\right]^{-1},
\end{equation}

\noindent where $a_i/R_1$ may be approximated at Roche Lobe overflow using equation (\ref{ch1.eq.RLobe}).

By observing post--common envelope systems we can use this model to reconstruct the possible values of $\alpha_{\rm CE}$ that may have produced the observed systems. An example of this, taken from \citet{Zorotovic10}, is shown in Fig.~\ref{ch1.fig.alpha}. We see that for most systems there is a wide range of possible values for the efficiency but most systems share a common value of $\alpha_{\rm CE}=0.2$.

The $\alpha$--model helps us to build up an idea for how common envelopes evolve but it is a very simple model. Over the past four decades, many attempts have been made to run numerical simulations of common envelopes at a range of complexities \citep[e.g.][]{Taam78,Meyer79,Bodenheimer84,Livio88,Terman94,Ricker08}. This has provided significant insight into how common envelopes evolve. However, these simulations often include very limiting simplifications or are only able to run over a fraction of the lifetime of the envelope. This is because of the huge range of length and time scales involved in the problem. For the moment, a fully detailed simulation of the entire common envelope phase of evolution is still highly impractical.

\section{Dissertation outline}

In chapter~\ref{ch2} we give the details of the physical models we use for simulating rotating stars with and without magnetic fields and how we implement those models numerically in the Cambridge stellar evolution code. We also discuss the generation of synthetic stellar populations from our stellar evolution models. In chapter~\ref{ch3} we look at the results of a number of common models of non--magnetic, rotating stars and examine where the greatest differences are in the predictions of each model. In chapter~\ref{ch4} we extend that analysis to synthetic stellar populations generated with the models of chapter~\ref{ch3}. We consider how the differences discussed in chapter~\ref{ch3} would appear in stellar populations and how this relates to the observations from the VLT--FLAMES survey of massive stars \citep{Dufton06, Hunter09}. In chapter~\ref{ch5} we look at how the introduction of magnetic fields into the physical model affects stellar evolution and its affect on our simulated populations. In chapter~\ref{ch6} we consider how white dwarfs with very strong surface fields might result from common envelope evolution in strongly interacting binary star systems. Finally, in chapter~\ref{ch7}, we summarize our conclusions from each of the chapters and present some suggestions for future work.

\begin{savequote}[60mm]
A theory has only the alternative of being right or wrong. A model has a third possibility: it may be right, but irrelevant. (Manfred Eigen)
\end{savequote}

\chapter{Modelling rotation and magnetic fields in stars}
\label{ch2}

Given the complicated and highly non--linear nature of the physics involved in rotation and magnetic fields, all but the simplest models must be solved numerically. The challenge for any model is choosing those elements of the physics which could have a strong effect on the results without creating a model whose complexity hinders simulation speeds and prevents progress in our understanding of the physical processes involved. In this chapter we discuss the physical models and numerical algorithms used throughout this dissertation.

\section{Stellar structure and evolution}
\label{ch2.sec.structure}

When we ignore secondary effects, such as rotation and magnetism, stars may be considered as one--dimensional objects. All of the fundamental properties such as pressure, density and temperature are determined according to the distance from the centre and are independent of latitude and longitude. In this case we refer to the star as being spherically symmetric. This approximation is extremely useful for numerical modelling. There have been attempts to create models of stars taking into account all three spatial dimensions \citep[e.g.][]{Bazan03}. However, these require huge amounts of computing power and are still only able to simulate stars on dynamical time scales which is impractical for studying the long--term behaviour of stars.

In one dimension we can formulate the equations for stellar evolution by defining variables according to $r$, the absolute distance from the centre of the star. In reality it is more convenient to work with the independent variable $m$, the mass enclosed within radius $r$. We can transform between the two coordinates with the mass conservation equation

\begin{equation}
\label{ch2.eq.mass}
\difft{m}{r}=4\pi r^2\rho,
\end{equation}

\noindent where $\rho$ is the local density.

There are some stages of stellar evolution that proceed on very short timescales. However, for the majority of a star's life we assume it exists in a state of hydrostatic equilibrium. In this state, the outward pressure is balanced by the inward pull of gravity and there are no bulk motions. This assumption breaks down in the presence of rotation as discussed in section~\ref{ch1.sec.vonzeipel}. The equation for hydrostatic equilibrium is

\begin{equation}
\label{ch2.eq.hydrostatic}
\difft{P}{m}=-\frac{Gm}{4\pi r^4},
\end{equation}

\noindent where $P$ is the local pressure and $G$ is the gravitational constant.

The equation for energy generation is

\begin{equation}
\label{ch2.eq.luminosity}
\difft{L}{m}=\epsilon,
\end{equation}

\noindent where $L$ is the local luminosity and $\epsilon$ is the energy generation rate. This term includes energy generation as a result of nuclear reactions, changes in gravitational energy and neutrino emission.

Energy is transported through the star according to the equation

\begin{equation}
\label{ch2.eq.temperature}
\difft{\log T}{m}=-\nabla\difft{\log P}{m}=-\frac{Gm}{4\pi r^4P}\nabla,
\end{equation}

\noindent where $T$ is the local temperature. The behaviour of $\nabla$ depends on whether energy is transported radiatively or convectively. Radiative transport is when energy is transported by repeated emission and absorption of photons. If the temperature gradient is sufficiently high or the opacity is sufficiently low then it is more efficient for energy to be transported by convection. In this case, material becomes unstable to vertical motion and circulation patterns of rising and falling fluid parcels develop. Strictly speaking our assumption of hydrostatic equilibrium breaks down in this case but we can still model the process within our hydrostatic framework because the perturbation to the underlying structure is small.

Where the star is radiative we take

\begin{equation}
\label{ch2.eq.nablar}
\nabla=\nabla_{\rm r}=\frac{3\kappa P L}{16\pi acgr^2T^4},
\end{equation}

\noindent where $\kappa$ is the opacity, $a$ is the radiation constant, $c$ is the speed of light and $g=Gm/r^2$ is the gravitational field strength.

The adiabatic gradient $\nabla_{\rm ad}=\left(\difft{\log T}{\log P}\right)_S$, where $S$ is the entropy of the stellar material. If $\nabla_{\rm ad}-\nabla_{\rm r}<0$, known as the Schwarzschild criterion, then the material is unstable to convective motions and we take $\nabla=\nabla_{\rm ad}+\Delta$, where $\Delta$ is the superadiabatic gradient. The adiabatic and superadiabatic temperature gradients can be calculated from the equation of state and mixing--length theory \citep{Bohm58}. This also gives us the convective diffusion coefficient for use in equation~(\ref{ch2.eq.composition}). The coefficient $D_{\rm con}=\left(C_{\rm con}\left(\nabla_{\rm r}-\nabla_{\rm ad}\right)^2m^2\right)/\tau_{\rm nuc}$ where $C_{\rm con}$ is a large constant calculated from mixing--length theory and $\tau_{\rm nuc}$ is the nuclear timescale. The coefficient $D_{\rm con}$ is non--zero only if the Schwarzschild criterion for convective instability is satisfied.

The evolution of the chemical composition of a star drives changes in the stellar structure and is responsible for the transition between the various phases of evolution. It is therefore of critical importance that we accurately track the changes that occur owing to nuclear fusion and chemical mixing. If $X_{i}$ is the mass fraction of element ${i}$ then it evolves according to the diffusion equation

\begin{equation}
\label{ch2.eq.composition}
\difft{X_{i}}{t}=\difft{}{m}\left(D\difft{X_{i}}{m}\right)-\sum_{j}R_{i,j}
\end{equation}

\noindent where $D$ is a diffusion coefficient which describes the transport of chemical composition. Typically for a non--rotating star this is only non--zero in convective zones although secondary effects such as overshooting, semi--convection, rotation and magnetism can all lead to mixing in radiative zones. The term $R_{i,j}$ is the rate of conversion of element ${i}$ to element ${j}$. This value can be either positive or negative depending on whether element ${i}$ is being created or destroyed.

In order to close these equations we need an equation of state which relates the temperature, pressure, entropy and density of the material and is a complex function of the physical variables. We also need to describe the opacity of the material and the reaction rates. We describe the methods used to determine these in section~\ref{ch2.sec.stars}.

\section{The Cambridge stellar evolution code}
\label{ch2.sec.stars}

This code was first developed by \citet{Eggleton71} and has undergone a number of significant revisions over the past 40 years \citep[e.g.][]{Pols95, Eldridge09}. The code is written in {\sc fortran 77} and, without the inclusion of rotation and magnetic fields, is approximately $7,000$ lines long. This is unusually short for a typical stellar evolution code. Its simplicity is one of the key reasons for the code's endurance and allows the modification of the internal physics with relative ease. 

The code splits the star into a sequence of mesh--points. These mesh--points are non--Lagrangian and are distributed according to a mesh--spacing function, $Q$. This is a complicated function of the pressure, temperature, mass, radius and density and can be easily modified depending on the needs of the user. The mesh points are distributed such that $\diffb{Q}{k}$ is constant throughout the star, where $k$ is the number of the mesh point. The function $Q$ is selected so that more mesh points are allocated to regions where higher resolution is needed such as the boundary of burning zones.

Prior to the inclusion of rotation and magnetic fields, the code solves $14$ finite--difference equations for the variables $r, m, T, L, \difft{Q}{m}, X_{\rm ^1H}, X_{\rm ^4He}, X_{\rm ^{12}C}, X_{\rm ^{14}N}, X_{\rm ^{16}O}, X_{\rm ^{20}Ne}, X_{\rm ^{24}Mg}$ and $X_{\rm ^{28}Si}$. The equations are solved by a Henyey technique \citep{Henyey59} which applies an iterative Newton--Raphson method. The solver uses an estimate for the solution of the equations based on the previous time step. This solution is then perturbed and the solver determines the relative fit between this solution and the previous solution. Based on this difference the code then produces a new trial solution. If this solution matches the equations sufficiently accurately then the code moves on to the next time step. If not then another iteration of this process is performed. If a solution with sufficient accuracy cannot be found or a maximal number of iterations is reached then the code terminates.

The difference equations are set up according to the structure equations described in section~\ref{ch2.sec.structure}. The distribution of mass is determined by

\begin{equation}
m_{k+1}-m_{k}=\Delta m_{k+1/2}
\end{equation}

\noindent where $\Delta m_{k}=\left(\diffb{m}{k}\right)_{k}$. The value of the derivative is determined by the mesh--spacing function. The mass conservation equation is

\begin{equation}
r^2_{k+1}-r^2_{k}=\left(\frac{1}{2\pi r \rho}\Delta m\right)_{k+1/2},
\end{equation}

\noindent where the right--hand side is the arithmetic mean of the function in brackets evaluated at mesh points ${k}$ and ${k+1}$. Similarly the equation for hydrostatic equilibrium is

\begin{equation}
(\log P)_{k+1}-(\log P)_{k}=-\left(\frac{Gm}{4\pi r^4 P}\Delta m\right)_{k+1/2}.
\end{equation}

The difference equation for energy transport is

\begin{equation}
(\log T)_{k+1}-(\log T)_{k}=-\left(\nabla\frac{Gm}{4\pi r^4 P}\Delta m\right)_{k+1/2},
\end{equation}

\noindent where $\nabla=\difft{\log T}{\log P}$ as described in section~\ref{ch2.sec.structure}. The difference equation for energy production is

\begin{equation}
L_{k+1}-L_{k}=\left(\epsilon\Delta m \right)_{k+1/2}.
\end{equation}

\noindent The rate $\epsilon$ contains terms for nuclear energy generation, energy changes owing to neutrino emission and changes in gravitational energy distribution. Finally, the evolution of the distribution of element ${i}$ is governed by

\begin{equation}
\begin{split}
\label{ch2.eq.chemdiff}
& \left(D\Delta m\right)_{k+1/2}\left(X_{i,k+1}-X_{{i,k}}\right)-\left(D\Delta m\right)_{k-1/2}\left(X_{i,k}-X_{i,k-1}\right)=\\ & \left(\left(\diffb{X_{i}}{t}+\sum_{j}R_{i,j}\right)\Delta m\right)_{k}+\left(X_{i,k+1}-X_{i,k}\right)\frac{\dot{m}_{k+1/2}}{2}+\left(X_{{i},k}-X_{i,k-1}\right)\frac{\dot{m}_{k-1/2}}{2}
\end{split}
\end{equation}

\noindent where $\dot{m}_{k}=\left(\diffb{m}{t}\right)_{k}$.

 The nuclear reaction rates were updated by \citet{Pols95} and \citet{Stancliffe05} and are based on the reaction rates of \citet{Caughlan88} and \citet{Gorres00}. The opacities were last updated by \citet{Eldridge04} and are based on the {\sc opal} opacities \citep{Iglesias96} at high temperatures and those of \citet{Ferguson05} at low temperatures. Also included are further corrections for major molecular opacities \citep{Stancliffe08}. The equation of state was last overhauled by \citet{Pols95} and includes the effects of Coulomb interactions and pressure ionization.

\section{Modelling stellar rotation}
\label{ch2.sec.rotation}

The centrifugal force caused by rotation affects the hydrostatic balance of the star, effectively reducing the local gravity. On a surface of constant radius the centrifugal force acts more strongly at the equator than the poles so the distortion of the star depends on co--latitude and our assumption of spherical symmetry is no longer valid. \citet{Tassoul78} showed that, except for stars that are close to critical rotation, the effect of rotational deformation remains axially symmetric. Enhanced mass loss from the surface because of rotation generally keeps stars rotating sufficiently below critical. Even when this is not the case it is only the outer--most layers that are affected. In models where the angular momentum distribution in convective regions is uniform the rotation rate may approach critical there but, because convective turbulence is already considered to be fully asymmetric, we do not need to consider further non--axial instabilities owing to the rotation. 

\subsection{Structure equations for rotating stars}
\label{ch2.sec.rotstructure}

We adopt similar adjustments to the stellar structure equations to those described by \citet{Endal76} and \citet{Meynet97}. First we define $S_P$ to be a surface of constant pressure, $P$. The volume contained within $S_P$ is $V_P$ and $r_P$ is the radius of a sphere with volume $V_P=4\pi r_P^3/3$. The mass conservation equation is then preserved in its non--rotating form,

\begin{equation}
\difft{m_P}{r_P}=4\pi r_P^2\rho,
\end{equation}

\noindent where $m_P$ is the mass enclosed within $S_P$ and $\rho$ is the density on the isobar which is assumed to be uniform. Owing to the strong stratification of stars, we expect hydrodynamic turbulence to be much stronger in the horizontal direction than in the vertical direction. This means that we expect variations in physical properties to be much smaller horizontally than vertically \citep{Zahn92}. This is known as the shellular hypothesis and is described further in section~\ref{ch2.sec.amtransport}. This applies to state variables such as pressure and temperature as well as composition variables. For a rotating star, the local gravity vector is

\begin{equation}
\label{ch2.eq.geff}
\bi{g}_{\rm eff}=\left(-\frac{Gm}{r^2}+\Omega^2r\sin^2\theta\right)\bi{e}_{r}+\left(\Omega^2r\sin\theta\cos\theta\right)\bi{e}_{\theta},
\end{equation}

\noindent where $\Omega$ is the local angular velocity. To proceed further we define the average of a quantity over $S_P$ as
\begin{equation}
\label{ch2.eq.average}
<q>\,\equiv\frac{1}{S_P}\oint_{S_P}q\, {\rm d}\sigma,
\end{equation}

\noindent where ${\rm d}\sigma$ is a surface element of $S_P$. Using this notation the equation of hydrostatic equilibrium becomes
\begin{equation}
\difft{P}{m_P}=-\frac{Gm_P}{4\pi r_P^4}f_P,
\end{equation}

\noindent where 

\begin{equation}
f_P=\frac{4\pi r_P^4}{Gm_PS_P}<g_{\rm eff}^{-1}>^{-1}
\end{equation}

\noindent and $g_{\rm eff}\equiv |\bi{g}_{\rm eff}|$. The derivation of $f_P$ is given in appendix~\ref{ap1.fp}. Hence with the new definition of variables we can retain the same 1D hydrostatic equilibrium equation as in non--rotating stars (c.f. equation \ref{ch2.eq.hydrostatic}) modified by a factor of $f_P$ which tends to unity for no rotation.

The equation for radiative equilibrium is similarly modified to

\begin{equation}
\difft{\log T}{\log P}=\frac{3\kappa P L_P}{16 \pi acGm_PT^4}\frac{f_T}{f_P},
\end{equation}

\noindent where $L_P$ is the total energy flux through $S_P$, $P$ is the pressure, $T$ is the temperature, $\kappa$ is the opacity, $a$ is the radiation constant, $c$ is the speed of light, $G$ is the gravitational constant and
\begin{equation}
f_T\equiv\left(\frac{4\pi r_P^2}{S_P}\right)\left(<g_{\rm eff}><g_{\rm eff}^{-1}>\right)^{-1}.
\end{equation}

\noindent Again, the non--rotating equation for stellar evolution has been preserved (c.f. equation~(\ref{ch2.eq.temperature})) except for the multiplication by $f_T/f_P$ (c.f. appendix~\ref{ap1.ft}). Of the two factors, $f_P$ deviates further from unity for a given rotation rate than $f_T$. Additional secondary effects of the reduced gravity must be taken into account when calculating quantities such as the pressure scale height and Brunt--V\"ais\"al\"a frequency. Hereinafter we drop the subscript $P$.

\subsection{Meridional circulation}
\label{ch2.sec.circulation}

The amount of thermal flux $F$ through a point in a star behaves as $F\propto g_{\rm eff}(\theta)$ (c.f. appendix~\ref{ap1.vonzeipel}). As seen in equation~(\ref{ch2.eq.geff}), $g_{\rm eff}$ is strongly dependent on co--latitude and so the radiative flux is greater at the poles than at the equator. This leads to a global thermal imbalance that drives a meridional circulation. The presence of meridional circulation has been considered for nearly a century since \citet{VonZeipel24}. In the past it has commonly been approximated by Eddington--Sweet circulation \citep{Sweet50}. \citet{Zahn92} proposed an alternative but similar treatment of the meridional circulation based on energy conservation along isobars and this is the formulation we are currently using. In spherical polar coordinates the circulation takes the form
\begin{equation}
\bi{U}=U(r)P_2(\cos\theta){\boldsymbol e}_r+V(r)\difft{P_2(\cos\theta)}{\theta}{\boldsymbol e}_{\theta},
\end{equation}

\noindent where $U$ and $V$ are linked by the continuity equation
\begin{equation}
V=\frac{1}{6\rho r}\difft{}{r}(\rho r^2U)
\end{equation}

\noindent and $P_2$ is the second Legendre polynomial $P_2(x)=\frac{1}{2}(3x^2-1)$. We currently use an approximate form for the meridional circulation by \citet{Maeder98}

\begin{equation}
\label{ch2.eq.circulation}
U=C_0\frac{L}{m_{\rm eff}g_{\rm eff}}\frac{P}{C_P\rho T}\frac{1}{\nabla_{\rm{ad}}-\nabla_{\rm r}+\nabla_{\mu}}\\\left(1-\frac{\epsilon}{\epsilon_m}-\frac{\Omega^2}{2\pi G\rho}\right)\left(\frac{4\Omega^2r^3}{3Gm}\right),
\end{equation}

\noindent where $m_{\rm eff}=m\left(1-\frac{\Omega^2}{2\pi G\rho}\right)$, $\epsilon=E_{\rm nuc}+E_{\rm grav}$, the total local energy emission, $\epsilon_m=L/m$, $C_P$ is the specific heat capacity at constant pressure, $\nabla_{\rm r}$ and $\nabla_{\rm{ad}}$ are the radiative and adiabatic temperature gradients respectively as described in section~\ref{ch2.sec.structure} and $\nabla_{\mu}=\diffb{\log\mu}{\log P}$ is the mean molecular weight gradient. For simplicity, the ratio of thermodynamic derivatives $\frac{\gamma}{\delta}$ used by \citet{Maeder98} is set to unity. As $\gamma=\left(\diffb{\log\rho}{\log\mu}\right)_{P,T}$ and $\delta=-\left(\diffb{\log\rho}{\log T}\right)_{P,\mu}$ this is correct for a perfect gas. We have also approximated the factor $\tilde{g}/g$ of \citet{Zahn92} by $\frac{4\Omega^2r^3}{3Gm}$. This is a suitable approximation throughout most of the star. The constant $C_0$ is included for aid of calibration and is discussed further in section~\ref{ch3.sec.testmodels}.

Meridional circulation receives very different treatments by different authors. More important than the precise formulation is the physical implementation. As discussed in section~\ref{ch2.sec.amtransport}, angular momentum is transported in stars by two main processes, advection and diffusion. Models such as those of \citet{Zahn92} and \citet{Maeder03} treat meridional circulation as an advective process where as \citet{Heger00} treat it as diffusive. This is an important distinction because an advective process can transport a variable in either direction with respect to the gradient of that variable. Therefore, advective transport of angular momentum can drive the generation of additional shear. Diffusion on the other hand can only transport a variable down the gradient of that variable and therefore can only act to reduce shear. Much emphasis is often placed on these two different treatments but shear can also be generated by structural changes from standard nuclear stellar evolution and mass--loss from the surface. However, the degree of shear is almost always significantly greater in models where meridional circulation is treated advectively.

\subsection{Mass loss with rotation}
\label{ch2.sec.massloss}

Observational evidence for enhanced mass loss is mixed \citep[see][]{Vardya85, Nieuwenhuijzen88} but, theoretically, near--critical rotation must drive additional mass loss to remove angular momentum and prevent the surface of the star from rotating super--critically \citep{Friend86}. We use the enhanced mass--loss rate of \citet{Langer98}
\begin{equation}
\dot{M}=\dot{M}_{\Omega=0}\left(\frac{1}{1-\frac{\Omega}{\Omega_{\rm crit}}}\right)^{\xi},
\end{equation}

\noindent where we take $\Omega_{\rm crit}=\sqrt{\frac{2GM}{3R^3}}$ and $\xi=0.45$. A more complete discussion of the critical rotation rates of stars is given by \citet{Maeder00} and \citet{Georgy11}. We use the non-rotating mass loss rates of \citet{Vink01}. 

\subsection{Rotation in convective zones}

Current 1D models of stellar evolution generally assume that convective zones are kept in solid body rotation. This may be caused by strong magnetic fields induced by dynamo action \citep{Spruit99} but there is no conclusive evidence that real fields generated by this mechanism are strong enough to enforce solid body rotation. Certainly in the Sun we see latitudinal variations in the angular velocity throughout the outer convective layer \citep{Schou98}. Standard mixing length theory suggests that a rising fluid parcel should conserve its angular momentum before mixing it with the surrounding material after rising a certain distance. This would lead to uniform specific angular momentum rather than solid body rotation. This is supported by a recent MLT--based closure model for rotating stars (Lesaffre et al., in prep.) and by 3D hydrodynamic simulations \citep{Arnett09}. In reality magnetic fields are likely to play some part in the transport of angular momentum but it is uncertain whether these are strong enough to affect the hydrodynamics. The asymptotic behaviour of the rotation profile in convective zones could have a profound effect on the evolution of the star, first because the total angular momentum content of a star for a given surface rotation increases dramatically for uniform specific angular momentum and secondly because uniform angular momentum in the convective zone produces a layer of strong shear at the boundary with the radiative zone and this drives additional chemical mixing. To explore the different possible behaviours we have introduced, in \textsc{rose} (section~\ref{ch2.sec.rose}), the capacity to vary the distribution of angular momentum in convective zones as discussed in section~\ref{ch2.sec.amtransport}.

\subsection{Angular momentum transport}
\label{ch2.sec.amtransport}

Differential rotation is expected to arise in stars because of hydrostatic structural evolution, mass loss and meridional circulation. Because of this stars are subject to a number of local hydrodynamic instabilities. These are expected to cause diffusion of radial and latitudinal variations in the angular velocity in order to bring the star back to solid body rotation, its lowest energy state. This occurs with characteristic diffusion coefficients $D_{\rm KH}$ and $D_{\rm h}$ respectively. \citet{Zahn92} proposed that, because of the strong stratification present in massive stars, the turbulent mixing caused by these instabilities is much stronger horizontally than vertically ($D_{\rm h}\gg D_{\rm KH}$). This leads to a situation where the angular velocity variations along isobars are negligible compared to vertical variations. Furthermore, all other state variables are assumed to be roughly constant over isobars and the mixing produces horizontal chemical homogeneity. This is referred to as shellular rotation, for which we describe the angular velocity by $\Omega=\Omega(r)$.

Taking into account all of the processes described in section~\ref{ch2.sec.rotation} we use the evolution equation for the angular velocity \citep{Zahn92}

\begin{equation}
\label{ch2.eq.amtransport}
\diff{r^2\Omega}{t}=\frac{1}{5\rho r^2}\diff{\rho r^4\Omega U}{r}+\frac{1}{\rho r^2}\frac{\partial}{\partial r}\left(\rho {D_{\rm KH} r^4\frac{\partial\Omega}{\partial r}}\right)+\frac{1}{\rho r^2}\frac{\partial}{\partial r}\left(\rho {D_{\rm{con}} r^{(2+n)}\frac{\partial r^{(2-n)}\Omega}{\partial r}}\right)
\end{equation}

\noindent and for the chemical evolution

\begin{equation}
\label{ch2.eq.chemtransport}
\frac{\partial X_{i}}{\partial t}=\frac{1}{r^2}\frac{\partial}{\partial r}\left(\left(D_{\rm KH}+D_{\rm{eff}}+D_{\Omega=0}\right)r^2\frac{\partial X_{i}}{\partial r}\right),
\end{equation}

\noindent where $X_{i}$ is the abundance of species ${i}$. The diffusion coefficient $D_{\rm con}$ is non--zero only in convective zones and the $D_{\rm KH}$ and $D_{\rm eff}$ are non--zero only in radiative zones. The parameter $n$ sets the steady state specific angular momentum distribution in convective zones, $n=2$ corresponds to solid body rotation and $n=0$ corresponds to uniform specific angular momentum. The coefficient $D_{\rm eff}$ describes the effective diffusion of chemical elements because of the interaction between horizontal diffusion and meridional circulation so

\begin{equation}
D_{\rm eff}=\frac{|rU|^2}{30 D_{\rm h}}.
\end{equation}

The main differences between models \citep[e.g.]{Talon97,Heger00,Meynet00,Maeder03} are the treatment of the meridional circulation, $U$, the diffusion coefficients, $D_{\rm KH}$, $D_{\rm eff}$ and $D_{\rm con}$, and the steady power--law distribution of angular momentum in convective zones determined by $n$. We describe many of these models which we have implemented with {\sc rose} in section~\ref{ch3.sec.testmodels}.

\subsection{Numerical implementation of rotation}
\label{ch2.sec.rose}

Rotation has been implemented in the {\sc stars} code described in section~\ref{ch2.sec.stars}. The new version of the code has been named {\sc rose} ({\sc ro}tating {\sc s}tellar {\sc e}volution). The code fundamentally remains the same but now solves an additional second--order difference equation based on equation~(\ref{ch2.eq.amtransport}) described in section~\ref{ch2.sec.amtransport}. Also included are modifications to the gravity and structure equations as described in section~\ref{ch2.sec.rotstructure}.

\subsubsection{Angular momentum transport}
\label{ch2.sec.amdifference}

Throughout the code we make the equivalence $J\equiv r^2 \Omega$ where $J$ is the new independent variable solved for by the code. In this sense $J$ is similar to specific angular momentum except that the specific angular momentum of a spherical shell rotating with angular velocity $\Omega$ is $\frac{2}{3}r^2\Omega$. We work with $J$ because it is more straightforward than including factors of $\frac{2}{3}$ in every instance. It also makes it easier to ensure angular momentum conservation. However, many processes are dependent on the amount of shear, $\diffb{\Omega}{r}$ and so we switch between the variables $\Omega$ and $J$ frequently.

The angular momentum evolves according to the difference equation

\begin{gather}
\label{ch2.eq.amdifference}
\Delta m_{k}\diffb{J_{{k}}}{t}+\left(J_{{k+1}}-J_{{k}}\right)\frac{\dot{m}_{k+1/2}}{2}+\left(J_{{k}}-J_{{k-1}}\right)\frac{\dot{m}_{k-1/2}}{2}=\nonumber \\
\frac{4\pi}{5}\left(\left(\rho r^2 J U\right)_{k+1/2}-\left(\rho r^2 J U\right)_{k-1/2}\right)+\nonumber \\
16\pi^2\left((D_{\rm KH}r^6\rho^2\Delta m^{-1})_{k+1/2}(\Omega_{k+1}-\Omega_{k})-(D_{\rm KH}r^6\rho^2\Delta m^{-1})_{k-1/2}(\Omega_{k}-\Omega_{k-1})\right)+\nonumber \\
16\pi^2\left(D_{\rm con}r^{4+n}\rho^2\Delta m^{-1}\right)_{k+1/2}\left(\left(r^{2-n}\Omega\right)_{k+1}-\left(r^{2-n}\Omega\right)_{k}\right)-\nonumber \\
16\pi^2\left(D_{\rm con}r^{4+n}\rho^2\Delta m^{-1}\right)_{k-1/2}\left(\left(r^{2-n}\Omega\right)_{k}-\left(r^{2-n}\Omega\right)_{k-1}\right).
\end{gather}

\noindent If we integrate equation~(\ref{ch2.eq.amtransport}) over the star and assume radiative behaviour exterior and interior the rate of change of total angular momentum $H_{\rm tot}$\footnote{Note that the variable $H_{\rm tot}$ is the total angular momentum as opposed to $J$ which is used to describe specific angular momentum, the angular momentum per unit mass.} 

\begin{equation}
\difft{H_{\rm tot}}{t}=4\pi r^4D_{\rm KH}\rho\diffb{\Omega}{r}{\bigg |}^R_0,
\end{equation}

\noindent where $R$ is the radius of the star. In other words, angular momentum conservation requires that we choose $\diffb{\Omega}{r}=0$ at the surface and the centre. We modify this condition in the presence of magnetic braking as discussed in section~\ref{ch2.sec.magnetism}. When mass is lost from the star through the action of stellar winds, it is effectively removed from the outer mesh point and so the total angular momentum of the star is reduced by the specific angular momentum of the outer mesh point multiplied by the total mass loss. Hence the total angular momentum of the star is reduced by the angular momentum lost in the stellar wind.

The difference equation for the evolution of the chemical composition is still given by equation~(\ref{ch2.eq.chemdiff}) except that the diffusion coefficient is modified to include the effects of rotation as described in equation~(\ref{ch2.eq.chemtransport}). The implementation of the difference equation for angular momentum transport and its boundary condition into {\sc rose} is fairly straightforward. The definition of suitable expressions for the diffusion coefficients is somewhat more complicated. The diffusion coefficients for purely rotating, non--magnetic stars are described in section~\ref{ch3.sec.testmodels}.

\subsubsection{Diffusion coefficient smoothing}

Numerical convergence becomes difficult at the surface and the centre of the star. Certain models, in which there is sudden change in behaviour at convective boundaries, also need special treatment. Much of this comes down to extreme changes of behaviour in angular momentum transport between radiative and convective zones. At these points, perturbations in one variable lead to large changes in another and this makes convergence very difficult. In order to avoid numerical instabilities, the diffusion coefficients are smoothed at the surface and the centre. The degree of smoothing is chosen so that a single choice of parameters allows all models with $3<M/{\rm M_{\odot}}<100$ with sub--critical initial rotation rate to reach the end of the main sequence. Many models are able to run with less aggressive smoothing but for consistency we choose parameters that do not require modification. 

At the centre of the star the situation is a little more complicated. During the main sequence, all of the stars considered in this dissertation have a convective core. When these stars reach the end of the main sequence their cores become radiative. From a numerical point of view, the value of $\nabla_{\rm ad}-\nabla_{\rm r}$ oscillates close to the centre and this again means that it is difficult to get convergent behaviour. To avoid this problem, a constant, uniform radiative diffusion coefficient is applied close to the centre and the convective diffusion coefficient goes to zero. This is also important where a uniform specific angular momentum distribution is assumed in convective zones because the rotation rate behaves as $\Omega\propto r^{-2}$ close to the centre. By assuming a small radiative region we prevent the singularity. In reality, viscous forces would eventually dominate here at length scales smaller than the convective length scale.

In the case of a constant, uniform convective diffusion coefficient, the behaviour changes suddenly at convective boundaries. Therefore, even small perturbations to the trial solution lead to large changes in the subsequent iteration. Convergence can be achieved by applying suitable smoothing to the convective diffusion coefficient at the convective boundary. This smoothing does not need to be applied in the case of a mixing--length theory based diffusion coefficient because the diffusion coefficient tapers off far more smoothly.

In order to apply smoothing to the diffusion coefficients we multiply them by variants of the function

\begin{equation}
\label{ch2.eq.smoothing}
f(x,x_0,\sigma)=\frac{1}{1+\exp(2(x-x_0)/\sigma)}=\frac{1}{2}\left(1-\tanh\left(\frac{x-x_0}{\sigma}\right)\right).
\end{equation}

\noindent Typically the arguments are taken to be some function of $m$, $k$ or $\nabla_{\rm ad}-\nabla_{\rm r}$. This function allows us to smoothly transition between different types of behaviour. At the surface the values are taken to be $x=k$, $x_0=0.2k_{\rm max}$ and $\sigma=10^{-2}$, where $k_{\rm max}$ is the number of mesh points used by the model, typically $399$. We note that the mesh points start at $1$ at the surface and increases to $k_{\rm max}$ at the centre. The radiative angular momentum transport diffusion coefficient tends to $10^{16}\,{\rm cm^2\,s^{-1}}$ which is a typical value calculated for the outer regions of massive stars. At the centre we take $x=k$, $x_0=0.99k_{\rm max}$ and $\sigma=10^{-3}$. At the centre of the star the diffusion coefficient again tends to $10^{16}\,{\rm cm^2\,s^{-1}}$. Even though the smoothing in the outer region covers around $20\%$ of the star by mesh point, this corresponds to a tiny fraction, typically around $10^{-3}$ of the star by mass. 

\subsubsection{The effective gravity and structure equations}
\label{ch2.sec.geff}

In the presence of rotation, the effective gravity is a complex function of the co--latitude, $\theta$. In order to simulate stars in one--dimension we have to perform suitable averages to remove this dependence. We use the framework set out in section~\ref{ch2.sec.rotstructure}. To calculate the value of $<g_{\rm eff}>$. We perform the average over spherical shells

\begin{equation}
<g_{\rm eff}>=\frac{1}{n_{\theta}+1}\sum_{j=0}^{n_{\theta}}g_{\rm eff}\left(r,\frac{2\pi j}{n_{\theta}}\right)
\end{equation}

\noindent and a similar average for $<g^{-1}_{\rm eff}>$. This allows us to calculate $f_P$ and $f_T$. We currently use $n_{\theta}=20$.

\section{Modelling stellar magnetism}
\label{ch2.sec.magnetism}

In order to simulate the magnetic field in stellar interiors we build on the code {\sc rose} described in section~\ref{ch2.sec.rotation}. In this section we describe a new physical model for magnetic field evolution in massive stars. It is based upon \citet{Spruit99} but encompasses many new elements. Most importantly, the magnetic field is evolved as an independent variable within the system, or rather two variables, one for the poloidal field and one for the toroidal field. The rotation is coupled to the magnetic field through the MRI instability and a simple $\alpha$--$\Omega$ dynamo 

\subsection{Magnetic field evolution}

We approach the evolution of the magnetic fields in a similar way to the evolution of the angular momentum distribution. In the radiative zones of stars, turbulence from purely rotational or magnetorotational instabilities leads to the generation of magnetic field by an $\alpha$--$\Omega$ dynamo mechanism. We assume a background velocity field of the form

\begin{equation}
\label{ch2.eq.velocity}
\bi{U}=\left\{U(r)P_2(\cos\theta),V(r)\difft{P_2(\cos\theta)}{\theta},\Omega(r)r\sin\theta\right\},
\end{equation}

\noindent where $P_2(x)$ is the second Legendre polynomial and $U(r)$ and $V(r)$ are the components of the meridional circulation and are related by the continuity equation

\begin{equation}
V=\frac{1}{6\rho r}\difft{}{r}(\rho r^2U).
\end{equation}

\noindent The radial component, $U(r)$, is taken to be the same as equation~(\ref{ch2.eq.circulation}) based on \citet{Maeder00}. It has been suggested that meridional circulation can be neglected in the presence of strong magnetic fields \citep{Maeder03b}. We discuss whether this is indicated by our model in section~\ref{ch5.sec.angmom}. For now we leave it in our equations for completeness.

The evolution of the large--scale magnetic field is described by the induction equation

\begin{equation}
\label{ch2.eq.induction}
\diffb{\bi{B}}{t}=\nabla\times(\bi{U} \times \bi{B})-\nabla\times(\eta\nabla\times\bi{B}).
\end{equation}

\noindent Assuming an azimuthal form for the mean field we may write $\bi{B}$ as

\begin{equation}
\label{ch2.eq.magfield}
\bi{B}=B_{\phi}(r,\theta)\bi{e}_{\phi}+\nabla\times(A(r,\theta)\bi{e}_{\phi}).
\end{equation}

\noindent Substituting equations (\ref{ch2.eq.velocity}) and (\ref{ch2.eq.magfield}) into (\ref{ch2.eq.induction}) gives

\begin{align}
\label{ch2.eq.toroidalfield}
\diffb{B_{\phi}}{t}=&\,rB_r\sin\theta\diffb{\Omega}{r}+B_{\theta}\sin\theta\diffb{\Omega}{\theta}-\nonumber\\
& \frac{1}{r}\diffb{}{\theta}\left(V(r)\difft{P_2(\cos\theta)}{\theta}B_{\phi}\right) -\frac{1}{r}\diffb{}{r}\left(rU(r)P_2(\cos\theta)B_{\phi}\right)-\nonumber \\
&\left(\nabla\times(\eta\nabla\times\bi{B})\right)_{\phi}
\end{align}
\noindent and
\begin{equation}
\label{ch2.eq.poloidalfield}
\diffb{A}{t}=-\frac{2V(r)}{r}\difft{P_2(\cos\theta)}{\theta}A\cot{\theta}-\frac{U(r)P_2(\cos\theta)}{r}\diffb{Ar}{r}\sin\theta+\alpha B_{\phi}-\nabla\times(\eta\nabla\times A\bi{e}_{\phi}),
\end{equation}

\noindent where we have introduced the $\alpha$--term in equation~(\ref{ch2.eq.poloidalfield}) to describe the regeneration of the poloidal field by the dynamo \citep{Schmalz91}. The radial and latitudinal components of the magnetic field are $B_r$ and $B_{\theta}$ respectively. Under the assumption of shellular rotation, the term $B_{\theta}\sin\theta\diffb{\Omega}{\theta}=0$.

In order to reduce the equations to one dimension we need to choose the $\theta$--dependence of the magnetic field and perform a suitable latitudinal average of equations (\ref{ch2.eq.toroidalfield}) and (\ref{ch2.eq.poloidalfield}). First we choose $A(r,\theta)=\tilde{A}(r)\sin\theta$ so that in the limit of no meridional circulation or magnetic stresses, the poloidal field tends towards a dipolar geometry. Under this assumption $B_r=\frac{2\tilde{A}\cos\theta}{r}$ and $B_{\theta}=-\difft{(r\tilde{A})}{r}\sin\theta$. We could equally choose a quadrupolar or higher order geometry but we start with this as the simplest case. The radial field has negative parity about the equator so this must also be true of the toroidal field. The toroidal field must also vanish at the poles to avoid singularities. We therefore choose $B_{\phi}=\tilde{B_{\phi}}(r)\sin(2\theta)$. Again, this is not a unique choice but is the lowest order Fourier mode that meets our requirements. Finally we take $\alpha=\tilde{\alpha}(r)$ and $\eta=\tilde{\eta}(r)$. 

We take the average\footnote{This is a different averaging process from that described in equation~(\ref{ch2.eq.average}).} of a quantity $q$ to be

\begin{equation}
\langle q \rangle_{\rm mag} = \int^{\pi/2}_0q\sin\theta{\rm d}\theta = -\int^{\pi}_{\pi/2}q\sin\theta {\rm d}\theta.
\end{equation}

\noindent The second identity holds because of our choice of parity for the various terms in equations~(\ref{ch2.eq.toroidalfield}) and~(\ref{ch2.eq.poloidalfield}). Hereinafter we drop the use of angled brackets and write $q=\tilde{q}$ for the radially--dependent components of the magnetic field and related quantities. Taking averages of equations~(\ref{ch2.eq.toroidalfield}) and~(\ref{ch2.eq.poloidalfield}) we get

\begin{equation}
\label{ch2.eq.toroidalfield2}
\diffb{B_{\phi}}{t}=A\diffb{\Omega}{r}-\frac{6}{5r}VB_{\phi}-\frac{1}{10r}UB_{\phi}+r\diffb{}{r}\left(\frac{\eta}{r^4}\diffb{}{r}(r^3B_{\phi})\right)
\end{equation}
\noindent and
\begin{equation}
\label{ch2.eq.poloidalfield2}
\diffb{A}{t}=\frac{3V}{2r}A-\frac{U}{8r}\diffb{}{r}(Ar)+\frac{8\alpha}{3\pi} B_{\phi}+\diffb{}{r}\left(\frac{\eta}{r^2}\diffb{}{r}(r^2A)\right).
\end{equation}

\noindent In the case where diffusion dominates, $A\to 1/r^2$ and $B_{\phi}\to 1/r^3$. This is what we expect for a dipolar field. Our boundary conditions are $B_{\phi}=0$ and $B_{\theta}\propto\frac{\pi}{2}\diff{rA}{r}=0$ at $r=0$ and $R$.

\subsection{Angular momentum evolution with a magnetic field}
\label{ch2.sec.ammagnetic}

In the Taylor--Spruit dynamo \citet{Spruit02} angular momentum transport is driven by the Maxwell stress produced by the magnetic field. This process is assumed diffusive and an effective diffusion coefficient is derived. We treat the angular momentum evolution in radiative zones by extending equation~(\ref{ch2.eq.amtransport}) to

\begin{equation}
\label{ch2.eq.magamtransport}
\diff{r^2\Omega}{t}=\frac{1}{5\rho r^2}\diff{\rho r^4\Omega U}{r}+\frac{3r}{8 \pi\rho}\langle\left(\nabla\times \textbf{\textit{B}}\right)\times\textbf{\textit{B}}\rangle_{\phi}+\frac{1}{\rho r^2}\frac{\partial}{\partial r}\left(\rho D_{\rm tot}r^4\frac{\partial\Omega}{\partial r}\right),
\end{equation}

\noindent where the pre--factor in the magnetic stress term comes from the combination of a factor of $\frac{1}{4\pi}$ for the permeability of free space and $\frac{3}{2}$ from the spherical average, $<r^2\sin^2\theta>_{\rm mag}$, on the left--hand side. The term $D_{\rm tot}$ is the total diffusion of angular momentum that arises from a combination of purely rotationally--driven turbulence, magneto--rotational turbulence and convection. Purely hydrodynamic turbulence comes from Kelvin--Helmholtz instabilities that are driven by shear. As in section~\ref{ch2.sec.rotation}, we refer to this diffusion coefficient as $D_{\rm KH}$. There are other sources of hydrodynamic turbulence, including an effective diffusion owing to the meridional circulation, but we group these all in $D_{\rm KH}$.  We use the formulation of \citet{Potter11}, based on that of \citet{Maeder03}, but other formulations may be used instead as described in chapters~\ref{ch3} and~\ref{ch4}. The diffusion by convective transport is $D_{\rm con}$ and is based on the effective diffusion from mixing--length theory \citep{Bohm58}. Finally the magnetic diffusion is $D_{\rm mag}$. With this notation $D_{\rm tot}=D_{\rm KH}+D_{\rm con}+D_{\rm mag}$. After averaging the magnetic stress term in equation~(\ref{ch2.eq.magamtransport}) over co--latitude we find

\begin{equation}
\diff{r^2\Omega}{t}=\frac{1}{5\rho r^2}\diff{\rho r^4\Omega U}{r}+\frac{3}{64 \rho r^3 B_\phi}\diffb{}{r}\left(r^3B^2_{\phi}A\right)+\frac{1}{\rho r^2}\frac{\partial}{\partial r}\left(\rho D_{\rm tot} r^4\frac{\partial\Omega}{\partial r}\right),
\end{equation}

\noindent where a factor of $\frac{8}{\pi}$ appears in the Maxwell stress term from the spherical average.

 We see that the Maxwell stress does not act diffusively as is often suggested. \citet{Spruit02} equates the Maxwell stress, $S_{\rm mag}$, to $r \rho\diffb{\Omega}{r}\nu_{\rm e}$, where $\nu_{\rm e}$ is some effective diffusivity. This automatically assumes that the large scale stresses lead to solid body rotation and is unjustified. It leads to a diffusion coefficient of the form $\nu_{\rm e}\propto(\diffb{\Omega}{r})^{-1}$ and so high diffusion rates for small shear. We could have equally assumed any similar relation such as $S_{\rm mag}=\frac{\rho}{r}\diff{r^2\Omega}{r}\hat{\nu}_{\rm e}$, where $\hat{\nu}_{\rm e}$ is now an effective diffusivity which drives the system towards uniform specific angular momentum. For \citet{Spruit02} this never becomes a problem because he assumes a steady--state saturated magnetic field but it does present a problem for systems where the magnetic field strength is independently derived. The magnetic stress term in fact acts advectively and so can increase the amount of shear in the system.

\subsection{Magnetic diffusion}

Instead of relying on the large scale Maxwell stress to redistribute angular momentum in radiative zones, we use the magnetic turbulence from the Tayler--instability \citep[][]{Tayler73}. Turbulent diffusion coefficients for this instability were proposed by \citet{Spruit02} and \citet{Maeder04}. We follow a similar method to derive the associated diffusion coefficients here. The main difference is that we solve for the magnetic field and hence the Alfv\'{e}n velocity independently instead of treating it as a function of the rotation rate.

First, the energy of the instability must be enough to overcome the restoring buoyancy force. This puts a limit on the vertical extent of the magnetic instability

\begin{equation}
\label{ch2.eq.lr}
l_r<\frac{r\omega_{\rm A}}{N},
\end{equation}

\noindent where $\omega_{\rm A}^2\approx \frac{B_{\phi}^2}{4\pi r^2 \rho}$ is the Alfv\'{e}n frequency and $N$ is the relevant buoyancy frequency. If this length scale is too small then the magnetic diffusivity damps the instability. \citet{Spruit02} takes this limit to be

\begin{equation}
\label{ch2.eq.lr2}
l_r^2>\frac{\eta\Omega}{\omega_{\rm A}^2}.
\end{equation}

\noindent When account is taken of the thermal diffusivity, the buoyancy frequency given by \citet{Maeder04} is

\begin{equation}
\label{ch2.eq.bvfrequency}
N^2=\frac{\eta/K}{\eta/K+2} N_T^2 + N_{\mu}^2,
\end{equation}

\noindent where $K$ is the thermal diffusivity, $N_T^2$ is the Brunt--V\"{a}is\"{a}l\"{a} frequency and $N_{\mu}^2$ is the frequency associated with the mean molecular weight gradient. Substituting equation~(\ref{ch2.eq.bvfrequency}) into equations (\ref{ch2.eq.lr}) and (\ref{ch2.eq.lr2}) gives a quadratic equation for $\eta$,

\begin{equation}
\label{ch2.eq.quadratic}
(N_T^2+N_{\mu}^2)\eta^2+\left(2KN_{\mu}^2-\frac{r^2\omega_{\rm A}^4}{\Omega}\right)\eta-2Kr^2\omega_{\rm A}^4=0.
\end{equation}

In the limit $N_{\mu}^2\gg N_T^2$ and $K\ll\eta$ we recover equation~(1) of \citet{Maeder04} and in the limit $N_{\mu}^2\ll N_T^2$ and $K\gg\eta$ we recover their equation~(2). In most cases we find that $K\gg\eta$ and $N_T^2\gg N_{\mu}^2$ in which case we get

\begin{equation}
\label{ch2.eq.etaapprox}
\eta\approx r^2\Omega\left(\frac{\omega_{\rm A}}{\Omega}\right)^2\left(\frac{\Omega}{N}\right)^{1/2}\left(\frac{K}{r^2 N_T}\right)^{1/2}.
\end{equation}

\noindent In equation~(\ref{ch2.eq.quadratic}) we make the substitution $\eta=C_{\rm m}\eta'$ where $C_{\rm m}$ is a calibration constant which we expect to be of order unity. The chemical composition of the star evolves in radiative zones according to the equation

\begin{equation}
\label{ch2.eq.magchemevolution}
\frac{\partial X_{i}}{\partial t}=\frac{1}{r^2}\frac{\partial}{\partial r}\left({\rm Pr_c}D_{\rm tot}r^2\frac{\partial X_{i}}{\partial r}\right),
\end{equation}

\noindent where ${\rm Pr_c}$ is the chemical Prandtl number and $X_{i}$ is the mass fraction of element ${i}$. Similarly we take the magnetic diffusivity to be $\eta={\rm Pr_m}\,D_{\rm mag}$ where ${\rm Pr_m}$ is the magnetic Prandtl number. We look at the effect of varying these two parameters in section~\ref{ch5.sec.calibration} but we expect the magnetic Prandtl number to be of order unity \citep{Yousef03}.

\subsection{Dynamo model}

We describe the dynamo generation parameter by taking $\alpha=\gamma r/\tau_{\rm a}$ where $\gamma$ is an efficiency parameter and $\tau_{\rm a}$ is the amplification time scale of the field. Following \citet{Maeder04} we take $\tau_{\rm a}=\frac{N}{\omega_{\rm A}\Omega q}$ where $q=\diffb{\log\Omega}{\log r}$. Combining these our dynamo efficiency is given by

\begin{equation}
\label{ch2.eq.alpha}
\alpha=\gamma\frac{r\omega_{\rm A}\Omega q}{N}.
\end{equation}

\subsection{Magnetic braking}
\label{ch2.sec.braking}

Strongly magnetic intermediate--mass stars typically have rotation rates much slower than other stars in their parent population \citep{Mathys04}. If the Alfv\'{e}n radius, the radius at which the magnetic energy density is the same as the kinetic energy density in the stellar wind, is larger than the stellar radius then magnetic braking allows additional angular momentum to be carried away by the stellar wind. Consider equation~(\ref{ch2.eq.magamtransport}). Writing $\int_0^m{\rm d}m=\int_0^R4\pi r^2 \rho {\rm d}r$ we obtain the boundary condition for angular momentum loss from the surface

\begin{equation}
\difft{H_{\rm tot}}{t}=4\pi R^4\rho D_{\rm tot}\left(\diffb{\Omega}{r}\right)_{R}
\end{equation}

\noindent where $\difft{H_{\rm tot}}{t}$ is the total rate of angular momentum loss from the star and is given by

\begin{equation}
\difft{H_{\rm tot}}{t}=R_A^2\Omega\dot{M}=\sigma^2J_{\rm surf}.
\end{equation}

\noindent The Alfv\'{e}n radius is $R_A$, $\sigma=\frac{R_A}{R}$ and $J_{\rm surf}$ is the specific angular momentum at the surface of the star. Following the analysis of \citet{ud-Doula02} we can calculate the magnetic efficiency

\begin{equation}
\label{ch2.eq.eta}
\phi(r)=\frac{B_*^2R^2}{\dot{M}v_{\infty}}\frac{(\frac{r}{R})^{-4}}{1-\frac{R}{r}},
\end{equation}

\noindent where $v_{\infty}=v_{\rm esc}=\sqrt{2g_{\rm eff}R}$ and $v_{\rm esc}$ is the escape velocity at the stellar surface. We have assumed that the external field is dipolar ($q=3$). The Alfv\'{e}n radius is typically taken where the dynamo efficiency equals unity. Rearranging equation~(\ref{ch2.eq.eta}), and setting $\phi=1$ and $\sigma=\frac{r}{R}=\frac{R_A}{R}$ at $r=R_A$ we find

\begin{equation}
\label{ch2.eq.sigma4}
\sigma^4-\sigma^3=\frac{B_*^2R^2}{\dot{M}v_{\rm esc}}.
\end{equation}

\noindent We assume $\sigma\gg 1$ so that

\begin{equation}
\label{ch2.eq.sigma2}
\sigma^2=\sqrt{\frac{B_*^2R^2}{\dot{M}v_{\rm esc}}}
\end{equation}

\noindent for the remainder of this dissertation. If $R_A<R$ then we take $\sigma=1$ so that, as a star loses mass, material carries away the specific angular momentum at the surface. When we approach this limit we should calculate $\sigma$ exactly from equation~(\ref{ch2.eq.sigma4}) but for now we assume that equation~(\ref{ch2.eq.sigma2}) remains valid. In section~\ref{ch5.sec.mvb} we typically find either strong fields where $\sigma\gg 1$ or very weak fields where we can safely take $\sigma=1$. So far we have been unable to produce a stable model for the mass--loss rates of \citet{Vink01} and so use the rate of \citet{Reimers75} in equation~(\ref{ch2.eq.sigma2}). For intermediate--mass stars on the main sequence this approximation is reasonably accurate.

\subsection{Free parameters}

Like most theories for stellar rotation and magnetic field evolution we have produced a closed model which depends on a number of free parameters. We look at typical physically motivated values for these parameters in section~\ref{ch5.sec.calibration} and also the effect of varying them. In total we have four free parameters. The parameter $C_{\rm m}$ affects the overall strength of the turbulent diffusivity. The magnetic and chemical Prandtl numbers, ${\rm Pr_m}$ and ${\rm Pr_c}$, describe how efficiently the turbulent diffusivity transports magnetic flux and chemical composition compared to angular momentum. And $\gamma$ affects the strength of the dynamo generation. Whilst ${\rm Pr_m}$ and $C_{\rm m}$ are both expected to be of order unity we have left them as free parameters for the moment to maintain generality.

\subsection{Numerical implementation}

The implementation of magnetic fields in the code is very similar to the implementation of angular momentum as described in section~\ref{ch2.sec.amdifference}. Before we proceed we consider the modification the angular momentum difference equation. With the inclusion of magnetic stress the equation is now

\begin{gather}
\label{ch2.eq.amdifferencemag}
\Delta m_{k}\diffb{J_{{k}}}{t}+\left(J_{{k+1}}-J_{{k}}\right)\frac{\dot{m}_{k+1/2}}{2}+\left(J_{{k}}-J_{{k-1}}\right)\frac{\dot{m}_{k-1/2}}{2}=\nonumber\\
\frac{4\pi}{5}\left(\left(\rho r^2 J U\right)_{k+1/2}-\left(\rho r^2 J U\right)_{k-1/2}\right)\nonumber
\\+16\pi^2\left(\left(D_{\rm KH}r^6\rho^2\Delta m^{-1}\right)_{k+1/2}\left(\Omega_{k+1}-\Omega_{k}\right)-\left(D_{\rm KH}r^6\rho^2\Delta m^{-1}\right)_{k-1/2}\left(\Omega_{k}-\Omega_{k-1}\right)\right)\nonumber\\
+16\pi^2\left(D_{\rm con}r^{4+n}\rho^2\Delta m^{-1}\right)_{k+1/2}\left(\left(r^{2-n}\Omega\right)_{k+1}-\left(r^{2-n}\Omega\right)_{k}\right)\nonumber\\
-16\pi^2\left(D_{\rm con}r^{4+n}\rho^2\Delta m^{-1}\right)_{k-1/2}\left(\left(r^{2-n}\Omega\right)_{k}-\left(r^{2-n}\Omega\right)_{k-1}\right)\nonumber\\ 
+\frac{3\pi}{16(rB_{\phi})_{k}}\left(\left(r^3B^2_{\phi}A\right)_{k+1/2}-\left(r^3B^2_{\phi}A\right)_{k-1/2}\right).
\end{gather}

With the inclusion of magnetic fields the code now solves for two additional second--order equations, making 16 in total. The difference equations for the evolution of the poloidal and toroidal field components are 

\begin{equation}
\label{ch2.eq.bpdifference}
\begin{split}
\Delta m_{k} \diffb{B_\phi}{t}\,=\,& (2\pi r^2\rho A_{k+1/2})\left(\Omega_{k+1}-\Omega_{k}\right)+(2\pi r^2 \rho A_{k-1/2})\left(\Omega_{k}-\Omega_{k-1}\right)+ \\
& 16\pi^2 r_{k}\left(\eta\rho^2\Delta m^{-1}\right)_{k+1/2}\left(\left(r^{3}B_{\phi}\right)_{k+1}-\left(r^{3}B_\phi\right)_{k}\right)-\\
& 16\pi^2 r_{k}\left(\eta\rho^2\Delta m^{-1}\right)_{k-1/2}\left(\left(r^{3}B_\phi\right)_{k}-\left(r^{3}B_\phi\right)_{k-1}\right)
\end{split}
\end{equation}

\noindent and

\begin{equation}
\label{ch2.eq.btdifference}
\begin{split}
\Delta m_{k} \diffb{A_\phi}{t}\,=\,& \frac{8\alpha_{k}}{3\pi}\Delta m_{k}B_{\phi,{k}}+ \\
& 16\pi^2\left(\eta\rho^2r^2\Delta m^{-1}\right)_{k+1/2}\left(\left(r^{2}A\right)_{k+1}-\left(r^{2}A\right)_{k}\right)-\\
& 16\pi^2\left(\eta\rho^2r^2\Delta m^{-1}\right)_{k-1/2}\left(\left(r^{2}A\right)_{k}-\left(r^{2}A\right)_{k-1}\right).
\end{split}
\end{equation}

\section{Simulating stellar populations}
\label{ch2.sec.starmaker}

Whilst creating models is extremely important for understanding how rotation and magnetic fields effect the evolution of stars, it is essential that we compare our results with observations. In order to do that we need to predict how a stellar population would look if stars obeyed our model. To create synthetic stellar populations we use the code {\sc starmaker}, described by \citet{Brott11}. It was originally designed to work with the evolutionary models of \citet{Brott11b}. We have adapted it for use with {\sc rose} stellar evolution models. 

Based on a grid of evolutionary models, {\sc starmaker} generates a population of millions of individual stars. The properties of each star are determined by generating random values for the initial mass, initial surface velocity and age. The initial mass is chosen so that the initial mass function, IMF, is $\xi(m){\rm d}m\propto m^{-2.35}{\rm d}m$, where $\xi(m){\rm d}m$ is the number of stars with mass between $m$ and $m+{\rm d}m$. The initial surface rotation velocity distribution is that of \citet{Dufton06} for Galactic B--type stars, a Gaussian function truncated at zero with mean $\mu = 175\,{\rm km\,s^{-1}}$ and standard deviation $\sigma=94\,{\rm km\,s^{-1}}$. The age is chosen from a uniform distribution that lies between pre--defined upper and lower bounds. The values of physical variables are then estimated from the stellar evolution models by interpolating between the four neighbouring models in initial mass and initial surface velocity space.

Models may be filtered depending on whether we would expect them to appear in an observed sample of stars. Stars must be sufficiently massive to produce enough light to pass the detection threshold of the telescope. As such, stars are excluded if their visual magnitude falls below a certain, cluster dependent, value. Stars that are too massive have luminosities so high that the surface chemical abundance cannot be resolved from stellar spectra. As such, stars hotter than 35,000$\,{\rm K}$ are excluded from the sample. Finally, stars with surface gravity  $g_{\rm eff}<10^{3.2}{\,\rm cm\,s^{-2}}$ are classed as giants and are also excluded. We can also apply randomised errors to the predictions to better reflect observational errors in the data. For example, the error in the nitrogen enrichment is approximately $\log_{10}[N/H]\approx 0.2$. These errors are chosen from a Gaussian distribution and applied to each star.

\begin{savequote}[60mm]
Remember that all models are wrong; the practical question is how wrong do they have to be to not be useful. (George Edward Pelham Box)
\end{savequote}

\chapter{Comparison of stellar rotation models}
\label{ch3}

\section{Introduction}

It has been known for many years that rapid rotation can cause significant changes in the evolution of stars \citep[e.g.][]{Kippenhahn70}. Not only does it cause a broadening of the main sequence in the Hertzsprung Russell diagram but it also produces enrichment of a number of different elements at the stellar surface \citep[e.g.][]{Hunter09, Frischknecht10}. With new large scale surveys, such as the VLT--FLAMES survey of massive stars, now reaching maturity, the data available for rotating stars are growing rapidly \citep[e.g.][]{Evans05,Evans06}. Any viable model of stellar rotation must be able to match the observed changes in surface chemical enrichment, temperature and luminosity of rotating stars. Whilst some useful conclusions about the internal structure of individual rotating stars can be derived from asteroseismology \citep[e.g.][]{Aerts03} it is still impractical to do this for large populations. The treatment of rotation and its induced chemical mixing in stars has changed dramatically over the past two decades. The model of \citet{Zahn92} has formed the basis for much of this work and many variations from the original have been used during this time to generate different predictions of stellar evolution with rotation \citep[e.g.][]{Talon97,Meynet00,Maeder03}. Alternative formalisms, such as that of \citet{Heger00}, treat the physical processes very differently. Particular emphasis is often placed on the treatment of meridional circulation. While those models based on that of \citet{Zahn92} treat meridional circulation as an advective process, \citet{Heger00} treat it as a diffusive process. This leads to a fundamentally different behaviour. This is just one feature of the model which may lead to significantly different results. Even so, both treatments are used and frequently quoted in the literature for their predictions of the effect of rotation on stellar evolution. 

One of the most poorly explored features of stellar rotation is the treatment of angular momentum transport within convective zones. Current 1D models treat convective zones as rotating solid bodies. This is not necessarily correct and there are strong reasons to explore alternatives such as uniform specific angular momentum \citep[e.g.][Lesaffre et al., in prep.]{Arnett09}. This potentially has dramatic consequences for stellar evolution because a star with uniform specific angular momentum through its convective core has much more angular momentum for a given surface rotation velocity. It also has a strong shear layer at the convective boundary that can drive additional transport of chemical elements.

Different models for stellar rotation have not been compared directly on a common numerical platform alongside otherwise identical input physics before. From the Cambridge stellar evolution code \citep{Eggleton71,Pols95} we have produced a code capable of modelling rotating stars in 1D under the shellular rotation hypothesis of \citet{Zahn92}. The code, {\sc rose} (Rotating Stellar Evolution), can easily be programmed to run with different physical models of stellar rotation and can model both radiative and convective zones under a range of different assumptions. This allows us to compare a variety of models for stellar rotation and determine what, if any, observable traits could be used to distinguish between them. We foresee two possibilities, either we can identify clear observational tests to eliminate certain models or the models show no testable difference in which case a simplified model could be formulated to provide the same results.

In section~\ref{ch3.sec.model} we outline the physical models implemented in {\sc rose}, in section~\ref{ch3.sec.results} we present a comparison of the evolutionary predictions for each model and in section~\ref{ch3.sec.conclusions} we present our conclusions.

\section{Models of stellar rotation}
\label{ch3.sec.model}

In section~\ref{ch2.sec.rotation} we introduced the basic physical ingredients for simulating the evolution of rotating stars using the code {\sc rose}. The structure equations are modified owing to the centrifugal force acting within rotating stars. Shear arises from processes such as circulation, mass loss and hydrostatic structural evolution. This causes Kelvin--Helmholtz instabilities that drive hydrodynamic turbulence. Because of the strong stratification of material in stars, horizontal turbulence along isobars is expected to be much stronger than vertical turbulence. As a result, horizontal variations in physical quantities are expected to be much smaller than vertical variations and we therefore approximate the star as retaining spherical symmetry according to the modifications to the structure equations described in section~\ref{ch2.sec.rotstructure}. This is the shellular hypothesis of \citet{Zahn92}.

\section{Angular momentum transport}
\label{ch3.sec.amtransport}

 As in section~\ref{ch2.sec.amtransport}, we model the evolution of the angular momentum distribution according to the equation

\begin{equation}
\label{ch3.eq.amtransport}
\diff{r^2\Omega}{t}=\frac{1}{5\rho r^2}\diff{\rho r^4\Omega U}{r}+\frac{1}{\rho r^2}\frac{\partial}{\partial r}\left(\rho {D_{\rm KH} r^4\frac{\partial\Omega}{\partial r}}\right)\\+\frac{1}{\rho r^2}\frac{\partial}{\partial r}\left(\rho {D_{\rm con} r^{(2+n)}\frac{\partial r^{(2-n)}\Omega}{\partial r}}\right),
\end{equation}

\noindent where $\Omega(r)$ is the angular velocity, $r$ is the radius from the centre of the star, $\rho$ is the density, $n$ determines the local power--law distribution of specific angular momentum in convective zones, $D_{\rm con}$ is the diffusion coefficient in convective zones and $D_{\rm KH}$ is the diffusion coefficient in radiative zones. The meridional circulation, $U$, is given by

\begin{equation}
U=C_0\frac{L}{m_{\rm eff}g_{\rm eff}}\frac{P}{C_P\rho T}\frac{1}{\nabla_{\rm{ad}}-\nabla+\nabla_{\mu}}\\\left(1-\frac{\epsilon}{\epsilon_m}-\frac{\Omega^2}{2\pi G\rho}\right)\left(\frac{4\Omega^2r^3}{3Gm}\right),
\end{equation}

\noindent where $L$ is the stellar luminosity, $m_{\rm eff}=m\left(1-\frac{\Omega^2}{2\pi G\rho}\right)$, $g_{\rm eff}$ is the effective gravity (c.f. equation \ref{ch2.eq.geff}), $P$ is the pressure, $C_P$ is the specific heat capacity at constant pressure, $T$ is the temperature, $G$ is the gravitational constant, $\epsilon=E_{\rm nuc}+E_{\rm grav}$, the total local energy emission $\epsilon_m=L/m$, $\nabla$ is the radiative temperature gradient, $\nabla_{\rm{ad}}$ is the adiabatic temperature gradient and $\nabla_{\mu}$ is the mean molecular weight gradient. It is similar to the formulation of \citet{Maeder98}. The variable $C_0$ is a calibration constant which is discussed further in section~\ref{ch3.sec.testmodels}. Meridional circulation is discussed in more detail in section~\ref{ch2.sec.circulation}.

\subsection{Test cases}
\label{ch3.sec.testmodels}

Since the shellular hypothesis of \citet{Zahn92} there have been many alternate prescriptions for stellar rotation \citep[e.g.][]{Zahn92, Meynet97, Talon97, Heger00, Maeder03}. Most of these are based on similar assumptions but vary in their implementation of the various diffusion coefficients for angular momentum and chemical transport because of rotation. 

{\sc rose} is able to simulate stars with a number of common models. The details of these models are summarized in Table \ref{ch3.table.models}. We examine the difference between these models in section~\ref{ch3.sec.results}. Unless otherwise stated, the metallicity is taken to be solar ($Z=0.02$). For each model we compute the stellar evolution for a range of masses between $3$\ms\ and $100$\ms\ and initial equatorial surface rotation velocities between $0$ and $600\,{\rm km\,s^{-1}}$, except where the initial surface rotation rate would be super--critical.

\begin{table}
\begin{center}
\label{ch3.table.models}
\caption[The diffusion coefficients used by each of the different cases examined in chapter~\ref{ch3}]{The diffusion coefficients used by each of the different cases examined in this chapter. For each case we give the value of $n$, the power--law profile for the distribution of specific angular momentum distribution in convective zones. In addition the table gives the source of the various diffusion coefficients.\\}

\begin{tabular}{ccccc}
\hline
Case & $n$ & $D_{\rm KH}$ & $D_{\rm h}$ & $D_{\rm con}$\\
\hline
1 & 2 & \citet{Talon97} & \citet{Maeder03} & $D_{\rm mlt}$\\
2 & 2 & \citet{Heger00} & N/A &  $D_{\rm mlt}$\\
3 & 2 & \citet{Zahn92} & \citet{Zahn92} & $D_{\rm mlt}$\\
4 & 0 & \citet{Talon97} & \citet{Maeder03} & $D_{\rm mlt}$\\
5 & 0 & \citet{Talon97} & \citet{Maeder03} & $10^{10}{\rm cm^2\,s^{-1}}$\\
6 & 0 & \citet{Talon97} & \citet{Maeder03} & $10^{14}{\rm cm^2\,s^{-1}}$\\
\hline
\end{tabular}
\end{center}
\end{table}

\subsubsection{Case 1:}
\label{ch3.sec.case1}

Here we use the formulation for $D_{\rm KH}$ of \citet{Talon97},
\begin{equation}
D_{\rm KH}=\frac{2 {\rm Ri}_{\rm c}\left(r\frac{d\Omega}{dr}\right)^2}{N_T^2/(K+D_{\rm h})+N_{\mu}^2/D_{\rm h}},
\end{equation}

\noindent where ${\rm Ri}_{\rm c}=(0.8836)^2/2$ is the critical Richardson number which we have taken to be the same as did \citet{Maeder03}.
\begin{equation}
N_T^2=-\frac{g_{\rm eff}}{H_P}\left(\frac{\partial\ln\rho}{\partial \ln T}\right)_{\!\!P,\mu}\left(\nabla_{\rm ad}-\nabla\right)
\end{equation}
\noindent and
\begin{equation}
N_{\mu}^2=\frac{g_{\rm eff}}{H_P}\left(\frac{\partial\ln\rho}{\partial\ln\mu}\right)_{\!\!P,T}\frac{d \ln\mu}{d \ln P}.
\end{equation}

\noindent Following \citet{Maeder03} we take
\begin{equation}
D_{\rm h}=0.134 r\left(r\Omega V\left(2V-\alpha U\right)\right)^{\frac{1}{3}}{,}
\end{equation} 

\noindent where
\begin{equation}
\alpha=\frac{1}{2}\diff{r^2\Omega}{r}.
\end{equation}

\noindent The differential equations derived by \citet{Zahn92} are fourth order in space. Our model differs in that third order derivatives and above cannot be reliably computed and must be ignored. The constant $C_0$ is included as a means of calibrating the model in light of this difference. Because our ultimate intention is to compare these models to data from the VLT--FLAMES survey of massive stars, we have chosen $C_0$ so that we reproduce the terminal--age main--sequence  (TAMS) nitrogen enrichment of a $40$\ms\ star initially rotating at $270{\,\rm km\,s^{-1}}$ and with Galactic composition given by \citet{Brott11}. This gives $C_0=0.003$. Whilst this is admittedly quite small, it is important to note that the non--linearity of the angular momentum transport equation means that a small change in the amount of diffusion corresponds to quite a large change in $C_0$.  We could have similarly adjusted the magnitude of $D_{\rm KH}$ in case~1 to give the desired degree of nitrogen enrichment. However, attempting to reduce the diffusion coefficient results in more shear which drives the diffusion back up so $D_{\rm KH}$ would need to be reduced by a large factor to give the desired effect. This would have the further consequence that the meridional circulation would have a much stronger effect on the system and produce stars with a very high degree of differential rotation between the core and the envelope. Hence we have chosen to adopt this method of calibration as the most physically reasonable. The diffusion of angular momentum in convective zones is determined by the characteristic eddy viscosity from mixing length theory such that $D_{\rm con}=D_{\rm mlt}=\frac{1}{3}v_{\rm mlt}l_{\rm mlt}$. This model takes $n=2$.

\subsubsection{Case 2:}

This is the model of \citet{Heger00}. In this case we set $U=0$ because circulation is treated as a diffusive process. The details of the various diffusion coefficients are extensive so we refer the reader to their original paper. With their notation the diffusion coefficients are
\begin{equation}
D_{\rm KH}=D_{\rm sem}+D_{\rm DSI}+D_{\rm SHI}+D_{\rm SSI}+D_{\rm ES}+D_{\rm GSF}
\end{equation}

\noindent and
\begin{equation}
D_{\rm eff}=(f_c-1)(D_{\rm DSI}+D_{\rm SHI}+D_{\rm SSI}+D_{\rm ES}+D_{\rm GSF}),
\end{equation}

\noindent where each of the $D_i$ corresponds to a different hydrodynamic instability. We note that our notation differs slightly from the original paper. The diffusion coefficients $\nu$ and $D$ used by \citet{Heger00} are equivalent to $D_{\rm KH}$ and $D_{\rm KH}+D_{\rm eff}$ respectively. \citet{Heger00} take $f_{\rm c}=1/30$ which is what we use here. The parameter $f_{\mu}$ used by \citet{Heger00} is taken to be zero. Mean molecular weight gradients play an important part in chemical mixing near the core however we have performed a number of test runs with $f_{\mu}=0.05$. Although there were some differences, they were not significant and could be largely masked by modifying the other free parameters associated with this case. The model differs from the original formulation of \citet{Heger00} in that we are unable to use {\sc stars} to consistently compute non--local quantities such as the spatial extent of instability regions used in some of the expressions for the various diffusion coefficients. We do not expect this to have a significant effect on the results because $D_{\rm ES}$ dominates the total diffusion coefficient and its limiting length scale is the pressure scale height rather than the extent of the unstable region.

This model is calibrated by modifying the dominant diffusion coefficient for transport owing to meridional circulation, $D_{\rm ES}$, by a constant of order unity to give the same TAMS nitrogen enrichment as case~1 for a $20$\ms\ star with initial surface angular velocity of $300\,{\rm km\,s^{-1}}$. This model has $n=2$ and $D_{\rm con}=D_{\rm mlt}$.

\subsubsection{Case 3:}

This is a reproduction of the original model of \cite{Zahn92} and is included as a baseline to highlight the differences in predictions of stellar rotation from the original model. In this model
\begin{equation}
D_{\rm KH}=\nu_{\rm v,h}+\nu_{\rm v,v},
\end{equation}

\noindent where
\begin{equation}
\nu_{\rm v,v}=\frac{8 C_1 {\rm Ri}_{\rm c}}{45}K\left(\frac{r}{N_T^2}\frac{d\Omega}{dr}\right)^2,
\end{equation}
\begin{equation}
\nu_{\rm v,h}=\frac{1}{10}\left(\frac{\Omega}{N_T^2}\right)\left(\frac{K}{D_{\rm h}}\right)^{\frac{1}{2}}r|2V-\alpha U|,
\end{equation}

\noindent $K$ is the thermal diffusivity and $\alpha$ is the same as in case~1. In this model
\begin{equation}
D_{\rm h}=C_2r|2V-\alpha U|
\end{equation}

\noindent and $C_1$ and $C_2$ are constants used for calibration. We constrain $C_1$ and $C_2$ by matching as closely as possible the TAMS nitrogen enrichment and luminosity of a $20$\ms\ case--1 star with initial surface rotation of $300\,{\rm km\,s^{-1}}$. We find that $C_1=0.019$ which is surprisingly small. This is because this model does not take into account mean molecular weight gradients and this leads to far more mixing between the convective core and the radiative envelope than in case~1. The TAMS luminosity is always greater than case~1 and so we minimize it with respect to $C_2$ so $C_2\to\infty$. This is realised by setting $D_{\rm eff}=0$ and is the case when the horizontal diffusion completely dominates over the meridional circulation and so is consistent with our assumption of shellular rotation. The constant $C_0$ is the same as in section~\ref{ch3.sec.case1}.

The main objection to this model, apart from the exclusion of mean molecular weight gradients, is that in the formulation of $D_{\rm h}$ we assume that, if the horizontal variation in the angular velocity along isobars takes the form $\tilde{\Omega}=\Omega_2(r)P_2(\cos\theta)$, then $\Omega_2(r)/\Omega(r)$ is constant and this is not physically justified. Again we set $n=2$ and $D_{\rm con}=D_{\rm mlt}$.

\subsubsection{Cases 4, 5 and 6:}

For these cases we use the same $D_{\rm KH}$ and $D_{\rm h}$ as in case~1 but we set $n=0$ to produce uniform specific angular momentum through the convective zones and test the effect of varying the convective diffusion coefficient $D_{\rm con}$,
\begin{equation}
D_{\rm con}=
\begin{cases}
D_{\rm mlt}& {\rm case\; 4,} \\10^{10}\,{\rm cm^2\,s^{-1}} & {\rm case\; 5,}\\10^{14}\,{\rm cm^2\,s^{-1}} & {\rm case\; 6.}
\end{cases}
\end{equation}

\subsubsection{Metallicity variation}

We have also calculated the evolution in cases 1, 2 and 3 for the same masses and velocities but with $Z=0.001$. We shall represent these cases with a superscript {\it Z} (case~${\rm 1}^Z$ is the low metallicity analogue of case~1 etc.).

\section{Results}
\label{ch3.sec.results}

Whilst there are many potential observables which may be used to distinguish between different models, it is important to choose the ones that are most easily compared with observational data. From our stellar evolution calculations we find a number of important differences between the test cases.

\subsection{Evolution of a ${\bf 20\,M}_{\odot}$ star in cases 1, 2 and 3}

\begin{figure}
\begin{center}
\includegraphics[width=0.99\textwidth]{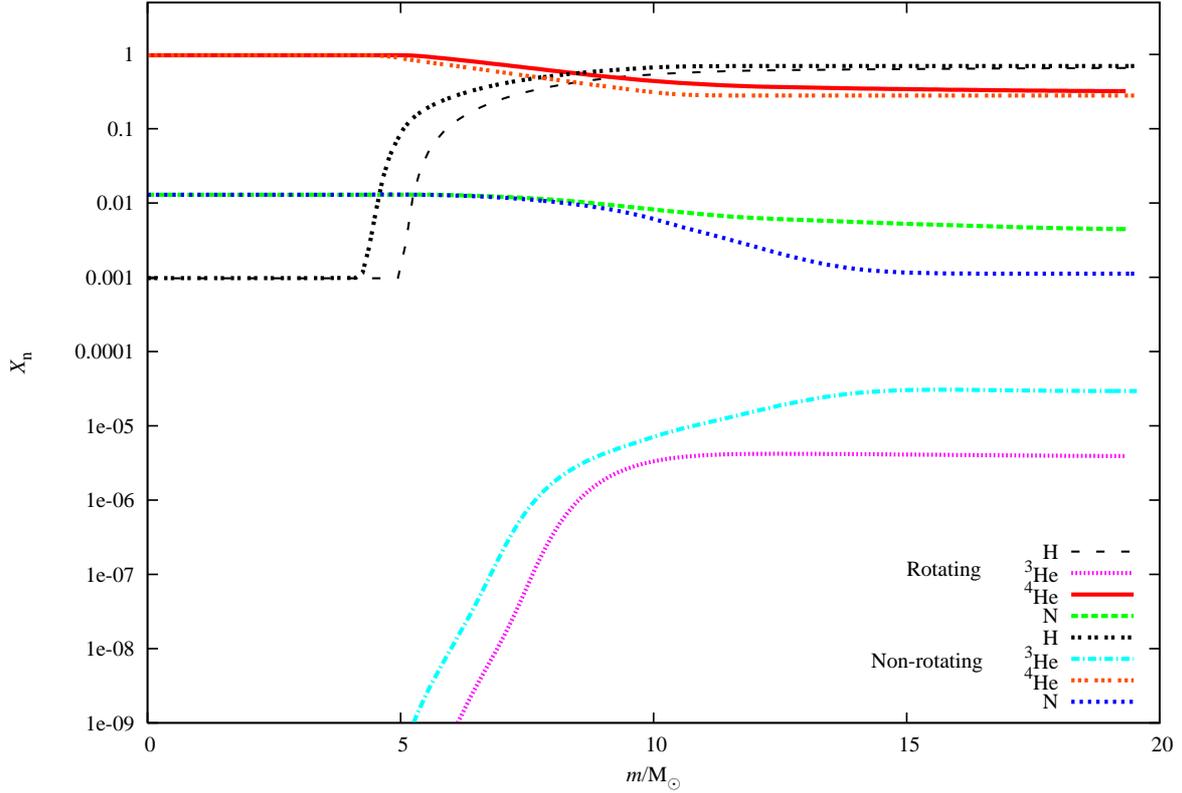}
\end{center}
\caption[Terminal age main sequence composition of a $20$\ms\ rotating star]{Terminal--age main sequence composition of  $20$\ms\ case--1 stars. The solid lines are for a star initially rotating at $300\,{\rm km\,s^{-1}}$ and the dashed lines are for a non--rotating star. Both stars have central hydrogen abundance $X_{\rm H}=10^{-3}$. Note that rotational mixing results in a larger core and mixing of helium and nitrogen throughout the radiative envelope.}
\label{ch3.fig.abundances}
\end{figure}

\begin{figure}
\begin{center}
\includegraphics[width=0.99\textwidth]{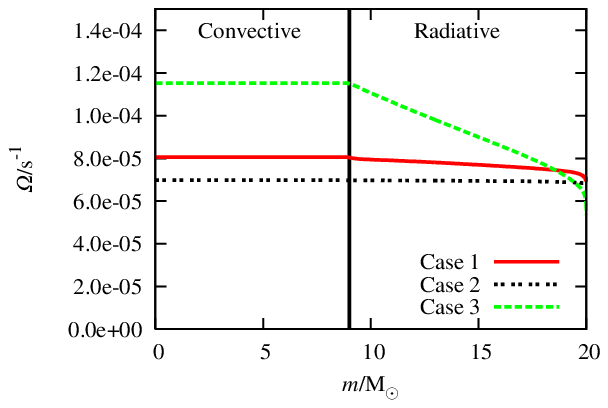} \\\includegraphics[width=0.99\textwidth]{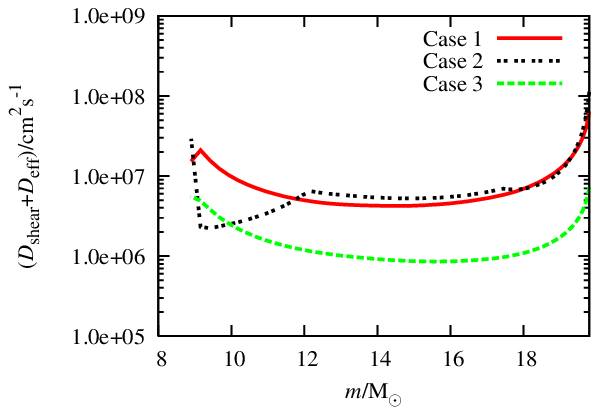}
\end{center}
\caption[Zero age main sequence properties for the angular momentum distribution in a $20$\ms\ rotating star]{Zero--age main sequence properties of a $20$\ms\ star initially rotating at $300\,{\rm km\,s^{-1}}$. The top panel shows the angular velocity through the star and the bottom panel shows the diffusion coefficient for chemical transport through the radiative envelope}
\label{ch3.fig.diff}
\end{figure}

First it is helpful to briefly examine the internal evolution that occurs. We consider here the main--sequence evolution of a $20$\ms\ star with initial surface rotation of $300\,{\rm km\,s^{-1}}$ for cases 1, 2 and 3. Although the centrifugal force causes some change in a star's structure, its evolution is most strongly affected by changes in the chemical composition. Fig.~\ref{ch3.fig.abundances} shows how the composition of a rotating $20$\ms\ case--1 star differs from a non--rotating $20$\ms\ star at the end of the main sequence. The difference in the rotation--induced mixing produces the variation in results between the various cases. Fig.~\ref{ch3.fig.diff} shows the angular velocity profile and the diffusion coefficient for vertical angular momentum transport in radiative zones predicted by each of cases 1, 2 and 3 at the zero--age main sequence (ZAMS). Note that, even though the stars have the same surface rotation, their core rotation and hence total angular momentum content can vary significantly between models

\begin{figure}
\begin{center}
\includegraphics[width=0.99\textwidth]{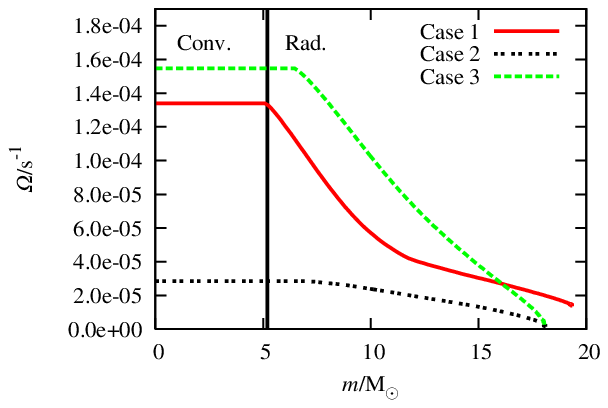} \\\includegraphics[width=0.99\textwidth]{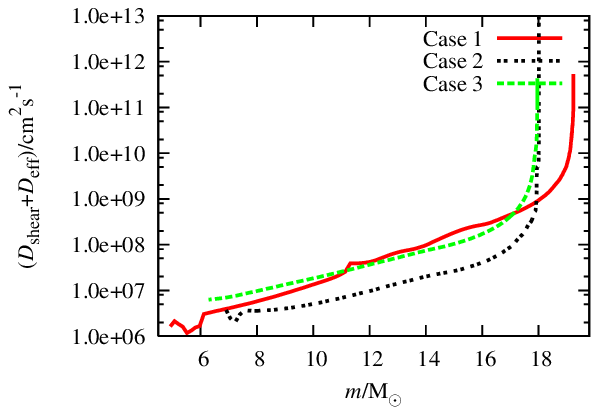}
\end{center}
\caption[Terminal age main sequence properties of the angular momentum distribution in a $20$\ms\ rotating star]{Terminal--age main sequence properties of a $20$\ms\ star initially rotating at $300\,{\rm km\,s^{-1}}$. The top panel shows the angular velocity through the star and the bottom panel shows the diffusion coefficient for chemical transport through the radiative envelope. The vertical line of the top panel separates the convective and radiative zones of the case--1 star.}
\label{ch3.fig.diff2}
\end{figure}

Despite their similar treatments, cases~1 and 3 have quite different initial rotation profiles. This is largely because of our choice of calibration. Because in case~3 we ignore mean molecular weight gradients, the overall efficiency of mixing must be reduced to match our calibration criterion (section~\ref{ch3.sec.testmodels}). This means that shear diffusion is much weaker relative to advection and so a profile with more differential rotation results. Had we chosen to calibrate the mixing by reducing $C_0$ instead of $C_1$ we would have found the opposite effect. This highlights one possible pitfall of including multiple free parameters within a given system.

 Case~2 is dominated by diffusion because of the diffusive treatment of meridional circulation. Recall that meridional circulation is treated advectively in cases 1 and 3 so is not included in the diffusion coefficient. In fact it is responsible for production of the shear at the ZAMS despite turbulence trying to restore solid body rotation. Because there is no perturbation to the rotation at the start of the main sequence, the star in case~2 rotates as a solid body. As the star evolves and mass is lost from its surface the solid body rotation is disturbed. Even so, because of the strong diffusion, case~2 stars never deviate far from solid body rotation as can be seen in Fig.~\ref{ch3.fig.diff2}. We note that case--1 stars reach the end of the main sequence with a higher mass than those in case~2 or case~3. This is because case--2 and case--3 stars have a longer main--sequence life owing to more efficient mixing at the core--envelope boundary. This allows hydrogen to be mixed into the core more rapidly than in case~1. This also leads to larger core masses in cases~2 and~3 compared to case~1.

We see in Fig.~\ref{ch3.fig.diff} that the predicted diffusion coefficients for cases~1 and~2 are similar throughout most of the envelope. By the TAMS, the diffusion predicted by case~2 is significantly lower than the other two cases. This is possibly because rising shear, owing to rapid hydrostatic evolution at the TAMS, causes the diffusion in cases~1 and 3 to increase while in case~2 diffusion is dominated instead by the circulation. Unsurprisingly, the diffusion coefficient in case~3 is similar in form to case~1 but significantly smaller, a result of our choice of $C_1$. However, we note that the diffusion coefficient predicted by the two cases is very similar by the end of the main sequence. Also, the diffusion coefficient at the core--envelope boundary at the end of the main sequence in case~1 is around an order of magnitude lower than in case~3. This is partially because of the core of the case--3 star rotating faster than in case~1 but is mostly because of the inclusion of the mean molecular weight gradient in the formulation for case~1. \citet{Frischknecht10}, who use a model very similar to our case~1, predict a much greater decline in the mixing near the core but we have been unable to reproduce this. It is likely that the difference is a result of the inhibiting effect of the mean molecular weight gradient on the rotational mixing. Whilst it is included in our models, the results of \citet{Frischknecht10} suggest that, towards the end of the main--sequence, its effect covers a much larger proportion of the radiative envelope than in our models.

\subsection{Effect on Hertzsprung--Russell diagram}

\begin{figure}
\begin{center}
\includegraphics[width=0.99\textwidth]{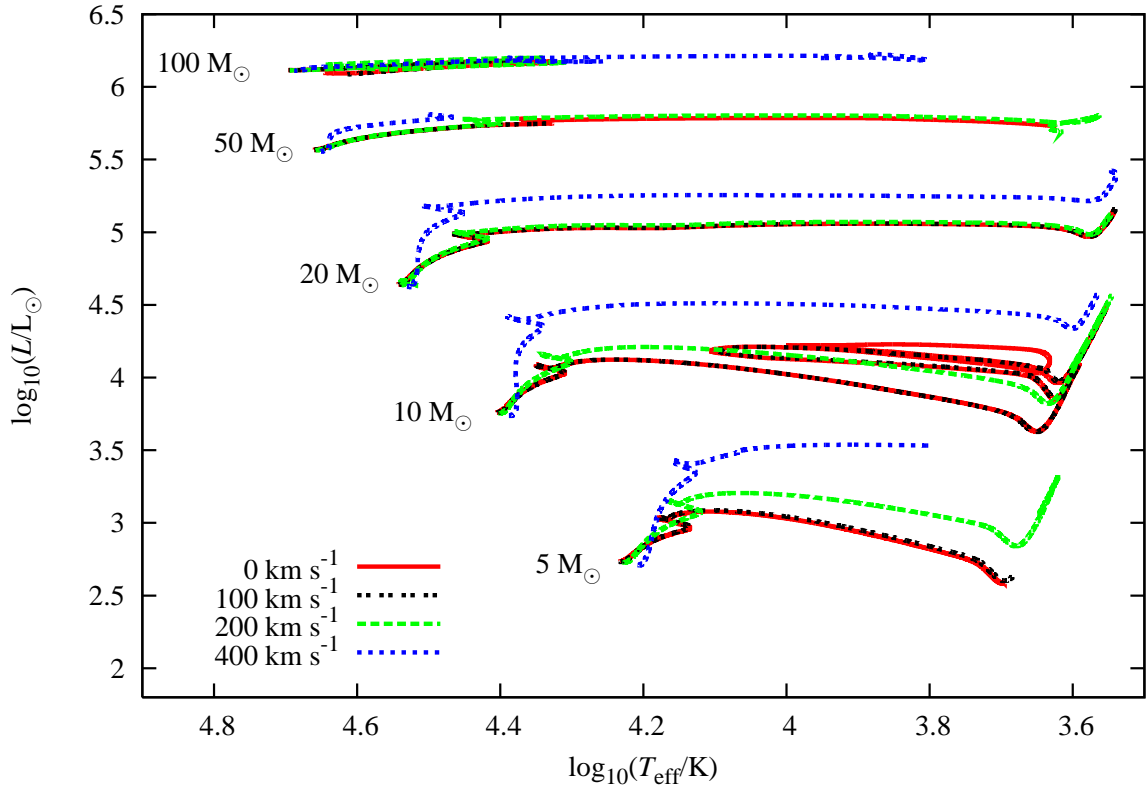}
\end{center}
\caption[Hertzsprung--Russell diagram for a range of stellar masses and rotation rates]{Stellar evolution tracks for stars between $5$\ms\ and $100$\ms\ calculated with {\sc rose} for case~1 with surface velocities of $0$ to $400\,{\rm km\,s^{-1}}$. The tracks for cases 4, 5 and 6, where the angular momentum transport in convective zones is varied, are almost indistinguishable from each other but produce more luminous stars than in case~1. There is significant difference between the predictions made for cases 2 and 3, where the angular momentum transport in radiative zones is varied.}
\label{ch3.fig.HR}
\end{figure}

As expected, the effects of rotation on the structure and chemical evolution of each star are significant across the HR diagram. We have plotted the case--1 models in Fig \ref{ch3.fig.HR}. Centrifugal forces cause the star to expand making it dimmer and redder. However, because of the additional chemical mixing, more hydrogen is made available during the main sequence so, by the TAMS, rotating stars are generally more luminous than similar mass non--rotating stars. This effect becomes more pronounced at higher rotation rates and is most apparent for stars with masses up to $20$\ms.

\begin{figure}
\begin{center}
\includegraphics[width=0.99\textwidth]{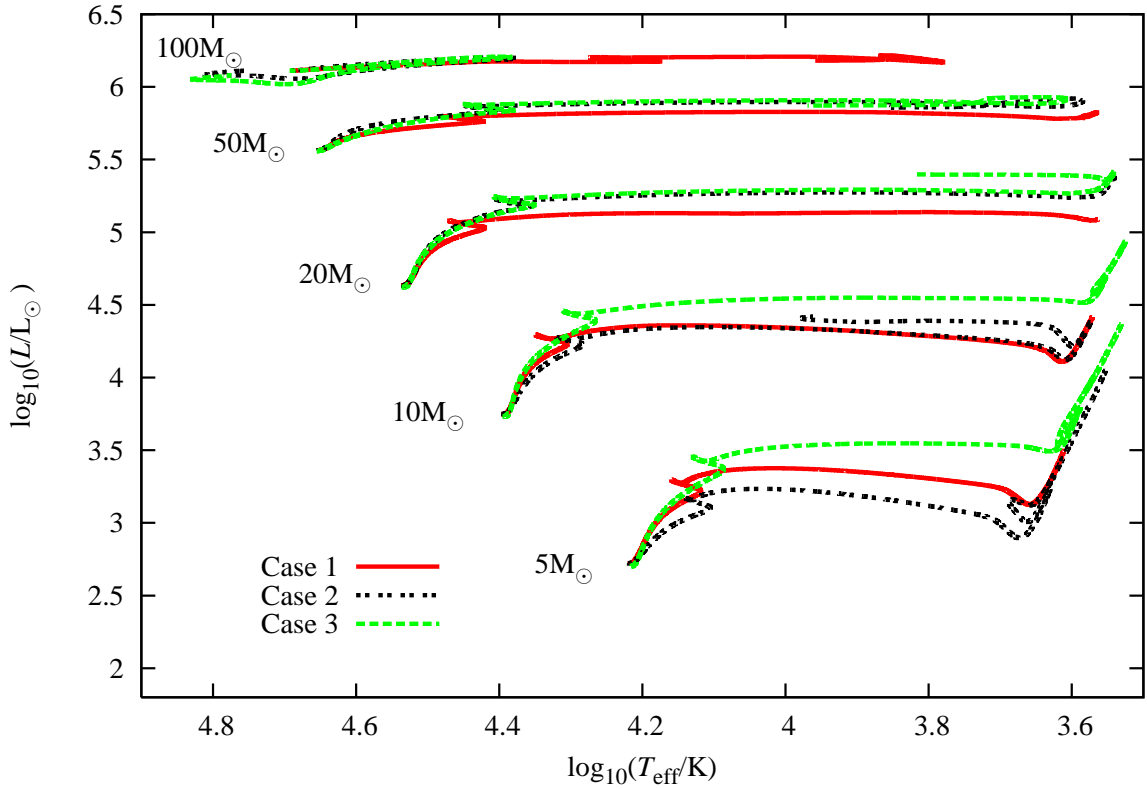}
\end{center} 
\caption[Hertzsprung--Russell diagram for various masses of star rotating at $300\,{\rm km\,s^{-1}}$ for a number of different models]{Stellar evolution tracks for stars between $5$\ms\ and $100$\ms\ with initial surface rotation of $300\,\rm{km}\,\rm{s}^{-1}$ for cases 1, 2 and 3. The model of \citet{Heger00} predicts a greater enhancement and higher surface temperatures compared to that based on \citet{Maeder03} for stars more massive than $10$\ms.}
\label{ch3.fig.HRrad1}
\end{figure}

There are clear differences between the predicted evolution of rotating stars in each of the three cases. Fig.~\ref{ch3.fig.HRrad1} shows the evolution of five different stellar masses in the HR diagram for cases 1, 2 and 3 with an initial surface rotation of $300\,\rm{km}\,\rm{s}^{-1}$. These cases are the three different models of rotational mixing in radiative zones at solar metallicity. Rotating case--1 stars appear to be the most luminous at low masses but least luminous at high masses. Case--2 stars are also consistently cooler than their case--1 and case--3 equivalents except below $10$\ms. This is because the strength of rotation--induced mixing increases rapidly with mass in case--2 stars unlike in cases~1 and~3, where the difference is more modest.

Although there is apparently a large difference between the three models in the HR diagram, to distinguish between them from stellar populations requires either a very large sample or accurate independent measurements of stellar masses and rotation velocities. Both of these are very challenging but, with the advent of large scale surveys, the former is quickly becoming a possibility.

\subsection{Nitrogen enrichment}

Currently the key test for any model of stellar rotation is how well it reproduces the spread of data in a Hunter diagram \citep{Hunter09}. Hunter plotted nitrogen enrichment against surface rotation. Large scale surveys, such as the VLT--FLAMES Tarantula survey \citep{Evans10}, will greatly increase the data available for surface rotation velocities and surface abundances over the coming decade. Thus, if different models can be distinguished in a Hunter diagram, this would form a key test for stellar rotation models.

\begin{figure}
\centering
\includegraphics[width=0.99\textwidth]{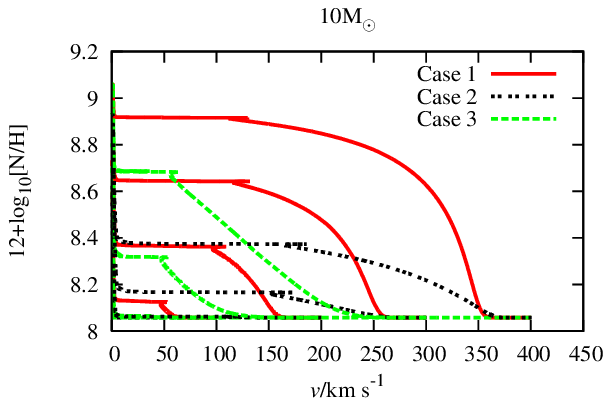} \\\includegraphics[width=0.99\textwidth]{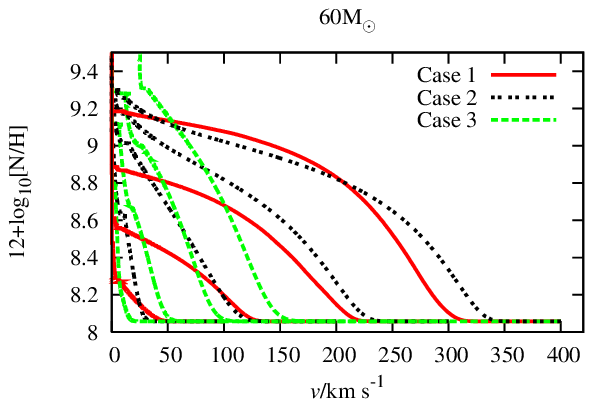}
\caption[Hunter diagrams for different stellar masses for a number of different physical models]{Nitrogen enrichment (by number of nuclei) variation with initial surface rotation for cases 1, 2 and 3. Stars start on the ZAMS with low nitrogen abundances and high velocities and evolve to higher abundances and lower velocities during the main sequence. The top panel is for $10$\ms\ stars and the bottom panel is for $60$\ms\ stars. Note the different scales for each panel. As expected the enrichment is much greater for more massive stars. Case--3 stars spin down more before any enrichment occurs. They give greater enrichment than either of the other two models at high masses but significantly less than case~1 for low masses. Cases~1 and 2 enrich to a similar degree for the high--mass stars but case~2 exhibits significantly less enrichment for low--mass stars than case~1.}
\label{ch3.fig.Nsol}
\end{figure}

In Fig.~\ref{ch3.fig.Nsol} we plot our theoretical Hunter diagram for $10$\ms\ and $60$\ms\ stars at different initial surface rotation velocities with the different radiative--zone models at solar metallicity. Each star begins at the bottom of the plot with the same nitrogen abundance but different initial surface velocities. There is an initial period where the star spins down before any enrichment has occurred. During this time the star moves straight to the left of the plot. Eventually the surface nitrogen abundance begins to increase because of rotation induced mixing from the burning region. At the same time the star continues to spin down because of mass loss and structural evolution. The net effect is that the star moves towards the upper left--hand region of the plot. At the end of the main sequence the star expands and rapidly spins down. This appears as a near--horizontal line as the star moves rapidly towards the left--hand edge of the plot. Some further enrichment may occur during the giant phase.

The evolution of the surface abundances is very model dependent. Case~3, which is based on the early model of \citet{Zahn92}, gives more nitrogen enrichment than case~1 at high masses ($M\geq 60$\ms) but significantly less at low masses ($M\leq 10$\ms). Most notably though, the case--3 stars spin down to a far greater degree before enrichment occurs. We attribute this to the neglect of mean molecular weight gradients. Mixing near the core is more efficient in case~3 but, owing to the overall calibration, is weaker near the surface.

For case~2 the amount of mixing in lower--mass stars is much less than for both cases 1 and 3. By comparison the enrichment of case--2 $60$\ms\ stars is greater than in the other two cases, particularly for slower rotators. This mass dependent behaviour of the rotating models could provide important clues to distinguish between the models.

At solar metallicity, owing to the enhanced mass loss, massive stars spin down before the end of the main sequence so, in this case, we would not observe the absence of the moderately rotating, highly enriched stars seen at low metallicity \citep{Hunter09}. Observations of multiple clusters at different ages at this metallicity would be a good test for rotating stellar models because the evolution across the Hunter diagram is significantly different in each case~even when they are calibrated to give the same level of enrichment at the end of the main sequence.

\subsection{Helium--3 enrichment}

Apart from the enrichment of nitrogen, rotation can have a large effect on the evolution of other elements. Changes in the carbon and oxygen abundances in rotating stars predicted from models have been considered but the accuracy of the data prohibits any strong conclusions from being made. \citet{Frischknecht10} discuss the effect rotation may have on the surface abundances of light elements. We consider here the evolution of the surface abundance of $^3$He. A similar analysis could be performed for other elements such as lithium and boron. The changes in the overall abundance of $^3$He because of rotation could partially explain the discrepancy between the predicted abundances produced by stars and the lack of enrichment of the inter--stellar medium compared to levels predicted by primordial nucleosynthesis \citep{Dearborn86,Hata95,Dearborn96}. This has been explained in the past by thermohaline mixing \citep{Stancliffe10} but, as our results show, the surface $^3$He abundance is strongly affected by rotation and so it is likely to make at least some contribution to this effect. We leave the issue of whether the total production increases or decreases over the stellar lifetime to future work. In either case, the evolution of helium--3 with respect to the surface rotation is very different between alternative models and so, as for nitrogen enrichment, this would form a useful comparison of stellar rotation models. Unlike nitrogen, helium--3 enrichment is stronger at low masses and so provides a greater number of candidate stars for observations.

\begin{figure}
\centering
\includegraphics[width=0.99\textwidth]{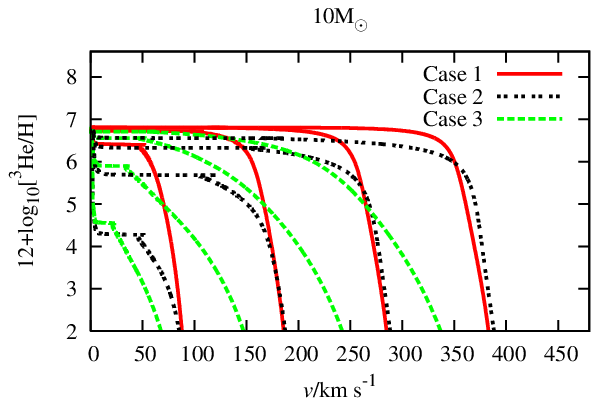} \\\includegraphics[width=0.99\textwidth]{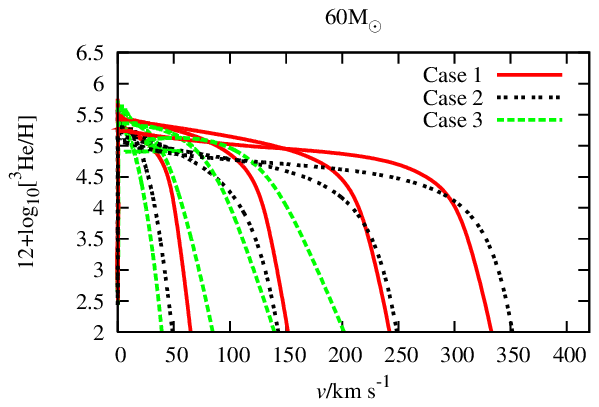}
\caption[Helium--3 enrichment variation with surface rotation for $10$\ms\ and $60$\ms\ stars for a number of different physical models]{Helium--3 enrichment variation with initial surface rotation for cases 1, 2 and 3. The top panel is for $10$\ms\ stars and the bottom panel is for $60$\ms\ stars. At high masses all three cases show a similar degree of TAMS enrichment though there is some variation in the evolutionary paths in each case. Case--3 stars spin down more before enrichment occurs and so lie to the left of the other two cases. At low masses, both case--2 and case--3 stars are less enriched at the TAMS than case--1 stars. This is especially true for slow rotators.}
\label{ch3.fig.He3sol}
\end{figure}

Fig.~\ref{ch3.fig.He3sol} shows the helium--3 enrichment for $10$\ms\ and $60$\ms\ stars of varying initial surface velocities for each of the different radiative zone models at solar metallicity. As for nitrogen, all three cases show comparable amounts of enrichment in high--mass stars. The amount of enrichment at the end of the main sequence is the same in each case~but case--3 stars are slightly more enriched at all rotation rates. Case--3 stars spin down much more before enrichment occurs so the paths for these stars lie to the left of case--1 and case--2 stars but the amount of enrichment at the end of the main sequence is comparable to, though slightly higher than, the other two cases.

The difference between the test cases is far greater at lower masses. Both case--2 and case--3 stars show substantially less enrichment during the main sequence especially for slow rotators and case--3 stars have much slower surface rotation at the end of the main sequence than in the other two cases. Both of these contribute to very different evolution which should be distinguishable observationally. Indeed, whilst the models may produce similar results for full population synthesis calculations, it has been found that there is often far less agreement when different mass ranges are considered separately \citep{Brott11b}.

\subsection{Metallicity dependence}

In order to compare stellar rotation models with data it is important to distinguish which effects are observable at different metallicities. Low--metallicity stars are particularly useful because they have significantly lower mass--loss rates \citep{Vink01}. For stars of metallicity $Z=0.001$ the mass--loss rate is roughly ten times lower than at solar metallicity. This allows us to rule out effects on the models produced by our prescription for mass loss.

\begin{figure}
\centering
\includegraphics[width=0.99\textwidth]{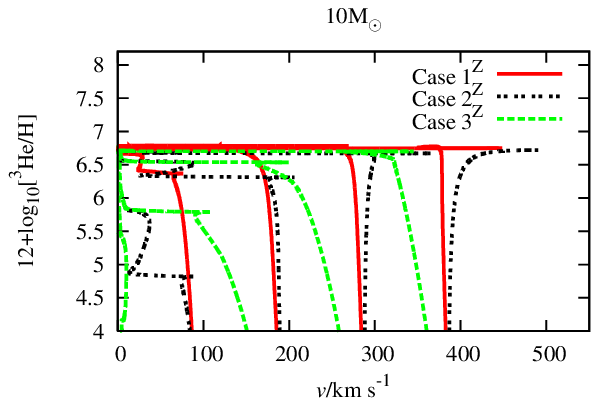}\\\includegraphics[width=0.99\textwidth]{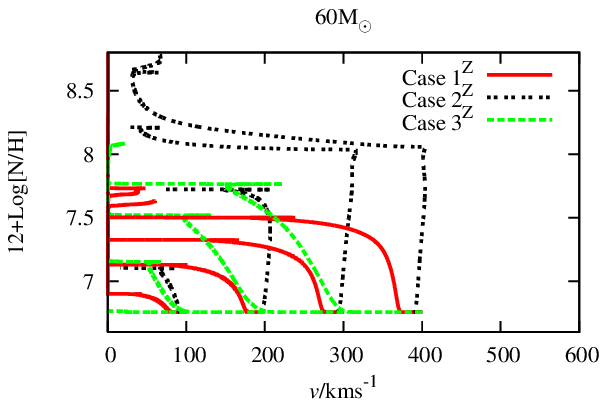}
\caption[Hunter diagrams at low metallicity for different stellar masses for a number of different physical models]{Chemical enrichment variation at low metallicity. The top panel shows the enrichment of helium--3 and the bottom panel shows the enrichment of nitrogen. The upper plot is for $10$\ms\ stars and the lower plot is for $60$\ms\ stars. Case~2$^Z$ now shows much higher nitrogen enrichment than case~1$^Z$, the opposite to what we found in the solar metallicity cases. The enrichment of helium--3 in case~2$^Z$ stars is still considerably less than in case~1$^Z$.}
\label{ch3.fig.Nz1}
\end{figure}

The low--metallicity cases show similar distinctions in the HR diagram to those at solar metallicity, although high--mass, rapidly rotating, case--2 stars are sufficiently well mixed to undergo quasi--homogeneous evolution. There are also significant differences in the nitrogen enrichment between the different models. Because the mass--loss rate is lower in low--metallicity stars they retain their surface rotation for much longer so the main sequence appears much more vertical in a Hunter diagram. Unlike the solar metallicity cases, case~2$^Z$ exhibits significantly more mixing than case~1$^Z$ particularly for slow and moderately rotating stars (Fig.~\ref{ch3.fig.Nz1}). This is the complete opposite of the results at solar metallicity and highlights the importance of testing different stellar environments to discover clues to distinguish between different models. In contrast, the enrichment of helium--3 in case~2$^Z$ stars is less than in case~1$^Z$.

\subsection{Surface gravity cut--off}

\begin{figure}
\centering
\includegraphics[width=0.99\textwidth]{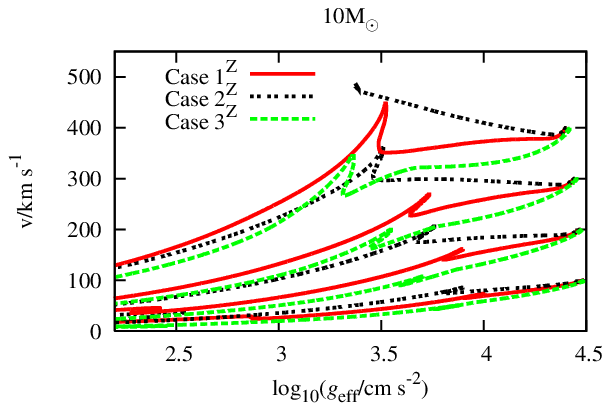}\\\includegraphics[width=0.99\textwidth]{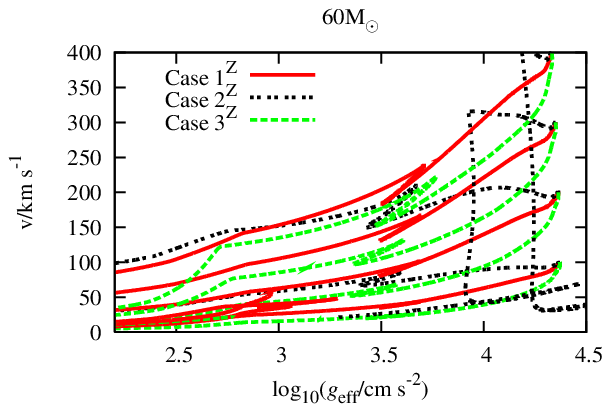}
\caption[Effective gravity variation with surface velocity for $10$\ms\ and $60$\ms\ low--metallicity stars for a number of different physical models]{The effective gravity variation with surface velocity for $10$\ms\ stars (top panel) and $60$\ms\ stars (bottom panel) for cases 1$^Z$, 2$^Z$ and 3$^Z$. The end of the main sequence occurs at the right--most cusp of each path. Although the surface gravity is similar for all the models and all initial rotation rates, the TAMS surface velocities are very different between the different cases. Exceptions to this are the rapidly--rotating, high--mass, case--2$^Z$ stars which undergo almost homogeneous evolution.}
\label{ch3.fig.gravz}
\end{figure}

As a consequence of increasing stellar radius and angular momentum conservation all models for stellar rotation predict a rapid decay in the surface rotation velocity after the end of the main sequence. Observations suggest that, even for rotating stars, there is a sharp cut--off in the effective gravity at $\log_{10}(g/{\rm cm^2s^{-1}})\approx 3.2$ when a star leaves the main sequence and moves over to the giant branch \citep{Brott11}. This effect depends on stars reaching the TAMS without spinning down too much during the main sequence. Therefore it is more easily seen at lower metallicities where the mass--loss rate is lower. The observed value for the TAMS gravity can be enforced in rotating models by including a degree of overshooting. However this simply introduces an additional free parameter into the models. In Fig.~\ref{ch3.fig.gravz} we show the different cut--offs in the effective gravity predicted by cases 1$^Z$, 2$^Z$ and 3$^Z$. The end of the main sequence is indicated by a distinct cusp in the path of the star in the range $3<\log_{10}(g/{\rm cm^2s^{-1}})< 4$. Note that the expected number of stars with significant rotation after the main--sequence cut--off is low because the evolution to a very slowly rotating giant is extremely rapid compared to the rest of the main sequence.

For higher--mass stars there is a sharp cut--off in the surface gravity at the end of the main sequence that is the same regardless of the model used although the TAMS surface rotation is somewhat higher in case~2. The mixing in low--metallicity, high--mass, rapidly rotating, case--2 stars is very efficient and so they evolve almost homogeneously and thus they appear differently in the plot. For lower--mass stars the change in surface gravity at the end of the main sequence is still clear but varies by around an order of magnitude across all rotation rates and test cases. Case--2 and case--3 stars have lower terminal--age main--sequence surface gravities than case--1 stars and show a tendency towards lower TAMS surface gravities for more rapid rotators. Case--3 stars generally have lower rotation rates at  the end of the main sequence for low--mass stars. This distinguishes them from case~2. Once again we see that the difference in stellar properties predicted by each model is strongly dependent on rotation rate, mass and metallicity suggesting that it is essential to explore populations covering as wide a range as possible in order to test rotating models.

\subsection{Alternative models for convection}

\begin{figure}
\centering
\includegraphics[width=0.99\textwidth]{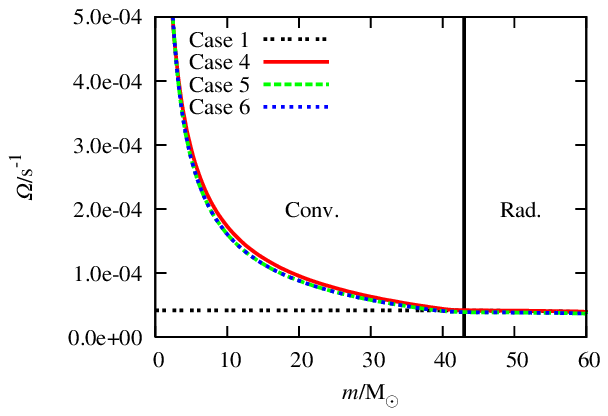}\\\includegraphics[width=0.99\textwidth]{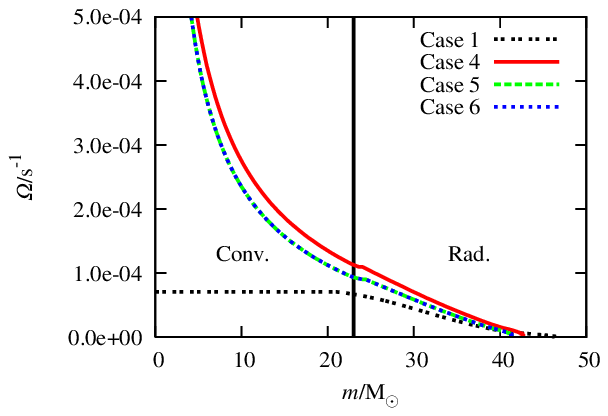}
\caption[Angular velocity distributions for $60$\ms\ stars initially rotating at $300\,{\rm km\,s^{-1}}$ for different models of angular momentum transport in convective zones]{Angular velocity distributions for $60$\ms\ stars initially rotating at $300\,\rm{km}\,\rm{s}^{-1}$ for different convective models. The top panel shows the ZAMS distributions and the bottom panel shows the TAMS distributions. Despite there being four orders of magnitude difference between the convective diffusion coefficients in cases~4, 5 and~6, there is very little difference between the angular velocity distributions they predict. In each case, the models with uniform specific angular momentum in convective zones predict more shear at the convective boundary than the models with solid body rotation in convective zones.}
\label{ch3.fig.convdist}
\end{figure}

In order to compare the difference in the evolution of a star owing to the details of the model for convective angular momentum transport we now focus on cases 1, 4, 5 and 6. Uniform specific angular momentum in the core causes more shear mixing near the core--envelope boundary than when the core is solid body rotating as shown in Fig.~\ref{ch3.fig.convdist}. This results in higher luminosity stars with similar temperatures. There is almost no difference between cases 4, 5 and 6 in the HR diagram.

\begin{figure}
\centering
\includegraphics[width=0.99\textwidth]{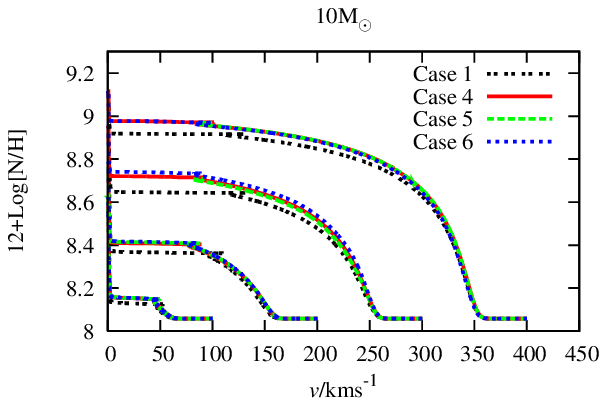}\\\includegraphics[width=0.99\textwidth]{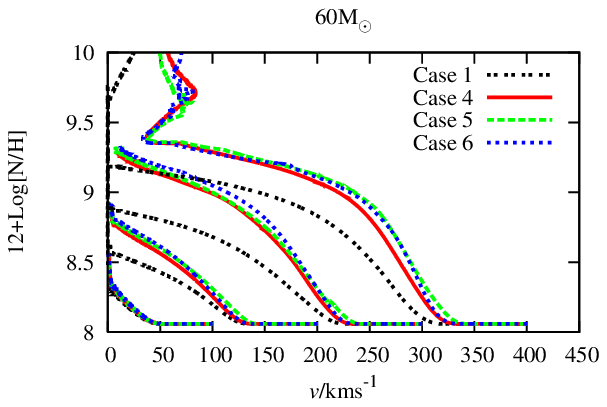}
\caption[Hunter diagrams for $10$\ms\ and $60$\ms\ stars for different models of angular momentum transport in convective zones]{Nitrogen enrichment variation with initial surface rotation for cases 4, 5 and 6. The top panel is for $10$\ms\ stars and the bottom panel is for $60$\ms\ stars. As expected the enrichment is much greater for more massive stars. There is much more enrichment for models in which the core has uniform specific angular momentum, an effect that becomes progressively more pronounced for higher masses and rotation rates}.
\label{ch3.fig.Nsol2}
\end{figure}

When we compare the different models for convection in a Hunter diagram we see that cases 4, 5 and 6, which have uniform specific angular momentum throughout their convective zones, have significantly more enrichment for all masses and rotation rates than case~1. The difference is more pronounced for higher--mass rapid rotators (Fig.~\ref{ch3.fig.Nsol2}). However, we note that it is more difficult to distinguish between cases 4, 5 and 6. For the highest mass stars we do find some difference in the enrichment of nitrogen and helium--3 between the models but recall that there is a difference in $D_{\rm con}$ of four orders of magnitude between cases~5 and 6.

Given the small magnitude of the change in enrichment and structure over such a range of diffusion coefficients it seems unlikely that these tests can adequately distinguish between convective models. In addition, adjusting the calibration of case~1 could produce a very similar effect making it difficult even to distinguish between $n=0$ and $n=2$ models from observations. At the same time though, it is interesting to note the significant change that modifying the core angular momentum distribution has had on the evolution of the surface composition.

\section{Conclusions}
\label{ch3.sec.conclusions}

Rotation in stars has a number of profound effects on their evolution. Not only are there significant changes in the hydrostatic structure \citep{Endal76} but this causes a thermal imbalance that can lead to a strong meridional circulation current \citep{Sweet50}. Meridional circulation leads to additional shear which induces a number of instabilities. The resulting turbulence leads to strong mixing of both angular momentum and chemical composition.

Although most modellers include all of these effects, the exact implementation of stellar rotation can vary dramatically. For example \citet{Heger00} use a model where meridional circulation is treated diffusively. The diffusion owing to shear is the linear combination of a number of coefficients based on different possible instabilities. On the other hand, modellers such as \citet{Zahn92}, \citet{Talon97} and  \citet{Maeder03} treat the circulation as advective and use a single diffusion coefficient based on the magnitude of Kelvin--Helmholtz instabilities induced by shear. In these models it is also necessary to define the magnitude of diffusion along isobars and this too varies between different models.

We have shown that different models generally give rise to similar qualitative conclusions but there are significant differences in the results based on mass, rotation rate and metallicity. There are also open questions about how angular momentum transport occurs in convective zones.

Comparing the models based on \citet{Talon97} and \citet{Maeder03}, case~1, and that based on \citet{Heger00}, case~2, we find that case~1 gives higher luminosity stars for masses less massive than $10$\ms\ and more luminous, hotter stars at higher masses than case~2. High--mass stars give similar levels of nitrogen enrichment in each case but case~2 produces far less enrichment for low--mass and intermediate--mass stars than case~1. The situation is similar for their helium--3 enrichment.

The predicted effects of rotation appear to be highly dependent on the metallicity. This is one of the clearest tests for different stellar rotation models. At metallicity of $Z=0.001$, case~2 actually produces significantly more nitrogen enrichment in high--mass stars although the degree of helium--3 enrichment is still lower in case~2 than case~1.  Additional effects such as the variation of the surface gravity with respect to surface rotation velocity are seen in lower--metallicity stars where the mass--loss rate is lower. We see a sharp cut--off in the effective gravity at the end of the main sequence but the TAMS rotation rates are very different between different models. Case--2 stars reach the end of the main sequence with higher rotation rates than the other two cases for both low--and high--mass stars. Case--1 stars reach the end of the main sequence with higher surface gravity than the other two cases but only for lower--mass stars.

All current models treat convective zones as rotating solid bodies. This may be justified if convective zones can generate sufficiently strong magnetic fields \citep[e.g.][]{Spruit99} but, if not, hydrodynamic models and calculations suggest that convective zones should tend towards uniform specific angular momentum \citep[Lesaffre et al., in prep.,][]{Arnett09}. Identifying whether this is the case or not is difficult from surface observations. There is no significant change in the paths of stars across the HR diagram between cases 4, 5 and 6. In fact, there is no test we have found that would adequately differentiate between these three cases  even though the diffusion coefficient in convective zones varies by four orders of magnitude. There are some minor differences in the amount of enrichment for massive stars but these would be hard to test with existing data. The models with uniform specific angular momentum generally produce slightly more luminous stars and higher surface chemical enrichment than those models with solidly rotating cores. However, slightly  less efficient mixing in the radiative zone could mask this difference.

We have thus far not included magnetic fields in the models. It has been suggested that the strong turbulence generated by rotation could result in a radiative magnetic dynamo \citep{Spruit99}. A sufficiently strong magnetic field can effectively suppress the meridional circulation and reduce the overall shear \citep{Maeder05}. It could also result in additional mass and angular momentum loss \citep[e.g.][]{Lau11, ud-Doula02}. As with rotation, there is little consensus on the details of magnetic field generation but it is generally accepted that the effects of rotation and magnetic fields cannot be considered in isolation. In chapter~\ref{ch5} we include magnetic fields in {\sc rose} in a similar manner in order to better explore this important feature of stellar evolution.

Owing to the range of available models, it is an extremely challenging problem to try to identify the one which best fits observed stellar populations. Now that large scale surveys are starting to produce data for many stars in different regions it is becoming possible to make progress and isolate which effects dominate. The key to distinguishing the relevant physics seems to be taking measurements of groups of stars at different masses, ages and metallicities. In individual clusters, models should be able to match not only the full distribution of observed stars but also the expected distribution in each mass range. Whilst most of the models can be calibrated to fit the data for a single cluster, we have shown that the behaviour of each model is highly dependent on mass and metallicity and so being able to fit data for a range of stellar environments is the true test of any model. 

\begin{savequote}[60mm]
A few observation and much reasoning lead to error; many observations and a little reasoning to truth. (Alexis Carrel)
\end{savequote}

\chapter{Model--dependent characteristics of stellar populations}
\label{ch4}

\section{Introduction}

The effect of rotation on the internal physics of stars has been considered for many years \citep[e.g.][]{Kippenhahn70}. Rotation causes significant changes in the hydrostatic balance of the star \citep{Endal76}, thermal imbalance causes a meridional circulation current \citep{Sweet50} and differential rotation leads to shear instabilities \citep[e.g.][]{Spiegel70}. These all result in mixing of angular momentum and chemical elements within the star leading to changes its surface properties such as the surface gravity, temperature, luminosity and chemical composition. Over the course of several decades, the physical formulations used to describe stellar rotation have proliferated \citep{Zahn92,Talon97,Meynet97,Heger00,Maeder05}. Whilst each new model has been suitably justified physically, there has been little observational data to back up claims of improved physical agreement. This leads to the possibility that any number of physical models can be chosen to produce a range of desired results which may or may not be accurate. This situation is worsened because the data required to constrain the models is scarce. However, with the observations of the VLT--FLAMES survey of massive stars \citep{Evans05,Evans06} and VLT--FLAMES Tarantula survey \citep{Evans10} it is now becoming possible to make such comparisons of different physical models and place some constraints on the formulations used. 

Comparing stellar models is still problematic because of the difficulty of isolating the effects of rotation from other physical and numerical differences in the results of other groups. In chapter \ref{ch3} we presented {\sc rose}, a code capable of performing stellar evolution calculations with a number of different models of stellar rotation, eliminating any differences owing to other numerical or physical effects between different codes. 

In chapter~\ref{ch3} we considered the main differences between the evolution of individual stars under the assumptions of several popular models. In this chapter we combine that analysis with the stellar population code, {\sc starmaker} \citep{Brott11b}, to determine the differences in stellar populations that arise from two physical models. One is based upon \citet{Heger00} and has solely diffusive transport of angular momentum. The other is based on \citet{Talon97} and \citet{Maeder03} and has both diffusive and advective transport of angular momentum. The two different models have very different diffusion coefficients and there are marked differences in the results for individual stars. It is possible to get better agreement between the models under different criteria by adjusting the associated unknown constants but this leads to poorer agreement elsewhere.

One particular consequence of different input physics that we found in chapter~\ref{ch3} is that the mass dependence of the mixing is very different in each case. The models agree in that the total enrichment in low--mass stars ($M < 20$\ms) is much less than in high--mass stars ($M > 20$\ms). However, the enrichment found with each model is very different for low--mass stars despite reasonable agreement for high--mass stars. In chapter~\ref{ch3} we also concluded that the difference between the two models varies for different metallicities. For $Z=0.001$, the model based on \citet{Heger00} actually produces significantly more nitrogen enrichment in high--mass stars, particularly for slow and moderate rotators. We shall explore all of these features further in this chapter.

In section~\ref{ch4.sec.physics} we briefly review the physics of the models. For full descriptions we refer the reader to chapter \ref{ch2} and \citet{Brott11}. We also describe the models under comparison. In section~\ref{ch4.sec.results} we compare the stellar population predictions and consider the similarities and differences between the models and in section~\ref{ch4.sec.conclusions} we present our summary and conclusions.

\section{Input physics}
\label{ch4.sec.physics}

Each grid of models is produced with the code {\sc rose} described in section~\ref{ch2.sec.rose}. The physical model for the evolution of rotating stars is described in section~\ref{ch2.sec.rotation}.  The stellar populations are then generated with the code {\sc starmaker} \citep{Brott11}, described in section~\ref{ch2.sec.starmaker}. It was originally designed to work with the evolutionary models of \citet{Brott11b}. We have adapted it for use with {\sc rose} stellar evolution models. Based on a grid of evolutionary models, {\sc starmaker} interpolates for stellar properties given an initial mass, initial surface velocity and age. These are chosen at random according to user--defined distribution functions. Each simulated star is assigned a random orientation in space. The newly generated sample can subsequently be filtered according to observational selection effects to enable comparison with observed samples.

\begin{figure}
\begin{center}
\includegraphics[width=0.99\textwidth]{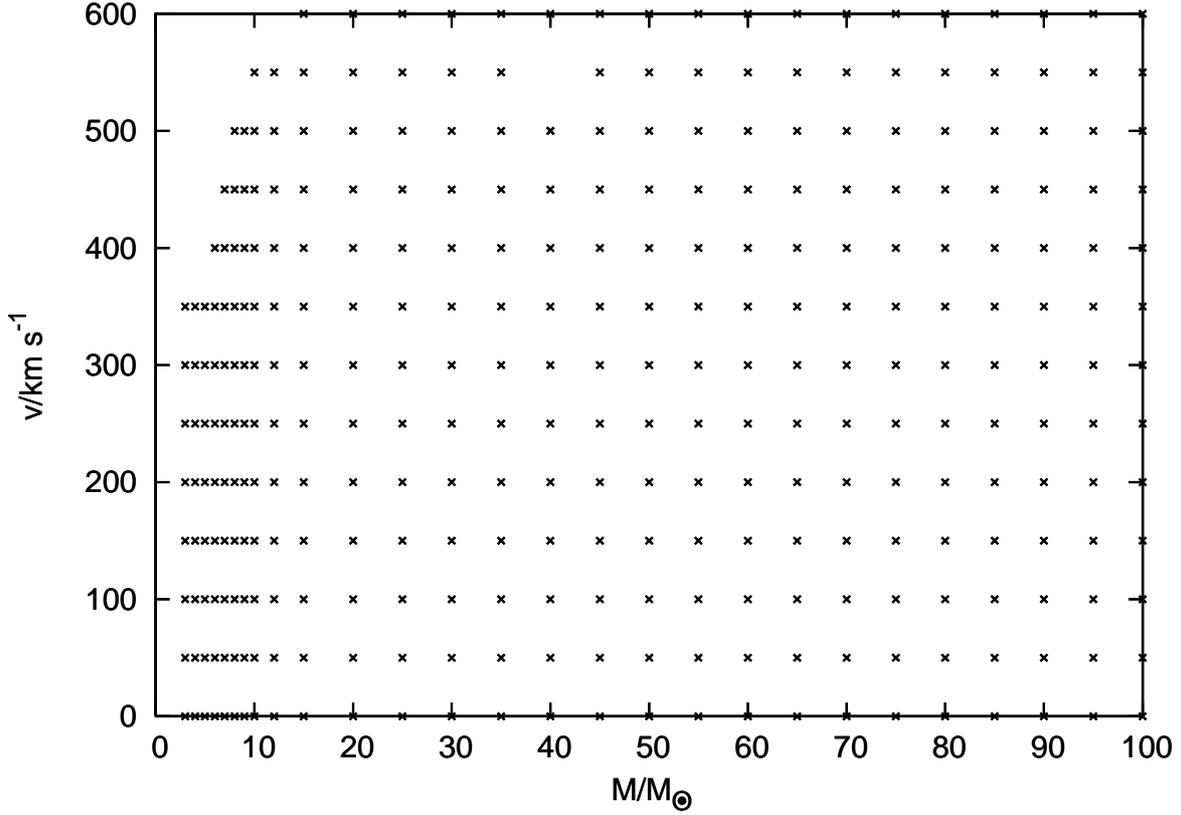}
\end{center}
\caption[Grid of models for simulating stellar populations in chapter~\ref{ch4}]{Grid of initial models, in initial mass--initial equatorial velocity space, used for simulating stellar populations.}
\label{ch4.fig.grid}
\end{figure}

In this chapter we consider two models for comparison, these are case~1 and case~2 from chapter~\ref{ch3}. For both models we evolve a grid of stars with masses between $3$\ms\ and $100$\ms\ and initial equatorial surface rotation velocities between $0$ and $600\,{\rm km\,s^{-1}}$. The zero age main sequence is taken to be the point of minimum luminosity at the onset of hydrogen burning. The masses computed are

\begin{align}
m/{\rm M}_\mathrm{\odot}\,\in\,&\{3,4,5,6,7,8,9,10,12,15,20,25,\nonumber\\ & 30,35,40,45,50,55,60,65,70,75,80,\nonumber\\ & 85,90,95,100\}
\end{align}

\noindent and for each mass the initial surface velocities used are

\begin{align}
v_{\rm ini}/{\rm km\,s^{-1}}\in\,&\{0,50,100,150,200,250,300,\nonumber\\ & 350,400,450,500,550,600\},
\end{align}

\noindent except when the rotation velocity would be too close to critical rotation to achieve numerical convergence. This becomes more difficult for stars less massive than $10$\Msun. Convergence can be achieved for a $10$\Msun\ star rotating faster than $95$\% of critical rotation, although the assumptions of the model are likely to become invalid this close to critical. For a $3$\Msun\ star the limit for convergence is close to $70$\% of critical rotation. Both the case~1 and case~2 models for each mass and initial surface velocity must reach the end of the main sequence for either of them to be used in the grid. The end of the main sequence is taken as the point of maximum temperature before a star moves onto the Hertzsprung gap. Each model evolved is plotted in Fig.~\ref{ch4.fig.grid}. For both models, the diffusion of angular momentum in convective zones is determined by the characteristic eddy viscosity given by mixing length theory such that $D_{\rm conv}=D_{\rm mlt}=\frac{1}{3}v_{\rm mlt}l_{\rm mlt}$. The position of the convective boundary is determined by the Schwarzschild criterion and although the code includes a model for convective overshooting, we do not use it in these simulations. Unlike for chapter~\ref{ch3}, we only consider the case in which the convective core tends to a state of solid body rotation. In section~\ref{ch4.sec.calibration} we examine the effects of changing the free parameters associated with the model. Models generated with this calibration are referred to as case~2$_{\rm b}$. In this chapter we generate models with two different metallicities, Galactic and LMC, as defined by \citet{Brott11b}. Other than the initial composition, the input physics is the same for both metallicities. However, for clarity, we distinguish models that use LMC metallicity by referring to them with a superscript `Z' (e.g. case~${\rm 1^Z}$).

The angular momentum evolves according to

\begin{equation}
\label{ch5.eq.amtransport}
\diff{r^2\Omega}{t}=\frac{1}{5\rho r^2}\diff{\rho r^4\Omega U}{r}+\frac{1}{\rho r^2}\frac{\partial}{\partial r}\left(\rho {D_{\rm KH} r^4\frac{\partial\Omega}{\partial r}}\right)+\frac{1}{\rho r^2}\frac{\partial}{\partial r}\left(\rho {D_{\rm{con}} r^{2}\frac{\partial \Omega}{\partial r}}\right).
\end{equation}

\noindent This is equation~(\ref{ch2.eq.amtransport}) with $n=2$. The terms are described in more detail in chapter~\ref{ch2}. The primary difference between the cases is the treatment of meridional circulation $U$ and the diffusion coefficient $D_{\rm KH}$.

\subsection{Case~1}

Our case~1 model uses the formulation for $D_{\rm KH}$ of \citet{Talon97},
\begin{equation}
D_{\rm KH}=C_0\frac{2 Ri_{\rm c}\left(r\frac{d\Omega}{dr}\right)^2}{N_T^2/(K+D_{\rm h})+N_{\mu}^2/D_{\rm h}},
\end{equation}

\noindent where

\begin{equation}
N_T^2=-\frac{g_{\rm eff}}{H_P}\left(\frac{\partial\ln\rho}{\partial \ln T}\right)_{\!\!P,\mu}\left(\nabla_{\rm ad}-\nabla\right)
\end{equation}
\noindent and
\begin{equation}
N_{\mu}^2=\frac{g_{\rm eff}}{H_P}\left(\frac{\partial\ln\rho}{\partial\ln\mu}\right)_{\!\!P,T}\frac{d \ln\mu}{d \ln P}.
\end{equation}

\noindent As in chapter~\ref{ch3} we follow \citet{Maeder03} by taking the critical Richardson number, $Ri_{\rm c}=(0.8836)^2/2$. We have also chosen $C_0$ so that we reproduce the terminal--age main--sequence (TAMS) nitrogen enrichment of a $40$\ms\ star initially rotating at $270{\,\rm km\,s^{-1}}$ with Galactic composition given by \citet{Brott11}. The effective diffusion coefficient $D_{\rm eff}$ is

\begin{equation}
D_{\rm eff}=\frac{|rU|^2}{30 D_{\rm h}}
\end{equation}

\noindent and we take
\begin{equation}
D_{\rm h}=0.134 r\left(r\Omega V\left(2V-\alpha U\right)\right)^{\frac{1}{3}},
\end{equation} 

\noindent where
\begin{equation}
\alpha=\frac{1}{2}\frac{d(r^2\Omega)}{dr}.
\end{equation}

\subsection{Case~2}

Our case~2 model is that of \citet{Heger00}. In this case $U=0$ because circulation is treated as a purely diffusive process. The details of the various diffusion coefficients are extensive so we refer the reader to the original paper. With their notation the diffusion coefficients are
\begin{equation}
D_{\rm KH}=D_{\rm sem}+D_{\rm DSI}+D_{\rm SHI}+D_{\rm SSI}+D_{\rm ES}+D_{\rm GSF}
\end{equation}

\noindent and
\begin{equation}
D_{\rm eff}=(f_{\rm c}-1)(D_{\rm DSI}+D_{\rm SHI}+D_{\rm SSI}+D_{\rm ES}+D_{\rm GSF}),
\end{equation}

\noindent where each $D_i$ corresponds to a different hydrodynamic instability. \citet{Heger00} take $f_{\rm c}=1/30$ and we use this too. We also use $f_\mu=0$. The consequences of this are discussed in chapter~\ref{ch3}. Unless otherwise stated, we calibrate this model by scaling $D_{\rm ES}$, the dominant diffusion coefficient, so that the nitrogen enrichment of a 20\ms, solar metallicity star with initial surface angular velocity of $300\,{\rm km\,s^{-1}}$ is the same as for case~1 at the TAMS.

\subsection{Stellar populations}

\begin{table}
\begin{center}
\begin{tabular}{cccc}
\hline
Age/yr&Number of&Maximum mass&Maximum mass\\
 &stars remaining&(case~1)/${\rm M}_\mathrm{\odot}$&(case~2)/${\rm M}_\mathrm{\odot}$\\
\hline
$5\times 10^6$&$9749375$ & $42.2$ & $41.6$\\
$10^7$&$9168582$&$21.6$ & $22.3$\\
$2\times 10^7$&$8318064$ & $14.1$ & $13.7$\\
$5\times 10^7$&$6449586$ & $8.2$ & $7.4$\\
\hline
\end{tabular}
\end{center}
\label{ch4.tab.pops}
\caption[The properties of the single--aged stellar populations used in section~\ref{ch4.sec.results}]{The properties of different single--aged stellar populations used in section~\ref{ch4.sec.results}. The original size of the population in each case is $10^7$ stars. Each population is generated by an instantaneous burst of star formation at $t=0$. The first column shows the age of the simulated population. The second column shows the number of stars remaining in the sample after stars that have reached the end of the main sequence are excluded. The third and fourth columns show the mass of the most massive star remaining in the sample at the given age for case~1 and case~2 respectively.}
\end{table}

Throughout this paper we use a variety of populations at different ages with different star formation histories. The main reason for this is that, as a population ages, the mass of the most massive stars remaining in the main--sequence population decreases, whilst stars much less massive than the maximum mass have not had sufficient time to produce significant nitrogen enrichment. The combination of these tendencies allows us to follow how the amount of enrichment varies with mass. This applies specifically to clusters in which we expect the range of ages of the stars to be small compared to the age of the cluster. It is important to note that rotation can significantly affect the upper bound to the mass of stars in the population. The populations used in this paper are listed in Table~\ref{ch4.tab.pops}. For the figures in section~\ref{ch4.sec.results} we compare the data using 2D histograms for which we have separated the data into a grid of $50 \times 50$ bins. The number of stars in each bin divided by the total number of stars. In case~1 this is $n_1$ and similarly $n_2$ for case~2. In each comparison we reduce the size of the larger population to be the same size as the other by randomly removing stars.

\section{Results}
\label{ch4.sec.results}

\begin{figure}
\begin{center}
\vspace{3cm}
\includegraphics[width=0.65\textwidth]{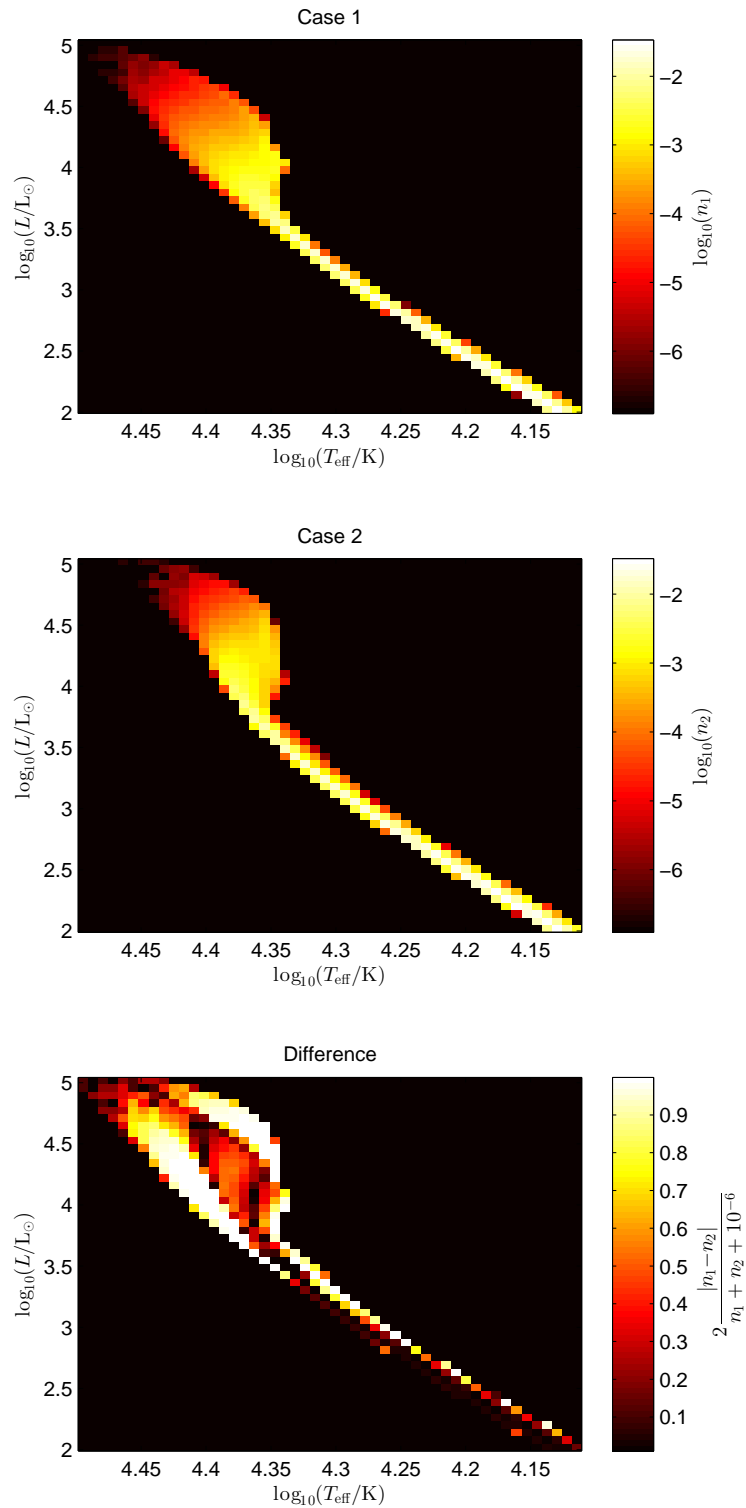}
\end{center}
\caption[Hertzsprung--Russell diagrams for stellar populations at age $2\times 10^7\,{\rm yr}$ for different physical models]{Hertzsprung--Russell diagrams for a population of stars at age $2\times 10^7{\rm \,yr}$. There is some slight variation between the two cases at the end of the main sequence (highest luminosity) but the effect is small. Otherwise there is no obvious difference between the results produced in cases~1~and~2. When comparing the two populations, the addition of $10^{-6}$ in the denominator is to avoid division by zero in unpopulated bins.}
\label{ch4.fig.2e7hr}
\end{figure}

We simulated stellar populations at a number of ages and metallicities for each model and found a number of significant differences.

\subsection{The Hertzsprung--Russell diagram}

When we look at the effect of the two models of rotation on stars in the Hertzsprung--Russell (HR) diagram we find that there is very little difference between them. Whilst there is variation in the TAMS temperatures and luminosities of the stars in each case, the difference is small and, for a single burst of star formation, only affects a handful of stars in the population at any given time. For most of a star's lifetime the predicted position in the HR diagram is sufficiently similar between the two cases that the difference in the population cannot be distinguished. Fig.~\ref{ch4.fig.2e7hr} shows the HR diagrams for cases~1 and~2 for simulated clusters with an age of $2\times 10^7{\rm \,yr}$. Apart from slightly different degrees of broadening at the main--sequence turn off, there is no difference between the two cases. This is true at all ages and if we simulate a population of stars with continuous star formation we still find only slight distinctions between the two cases. To distinguish these differences in real stellar populations would be made even more difficult because the broadening of the high--luminosity end of the main sequence in the HR diagram owing to rotation is similar to the effect of including binary stars \citep{Eldridge08}. This doesn't mean that mass determinations of rotating stars from their surface rotation, temperatures and luminosities are unaffected by the specific physics of the model. For individual models, the difference can be significant but the cumulative effect has little impact on the population as a whole. It is also important to note that the maximum mass of stars remaining in the sample varies between the cases because the main--sequence lifetimes are different. This means that care must be taken when identifying a cluster's age with respect to its most massive members if the cluster contains rapid rotators.

\subsection{Velocity distribution evolution}
\label{ch4.sec.veldist}

\begin{figure}
\begin{center}
\includegraphics[width=0.7\textwidth]{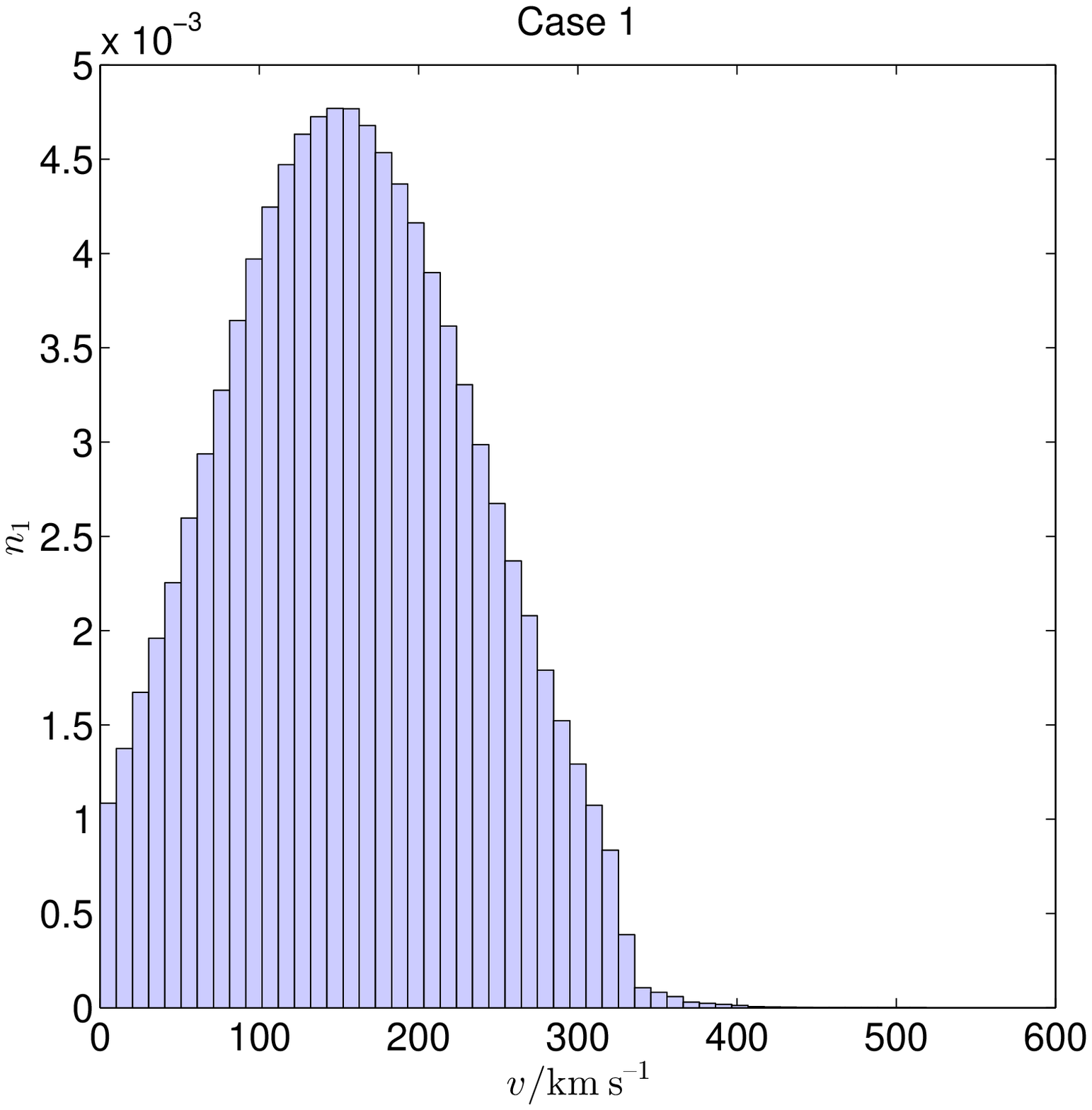}
\includegraphics[width=0.7\textwidth]{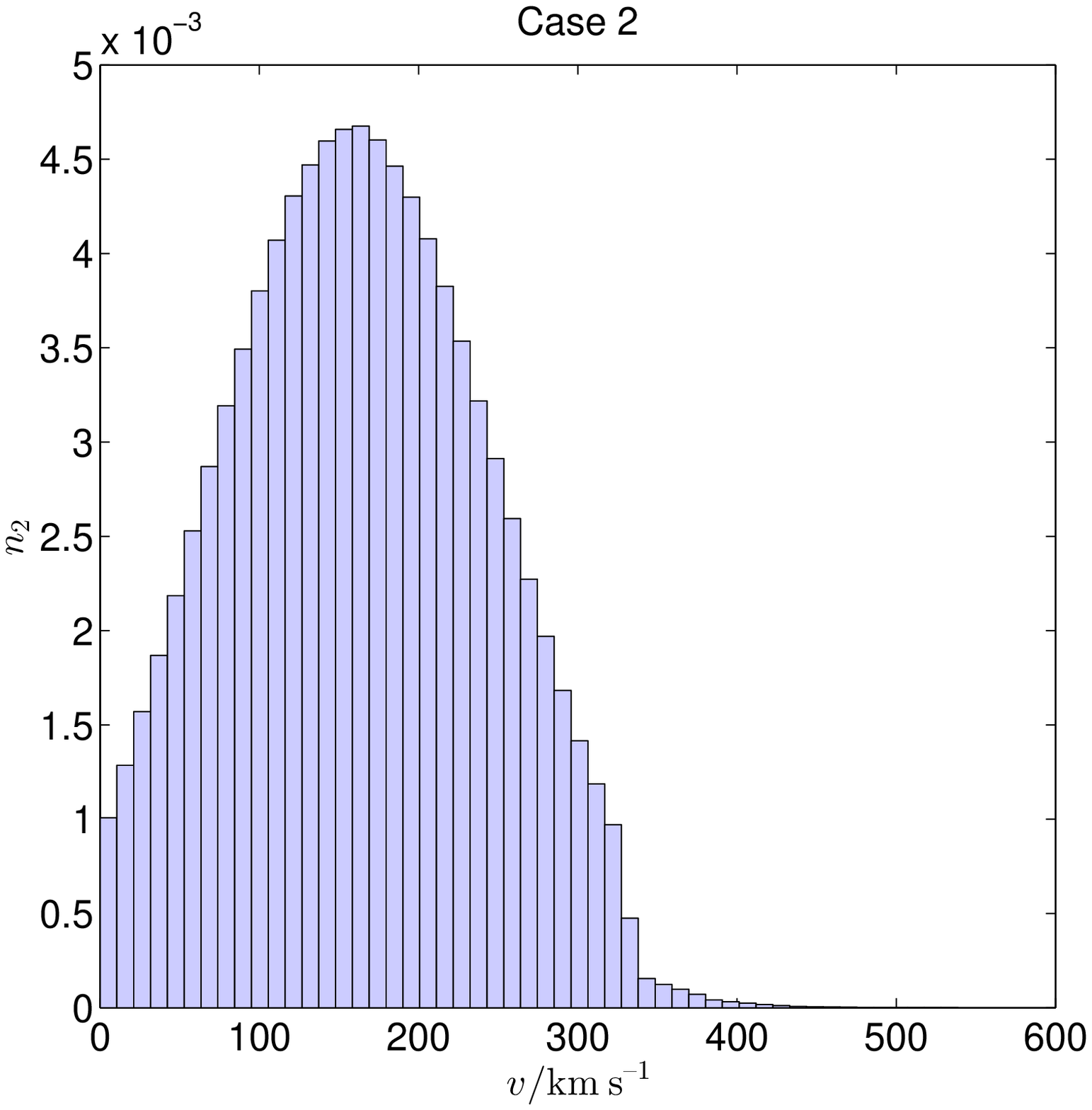}
\end{center}
\caption[Velocity distributions of stellar populations at age $2\times 10^7\,{\rm yr}$ for different physical models]{Velocity distribution of stars in cases~1 and~2 both at an age of $2\times 10^7{\rm \,yr}$ for the populations described in table \ref{ch4.tab.pops}. For this and subsequent figures, $v$ is the projected surface rotation velocity. There is no marked difference between the two distributions.}
\label{ch4.fig.2e7v}
\end{figure}

Because of variations in the amount of mixing and the evolutionary timescale between the two cases, we might expect differences between the distribution of rotation rates as the populations evolve. In Fig.~\ref{ch4.fig.2e7v} we plot the velocity distribution of the remaining stars in the single--aged populations at $2\times 10^7\,{\rm yr}$. This is the typical shape of the distribution at all ages considered and we see that there is very little difference between the two cases. 

\subsection{The Hunter diagram}
\label{ch4.sec.hunter}

\begin{figure}
\begin{center}
\hspace*{-2.2cm}\includegraphics[width=1.2\textwidth]{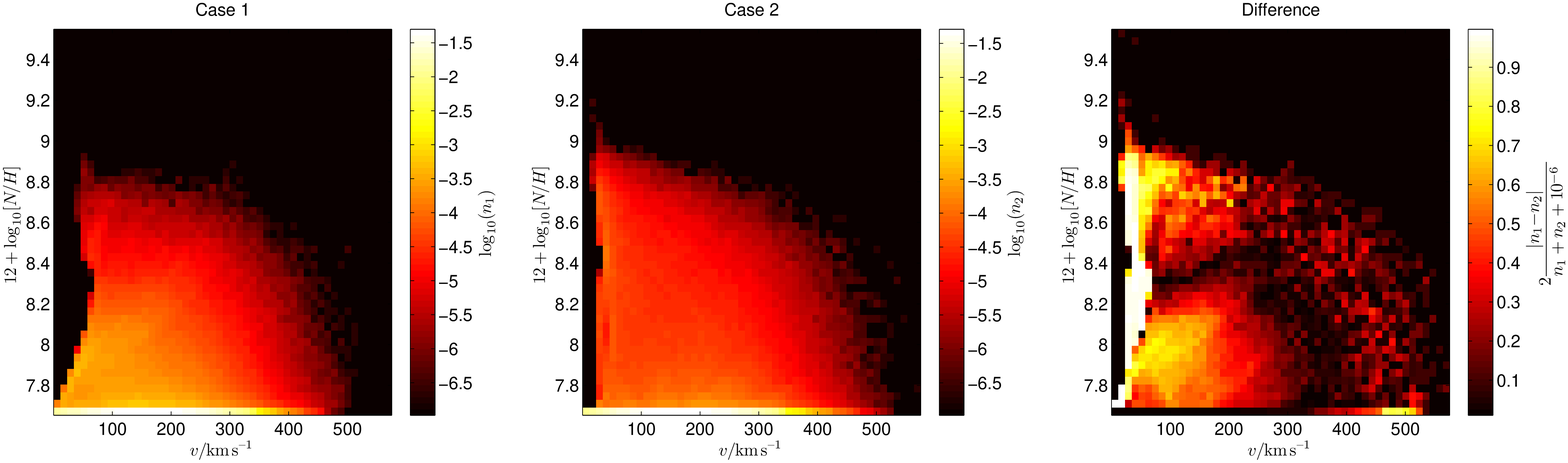}
\hspace*{-2.2cm}\includegraphics[width=1.2\textwidth]{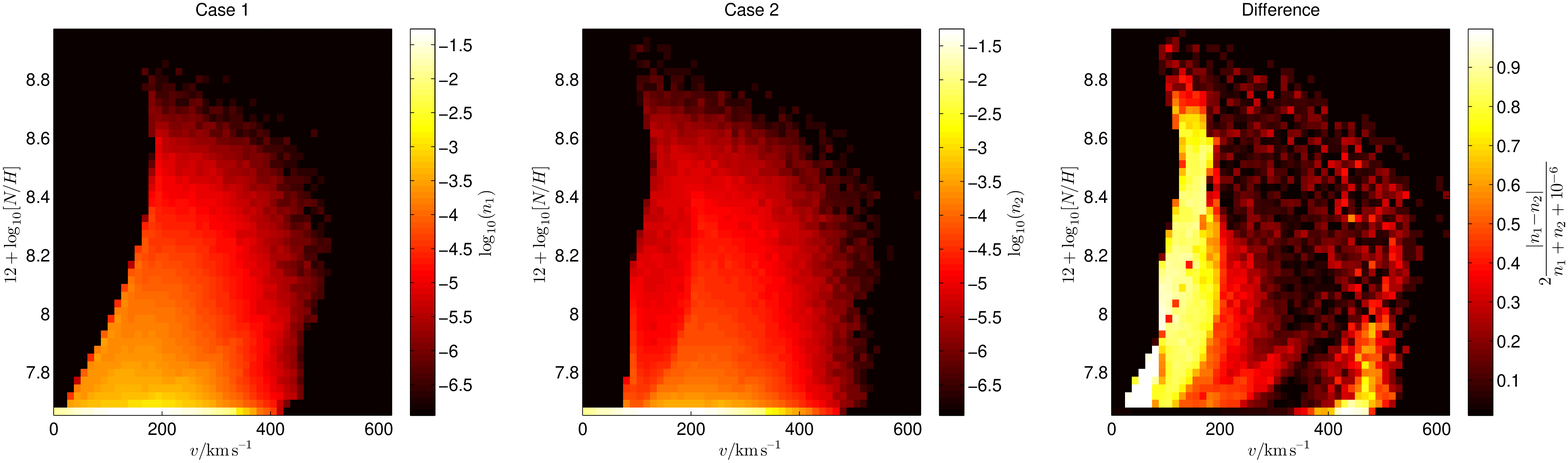}
\hspace*{-2.2cm}\includegraphics[width=1.2\textwidth]{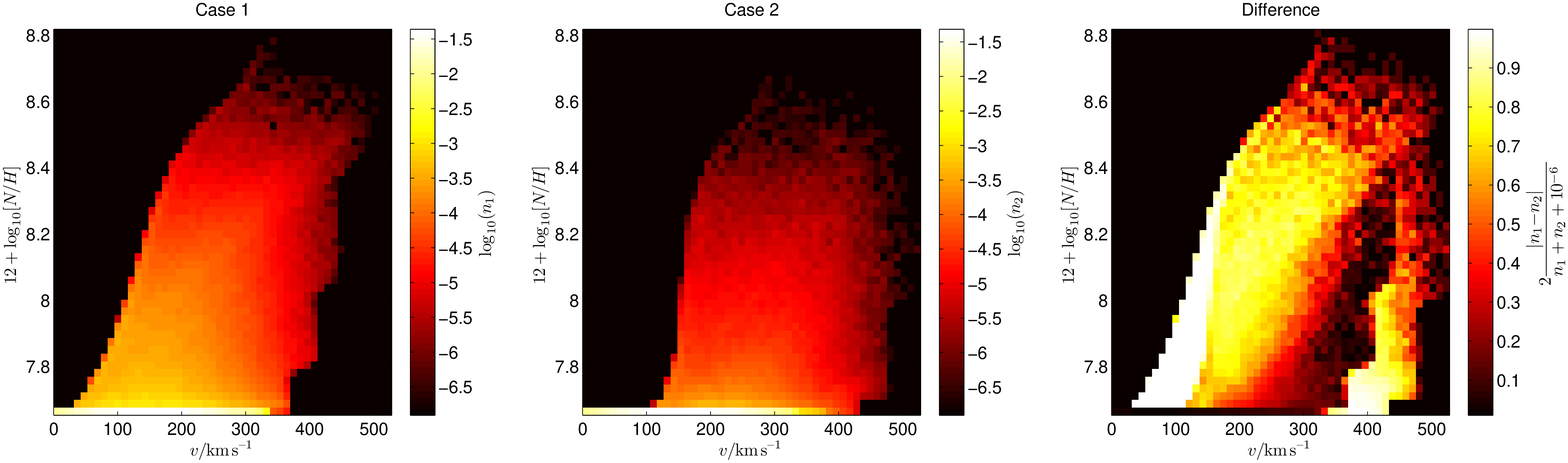}
\hspace*{-2.2cm}\includegraphics[width=1.2\textwidth]{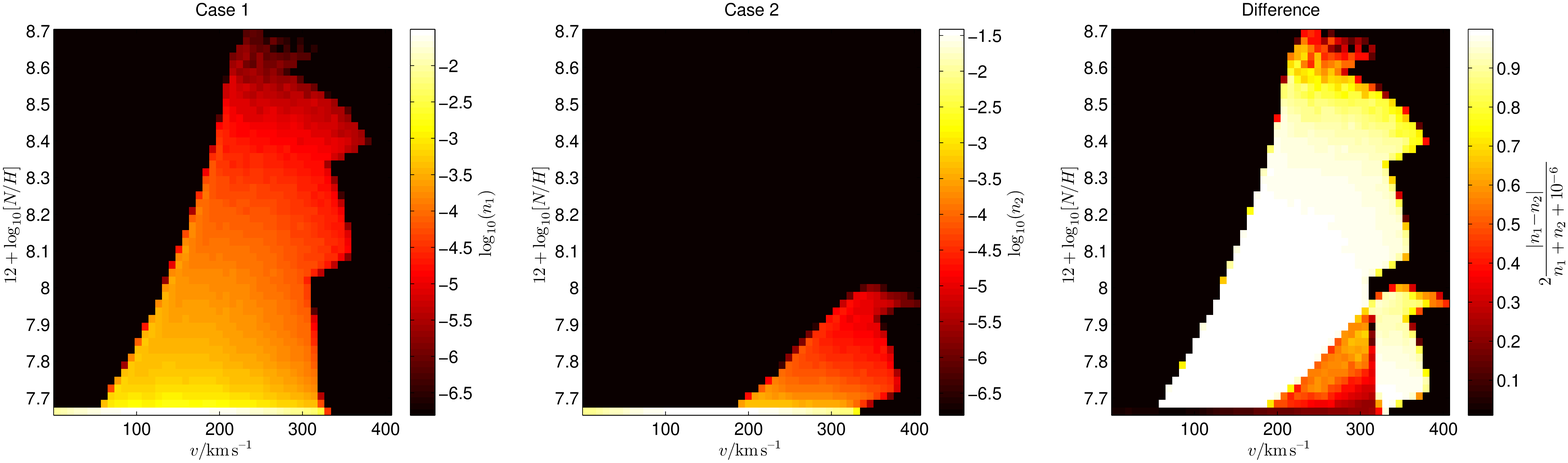}
\end{center}
\caption[Hunter diagrams for stellar populations at various ages for different physical models]{Hunter diagrams for single--aged populations of stars. From top to bottom, the four rows of figures correspond to $5\times 10^6\,{\rm yr}$, $10^7\,{\rm yr}$, $2\times 10^7\,{\rm yr}$ and $5\times 10^7\,{\rm yr}$. At early times the two cases are similar with slightly more enrichment of the fastest rotators in case~2 and more enrichment of stars rotating more slowly than $200\,{\rm km\,s^{-1}}$ in case~1. By $2\times 10^7\,{\rm yr}$ we see many more enriched stars in case~1 and by $5\times 10^7\,{\rm yr}$ the amount of mixing in case~2 has dropped off dramatically. The jagged right hand edge of the populations is a result of the grid geometry and the mass--independence of the initial rotation velocity distribution. Neither affects the large difference we see in the populations at late times.}
\label{ch4.fig.hunt}
\end{figure}

The surface abundance of various isotopes is an extremely important tracer of the effects of rotation. In section~\ref{ch4.sec.veldist} we showed that, using alternate models of rotation, we find only small effects on the velocity distribution in stellar populations. We now consider the effect of rotation on the surface abundance of nitrogen. We could make similar conclusions about other chemical elements but their usefulness depends on the accuracy to which they can be measured and the availability of data. For example in chapter~\ref{ch3} we discussed the effect of rotation on the surface abundance of helium--3 but this is difficult to measure and so is not particularly useful in this discussion. \citet{Frischknecht10} and \citet{Brott11} also consider how rotation is likely to affect the surface abundances of light elements. If we look at a plot of the surface abundance against surface rotation rate, commonly referred to as the Hunter diagram \citep{Hunter09}, for different ages (Fig.~\ref{ch4.fig.hunt}) we see that there are some very clear differences between the two cases. At each age, the most massive stars remaining in the population dominate the enriched stars. Stars more massive than this have already evolved off the main sequence. The less massive stars in the population evolve more slowly and so have not had enough time to become enriched. At early times the populations are very similar except that case~1 predicts rather more enrichment for stars rotating slower than $200\,{\rm km\,s^{-1}}$ while case~2 predicts more enrichment of the most rapidly rotating stars. As the population ages, the amount of enrichment in case~1 stays roughly the same but the amount of enrichment in case~2 drops off slowly followed by a large drop between $2\times10^7\,{\rm yr}$ and $5\times 10^7 \,{\rm yr}$. It is here, where only stars less massive than $8.2$\ms\ remain, that the difference between the models is clearest. However, even at $2\times10^7\,{\rm yr}$, we can see that there are far more enriched stars in case~1 than case~2 compared with earlier times.

\begin{figure}
\begin{center}
\vspace{3cm}
\includegraphics[width=0.65\textwidth]{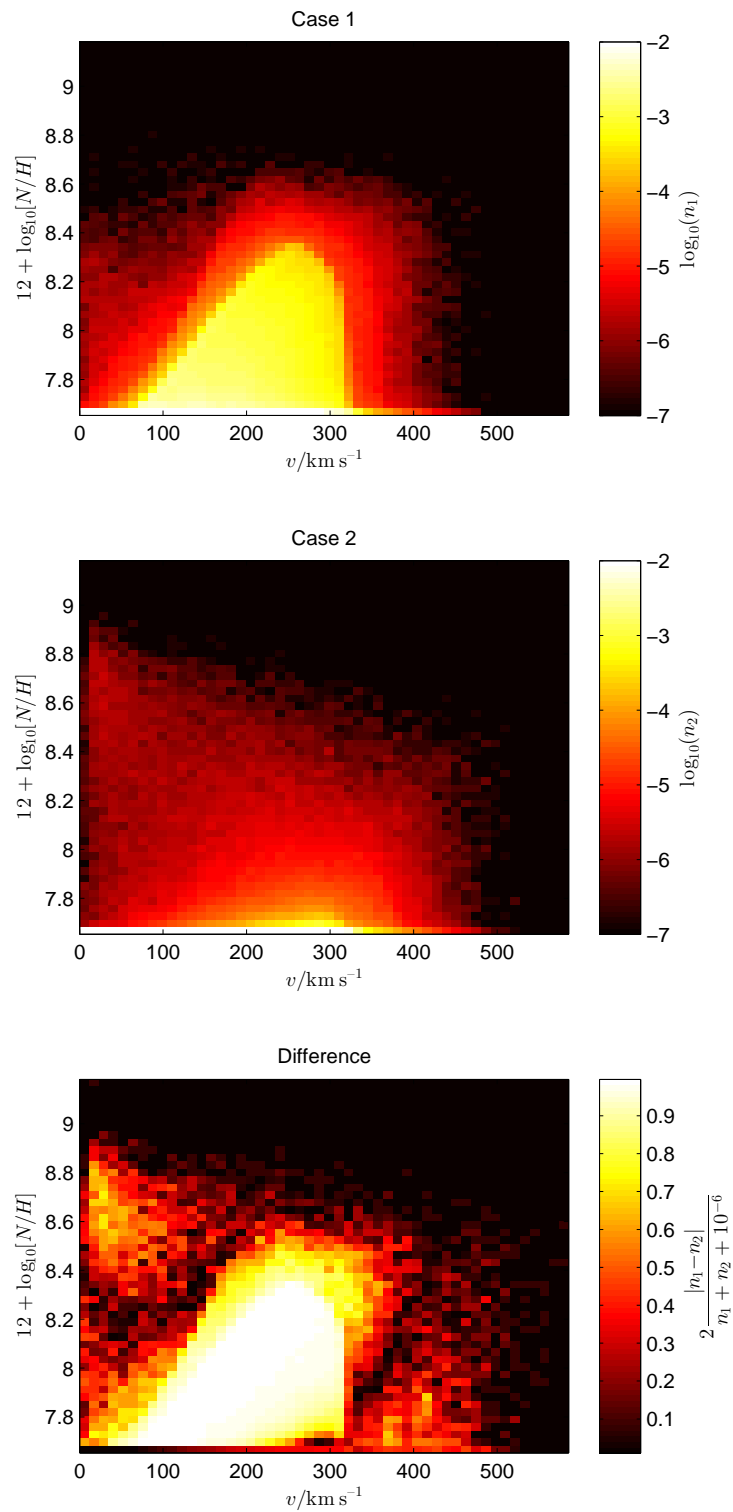}
\end{center}
\caption[Hunter diagrams for stellar population under the assumption of continuous star formation for different physical models]{Hunter diagrams for a population of stars with continuous star formation. Despite showing good agreement at low surface rotation rates, case~1 has many more fast--rotating highly enriched stars. However, this difference can often be accounted for by recalibration of the mixing coefficients and is difficult to observe owing to the rarity of rapidly rotating high--mass stars which occupy this region of the plot.}
\label{ch4.fig.huntcont}
\end{figure}

If we consider the case--dependence of the Hunter diagrams for a population of stars with a continuous star formation history we find much the same thing. Fig.~\ref{ch4.fig.huntcont} shows that, although both cases follow a trend of more frequent enrichment in stars with higher rotation rates, case~1 produces far more enriched stars than case~2. This is not surprising because we found far less mixing in case--2 stars at lower masses than in case~1 and it is these stars that dominate the population. We could have instead chosen to calibrate case~2 so that there were more mixing in low--mass stars but this would inevitably lead to a worse match in the populations elsewhere, perhaps in the enrichment of rapidly rotating very massive stars ($M>40$\ms). We discuss this further in section~\ref{ch4.sec.calibration}. Also important to consider is the effect of metallicity. In chapter~\ref{ch3} we found a reversal in the trend of less mixing in case--2, low--mass stars ($M<20$\ms). Increasing the mixing here to bring the two populations in line would make the low--metallicity agreement far worse. We discuss this in section~\ref{ch4.sec.metal}.

\subsection{Effective surface gravity and enrichment}
\label{ch4.sec.grav}

\begin{figure}
\begin{center}
\includegraphics[width=0.99\textwidth]{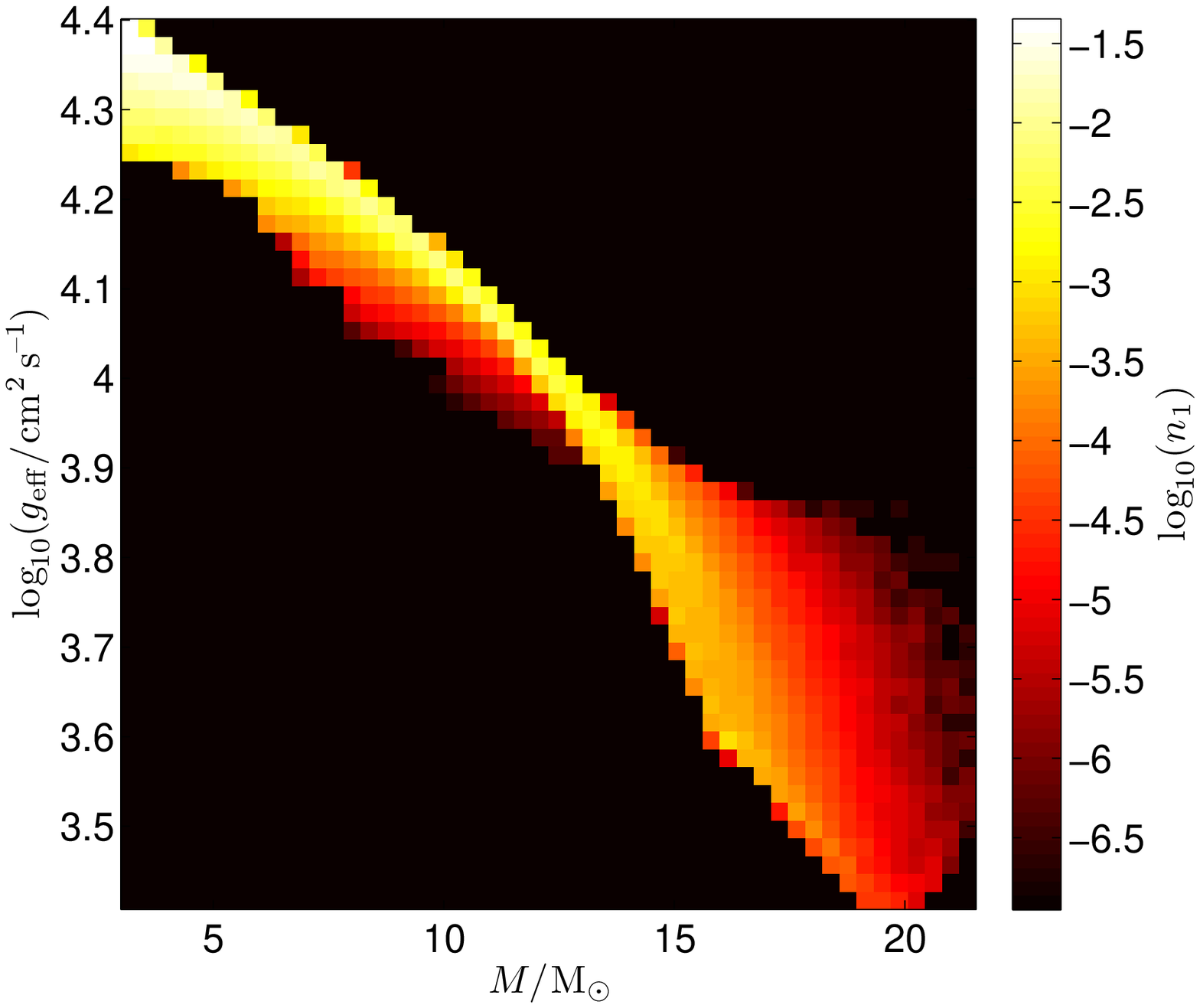}
\end{center}
\caption[Relation between the mass and surface gravity in a simulated stellar population of age $10^7\,{\rm yr}$]{Simulated single--aged stellar populations at $10^7{\rm \,yr}$ in case~1. The plot shows the strong correlation between mass and surface gravity. This relation only holds when the population has a single age. Because of rotation, the relation is degenerate and a measurement of the surface gravity corresponds to stellar masses with a range of up to $7$\ms.}
\label{ch4.fig.grav}
\end{figure}

\begin{figure}
\begin{center}
\hspace*{-2.2cm}\includegraphics[width=1.2\textwidth]{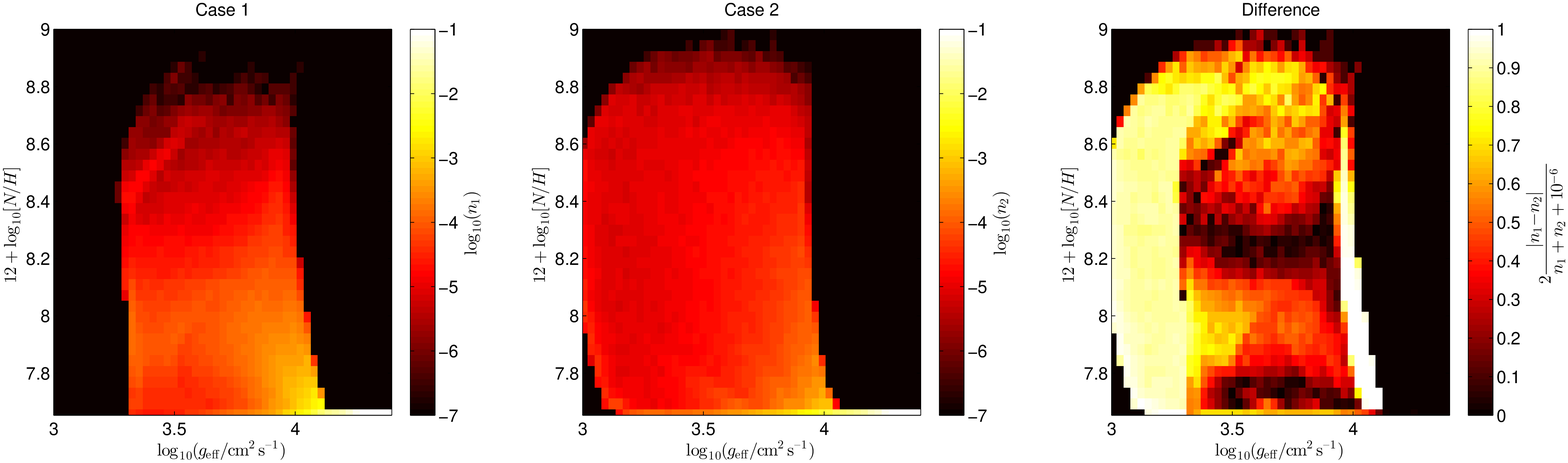}
\hspace*{-2.2cm}\includegraphics[width=1.2\textwidth]{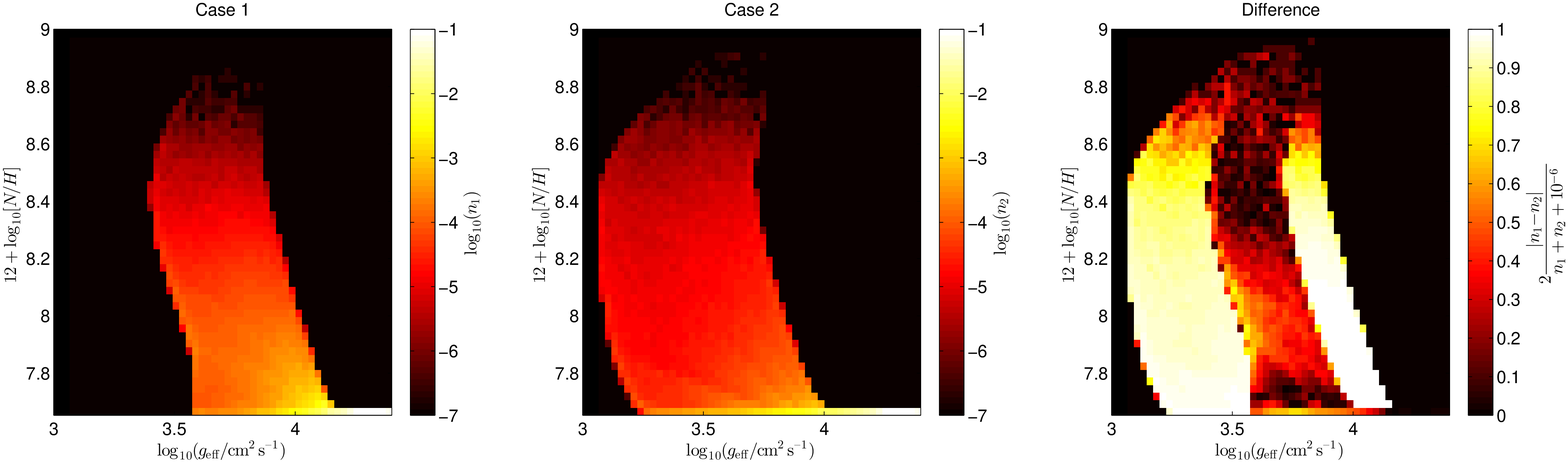}
\hspace*{-2.2cm}\includegraphics[width=1.2\textwidth]{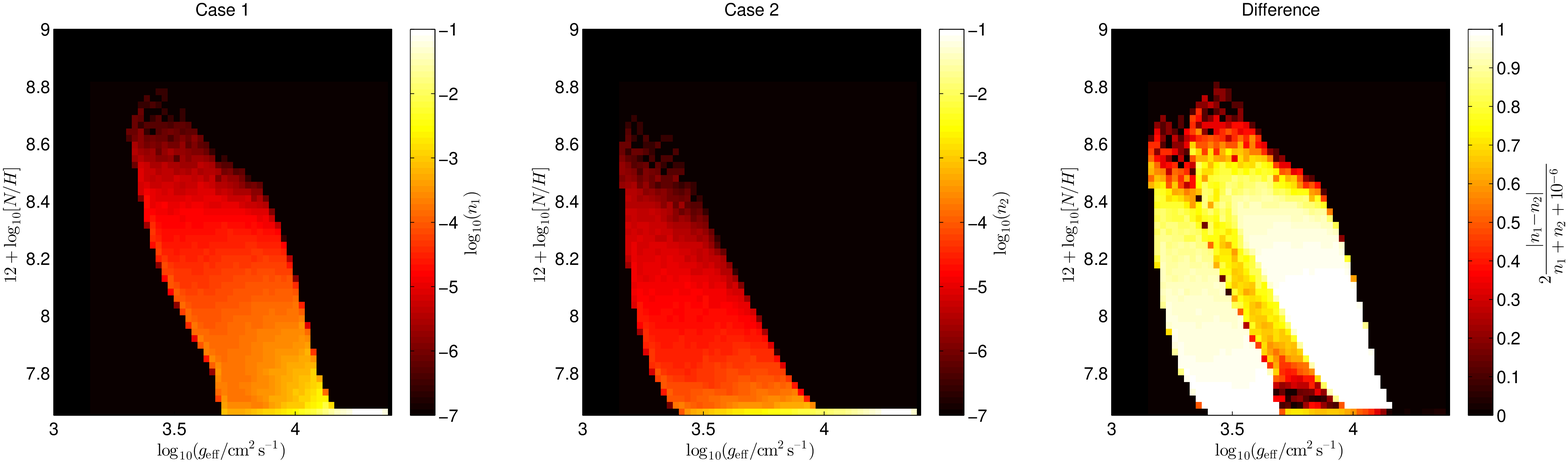}
\hspace*{-2.2cm}\includegraphics[width=1.2\textwidth]{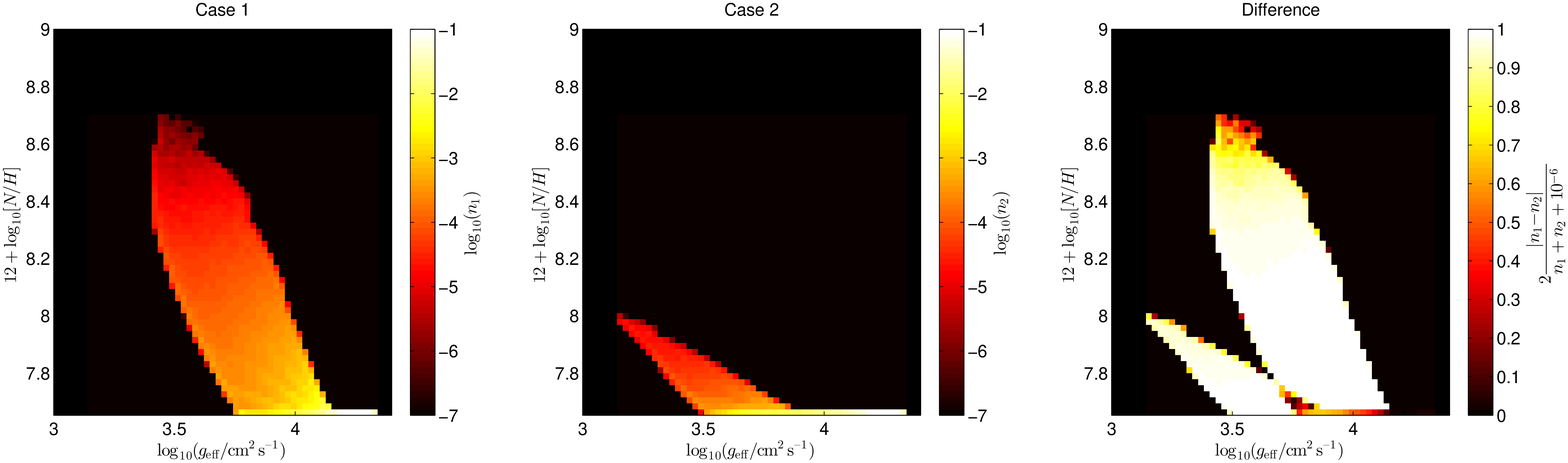}
\end{center}
\caption[Surface nitrogen enrichment against effective surface gravity for populations at a range of ages for different physical models]{Distribution of the surface nitrogen enrichment against effective surface gravity in single--aged stellar populations. From top to bottom, the four rows of figures correspond to $5\times 10^6\,{\rm yr}$, $10^7\,{\rm yr}$, $2\times 10^7\,{\rm yr}$ and $5\times 10^7\,{\rm yr}$. At early times case~2 gives a larger spread of effective surface gravities and higher enrichment of the fastest rotators. At later times the maximum enrichment is similar but case~2 predicts overall lower surface gravities than case~1. Finally at late times when only stars with mass smaller than $8.2$\ms$\ $ remain, case~2 predicts far less mixing than case~1 as well as much lower surface gravity for its fastest rotators. }
\label{ch4.fig.gn}
\end{figure}

As we suggested in chapter~\ref{ch3}, the difference between the two cases can be seen most clearly by considering different masses of stars. In our discussion of the Hunter diagram we have appealed to the single--aged population of stars to differentiate between stellar masses as the population ages. Unfortunately, determining the mass of rotating stars self--consistently is difficult because of the degeneracy that arises owing to rotation. Fig.~\ref{ch4.fig.grav} shows the typical relationship between mass and effective surface gravity in a simulated population. There is a strong correlation between the two but rotation causes degeneracy so estimates of the mass from effective gravity alone could be wrong by up to $7$\ms\ in this case. The correlation does not persist in the case of continuous star formation. Use of the effective surface gravity is also advantageous because it can be directly determined spectroscopically. However, caution is necessary for rapid rotators because the effective gravity is not uniform across the stellar surface \citep{VonZeipel24}. The Hunter diagram suffers from the problem that, even for simple stellar populations like this one, stars exist in all regions of the diagram and the population has few clear boundaries. If we look at the variation of effective surface gravity with nitrogen enrichment the difference between the models becomes very clear (Fig.~\ref{ch4.fig.gn}). There are sharp curves that bound the upper and lower effective surface gravities of the population. The lower bound occurs because stars evolve rapidly into giants with much lower surface gravity after this limit. The upper bound occurs because younger stars with higher surface gravities haven't evolved to the point where their surface nitrogen is enriched. There are features which distinguish the two populations at each age. For young populations ($5\times 10^6{\rm \,yr}$) case~2 has a higher upper bound for nitrogen enrichment and there is a much broader range of surface gravities than in case~1. For older populations ($10^7{\rm \,yr}$ and $2\times 10^7{\rm \,yr}$) case~2 predicts generally lower values for the surface gravity. Finally for old populations ($5\times 10^7{\rm \,yr}$) the difference becomes very stark. The amount of mixing in case~2 drops off dramatically compared to case~1 while we still predict much lower values for the surface gravity in rapid rotators.

\begin{figure}
\begin{center}
\vspace{3cm}
\includegraphics[width=0.65\textwidth]{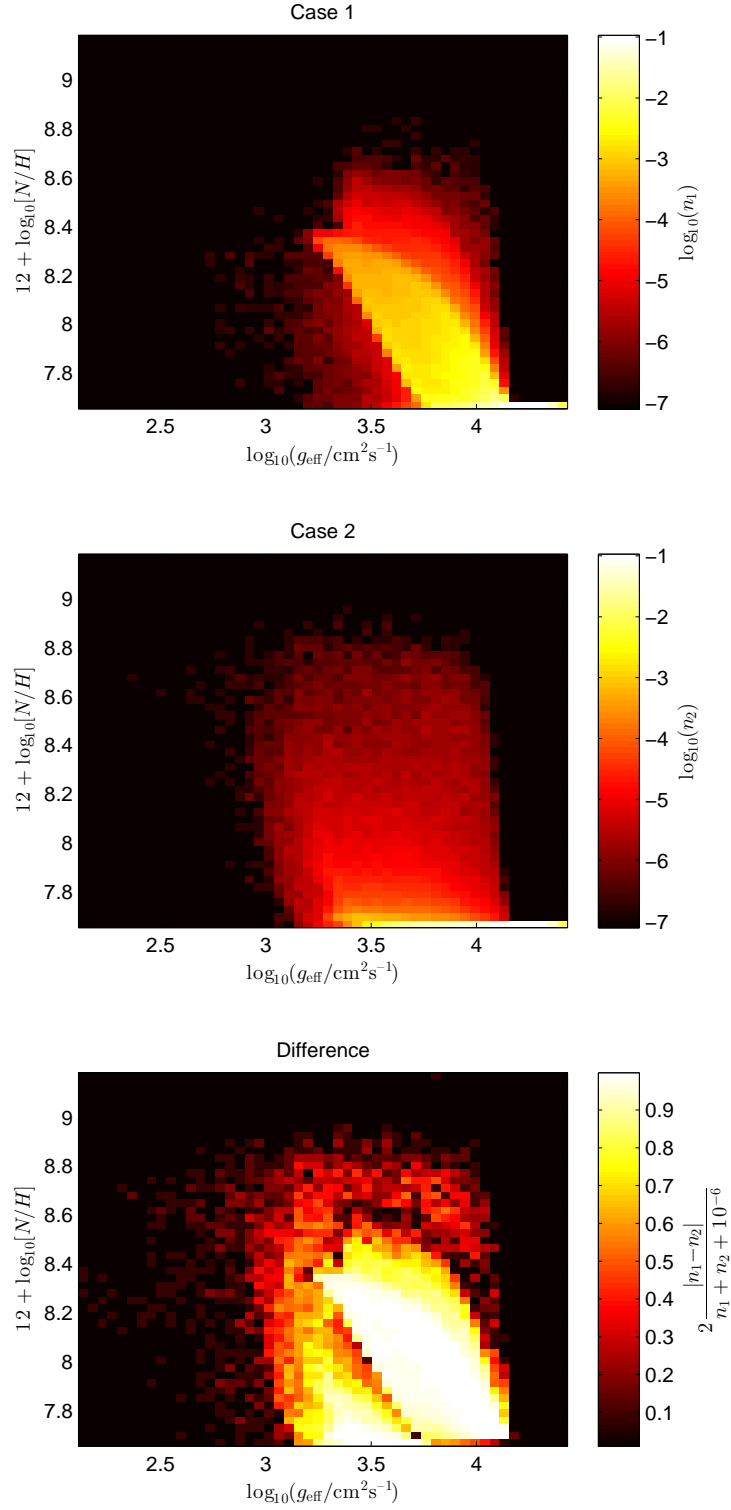}
\end{center}
\caption[Surface nitrogen enrichment against effective surface gravity of simulated populations of stars with continuous star formation for different physical models]{Surface nitrogen enrichment against effective surface gravity of simulated populations of stars with continuous star formation. The distributions are both confined to a narrow band and have similar ranges for enrichment, though slightly higher in case~2. However, case~1 produces many more enriched stars than case~2 when they are for the large part confined to a narrow band. There are some edge--of--grid effects that arise because the initial rotation function is mass independent and so produces many more low--mass stars close to their critical limit.}
\label{ch4.fig.contgn}
\end{figure}

When we consider a population of stars with continuous star formation history, the difference in the populations is still clear. Interestingly, unlike the Hunter diagram, this visualisation actually highlights the similarities between the two cases as well the differences. Fig.~\ref{ch4.fig.contgn} shows that stars in both models are confined to a similar band of effective gravities and their range of surface abundances are very similar. The main difference between the two cases, apart from the increased frequency of enriched stars in case~1 which we saw in section~\ref{ch4.sec.hunter}, is the confinement of the enriched case--1 stars to a distinct band. This contrasts to case~2 for which the stars are spread much more evenly across their range of enrichment.

\subsection{Recalibration}
\label{ch4.sec.calibration}

\begin{figure}
\begin{center}
\vspace{3cm}
\includegraphics[width=0.65\textwidth]{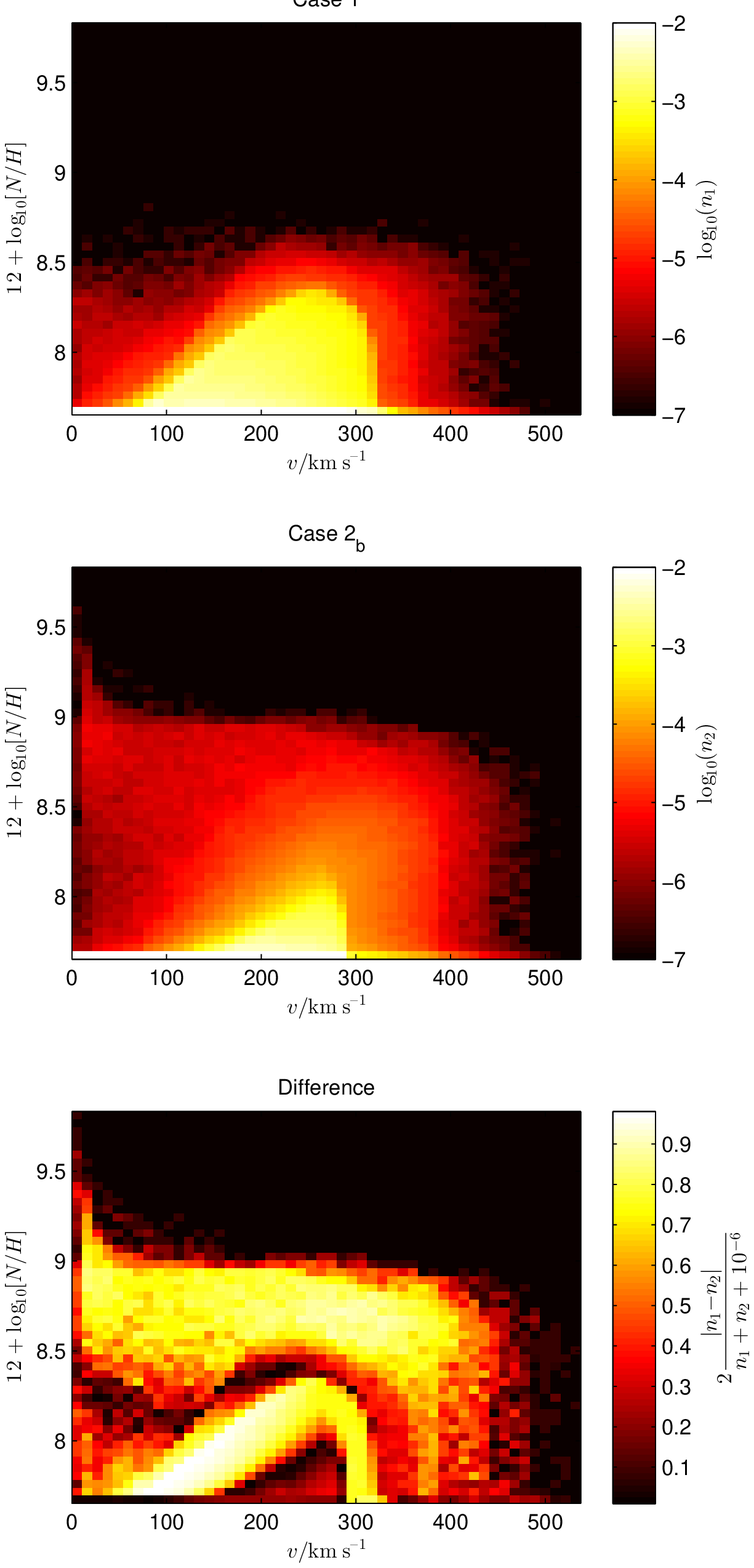}
\caption[Hunter diagrams for simulated stellar populations in case~2$_{\rm b}$]{Hunter diagrams for a population of stars undergoing continuous star formation with case~2 calibrated to give the same TAMS nitrogen enhancement as case~1 for a star of mass $10$\ms\ initially spinning with $v=200\,{\rm km\,s^{-1}}$ (case~2$_{\rm b}$). There are now more moderately enriched stars in case~2$_{\rm b}$ but still far fewer than in case~1 and the upper limit for enrichment in case~2$_{\rm b}$ is now far greater than case~1.}
\label{ch4.fig.huntcontb}
\end{center}
\end{figure}

We have thus far described the differences that arise between the two test cases under a specific calibration of the mixing. However, within each case is the flexibility to calibrate to some degree the amount of mixing that arises because of rotation. We chose in our initial calibration to match the TAMS nitrogen enrichment of $20$\ms$\ $ stars initially rotating at $v=300\,{\rm km\,s^{-1}}$. This is a reasonably good fit for stars of $M>15$\ms\ but for smaller masses the amount of mixing in case~2 drops off rapidly. Now suppose instead we had chosen to match the TAMS nitrogen enrichment of a star with initial mass $M=10$\ms\ and $v=200\,{\rm km\,s^{-1}}$. We refer to this model as case 2$_{\rm b}$. This is more representative of the stars observed in the VLT--FLAMES survey \citep{Dufton06} and so should produce mixing in line with the bulk of the population. We discuss the VLT--FLAMES survey data in relation to our simulated populations in section~\ref{ch4.sec.flames}.  Fig.~\ref{ch4.fig.huntcontb} shows the Hunter diagram for the new sample. We see that the agreement is better in the Hunter diagram but case~2$_{\rm b}$ still can't produce the tightly confined bulk of enriched stars seen in case~1. Also, the maximum enrichment observed in case~2$_{\rm b}$ is now far greater than in case~1.

\begin{figure}
\begin{center}
\hspace*{-2.2cm}\includegraphics[width=1.2\textwidth]{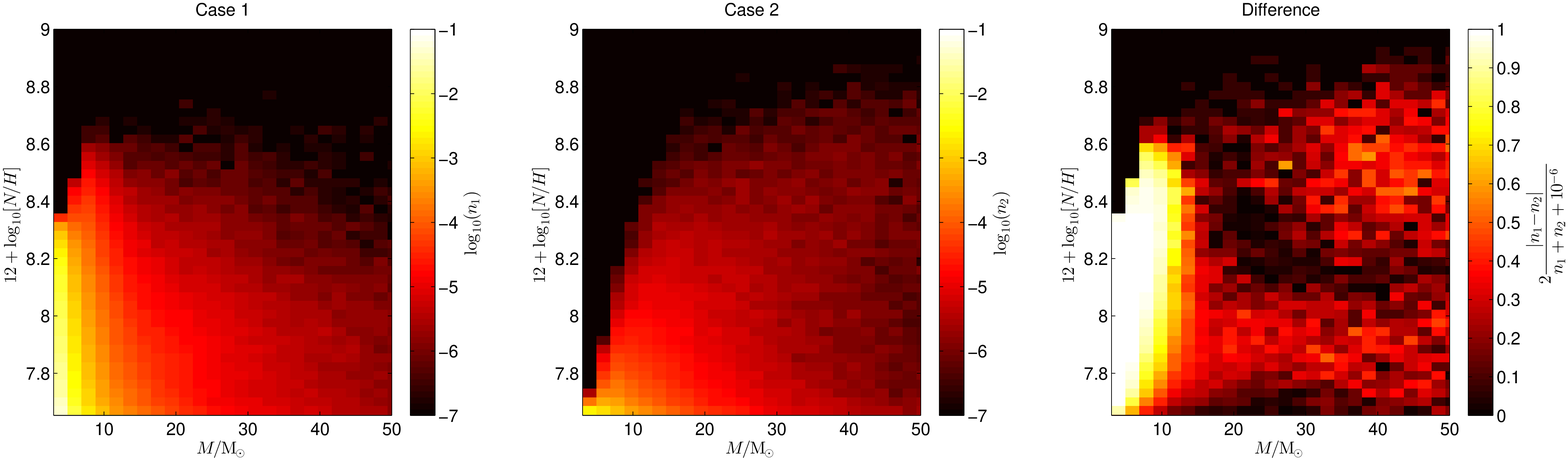}
\hspace*{-2.2cm}\includegraphics[width=1.2\textwidth]{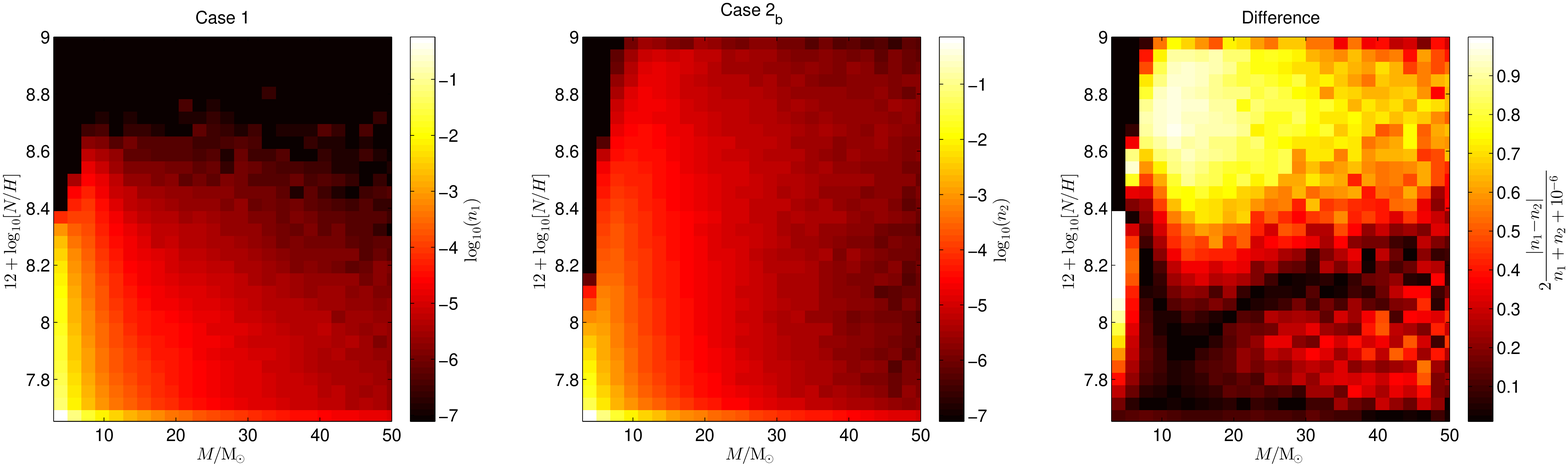}
\end{center}
\caption[Surface nitrogen enrichment against mass for simulated stellar populations with continuous star formation for different physical models]{Distribution of the masses of stars in case~1 and case~2 for a population with continuous star formation history. The top row shows the models with the calibration of case~2 to give the same TAMS nitrogen enhancement as case~1 for a star of $20$\ms\ initially spinning with $v=300\,{\rm km\,s^{-1}}$. The second row shows the same but with case~2 calibrated to give the same TAMS nitrogen enrichment as as star with initial mass $M=10$\ms$\ $ and surface rotation $v=200\,{\rm km\,s^{-1}}$ (case~2$_{\rm b}$). The two cases agree for masses greater than  $15$\ms$\ $ for the original calibration. The agreement continues to lower masses for the second calibration but now there are more highly enriched stars in case~2$_{\rm b}$ in the mass range $5$\ms$<M<20$\ms.}
\label{ch4.fig.mn}
\end{figure}

Although we can't directly measure the mass of stars, it is instructive to examine where the main differences in our sample arise. Fig.~\ref{ch4.fig.mn} shows the distribution of nitrogen enrichment by mass in case~1, case~2 and case~2$_{\rm b}$. In the first instance, the agreement between the two models is reasonable for stars more massive than around $15$\ms$\ $ but the mixing in case~2 drops off rapidly for lower masses as we observed in section~\ref{ch4.sec.grav}. For case~2$_{\rm b}$ the agreement holds to much lower masses except now there are far more highly enriched stars with mass $5$\ms$<M<20$\ms\ when compared to case~1.

\subsection{Effects of metallicity}
\label{ch4.sec.metal}

\begin{figure}
\begin{center}
\vspace{3cm}
\includegraphics[width=0.65\textwidth]{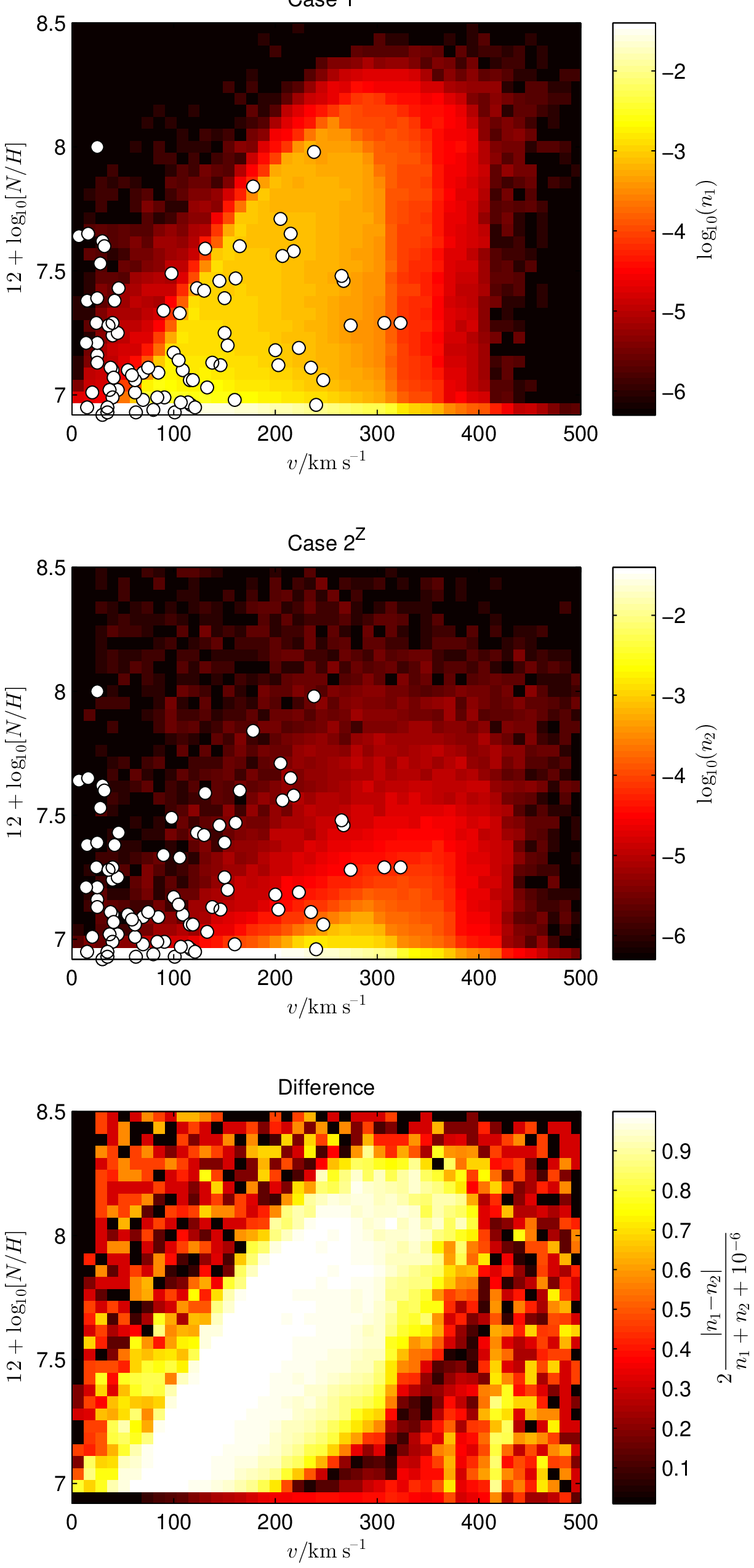}
\end{center}
\caption[Hunter diagrams for simulated stellar populations at LMC metallicity]{Hunter diagrams for populations of stars undergoing continuous star formation at LMC metallicity. The two populations are qualitatively similar to the populations at solar metallicity. We have also plotted the LMC stars observed by \citet{Hunter09}. Case~1$^{\rm Z}$ predicts a very confined distribution of enriched stars, whereas case~2$^{\rm Z}$ predicts a much wider spread of enrichment. The stars at the left--hand edge of the diagram can't be explained by rotational mixing alone.}
\label{ch4.fig.huntcontz}
\end{figure}

In chapter~\ref{ch3} we concluded that there was significant variation in the differences between the two cases at different metallicities. We simulated a grid of models at LMC metallicity, composition and initial velocity distribution, as given by \citet{Brott11}. This is not as low as the low--metallicity case described in chapter~\ref{ch3} but it does allow us to compare our results with the data from the LMC observations in the VLT--FLAMES survey of massive stars. We simulated a population in case~1$^{\rm Z}$ and case~2$^{\rm Z}$ at this composition with continuous star formation history. The Hunter diagram for this population is shown in Fig.~\ref{ch4.fig.huntcontz}.

We see that the qualitative distribution of stars in the simulated population is similar to the solar metallicity populations. Case~1$^{\rm Z}$ produces a much more well--defined band of enriched stars whereas case~2$^{\rm Z}$ produces fewer enriched stars that have a much greater spread in abundance. If we consider the mass--dependence of the rotational mixing we find a similar decline in the amount of mixing in case~2$^{\rm Z}$ compared to case~1$^{\rm Z}$ for stars less massive than $20$\ms. As in chapter~\ref{ch3}, we find that the amount of mixing in stars above this mass is higher in case~2$^{\rm Z}$ than in case~1$^{\rm Z}$. In fact, the mixing in case~1$^{\rm Z}$ decreases slightly for higher--mass stars. This means that, as in section~\ref{ch4.sec.calibration}, an increase in the mixing in case~2$^{\rm Z}$ is unlikely to produce a better correlation between the two cases. However, owing to the IMF, there are many more stars less massive than $20$\ms\ in the population and so case~1 produces many more enriched stars than case~2$^{\rm Z}$.

In Fig.~\ref{ch4.fig.huntcontz} we have also plotted those LMC stars for which the nitrogen abundances have been determined by \citet{Hunter09}. As remarked by \citet{Hunter09}, there are many highly enriched, slowly rotating stars that are not explained by either model of rotational mixing. These are addressed in chapter~\ref{ch5}. For the remainder of observed stars we see a trend of increasing enrichment for higher rotation rates. On initial inspection, case~1$^{\rm Z}$ fits the VLT--FLAMES data much more closely than case~2$^{\rm Z}$. However, selection effects are important and we consider these in section~\ref{ch4.sec.flames}.

\subsection{Selection effects in the VLT--FLAMES survey}
\label{ch4.sec.flames}

\begin{figure}
\begin{center}
\vspace{3cm}
\includegraphics[width=0.65\textwidth]{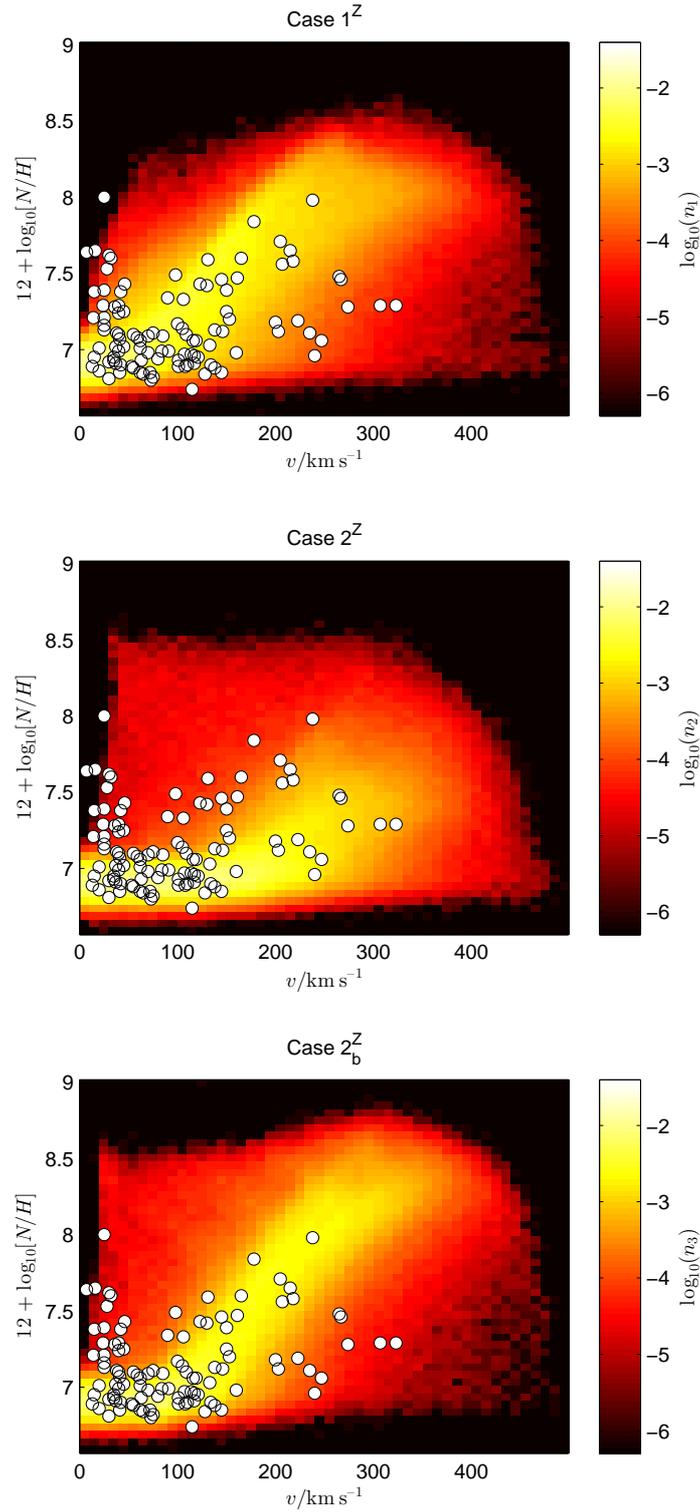}
\end{center}
\caption[Hunter diagrams for simulated stellar populations undergoing continuous star formation with different physical models]{Hunter diagrams for populations of stars undergoing continuous star formation at LMC metallicity for a number of different models with selection effects applied as described in section~\ref{ch4.sec.flames}. Populations have been simulated for cases~1$^{\rm Z}$, 2$^{\rm Z}$ and~2$_{\rm b}^{\rm Z}$. The populations are qualitatively similar to the populations at solar metallicity. We have also plotted the LMC stars observed by \citet{Hunter09}. The slowly--rotating, highly--enriched stars at the left--hand edge of the diagram cannot be explained by rotational mixing alone.}
\label{ch4.fig.huntcontobsz}
\end{figure}
\begin{figure}
\begin{center}
\includegraphics[width=0.99\textwidth]{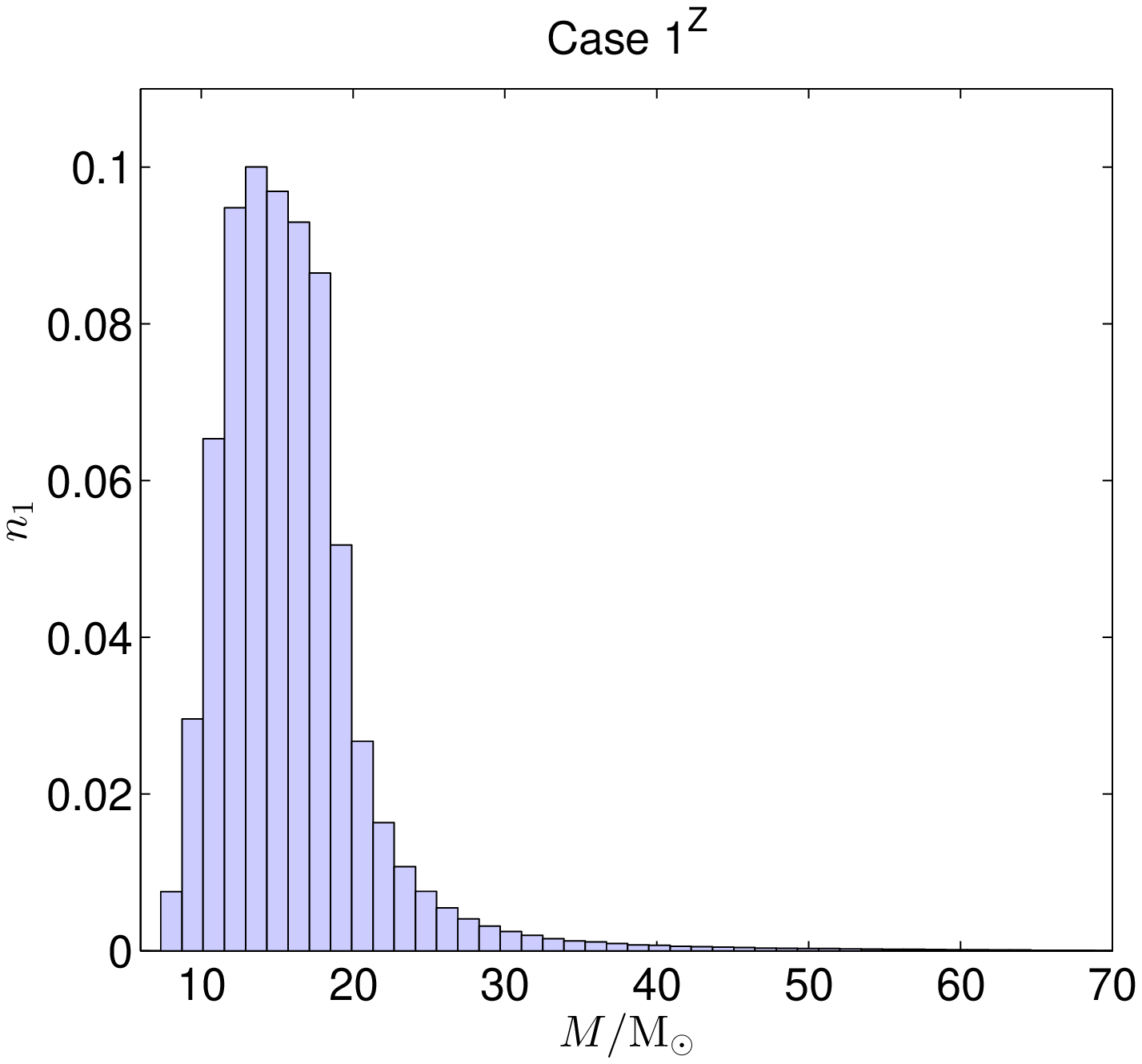}
\end{center}
\caption[Distribution of masses in a simulated stellar population at LMC metallicity]{Distribution of masses in a simulated population of case--1$^{\rm Z}$ stars at LMC metallicity with the inclusion of selection effects. The distribution is very similar in each case. We see that the distribution peaks strongly around $12$\ms.}
\label{ch4.fig.masscontobsz}
\end{figure}

The most common data set used to test rotational mixing in massive stars is the VLT--FLAMES survey of massive stars \citep{Evans05,Evans06,Dufton06} owing to the number of stars sampled and the detailed determination of surface composition. We repeated our population synthesis as in section~\ref{ch4.sec.metal} for a continuous population of LMC--metallicity stars but we have included the selection effects which affect the stars in the VLT--FLAMES survey so that we may compare the distributions more directly with those found by \citet{Hunter09}. The selection criteria we used are that of the cluster N11 in the LMC. For a detailed description see \citet{Brott11}. Stars are excluded if their visual magnitude is greater than $15.34$, if they are hotter than 35,000\,K, if their surface gravity is less than $10^{3.2}{\,\rm cm\,s^{-2}}$ or they are rotating faster than $90\%$ of their critical rotation rate. In addition, a random error is applied to $\log_{10}[N/H]$ selected randomly from a Gaussian with standard deviation $\sigma = 0.2$.

The simulated population produced after we apply the selection effects is shown in Fig.~\ref{ch4.fig.huntcontobsz} along with the LMC stars data of \citet{Hunter09}. We show the population in cases~1$^{\rm Z}$, 2$^{\rm Z}$ and~2$_{\rm b}^{\rm Z}$ (the LMC analogue of case~2$_{\rm b}$ with the calibration described in section~\ref{ch4.sec.calibration}). We see that, contrary to our discussion in section~\ref{ch4.sec.metal}, the differences between the various models are now far less apparent. Compared with the other two cases, case~2$^{\rm Z}$ predicts many more less--enriched fast rotators than are observed and the amount of mixing is insufficient to match the observed band of enriched stars. Compared with section~\ref{ch4.sec.metal}, cases~1$^Z$ and~2$^Z_b$ now both show a similarly good fit to the data. In both cases the band of predicted enriched stars is matched well by the observations. However, we note that we would expect to see a number of stars with $12+\log_{10}[N/H]>8$. The number of predicted stars in this range is greater for case~2$_{\rm b}^{\rm Z}$ than case~1$^Z$. If we were to reduce the amount of mixing we would get too little nitrogen enrichment in stars with $v<200{\rm \,km\,s^{-1}}$. This effect may be a result of the difficulty of measuring nitrogen abundances in this region but is otherwise difficult to resolve. A further increase in the mixing would exacerbate the problem that the upper bound to enrichment at rapid rotation is too high. A decrease in the mixing would mean that the band of enriched stars in the simulated population is less likely to produce a good fit to the model and would leave observed slowly--rotating, moderately--enriched stars that cannot be explained through the theoretical models.

Unfortunately, the mass dependence of the models is not well reflected in the VLT--FLAMES populations. With the selection effects described, the masses of the LMC--metallicity sample are confined between  $10$\ms\ and  $20$\ms\ as shown in Fig. \ref{ch4.fig.masscontobsz}. We find a similarly narrow range when we consider Galactic stars under similar selection effects. Therefore, any simulated population where we include the selection effects of VLT--FLAMES captures the rotational--dependence of the model but for only a small fraction of the mass--dependence which is where we have found the biggest differences between our two cases.

\section{Conclusions}
\label{ch4.sec.conclusions}

Rotation has many effects on stellar evolution. Some of these, such as that on the surface temperature, are because of rotation but only vary significantly between different models towards the end of the main sequence. Others, such as that on the surface rotation velocity, may evolve differently throughout the main sequence according to different models but do not produce significant changes in the distribution of stars in a simulated population. Therefore these properties alone are largely unhelpful to distinguish between the different implementations of rotation in stellar models. 

It has been observed that the surface abundances of several chemical elements change significantly because of rotation. The degree to which this happens in the theoretical models strongly depends on which particular model for stellar rotation is used and which constraints are used to calibrate them. Recently, the Hunter diagram has been the favoured diagnostic tool for analysing stellar rotation because it shows a clear connection between the surface rotation of a star and its surface enrichment. Our model based on that of \citet{Talon97}, case~1, shows a similar order of magnitude enrichment for all masses. On the other hand our model based on that of \citet{Heger00}, case~2, shows a steep decline in the amount of enrichment around $15$\ms. We can account for this to a degree by adjusting the calibration of the models but we see that increasing the mixing in case~2, so that low--mass stars show similar enrichment to those of case~1, then leads to a much higher maximum enrichment for a case--2 population than a case--1 population. This suggests that studies should focus on stars either side of this mass limit.

Because the two models are very different for different mass ranges, the effective gravity is a sensible tool to investigate the mass--dependence of the mixing strength. It is very difficult to self--consistently infer the mass of a star from a luminosity--temperature--rotation relation because any such relation depends on the model used for rotation. The effective gravity however can be derived directly from spectra and, despite some degeneracy, there is a strong relation between it and the stellar mass. This can be usefully applied to the study of rotational mixing. We have shown that there are very clear differences between the synthetic populations produced by each model when nitrogen enrichment is plotted against surface gravity. In the case of a population in which all the stars have the same age, they are confined to a very specific region. Stars with surface gravity below a certain limit evolve into giants and beyond on a short timescale compared to their main--sequence lifetime. Stars with surface gravity above another limit are less massive and have not had long enough to become enriched. This effect persists even in the case of a population of stars with continuous star formation. Case~1 predicts that nitrogen is enriched over a narrower range of surface gravities than case~2. We see that case~1 produces many more moderately enriched stars than case~2 but the maximum enrichment in case~1 is lower than in case~2.

We have also shown that similar trends appear at different metallicities. We simulated populations in both cases with continuous star formation history at LMC metallicity. The qualitative distribution of stars in each case was similar to that at Galactic metallicity. Case~1$^{\rm Z}$ still produced a confined band of enriched stars in the Hunter diagram, whereas case~2$^{\rm Z}$ produced a much greater spread. Case~1$^{\rm Z}$ also produced many more highly--enriched stars than case~2$^{\rm Z}$ although the maximum enrichment in case~2$^{\rm Z}$ was higher than in case~1$^{\rm Z}$. Similarly to solar metallicity, this is because of the mass dependence of the mixing in each case. The decline in the amount of mixing in case~2$^{\rm Z}$ begins at higher masses at low metallicity ($20$\ms\ at LMC rather than $15$\ms\ at solar metallicity) and the mixing in stars above this mass is relatively constant for stars in case~2$^{\rm Z}$, whereas case--1$^{\rm Z}$ stars show slightly less enrichment as the mass of stars increases. This may not be indicative of the strength of the rotational mixing but more that, because the main--sequence lifetime of the stars decreases with increasing mass, there is less time to transport nitrogen from the core to the surface. 

When we compare the simulated populations to the LMC--metallicity stars in the VLT--FLAMES survey we find that both cases~1$^{\rm Z}$ and~2$_{\rm b}^{\rm Z}$, which uses the second calibration for case~2 described in section~\ref{ch4.sec.calibration}, give a reasonable fit to the observed data and it is difficult to determine which fits the data more closely. Unfortunately the range of initial masses that remain in the simulated samples after we apply the selection effects is extremely narrow, between $10$\ms\ and $20$\ms\ in the LMC populations. This means that it is difficult to observe the large difference in the mass--dependence of the models. Hence, a close fit between the Hunter diagrams for a simulated population and the data of the VLT--FLAMES survey is a useful test of models for rotational mixing but cannot establish the validity of a model by itself. As we have seen, models for rotational mixing with very different mass--dependencies can reproduce similarly good fits to the current data. Whilst, in the future, determinations of the nitrogen abundance in a wider mass--range of stars may help solve this problem, it is likely that examining the enrichment and depletion of other elements will be necessary. For example, \citet{Brott11b} look at the effect of rotation on the surface Boron abundance. Different initial rotation velocity distributions may also have a significant effect on the simulated populations.

Despite sharing many similar features, it is unreasonable to expect that two different models for stellar rotation can produce identical qualitative results for an extended range of masses, rotation rates and metallicities. We have shown that, whilst the two models agree for stars more massive than $15$\ms, there is much less agreement for less massive stars. Furthermore, we have only thus far made a comparison of two particular models for stellar rotation chosen from the many available. In particular we haven't included models based around the Taylor--Spruit dynamo \citep{Spruit02} such as that investigated by \citet{Brott11} up to this point. Whilst this is a similar model to that of \citet{Heger00} it produces very different results owing to the inclusion of magnetic fields. We examine the effect of magnetic fields in chapter~\ref{ch5}. Whilst the marginally better fit of case~1 suggests that models in which meridional circulation is treated advectively and diffusion comes solely from hydrodynamic instabilities are more realistic, the difference is still not convincing and the results of \citet{Brott11} also produce a reasonable fit to observed LMC stars. To distinguish between these models, and others, it is necessary to continue with this analysis and extend it to different masses and metallicities as more data becomes available.

\begin{savequote}[60mm]
Magnetism, as you recall from physics class, is a powerful force that causes certain items to be attracted to refrigerators. (Dave Barry)
\end{savequote}

\chapter[Stellar evolution with an alpha--omega dynamo]{Stellar evolution of massive stars with a radiative alpha--omega dynamo}
\label{ch5}

\section{Introduction}
\label{ch5.sec.introduction}

The study of rotation in the radiative zones of stars is strongly coupled with the evolution of magnetic fields. Observation of stellar magnetic fields is difficult but a number of magnetic O and B stars have been discovered \citep{Donati01,Donati02,Neiner03,Donati06,Donati06b,Grunhut11}. Combined with this, a number of chemically peculiar A and B stars (known as Ap and Bp stars respectively) with surface field strengths up to $20$\,kG have been identified \citep[e.g.][]{Borra78,Bagnulo04,Hubrig05}. We direct the reader to \citet{Mathys09} for a review. These large--scale fields tend to have simple geometries and there is debate over whether they arise from fossil fields present during a star's formation \citep{Cowling45, Alecian08} or from a rotationally--driven dynamo operating in the radiative zone of the star \citep{Spruit99,Maeder04}. In this chapter we focus on the latter but we give consideration to whether a fossil field can be sustained throughout the stellar lifetime. 

In low--mass stars, where the outer region is convective, magnetic fields are expected to be formed in a strong shear layer at the base of the convection zone and then transported to the surface by convection and magnetic buoyancy \citep{Nordhaus10}. In radiative zones there is no strong bulk motion to redistribute magnetic energy. In most dynamo models, magnetic flux is redistributed by  magneto--rotational turbulence \citep{Spruit02}. This turbulence is also responsible for driving the generation of large--scale magnetic flux. This is the $\alpha$--effect \citep[e.g.][]{Brandenburg01} which applies to both poloidal and toroidal components, although in rotating systems shear is generally more effective at producing toroidal field from the poloidal component and so the $\alpha$--effect is needed for the poloidal field only. The toroidal field is instead maintained by the conversion of poloidal field into toroidal field by differential rotation. This is commonly referred to as an $\alpha$--$\Omega$ dynamo \citep{Schmalz91}.

Because observed fields are potentially strong enough to affect chemical mixing and angular momentum transport, their inclusion in stellar evolution models is essential. Rotation itself is a likely candidate to drive dynamo mechanisms within a star and theoretical models \citep[e.g.][]{Spruit99} have predicted magnetic fields that can produce turbulent instabilities which dominate the transport of angular momentum. Whilst the purely hydrodynamic evolution of the angular momentum distribution in main--sequence stars has been considered extensively in the framework of one--dimensional stellar evolution calculations \citep[e.g.][]{Meynet00,Heger00}, magnetic fields have received far less attention \citep{Maeder04, Brott11b}. The evolution of the angular momentum distribution and magnetic field strength have a significant effect on the final fate of a star and its ejecta. 

Apart from causing chemical mixing, sufficiently strong magnetic fields are expected to cause magnetic braking that results in the rapid spin down of rotating magnetic stars \citep{Mathys04}. It has been suggested that magnetic fields might explain the existence of slowly--rotating, chemically peculiar stars in surveys of rotating stars \citep{Hunter09}. We include a model for magnetic braking based on that of \citet{ud-Doula02} and show the effects it has on the models of magnetic stars.

Many studies of magnetic fields in massive main--sequence stars consider the Tayler--Spruit dynamo mechanism \citep{Spruit02}. This model asserts that pinch--type instabilities \citep{Tayler73, Spruit99} arise in toroidal fields that drive magnetic turbulence that enforces solid--body rotation. The growth of instabilities is controlled by magnetic diffusion which ultimately determines the equilibrium strength of the field. This idea was built upon by \citet{Maeder04} who found that the Tayler--Spruit dynamo did indeed result in far less differential rotation than in solely hydrodynamic models. It was also incorporated in the work of \citet{Brott11b} who compared stellar evolution calculations based on the Tayler--Spruit dynamo with the data from the VLT--FLAMES survey of massive stars \citep[e.g.][]{Evans05,Evans06}. They found reasonable agreement between the observed and simulated samples \citep{Brott11}. However, in chapter~\ref{ch4} we found equally good agreement between the data from the VLT--FLAMES survey and purely hydrodynamic models based on models of \citet{Heger00} and \citet{Meynet00}.

In the models of of \citet{Spruit02} and \citet{Maeder04}, the magnetic field is purely a function of the stellar structure and rotation. Whilst it feeds back on the system via turbulent diffusivities, the magnetic field doesn't appear as an independent variable within the system. In this work we have continued along similar lines to \citet{Spruit02} but have developed a magnetic model where the poloidal and toroidal components are evolved via advection--diffusion equations derived from the induction equation. These are similar in form to the angular momentum evolution equation. The magnetic field and angular momentum evolution are coupled by turbulent diffusivities, magnetic stresses and conversion of poloidal field into toroidal field by differential rotation. The dynamo is completed by regeneration of magnetic flux by a simple $\alpha$--$\Omega$ dynamo. We look at how the predicted surface magnetic field varies with age and rotation rate for a range of initial masses and how a simulated population of magnetic stars compares to the data from the VLT--FLAMES survey of massive stars \citep{Dufton06}. We also consider how our model behaves with a strong initial fossil field but without the action of a dynamo.

In section~\ref{ch5.sec.model} we briefly review the model we use to simulate the magnetic fields including the equations for the $\alpha$--$\Omega$ dynamo and magnetic braking. For the full details we direct the reader to section~\ref{ch2.sec.magnetism}. In section~\ref{ch5.sec.results} we look at the predictions of the model for a range of stellar masses and initial rotation rates and how simulated populations compare with observations and in section~\ref{ch5.sec.conclusions} we present a discussion of the results and our conclusions.

\section{Magnetic rotating model}
\label{ch5.sec.model}

In order to simulate the magnetic field in stellar interiors we build on the code {\sc rose} \citep{Potter11} described in section~\ref{ch2.sec.rose}. We briefly review our model for magnetic fields in rotating stars in this section. For full details we refer the reader to section~\ref{ch2.sec.magnetism}. We assume the magnetic field, $\bi{B}$ takes the form

\begin{equation}
\label{ch5.eq.b}
\bi{B}=B_{\phi}(r,\theta)\bi{e}_{\phi}+\nabla\times(A(r,\theta)\bi{e}_{\phi}).
\end{equation}

\noindent By substituting this into the induction equation (c.f. equation~(\ref{ch2.eq.induction})) and performing a suitable spherical average we get the equations

\begin{equation}
\label{ch5.eq.toroidal}
\diffb{B_{\phi}}{t}=A\diffb{\Omega}{r}-\frac{6}{5r}VB_{\phi}-\frac{1}{10r}UB_{\phi}+r\diffb{}{r}\left(\frac{\eta}{r^4}\diffb{}{r}(r^3B_{\phi})\right)
\end{equation}
\noindent and
\begin{equation}
\label{ch5.eq.poloidal}
\diffb{A}{t}=\frac{3V}{2r}A-\frac{U}{8r}\diffb{}{r}(Ar)+\frac{8\alpha}{3\pi} B_{\phi}+\diffb{}{r}\left(\frac{\eta}{r^2}\diffb{}{r}(r^2A)\right),
\end{equation}

\noindent where the $\alpha$ term represents the regeneration of the poloidal field by magnetic turbulence. The turbulent diffusion coefficient $D_{\rm mag}$ represents transport of magnetic flux by turbulence owing to magnetic instabilities. With the inclusion of magnetic forces, the angular momentum evolves according to

\begin{equation}
\diff{r^2\Omega}{t}=\frac{1}{5\rho r^2}\diff{\rho r^4\Omega U}{r}+\frac{1}{4\pi r}\diffb{}{r}\left(rAB_{\phi}\right)+\frac{1}{\rho r^2}\frac{\partial}{\partial r}\left(\rho D_{\rm tot} r^4\frac{\partial\Omega}{\partial r}\right).
\end{equation}

\noindent The diffusion coefficient $D_{\rm tot}$ includes the diffusion of angular momentum by Kelvin--Helmholtz instabilities and convection as discussed in chapters~\ref{ch3} and~\ref{ch4} as well as additional turbulent transport as a result of magnetic instabilities. The magnetic diffusivity, $\eta$, is

\begin{equation}
\label{ch5.eq.eta}
\eta= r^2\Omega\left(\frac{\omega_A}{\Omega}\right)^2\left(\frac{\Omega}{N}\right)^{1/2}\left(\frac{K}{r^2 N_T}\right)^{1/2}.
\end{equation}

\noindent where $\omega_A$ is the Alfv\'{e}n frequency and $N$ is the Brunt--V\"{a}is\"{a}l\"{a} and $K$ is the thermal diffusivity. The chemical composition of the star evolves in radiative zones according to the equation

\begin{equation}
\label{ch5.eq.chemtransport}
\frac{\partial X_i}{\partial t}=\frac{1}{r^2}\frac{\partial}{\partial r}\left({\rm Pr_c}D_{\rm tot}r^2\frac{\partial X_i}{\partial r}\right),
\end{equation}

\noindent where ${\rm Pr_c}$ is the chemical Prandtl number. Similarly we take the magnetic diffusivity to be $\eta={\rm Pr_m}\,D_{\rm mag}$ where ${\rm Pr_m}$ is the magnetic Prandtl number.

The dynamo efficiency is given by

\begin{equation}
\label{ch5.eq.alpha}
\alpha=\gamma\frac{r\omega_A\Omega q}{N},
\end{equation}

\noindent where $q=\diffb{\ln \Omega}{\ln r}$. Owing to magnetic braking, angular momentum is lost from the star at a rate given by

\begin{equation}
\difft{H_{\rm tot}}{t}=4\pi R^4\rho D_{\rm tot}\left(\diffb{\Omega}{r}\right)_R,
\end{equation}

\noindent where $\difft{H_{\rm tot}}{t}$ is the total rate of angular momentum loss from the star and is given by

\begin{equation}
\difft{H_{\rm tot}}{t}=R_A^2\Omega\dot{M}=\sigma^2J_{\rm surf}.
\end{equation}

\noindent The Alfv\'{e}n radius is $R_A$, $\sigma=\frac{R_A}{R}$ and $J_{\rm surf}$ is the specific angular momentum at the surface of the star. The parameter $\sigma$ is evaluated by

\begin{equation}
\label{ch5.eq.sigma2}
\sigma^2=\sqrt{\frac{B_*^2R^2}{\dot{M}v_{\rm esc}}}.
\end{equation}

There are four free parameters in this model. The parameter $C_{\rm m}$ affects the overall strength of the turbulent diffusivity. The magnetic and chemical Prandtl numbers, ${\rm Pr_m}$ and ${\rm Pr_c}$ respectively, define the ratio between the diffusion coefficient for angular momentum transport and the diffusion coefficients for the magnetic field and chemical transport. Finally, $\gamma$ affects the efficiency of the magnetic dynamo.

\section{Results}
\label{ch5.sec.results}

\begin{figure}
\begin{center}
\includegraphics[width=0.99\textwidth]{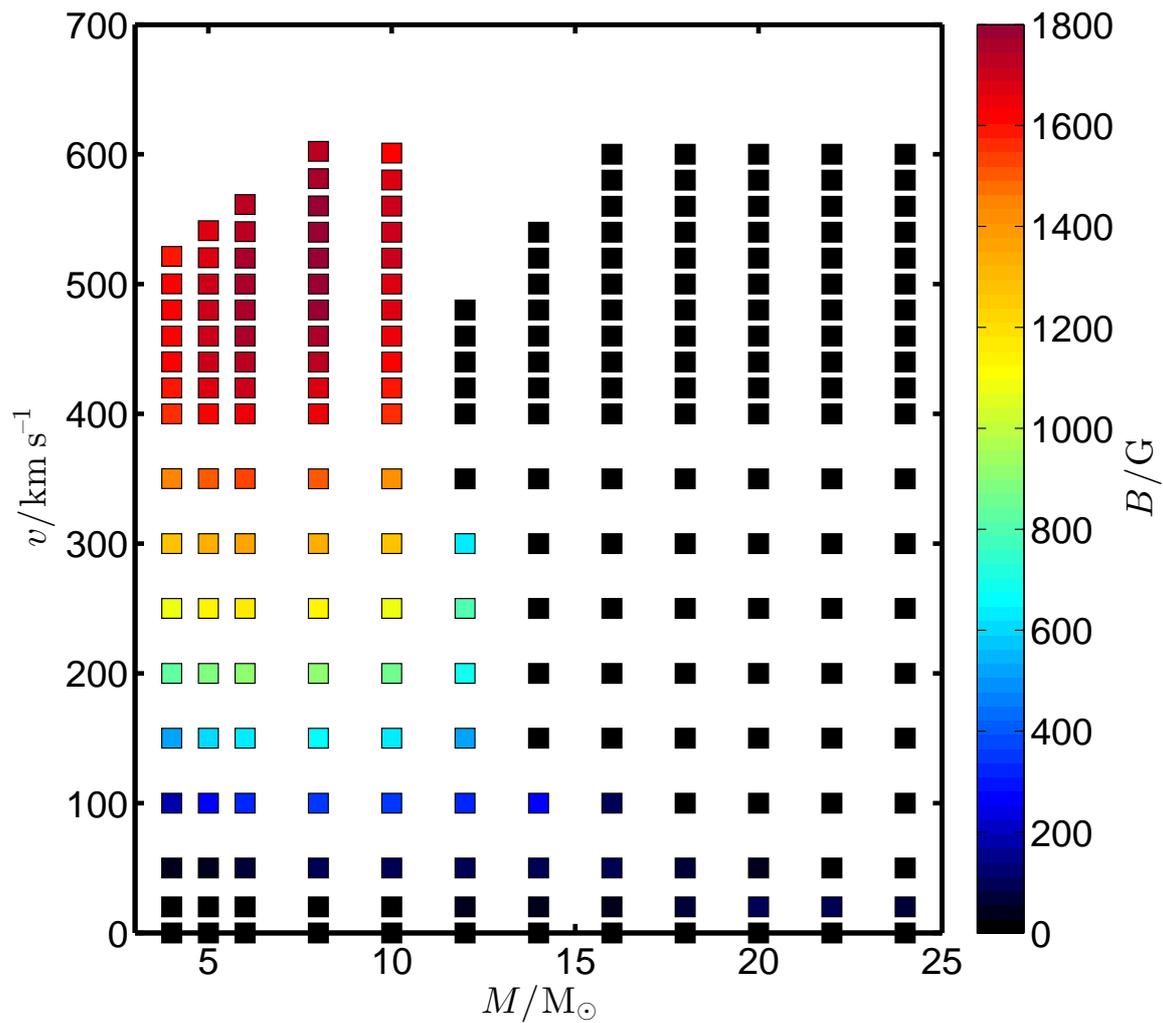}
\end{center}
\caption[Grid of models used in section~\ref{ch5.sec.results}]{Grid of models considered in section~\ref{ch5.sec.results}. The colour of each point indicates the surface field strength at the zero--age main sequence.}
\label{ch5.fig.grid}
\end{figure}

We simulated a grid of models with masses $4<M/$\ms$<24$ and initial rotation rates $0<v_{\rm ini}/{\rm km\,s^{-1}}<600$, except where the initial rotation rate is greater than the critical rotation rate of the star. All of the models described are at LMC metallicity as used by \citet{Brott11b}. We set $C_{\rm m}=1$ and ${\rm Pr_m}=1$. We also set $\gamma = 10^{-15}$ which results in a maximum field strength across the whole population of $B\approx 20\,{\rm kG}$. The maximum terminal--age main--sequence (TAMS) nitrogen enrichment in the simulated magnetic population, including observational constraints, is matched with the maximum enrichment in the slowly rotating population of \citet{Hunter09}. This gives ${\rm Pr_c}=0.01$. In each model the rotation and magnetic field were allowed to relax to equilibrium at the zero--age main sequence (ZAMS). The grid of initial models is shown in Fig.~\ref{ch5.fig.grid} which also shows the ZAMS surface field strength in each model. We will look at this in more detail in section~\ref{ch5.sec.mvb}.

\subsection{Magnetic field evolution}
\label{ch5.sec.evolution}

\begin{figure*}
\begin{center}
\includegraphics[width=0.99\textwidth]{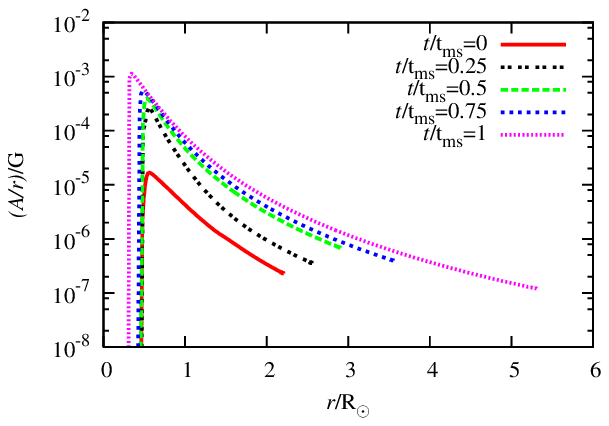} \\\includegraphics[width=0.99\textwidth]{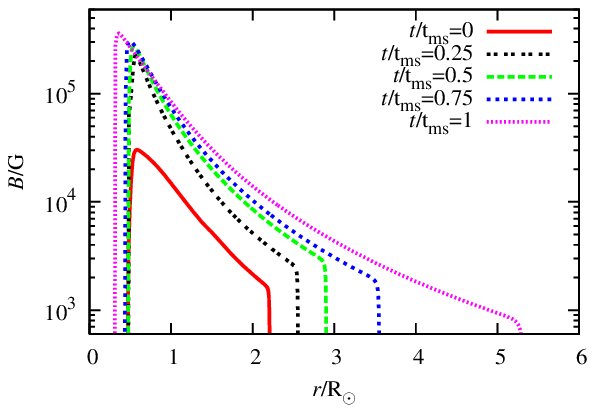}
\end{center}
\caption[Magnetic field evolution in a $5$\ms\ star initially rotating at $300\,{\rm km\,s^{-1}}$ without magnetic braking]{Evolution of the magnetic field in a $5$\ms\ star initially rotating at $300\,{\rm km\,s^{-1}}$ without magnetic braking. The top plot shows the magnetic potential and the bottom plot shows the toroidal field. The $\alpha$--effect produces a weak poloidal field which is efficiently converted into toroidal field by differential rotation. For each component, the field strength is approximately three orders of magnitude smaller at the surface than the core. The ratio of the toroidal and poloidal field strengths is of the order~$10^9$.}
\label{ch5.fig.mag1}
\end{figure*}

Owing to the strong magnetically--induced turbulence, the toroidal field behaves roughly as $B_{\phi}\propto r^{-3}$ and the poloidal field behaves as $A\propto r^{-2}$ so both are much stronger towards the core than at the surface of the star as shown in Fig.~\ref{ch5.fig.mag1}. The toroidal field falls to zero within a very narrow region near the surface of the star to meet the boundary conditions. The strength of the toroidal field predicted is around nine orders of magnitude larger than the poloidal field. This is because the $\Omega$--effect, the conversion of poloidal field into toroidal field by differential rotation, is much stronger than the $\alpha$--effect which regenerates the poloidal field. We take the surface value of the field to be the strength of the toroidal field just below the boundary layer. If we were instead to take the poloidal field, we would need a larger value of $\gamma$ to produce a stronger field. In this case the toroidal field is around six orders of magnitude larger than the poloidal field. So a surface poloidal field of $10^3\,{\rm G}$ would correspond to a toroidal field of $10^9\,{\rm G}$ just below the surface. The fields then increase by several orders of magnitude towards the core. Not only do these field strengths seem unreasonably energetic but also the magnetic stresses result in cores that are spinning near or above break--up velocity. However, spectropolarimetric observations have concluded that the large--scale structure of the external magnetic fields of massive stars are largely dipolar so there must be some mechanism for converting the toroidal field into poloidal field at the surface. It is likely that the stellar wind stretches the field lines in the radial direction, changing the toroidal field to a radial geometry as material is ejected from the stellar surface \citep{Parker58}.

Owing to the very large value for $D_{\rm con}$ predicted from mixing--length theory, the predicted field is extremely weak within the convective core. This is somewhat at odds with our observations in the Sun where large--scale magnetic flux can be transported through a convective region without being destroyed. It may be that convection is better treated by an anisotropic diffusivity. It has little effect on the magnetic field in the radiative zone though so does not strongly affect our model for massive stars. We therefore leave further consideration of this effect for future work.

\begin{figure}
\begin{center}
\includegraphics[width=0.99\textwidth]{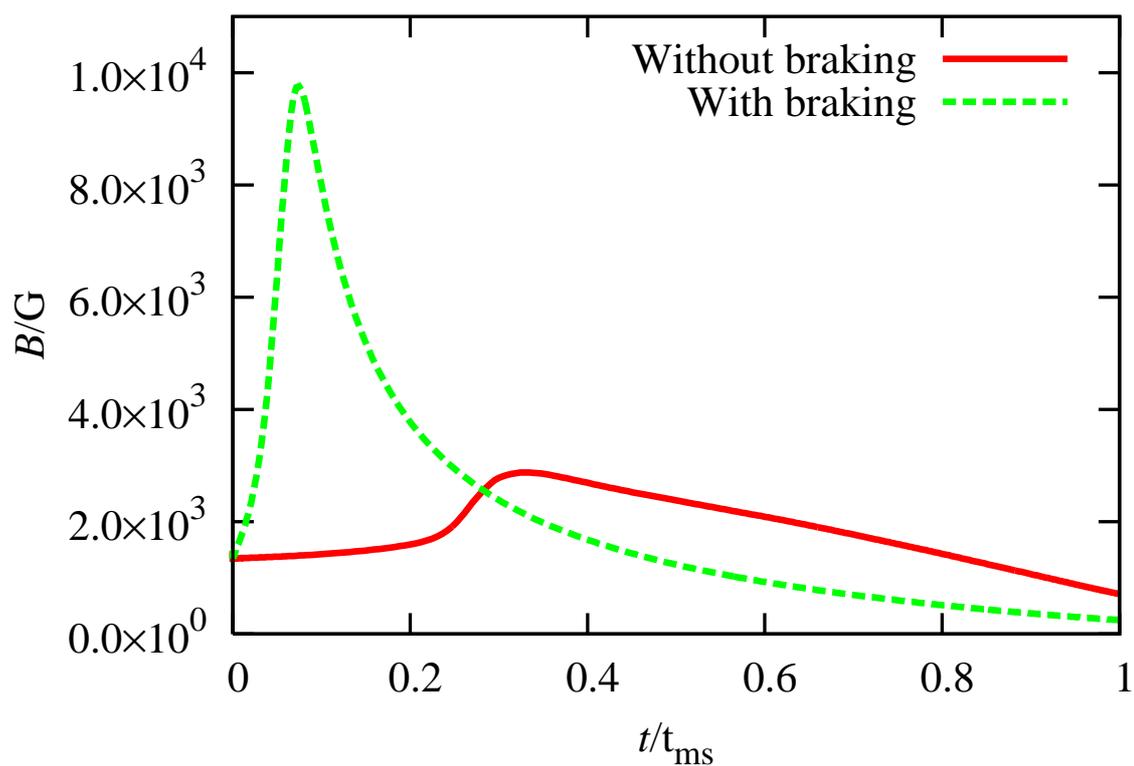}
\end{center}
\caption[Evolution of the magnetic field in a $5$\ms\ star initially rotating at $300\,{\rm km\,s^{-1}}$ with magnetic braking]{Evolution of the surface magnetic field strength in a $5$\ms\ star initially rotating at $300\,{\rm km\,s^{-1}}$ with and without magnetic braking. The surface field strength shows only a slight degree of variation during the main sequence when there is no magnetic braking. When magnetic braking is included the field strength peaks sharply after the ZAMS and then decays away rapidly. However, the field strength at the end of the main sequence is still several hundred Gauss.}
\label{ch5.fig.mag2}
\end{figure}

We first consider models in the absence of magnetic braking in order to distinguish evolutionary effects owing to the dynamo from those caused by braking. In this case, although the surface field only exhibits a small degree of variation (Fig.~\ref{ch5.fig.mag2}), the magnetic field inside the star becomes significantly stronger during the course of the main sequence. The surface magnetic field reaches a peak strength and then weakens towards the end of the main sequence. However, this change is always within a factor of three of the ZAMS value.  This is consistent with the model of \citet{Tout96} in which Ae/Be stars tap rotational energy early in their lives. The enhancement of the field inside the star is largely because the Brunt--V\"{a}is\"{a}l\"{a} frequency decreases as the star expands during the main sequence. It is also partly because the amount of differential rotation increases as a result of the changing hydrostatic structure of the star.

We might intuitively expect that the spin down of the star owing to magnetic braking would cause the magnetic field to decay rapidly and this is true later in the life of the star. However, the inclusion of magnetic braking first leads to a significant enhancement of the magnetic field shortly after the ZAMS. When braking is included, the loss of angular momentum from the surface is so fast that diffusion of angular momentum cannot prevent a build up of shear within the radiative envelope. This drives additional generation of magnetic flux through the $\alpha$--$\Omega$ dynamo and actually causes a much stronger peak field than without magnetic braking. The magnetic diffusion eventually reduces the amount of differential rotation and the magnetic spin down results in a weaker dynamo and faster rate of field decay. However, the field remains sufficiently large throughout the main sequence that the rate of chemical transport is still large enough to cause a significant amount of nitrogen enrichment. We discuss this further in section~\ref{ch5.sec.hunter}. Although the eventual decay of the surface field in the presence of magnetic braking is quite rapid, the field strength at the end of the main sequence is still several hundred Gauss. This is consistent with the observation that all chemically peculiar Ap stars have strong fields \citep{Auriere07}.

\subsection{Effect on angular momentum distribution}
\label{ch5.sec.angmom}

\begin{figure}
\begin{center}
\includegraphics[width=0.99\textwidth]{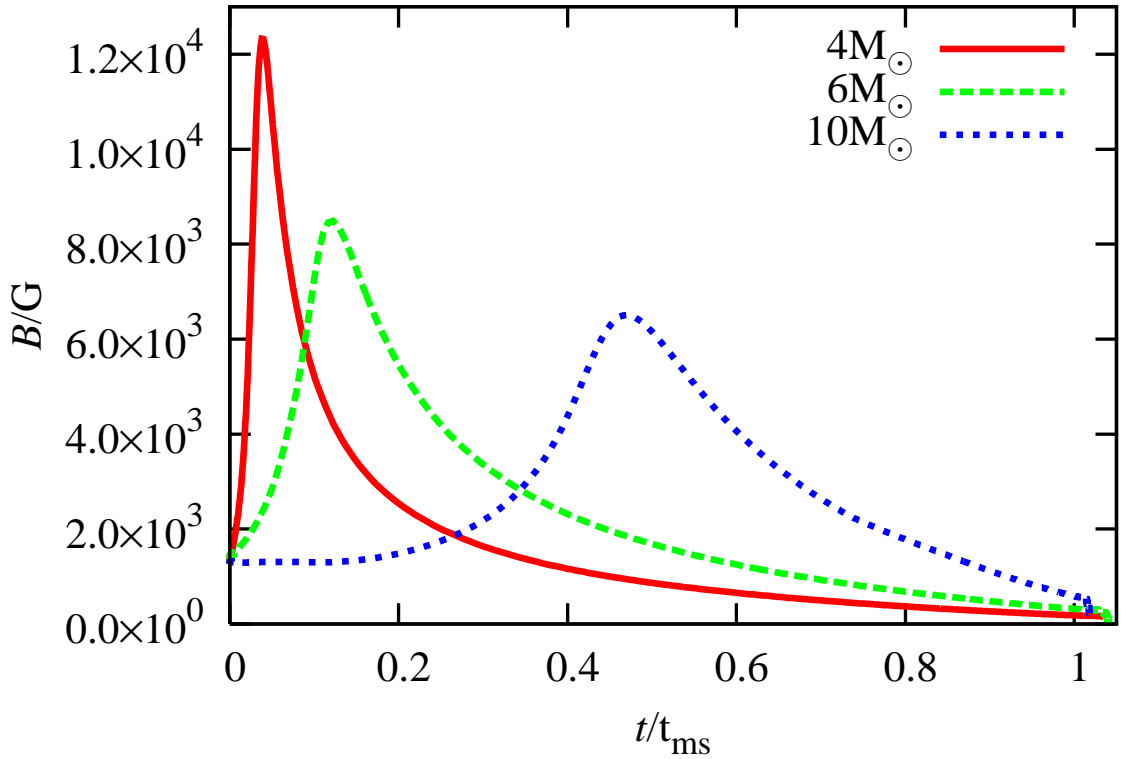}
\end{center}
\caption[Evolution of the surface magnetic field strengths in rotating stars of various masses]{Evolution of the surface magnetic field strengths in $4$\ms, $6$\ms\ and $10$\ms\ stars initially rotating at $300\,{\rm km\,s^{-1}}$ with an $\alpha$--$\Omega$ dynamo and magnetic braking. The maximum field strength is much greater in less massive stars. For all masses of star, the field strength increases sharply at the start of the main sequence owing to the rapid loss of angular momentum at the surface because of magnetic braking. This causes differential rotation which drives additional flux generation by the dynamo. This peak occurs later for more massive stars both in absolute time and as a fraction of their main--sequence lifetime. Following this, the field decays rapidly over the remainder of the main sequence.}
\label{ch5.fig.magstrength}
\end{figure}

\begin{figure}
\begin{center}
\includegraphics[width=0.95\textwidth]{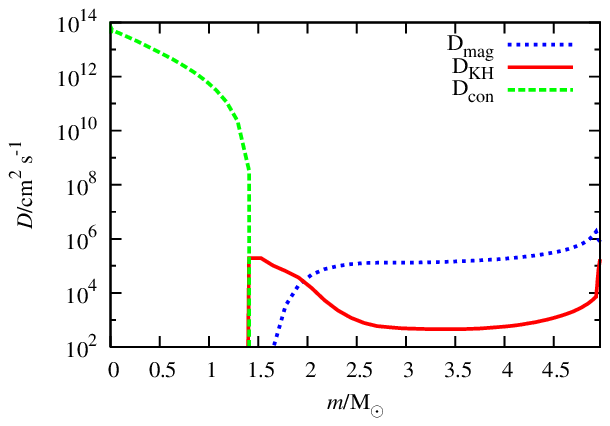}\\\includegraphics[width=0.95\textwidth]{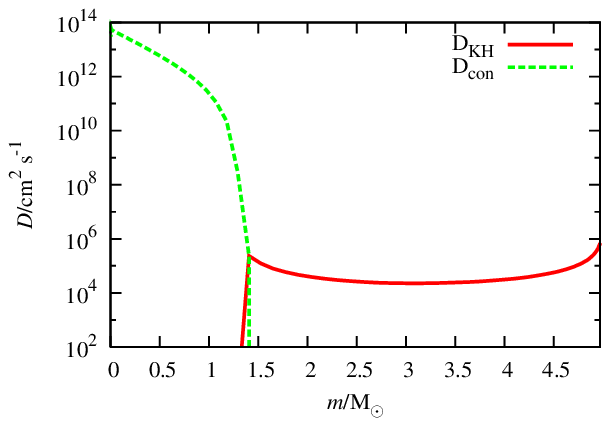}
\end{center}
\caption[Turbulent diffusivities governing angular momentum transport in a magnetic, rotating $5$\ms\ star]{Diffusivities for the angular momentum resulting from convection, hydrodynamic and magnetohydrodynamic effects in a $5$\ms\ star initially rotating at $300\,{\rm \, km\,s^{-1}}$. The top plot is for a magnetic star whereas the bottom plot is for a non--magnetic star. We note that the model predicts more efficient transport by magnetic effects compared to purely hydrodynamic effects. We also note that in the magnetic star, the diffusion of angular momentum by hydrodynamic turbulence is greatly reduced because the magnetic field reduces shear. There is a small region near the convective core where the magnetic diffusion becomes much smaller owing to mean molecular weight gradients. In this region the hydrodynamic turbulence dominates. This region only exists at the start of the ZAMS because the field becomes much stronger shortly after and the effects of rotation decrease as magnetic braking spins the star down.}
\label{ch5.fig.diffusion}
\end{figure}

\begin{figure}
\begin{center}
\includegraphics[width=0.6\textwidth]{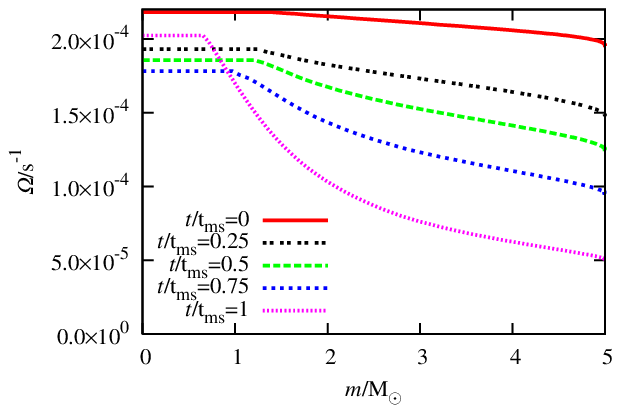}\\\includegraphics[width=0.6\textwidth]{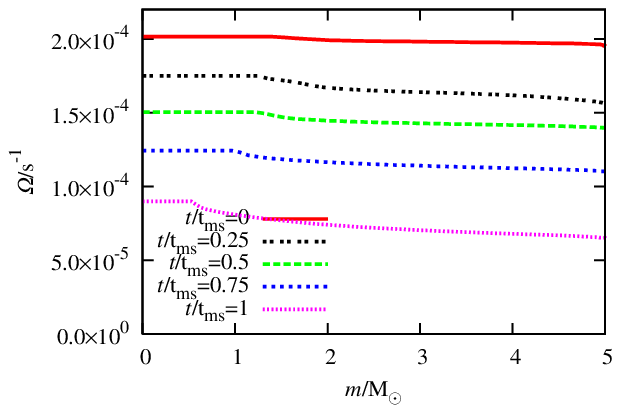}\\\includegraphics[width=0.6\textwidth]{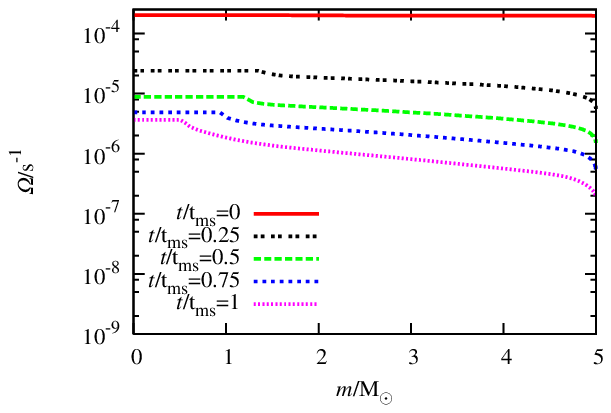} 
\end{center}
\caption[Evolution of the angular momentum distribution in a rotating, magnetic $5$\ms\ star]{Evolution of the angular momentum distribution in a $5$\ms\ star initially rotating at $300\,{\rm \, km\,s^{-1}}$. The top plot is for a non--magnetic star, the middle plot is for a magnetic star without braking and the bottom plot is for a magnetic star with braking. In magnetic stars without braking, the strong magnetic turbulence results in much less shear than the non--magnetic equivalent. Stronger diffusion in the magnetic stars also leads to far less differential rotation between the core and the envelope. This causes higher surface rotation in the non--braked magnetic star compared to the non--magnetic star. When braking is introduced to the magnetic star it spins down rapidly. The angular momentum loss from the surface leads to a much higher degree of differential rotation in the magnetic star with braking compared to the magnetic star without braking.}
\label{ch5.fig.omega}
\end{figure}

The effect of the magnetic field on the angular momentum distribution of a star has profound implications for its chemical evolution and the properties of its remnant. Shear arises in stars mostly as a result of changes in the structure from ongoing evolution and mass loss. Rotation also causes meridional circulation in stars. This contributes to the shear as we discussed in section~\ref{ch2.sec.circulation}. In the magnetic case, where magnetic braking is included, meridional circulation dominates over the magnetic stresses at the ZAMS for almost the entire star. For a $5$\ms\ star initially rotating at $300{\rm \,km\,s^{-1}}$ the meridional circulation is approximately six orders of magnitude stronger in the outer layers than the magnetic stresses at the ZAMS. Through most of the envelope the difference is between one and three orders of magnitude. However, when the magnetic field grows rapidly shortly after the ZAMS and magnetic braking begins to rapidly spin down the star this reverses and the magnetic stresses become much more important than the meridional circulation for the remainder of the main sequence. As we see in Fig~\ref{ch5.fig.magstrength}, this peak occurs later for more massive stars as a fraction of main--sequence lifetime and so the meridional circulation can dominate for longer. Therefore, whilst it is true that the meridional circulation has little effect on the evolution of magnetic stars for most of the main sequence, it is not necessarily true close to the ZAMS.

Apart from the physical effects that produce shear within the radiative envelope the other major factor that affects the angular momentum distribution is the strength of the turbulent diffusion. We have plotted the major diffusion coefficients at the ZAMS for a $5$\ms\ star initially rotating at $300,{\rm km\,s^{-1}}$ in Fig.~\ref{ch5.fig.diffusion}. The overall diffusion coefficient predicted by the magnetic model is significantly larger than produced by hydrodynamic turbulence alone. We note that $D_{\rm KH}\propto\left(\diffb{\Omega}{r}\right)^2$ predicted in the magnetic model is significantly lower than in the non--magnetic model. Whilst magnetic stresses should produce more shear than in the non--magnetic model, the diffusion coefficient is sufficiently high to cause an overall reduction in shear. This is illustrated in  Fig.~\ref{ch5.fig.omega} where we have plotted the evolution of the angular momentum distribution for the same star without a magnetic field, with a magnetic field but without braking and with both a magnetic field and braking. There is a small region near the convective core in the magnetic star where the magnetic diffusion becomes much smaller owing to mean molecular weight gradients. In this region the hydrodynamic turbulence dominates. This region only exists at the start of the ZAMS because the field becomes much stronger shortly after and the effects of rotation decrease as magnetic braking spins the star down.

In Fig.~\ref{ch5.fig.omega} we see that, in the magnetic star without braking, there is far less differential rotation throughout the star than in the non--magnetic star. This also means that the cores of magnetic stars are likely to be rotating more slowly than non--magnetic stars even before the effects of braking are included. When braking is included we see much the same trend except, in the model with magnetic braking, the whole star spins down rapidly. The typical Alfv\'{e}n radius for this star is approximately $50$, meaning that the rate of angular momentum loss is several thousand times faster than without a magnetic field. We note that there is far more differential rotation in this star compared with the magnetic star without braking. This is because of the rapid loss of angular momentum from the surface of the star.

\subsection{Mass--rotation relation of the main--sequence field strength}
\label{ch5.sec.mvb}

\begin{figure}
\begin{center}
\includegraphics[width=0.99\textwidth]{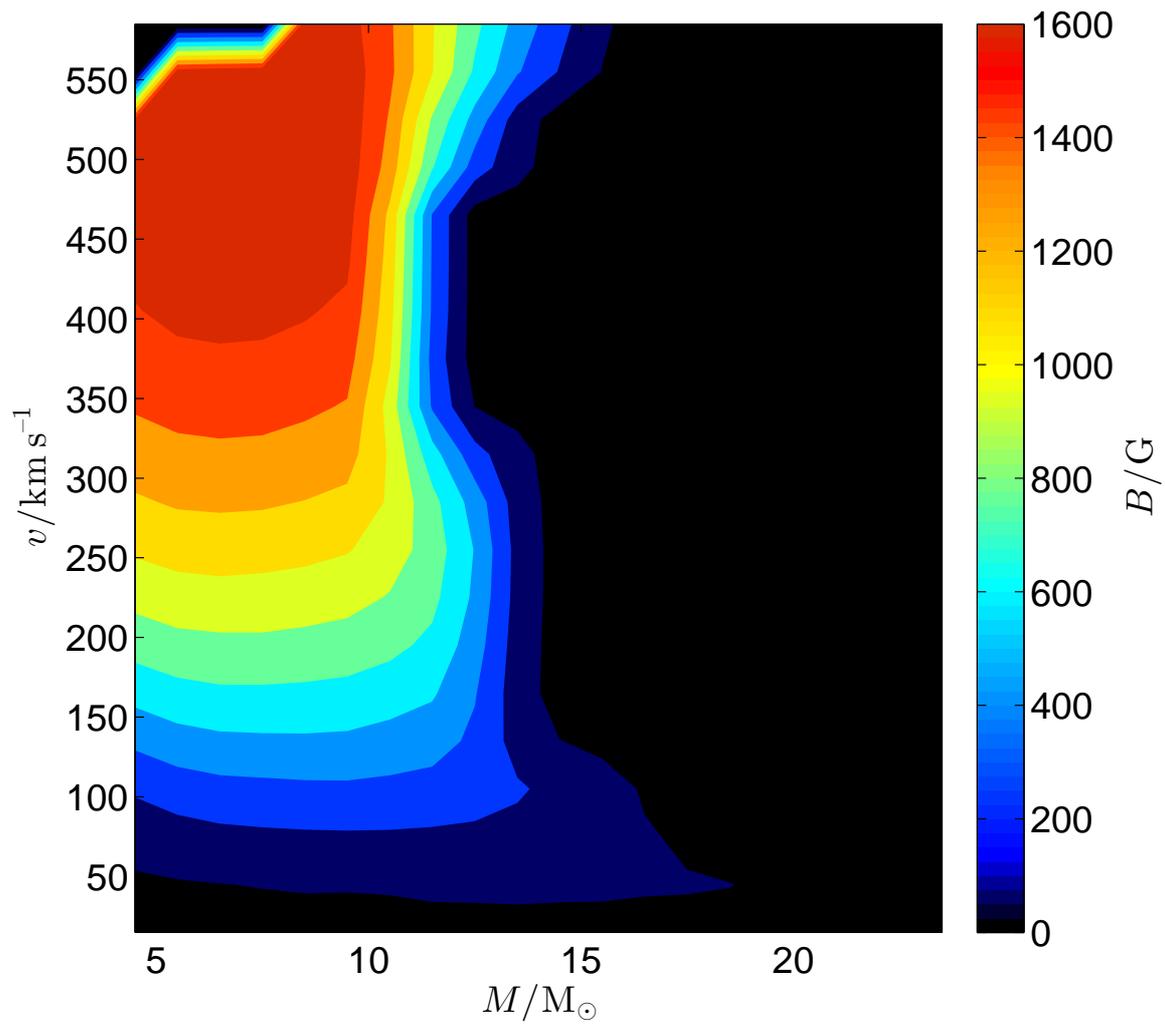}
\end{center}
\caption[Main--sequence magnetic field strengths for ZAMS stars with varying masses and rotation rates]{Main--sequence magnetic field strengths for intermediate--mass ZAMS stars at different rotation rates. Stars more massive than $15$\ms\ have almost no magnetic activity except for a weak field in slow rotators. The strongest fields occur in the most rapidly rotating stars with $4<M/$\ms$<10$.}
\label{ch5.fig.magevol}
\end{figure}

\begin{figure}
\begin{center}
\includegraphics[width=0.99\textwidth]{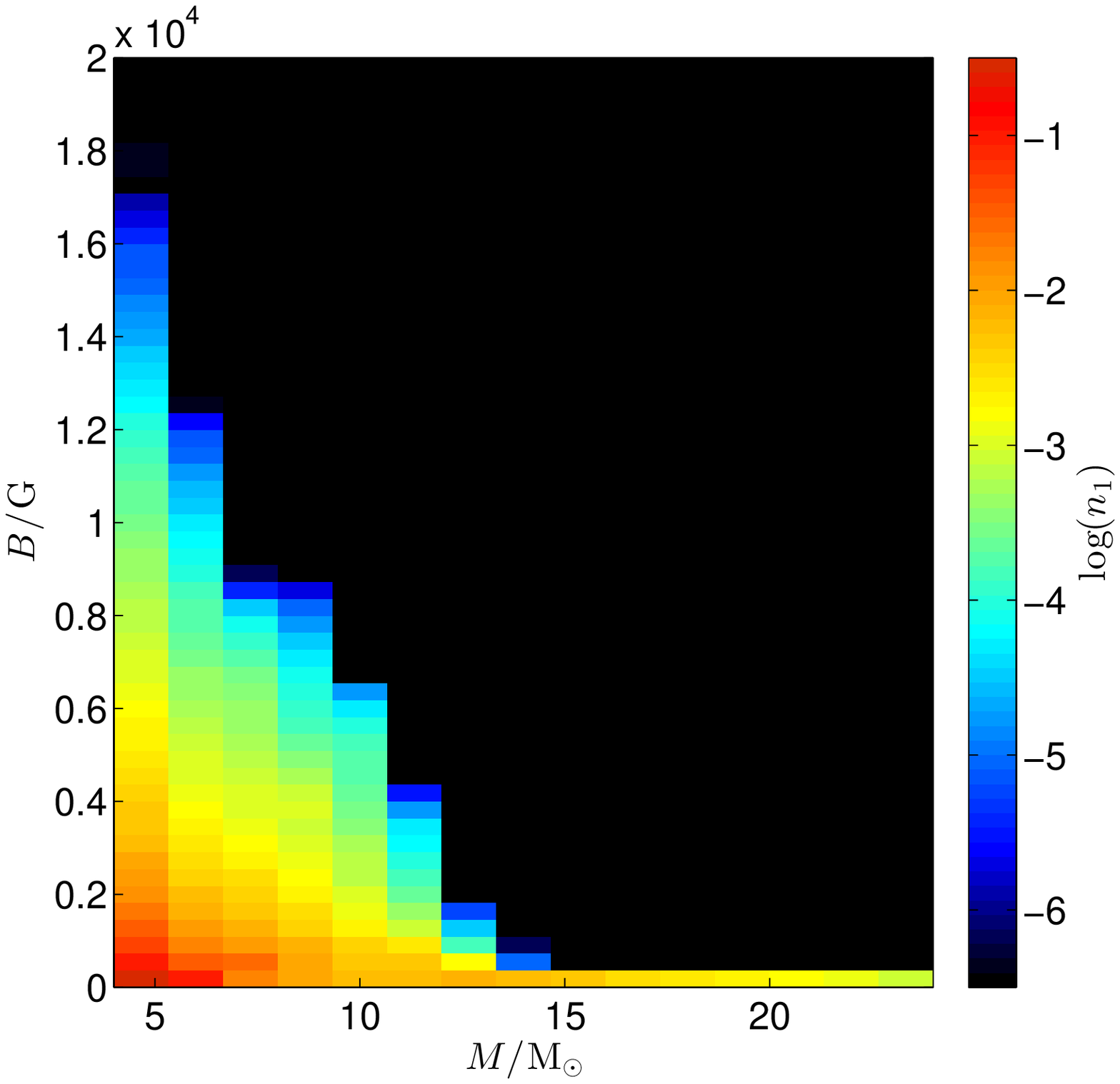}
\end{center}
\caption[Masses and magnetic field strengths of stars in a synthetic stellar population with ongoing star formation]{Simulated masses and magnetic field strengths for a population of stars drawn from the grid of models shown in Fig.~\ref{ch5.fig.grid}. The population undergoes continuous star formation, is drawn from a Salpeter IMF and the velocity distribution is Gaussian with mean $\mu=145\,{\rm \, km\,s^{-1}}$ and standard deviation $\sigma=94\,{\rm \, km\,s^{-1}}$. The number of stars in each bin as a fraction of the total number of stars is $n_1$. We see that lower--mass stars support much stronger fields. There is very little magnetic activity in stars more massive than around $15$\ms.}
\label{ch5.fig.mb}
\end{figure}

Historically the presence of strong magnetic fields in massive stars has been thought to be mainly confined to A~stars and perhaps some of the lower--mass B stars \citep{Mathys09}. This may have been because of the difficulty in observing magnetic effects in the broad absorption features of more massive stars \citep{Petit11}. However, as the amount of available data has grown, thanks to surveys such as MiMeS project \citep[e.g.][]{Wade09}, it has become clear that this is not caused simply by selection effects.

By applying our model to the grid shown in Fig.~\ref{ch5.fig.grid} we are able to track the evolution of the surface field strength of stars and, in particular, how it varies with mass and rotation rate. We show this dependence at the ZAMS in Fig.~\ref{ch5.fig.magevol}. It is immediately apparent that although stars less massive than around $15$\ms\ are able to sustain significant fields, no significant field is predicted in more massive stars except in very slow rotators. Even for high--mass, slow rotators the field doesn't exceed $200{\rm \, G}$. The transition between a strong ZAMS field and no field is sharpest in rapid rotators. This transition is caused by the interaction between hydrodynamic and magnetic turbulence. If $D_{\rm KH}$ exceeds $D_{\rm mag}$ for a sufficiently large region of the radiative envelope, the magnetic field decays exponentially and cannot be sustained by the dynamo. Because $\alpha\propto\omega_A$, the strength of the dynamo weakens with the magnetic field. In the case where $D_{\rm mag}$ is the dominant turbulent process, this is matched by a greater reduction in the turbulent diffusivity because, for most of the envelope, $D_{\rm mag}\propto\omega_A^2$. As the diffusivity drops, the field is less efficiently dissipated and so an equilibrium is reached. When $D_{\rm KH}$ dominates and the field decays the diffusivity is largely unaffected and so the dynamo continues to weaken causing the field to completely disappear. At higher masses and rotation rates $D_{\rm KH}$ is larger and so catastrophic quenching occurs for lower dynamo efficiencies. Assuming that both instabilities act in the radiative envelope, this explains why magnetic fields are more likely to be observed in A~stars and less frequently in O and B stars. 

Given that the magnetic field strength increases sharply after the main sequence before decaying away exponentially as discussed in section~\ref{ch5.sec.evolution}, we consider the distribution of magnetic field strengths in a population of stars with a continuous distribution of ages. The population is shown in Fig.~\ref{ch5.fig.mb}. The population undergoes continuous star formation, is drawn from a Salpeter IMF and a Gaussian velocity distribution with mean $\mu=145\,{\rm km\,s^{-1}}$ and standard deviation $\sigma=94\,{\rm km\,s^{-1}}$. We see that magnetic activity is highest in the least massive stars. As before, stars more massive than around $15$\ms\ show no magnetic activity. We note that the stars with the strongest fields fall outside the observational limits of the VLT--FLAMES survey of massive stars \citep{Dufton06}. We discuss this further in section~\ref{ch5.sec.hunter}.

 We therefore predict two distinct populations of stars. The first is a population of slowly rotating, magnetic and chemically peculiar stars with masses less than $15$\ms. The second is a population of more massive stars that are non--magnetic and follow the trend discussed by \citet{Hunter09} and chapter~\ref{ch4}, where rotation and nitrogen enrichment have a strong positive correlation. This is precisely what we observe \citep{Hunter09}. We may still observe A~stars that are rapidly rotating but not highly enriched. These stars should still support a strong magnetic field but are sufficiently young that no chemical enrichment has occurred. These rapidly rotating stars would be very infrequent owing to the efficient spin down by the magnetic braking. A rapidly rotating, highly magnetic massive star was observed by \citet{Grunhut12}. The star has a mass of $5.5$\ms\ and a surface rotation velocity of $290\,{\rm km\,s^{-1}}$ but has a surface field strength in excess of $10\,$kG. 

\subsection{Effect on the Hertzsprung--Russel diagram}

\begin{figure}
\begin{center}
\includegraphics[width=0.99\textwidth]{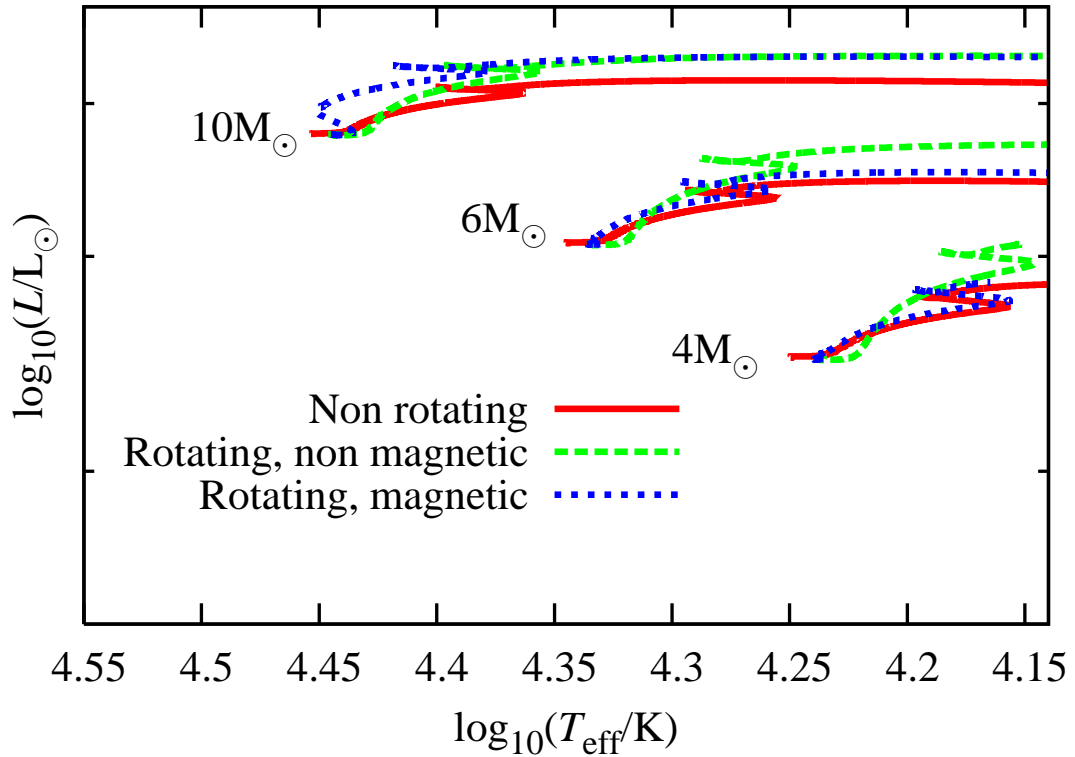}
\end{center}
\caption[Hertzsprung--Russell diagram for magnetic, rotating stars of various masses]{Hertzsprung--Russell diagram for stars with mass $4$\ms, $6$\ms\ and $10$\ms. The plot shows the predicted evolution for non--rotating stars, stars initially rotating at $300\,{\rm km\,s^{-1}}$ but with no magnetic field (c.f. case~1 from chapters~\ref{ch3} and~\ref{ch4}) and magnetic stars initially rotating at $300\,{\rm \, km\,s^{-1}}$. In less massive stars magnetic braking rapidly spins down the star so the structural effects of rotation are much less apparent. In more massive stars the effect of braking is much weaker and so the evolution is much closer to the rotating, non--magnetic model.}
\label{ch5.fig.hr}
\end{figure}

Because less massive stars have stronger fields, both magnetically induced mixing and magnetic braking are much more effective in these stars. Owing to the stronger magnetic mixing, chemical transport is more efficient in less massive stars as discussed in section~\ref{ch5.sec.hunter}. As a result, more hydrogen is mixed down into the core of less massive stars. However, because magnetic braking causes lower--mass stars to spin down very rapidly, the effects on brightness and temperature that arise from changes in the stellar structure in rotating stars are far less apparent when magnetic fields are introduced, as shown in Fig.~\ref{ch5.fig.hr}. In the $10$\ms\ model we see that the difference between the magnetic and non--magnetic rotating models is smaller owing to the much weaker field and hence less rapid spin down. However, in the evolution of the $4$ and $6$\ms\ models the magnetic stars remain barely distinguishable from the evolution of the non--rotating stars.

\subsection{The lifetime of fossil fields}
\label{ch5.sec.fossilfield}

\begin{figure}
\begin{center}
\includegraphics[width=0.99\textwidth]{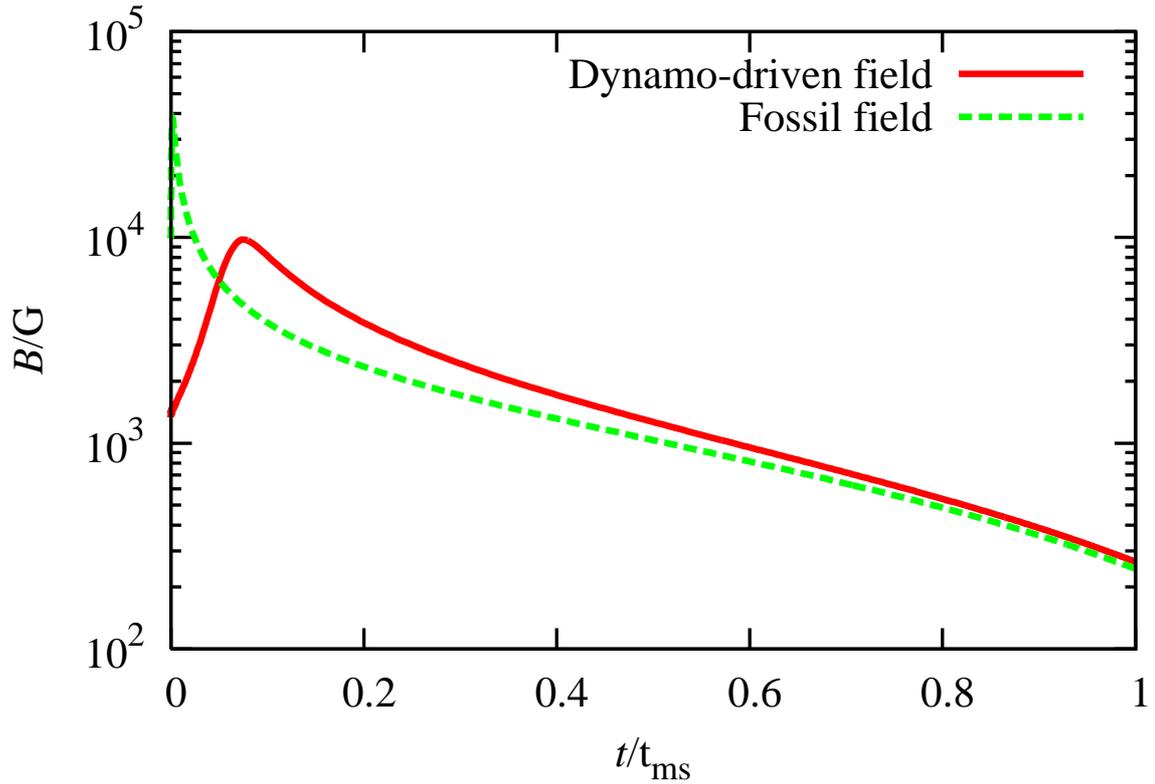}
\end{center}
\caption[Evolution of the magnetic field strength of $5$\ms\ in the absence of dynamo regeneration]{Evolution of two $5$\ms\ stars with different magnetic field models. The first star has no magnetic dynamo ($\gamma=0$) but starts with a very strong initial field ($B=10\,{\rm kG}$). The second star uses the same dynamo model and parameters as described in section~\ref{ch5.sec.model}. The star with an active dynamo is able to sustain the field for longer than the star with a fossil field but, owing to magnetic braking, both fields eventually decay exponentially. The two stars have similar field strengths at the end of the main sequence.}
\label{ch5.fig.fossil}
\end{figure}

An alternative to the radiative--dynamo model is that the magnetic field originates in the material that formed the star. If the protostellar cloud which forms a star is weakly magnetic, conservation of magnetic energy would result in a very strong main--sequence field. We call these fossil fields \citep{Braithwaite04}. In order for the fossil field model to work, the field must be able to survive the collapse of the protostellar cloud during the star formation process. The fossil field argument also relies on a stable field configuration being reached that would avoid destruction on main--sequence lifetimes. Certain stable configurations have been found in recent years \citep{Braithwaite06} and simulations have suggested that arbitrary field configurations do relax to these stable states \citep{Mathis11}. However, simple field configurations are still subject to the same instabilities as the fields we have generated by dynamo action, in particular the Tayler instability \citep{Tayler73}. There are a number of other instabilities that could occur in simple field configurations \citep{Parker66} but for now we consider only the Tayler instability.

We consider two stars, both initially rotating at $300\,{\rm km\,s^{-1}}$. The first star starts on the ZAMS with a magnetic field of $10\,{\rm kG}$ but $\gamma=0$ so no dynamo operates. The second star is a rotating magnetic star with dynamo parameters described in section~\ref{ch5.sec.model}. The evolution of the magnetic fields is shown in Fig.~\ref{ch5.fig.fossil}. In each case, the initial field undergoes some amplification at first owing to the onset of mass and angular momentum loss and the subsequent redistribution of angular momentum through the envelope. This is much more rapid in the case of the fossil field and does not appear in Fig.~\ref{ch5.fig.fossil}. The field then decays exponentially during the main sequence. We note that although the star with an operating dynamo is able to prevent the field from decaying for a short time, once magnetic braking has spun the star down sufficiently, the dynamo can no longer maintain the field which then decays exponentially. The final field strength is similar in each case.

Because the fossil field model predicts field evolution similar to that of the dynamo model it is difficult to argue which model is more physically accurate. However, we note that the fossil field strength has to be several orders of magnitude larger than the initial field in the case of a magnetic dynamo in order to reproduce the same final field. The question remains whether the fossil field argument can produce stars with strong enough initial fields so that they remain strong enough to influence chemical mixing in the star during the main sequence. \citet{Moss03} examined how much magnetic flux could potentially survive to the ZAMS from the pre--main sequence. He found that a significant fraction of flux could survive but only if the magnetic diffusivity was sufficiently low. Above this limit, no flux was expected to survive. The fossil field must also reproduce the two distinct observed populations in the Hunter diagram, shown in Fig.~\ref{ch5.fig.dufton}, discussed further in section~\ref{ch5.sec.hunter}. One could argue that this depends on the distribution of magnetic field strengths in protostellar clouds but the fossil field model must then also explain the mass--dependent distribution of field strengths observed in massive stars. Thus far we have come across no arguments that accurately reproduce these features of observed populations for fossil fields.

\subsection{Effect on surface composition}
\label{ch5.sec.hunter}

\begin{figure}
\begin{center}
\includegraphics[width=0.99\textwidth]{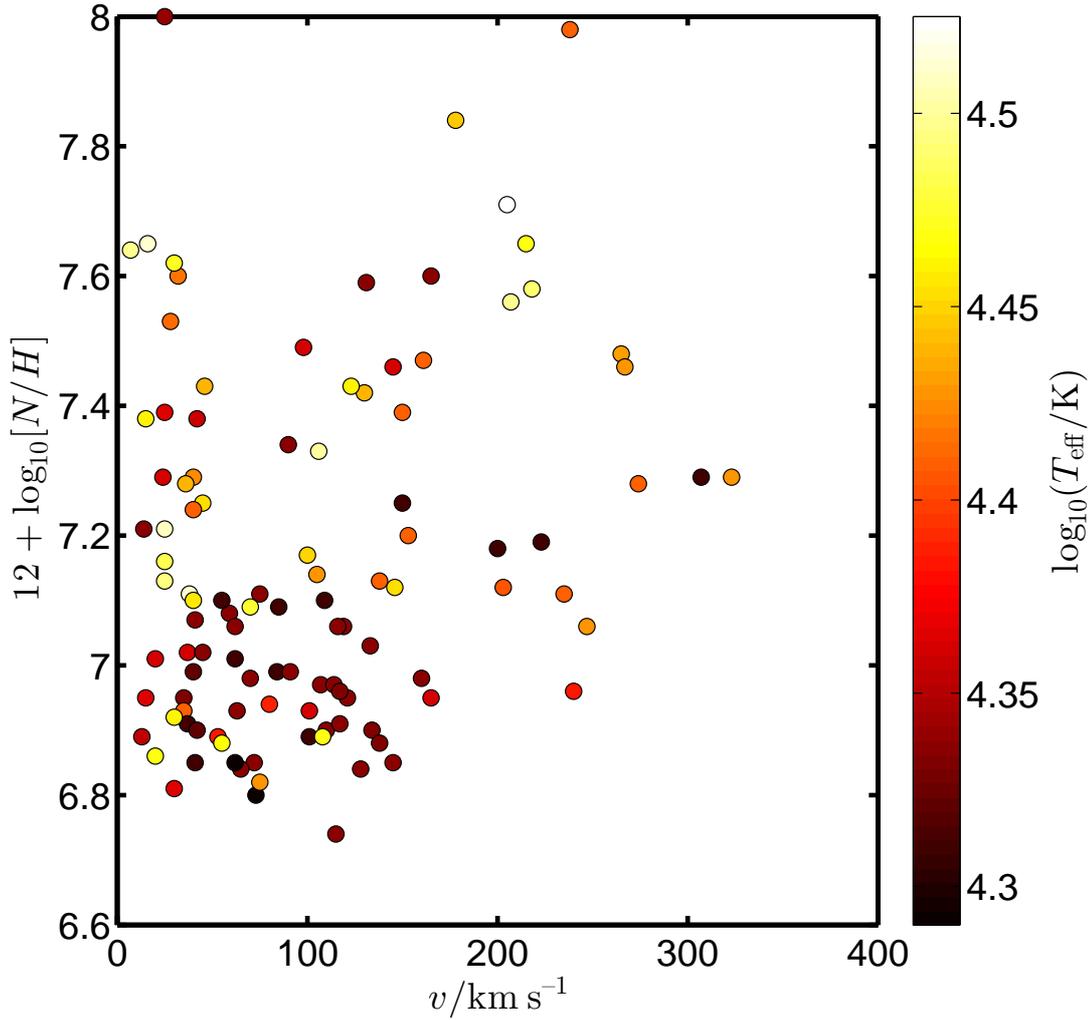} 
\end{center}
\caption[Hunter diagram for LMC stars observed in the VLT--FLAMES survey of massive stars]{Hunter diagram for the LMC stars observed in the VLT--FLAMES survey of massive stars \citep{Hunter09}. Stars with surface gravity smaller than $\log_{10}(g_{\rm eff}/{\rm cm^2\,s^{-1}})=3.2$ are classified as giants and have been excluded. The effective temperature of each star is also shown.}
\label{ch5.fig.dufton}
\end{figure}

\begin{figure}
\begin{center}
\includegraphics[width=0.99\textwidth]{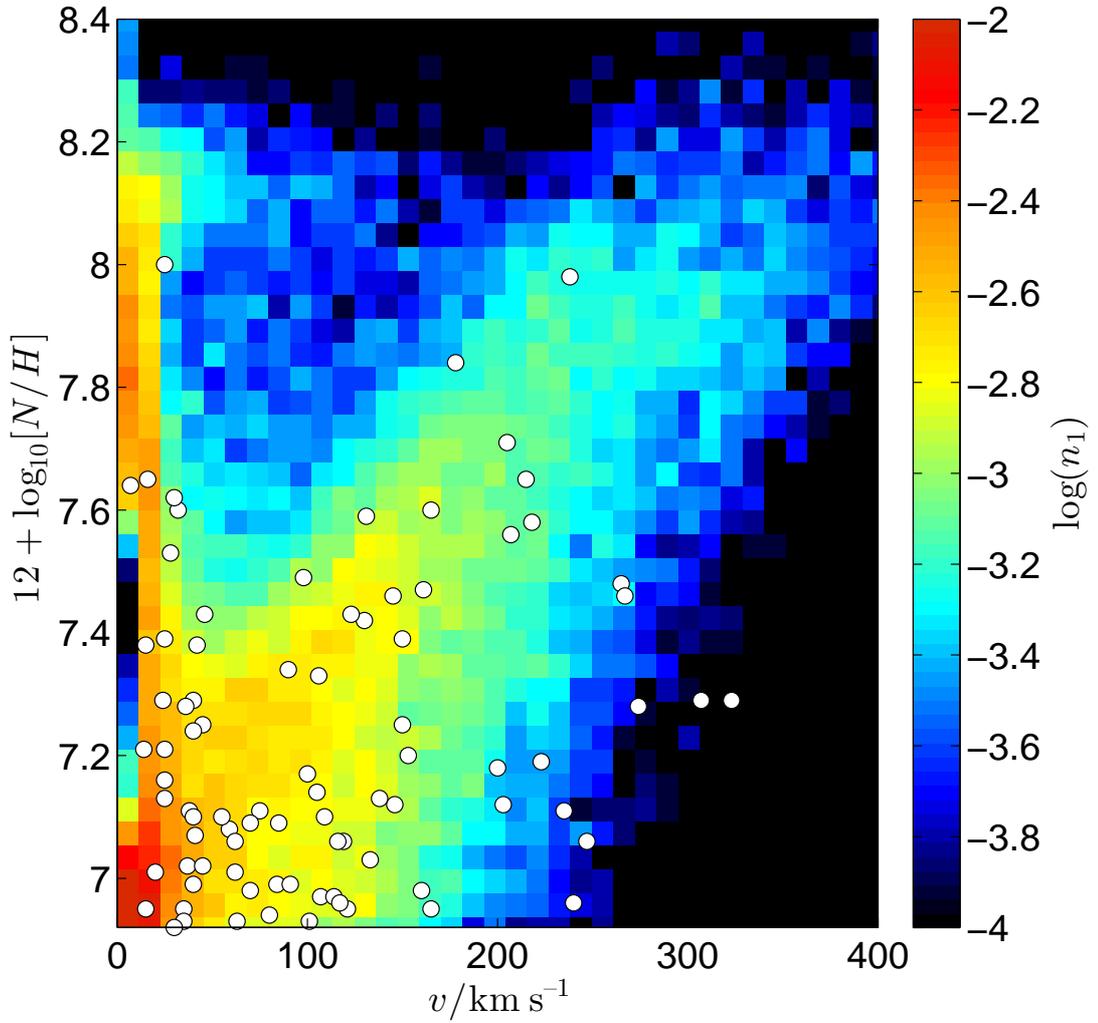} 
\end{center}
\caption[Hunter diagram for a population of stars drawn from the grid of models shown in Fig.~\ref{ch5.fig.grid}]{Hunter diagram for a population of stars drawn from the grid of models shown in Fig.~\ref{ch5.fig.grid}. The population undergoes continuous star formation, is drawn from a Salpeter IMF and the velocity distribution is Gaussian with mean $\mu=145\,{\rm km\,s^{-1}}$ and standard deviation $\sigma=94\,{\rm km\,s^{-1}}$. The number of stars in each bin as a fraction of the total number of stars is $n_1$. The magnetic model reproduces well the two distinct populations of stars observed in the VLT--FLAMES survey. More massive stars which cannot support a dynamo are enriched by rotational mixing whereas lower--mass stars are spun down rapidly and are enriched by magnetic mixing.}
\label{ch5.fig.hunter}
\end{figure}
\begin{figure}
\begin{center}
\includegraphics[width=0.99\textwidth]{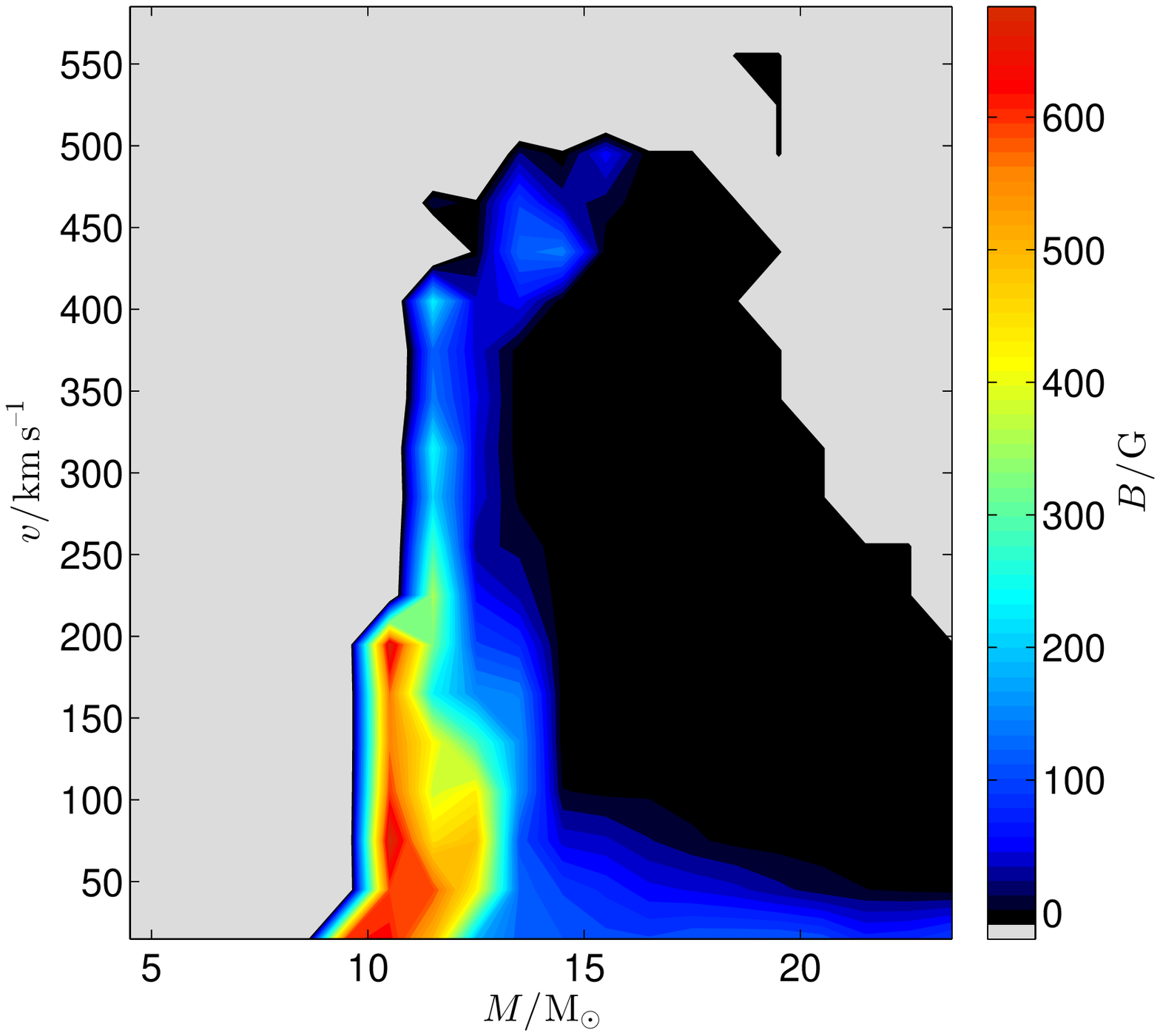} 
\end{center}
\caption[Relation between field strength, mass and rotation rate in a simulated population of stars according to the observational constraints of the VLT--FLAMES survey]{Distribution of magnetic field strengths with respect to mass and rotation rate for a population of stars undergoing continuous star formation. The population is drawn from a Salpeter IMF and the velocity distribution is Gaussian with mean $\mu=145\,{\rm km\,s^{-1}}$ and standard deviation $\sigma=94\,{\rm km\,s^{-1}}$. The gray region is where stars are not observed in the simulated population. Less massive stars are eliminated from the sample because they have insufficient magnitude for detection. The black region is for stars that appear in the simulated population but have no discernible field. We see that the magnetic stars in the sample, responsible for producing the slowly--rotating, enriched stars in Fig.~\ref{ch5.fig.hunter} come from a narrow region around $M=10$\ms.}
\label{ch5.fig.mvb2}
\end{figure}

The Hunter diagram \citep{Hunter09} is a plot of the surface nitrogen abundance in a star against surface velocity. The VLT--FLAMES survey of massive stars \citep{Evans05,Evans06,Dufton06} resulted in a significant amount of data on the nitrogen abundances in rotating stars in a number of samples from the Milky Way and Magellanic Clouds \citep{Hunter09}. In particular it was observed that there exists a class of stars that are slowly rotating ($v<60\,{\rm km\,s^{-1}}$) but exhibit significant nitrogen enrichment. It was suggested that these stars are, or once were, magnetic stars. If we extend the Hunter diagram to consider the effective temperature of each star as shown in Fig.~\ref{ch5.fig.dufton} we do not see a significant temperature variation between the two groups but we note that the mass range of stars in this sample is only $8<M/{\rm M_{\odot}}<20$ and so we cannot draw any strong conclusions about the relative mass distribution of the two enriched populations.

The observed distribution of surface abundance anomalies are well reproduced by our model which predicts magnetic fields only in stars less massive than around $15$\ms. The rest of the stars in the sample continue to evolve as non--magnetic stars as described in chapter~\ref{ch4}. The two distinct populations that we see in Fig.~\ref{ch5.fig.dufton} are reflected by the predictions made in section~\ref{ch5.sec.mvb}, shown in Fig.~\ref{ch5.fig.hunter}. This shows a simulated population of stars between $8$\ms\ and $20$\ms\ with our radiative--dynamo model and with magnetic braking. The mass range is smaller than that of the full grid shown in Fig.~\ref{ch5.fig.grid} owing to the removal of the least and most massive stars because of observational effects. The population was generated with the population synthesis code {\sc starmaker} \citep{Brott11}. It behaves exactly as we would expect from the VLT--FLAMES data. The stars initially have a full spread of rotation rates but the magnetic population spins down rapidly owing to the effects of magnetic braking. The magnetic field continues to affect the mixing and these stars become enriched as they age producing a population of magnetic, slowly rotating, chemically peculiar stars. More massive stars, where an equilibrium field cannot be supported by the dynamo, evolve as non--magnetic stars with hydrodynamic turbulence driving the mixing. This produces a second population whose enrichment increases with rotation rate as modelled in chapter~\ref{ch4}. The two populations are also highlighted in Fig.~\ref{ch5.fig.mvb2} which shows the relationship between field strength, mass and rotation rate in the simulated population. Most of the stars evolve without magnetic fields but there is a small region, at the lower mass limit of the sample ($M\approx 10$\ms), where stars are predicted to be magnetic. We note that, because this region is very narrow, small changes in the boundary between magnetic and non--magnetic evolution have a significant effect on the number of magnetic stars in the observed sample. It is possible this effect could be produced by fossil fields as discussed in section~\ref{ch5.sec.fossilfield} but thus far there is no way to explain why we see such distinct populations in the VLT--FLAMES data or why magnetic fields have a higher incidence rate amongst less massive stars.

We also note that those stars in Fig.~\ref{ch5.fig.dufton} with nitrogen enrichment $6.8<\log_{10}[N/H]<7.1$ and $0<v/{\rm km\,s^{-1}}<150$ cannot easily be categorized into either group of stars. They may be low--mass, fast rotators that have been partially spun down by magnetic braking, low--mass stars that are born with slow rotation or high--mass stars that are born with slow rotation. These stars evolve along a relatively similar path in the Hunter diagram.

\subsection{Variation with different parameters}
\label{ch5.sec.calibration}

\begin{table*}
\begin{center}
\begin{tabular}{cccccc}
\hline
$C_{\rm m}$&${\rm Pr_m}$&$\gamma$&$\max\left(\frac{rB_{\phi}}{A}\right)$&$\max(q)$&$\max(B_{\rm surf}/{\rm G})$\\
\hline
$1$&$1$&$1.93\times 10^{-16}$&$4.2\times 10^9$&$1.39$&$3.06\times 10^3$\\
$0.1$&$1$&$2.40\times 10^{-18}$&$1.1\times 10^{11}$&$0.96$&$7.10\times 10^2$\\
$10$&$1$&$3.53\times 10^{-14}$&$1.2\times 10^{8}$&$0.019$&$1.53\times 10^3$\\
$1$&$0.1$&$1.52\times 10^{-16}$&$3.74\times 10^{9}$&$1.04$&$7.40\times 10^{2{\rm (2)}}$\\
$0.1$&$0.1$&$3.00\times 10^{-17{\rm (3)}}$&&&\\
$10$&$0.1$&$1.81\times 10^{-14}$&$4.4\times 10^{8}$&$0.25$&$1.27\times 10^3$\\
$1$&$10$&$1.04\times 10^{-14}$&$8.13\times 10^{8}$&$0.0089$&$1.55\times 10^3$\\
$0.1$&$10$&$4.47\times 10^{-16}$&$1.95\times 10^{10}$&$0.14$&$8.92\times 10^2$\\
$10$&$10$&$1.58\times 10^{-11}$&$2.1\times 10^{7}$&$0.033$&${\rm N/a}^{(1)}$\\
\hline
\hline
\end{tabular}
\caption[The variation of magnetic stellar evolution owing to variation in parameters for magnetic field evolution]{The variation of magnetic and stellar parameters with different values for $C_{\rm m}$ and ${\rm Pr_m}$ for a $5$\ms\ star initially rotating at $200\,{\rm km\,s^{-1}}$ with magnetic braking. Each model was taken to have the same equilibrium ZAMS field. The table shows the values of $C_{\rm m}$, ${\rm Pr_m}$ and $\gamma$ used for each model as well as the maximum internal value of the ratio of the poloidal and toroidal field, $\frac{rB_{\phi}}{A}$ and $q=\diffb{\Omega}{r}$, taken at $5 \times 10^7 {\rm yr}$. Finally the table shows the maximum value of the surface field during the main sequence. We note three special entries in the table. (1) This star evolved quasi--homogeneously and produced a monotonically increasing field well beyond the normal main--sequence lifetime. Therefore defining a maximum main--sequence field was inappropriate. (2) This star evolved normally but we note that for a slightly smaller value of $\gamma$ we were unable to maintain an equilibrium field. This effect was discussed in section~\ref{ch5.sec.mvb}. (3) This star is similar to (2) but in this case we were totally unable to maintain an equilibrium field at the desired strength. We note that for stars (2) and (3), a stronger field can be maintained provided the dynamo--efficiency is sufficiently large.}
\label{ch5.tab.calibration}
\end{center}
\end{table*}

The model currently contains four parameters which we may vary independently. If we include possible recalibration of the Alfv\'{e}n radius by constants of order unity then this increases to five. We may fix the Alfv\'{e}n radius by ensuring that the population of enriched magnetic stars is confined to the appropriate band of rotation rates as discussed in section~\ref{ch5.sec.hunter}. We can also set ${\rm Pr_c}$ by ensuring that the maximal enrichment of magnetic stars is the same as in the VLT--FLAMES data also discussed in section~\ref{ch5.sec.hunter}. The remaining three parameters may then be varied so that typical field strengths are of the order $10\,{\rm kG}$, as observed in magnetic Ap stars \citep{Mathys09}. This value is subject to change though given the scarcity of observations of magnetic stars. This still leaves a high degree of freedom within the model. Up to this point we have used $C={\rm Pr_{\rm m}}=1$ and $\gamma=10^{-15}$ but we consider the effect of varying $C_{\rm m}$ and ${\rm Pr_m}$ by an order of magnitude in either direction. We ran our $5$\ms\ star initially rotating at $300\,{\rm \, km\,s^{-1}}$ with magnetic braking. The effect on a number of parameters is shown in Table~\ref{ch5.tab.calibration}.

For low magnetic Prandtl numbers it is much more difficult to sustain the dynamo. The same surface field is reproduced with smaller dynamo efficiencies but the minimum sustainable field strength is larger. In the case of small $C_{\rm m}$ and small ${\rm Pr_m}$, the field was completely quenched by the hydrodynamic turbulence as described in section~\ref{ch5.sec.mvb}. A dynamo could be sustained for stronger surface fields but only by increasing the dynamo efficiency significantly. Even for $C_{\rm m}=1$ we found that for a small reduction in $\gamma$ the ZAMS field collapsed.

For higher values of $C_{\rm m}$, the diffusion of the magnetic field requires a larger dynamo efficiency in order to maintain the same strength field and vice--versa for smaller values of $C_{\rm m}$. For simultaneously large values of $C_{\rm m}$ and ${\rm Pr_m}$ the field keeps growing monotonically with time during quasi--homogeneous evolution. This is to be expected when the dynamo--driven mixing becomes very high. Typically we could adjust $\rm Pr_c$ to compensate.

Regardless of our choice of $C_{\rm m}$ and ${\rm Pr_m}$, the ratio of the poloidal and toroidal field strength is well correlated with the dynamo efficiency. Larger values of the dynamo efficiency lead to a smaller ratio between the two field strengths. This is because of the form of equations (\ref{ch5.eq.toroidal}) and (\ref{ch5.eq.poloidal}). Because the two fields have the same diffusion timescales, their equilibria depend on the regeneration terms. In the case of the poloidal field this comes from the $\alpha$--effect and for the toroidal field it comes from the shear. In all of our models, the $\alpha$--effect is much weaker than the effect of shear and so the poloidal field strength is much smaller. However if $\gamma$ is increased, increasing the regeneration of the poloidal field but having little direct effect on the toroidal field, the ratio between the two becomes much smaller.

There are other aspects of the evolution that are much more difficult to explain and are related to the non--linearities in the model and their coupling to the effects of stellar evolution on nuclear timescales. We might expect the maximum value of the shear to always be smaller with higher values of $C_{\rm m}$ because the angular momentum transport is more efficient but, while this is true in general, it isn't a simple relationship. Likewise the maximum main--sequence surface field doesn't seem to correlate with either free parameter. 

In particular, the relative abundance of slow and fast rotating chemically peculiar stars may be explained by a shift in the position by mass of the cut--off between magnetic stars and non--magnetic stars discussed in section~\ref{ch5.sec.mvb}. The effect of these free parameters on the position of the cut--off is something we leave for future work.

\section{Conclusions}
\label{ch5.sec.conclusions}

Magnetic fields are one of the most mysterious and least understood aspects of stellar evolution. The first magnetic massive star was discovered over 65 years ago \citep{Babcock47} and yet debate still rages about whether these fields have primordial origin or are generated by a radiative dynamo acting within the stellar envelope. Models of magnetic stars must reproduce the observed phenomenon of magnetic A stars with unusual surface compositions that have much slower rotation rates than the rest of their  population \citep{Mathys04}. The data from the VLT--FLAMES survey of massive stars \citep{Evans05, Evans06} also supports the idea that there exists a population of stars that are slowly rotating but have a high degree of nitrogen enrichment \citep{Hunter09}.

We have presented a simple radiative dynamo model that arises because of the Tayler pinch--type instability \citep{Tayler73} and is based on the model of \citet{Spruit99} which was further developed by \citet{Spruit02} and \citet{Maeder04}. Unlike previous work, we have evolved both the poloidal and toroidal fields as independent variables at each radius in the star coupled to the angular momentum distribution of the star. The magnetic fields evolve according to a latitudinally--averaged induction equation with the inclusion of an $\alpha\Omega$--dynamo mechanism derived from mean--field magnetohydrodynamics \citep{Schmalz91}. We introduce a model for magnetic braking similar to that of \citet{ud-Doula02}. The model depends on a number of parameters, the overall strength of the magnetic turbulence, the magnetic Prandtl number, the chemical Prandtl number, the dynamo efficiency and the critical ratio of the kinetic energy to the magnetic energy, which defines the Alfv\'{e}n radius. The choices of $C_{\rm m}$, which affects the strength of the magnetic turbulence, and ${\rm Pr_c}$ have a strong effect on the dynamo efficiency needed to sustain the field but the relation between these parameters and the internal evolution of the models is complicated.

In models of the magnetic field, when magnetic braking is not included, the field varies only by a factor of a few during the main sequence. When we include magnetic braking, the Alfv\'{e}n radius is typically between $10$ and $100$ times greater than the stellar radius and so angular momentum loss is some $10^3$ times greater than from non--magnetic mass loss alone. The rapid angular momentum loss from the surface drives additional shear that leads to increased field generation. In magnetic stars with magnetic braking, the field increases rapidly at the start of the main sequence before decaying exponentially. The field strengths at the end of the main sequence are predicted to be of order $100{\rm \, G}$.

We consider a population of stars with this magnetic model and find two distinct types of behaviour. For stars more massive than around $15$\ms\ the Kelvin--Helmholtz turbulence dominates over the magnetic turbulence and a stable field cannot be sustained by the dynamo. In these cases we see no appreciable field strength during the main sequence so the stars evolve according to our normal prescription for non--magnetic, rotating stars. The predicted field strength is stronger for rapid rotators but the overall strength does not depend strongly on the stellar mass except near the limit at which the dynamo can sustain the field. Although the magnetic field decays exponentially after an initial peak, it remains strong enough to have a significant effect on the chemical evolution of the star. Though the actual mass at which this dichotomy sets in depends on parameters, the fact it exists is an important consequence of our model.

If we look at the evolution of an artificially strong initial field in the absence of any dynamo action, but subject to the diffusion that arises from the \citet{Tayler73} instability, we find that reproducing the same TAMS field requires an initial field several orders of magnitude larger than in the presence of a dynamo because any fossil field is predicted to decay exponentially. The fossil field hypothesis suffers from the problem that we expect the fields in low--mass stars to decay more than in more massive stars, likely because of their much longer main--sequence lifetimes. This is opposite to observed trends which suggest that less massive stars are more likely to support strong fields than more massive stars \citet{Grunhut11}. This model also offers no explanation as to why we see two distinct groups in the Hunter diagram. Both of these issues are well resolved by our $\alpha\Omega$--dynamo model.

We created an artificial population of stars with the population synthesis code {\sc starmaker} \citep{Brott11}, including the effects of the $\alpha$--$\Omega$ dynamo and magnetic braking. The population reflects well the observations of the VLT--FLAMES survey of massive stars. The survey observed two distinct populations of stars. The first shows increasing nitrogen enrichment with rotation rate, the second is a class of slow--rotating stars that exhibit unusually high nitrogen abundances compared to the rest of the population. This distribution of stars is well reproduced by the magnetic model. The fact that the two very different evolutionary paths arise naturally from the model is very encouraging to explain why we observe these two classes of star without having to appeal to the fossil fields argument.

There are still a number of open questions and further refinements that need to be made to the model. We have evolved a magnetic population of stars with the same initial velocity distribution as the non--magnetic stars. If the radiative dynamo has a strong effect on the pre--mainsequence evolution then magnetic braking causes magnetic stars to reach the ZAMS with significantly slower rotation rates than stars with no significant field. This is indeed observed in stellar populations \citep{Mathys04}. \citet{Alecian08} also discovered a number of stars on the pre--main sequence which exhibited significant magnetic activity. They attribute these to fossil fields by eliminating the possibility that the fields could be generated by a convective dynamo. However, if a radiative dynamo operates in these stars it could also be responsible for the generation of the observed fields. By comparison, the observations of \citet{Grunhut12} suggest that magnetic stars may reach the main sequence with significant rotational velocities. If magnetic stars were born with slower rotation rates than their non--magnetic counterparts then this would partly explain why the required dynamo efficiency is so small and why the predicted ratio between the poloidal and toroidal fields is so large. If magnetic stars were born with lower surface rotation rates then a higher dynamo efficiency would be needed to produce observed magnetic field strengths. This would reduce the difference between the $\alpha$--effect and the $\Omega$--effect and so the ratio of the strengths of the poloidal and toroidal fields would be closer to unity. Another possible explanation for why the predicted dynamo efficiency is so small is that we chose the radial coordinate as the length scale for the dynamo action. In reality a more sensible choice may have been the length scale of the saturated magnetic instabilities, $l_r$ (c.f. equations~(\ref{ch2.eq.lr}) and~(\ref{ch2.eq.lr2})). A shorter length scale would result in a weaker dynamo and therefore the dynamo efficiency would need to be higher to sustain the same field strength.

The observed proportion of Ap stars as a fraction of the whole population of A stars is roughly $10\%$ \citep{Moss01}. Our grid of models does not yet extend down to the mass range for A stars ($1.4<M/$\ms$<2.1$) and so we cannot yet say whether our population matches this statistic. We do expect that, given the predicted initial velocity distribution of massive stars, the population of A stars should still be dominated by slow rotators that do not support a radiative dynamo. In the mass range of our simulated population, over $90$\% of stars in the sample have surface field strength less than $187$\,G. This is well below the limit of $300$\,G anticipated for the transition to Ap classification \citep{Auriere07}. Although our population contains some very massive stars where we expect smaller field strengths, the form of the IMF ensures that the population is still dominated by intermediate--mass stars so the figure for A stars is likely to be similar. It is also likely that below a certain amount of shear, a dynamo does not operate. We have not taken this into account in our simple model. If it is the case, there may also be a sharp transition between magnetic and non--magnetic behaviour at low rotation rates.

In our models we have assumed a simple magnetic field geometry. Even if real fields are generated by dynamo action then they may still relax to stable field configurations such as those suggested by \citet{Braithwaite06,Mathis11}. Further work is needed to determine how the model might behave differently under these conditions. Further consideration must also be given to the action of convection on the magnetic field. Does our diffusive model apply in convective zones and if so is it anisotropic? Furthermore, can we better constrain the free parameters in the system, including the efficiency of magnetic braking? Although data on magnetic stars is scarce, a great deal of progress has been made possible by surveys such as the VLT--FLAMES survey of massive stars and the MiMeS project. These provide sufficient clues to further constrain our existing models. Additional progress will no doubt be possible thanks to ongoing developments in stellar observations from the MiMeS project \citep{Wade09, Grunhut11} and additional data on stellar surface compositions through projects such as the VLT--FLAMES tarantula survey \citep{Evans11}.

\begin{savequote}[60mm]
All things change; nothing perishes. (Ovid)
\end{savequote}

\chapter[WD magnetic fields in interacting binaries]{White dwarf magnetic fields in strongly interacting binaries}
\label{ch6}

\section{Introduction}
\label{ch6.sec.introduction}

Surveys of the galactic white dwarf (WD) population have discovered magnetic field strengths ranging up to about $10^9\,\rm{G}$ \citep{Schmidt03}. Typically WDs fall into two categories, those with field strengths of the order $10^6\,\rm{G}$ or higher and those with fields weaker than around $10^5\,\rm{G}$. We focus here on highly magnetic WDs (hereinafter MWDs) with field strengths greater than $10^6\,\rm{G}$. \citet{Landstreet71} proposed a fossil field mechanism for the origin of these fields based on the evolution of magnetic Ap/Bp stars. We discussed the fossil field argument in chapter~\ref{ch5} and concluded that there are many outstanding issues with this argument. In particular it is likely that a large fraction of the magnetic flux would be dissipated before reaching the terminal--age main sequence (TAMS). Here we build on the proposal of \citet{Tout08} that the origin of the strong fields of MWDs lies in the interaction between the WD and its companion star in a binary system.

This assertion is based on observations from the SDSS that approximately $10\,\%$ of isolated WDs are highly magnetic \citep{Liebert05} as are $25\,\%$ in cataclysmic variables \citep{Wickramasinghe00} while there are none to be found in wide detached binary systems. Of the 1,253 binary systems comprising a WD and non--degenerate M--dwarf star surveyed in the SDSS Data Release Five \citep{Silvestri07} none have been identified with magnetic fields greater than the detection limit of about $3\,{\rm MG}$. The relatively high occurrence of MWDs in strongly interacting binaries compared with elsewhere suggests that the generation of their strong fields is likely the result of the interaction between the binary components. The MWDs observed in isolated systems may be explained by either the total disruption of the companion star during unstable mass transfer or the coalescence of the MWD and the core of its companion following loss of sufficient orbital energy to the common envelope (hereinafter CE) or via gravitational radiation.

When a giant with a degenerate core expands beyond its Roche Lobe (c.f. section~\ref{ch1.sec.ce}) mass transfer may proceed on a dynamical time scale.  A dense companion, typically a main--sequence star, cannot accrete the overflowing material fast enough and so instead swells up to form a giant CE. As a result of energy and angular momentum transfer to the CE during orbital decay of the dense companion and the remnant core, strong differential rotation is established within the envelope. Also, owing to its size and thermal characteristics together with the nuclear energy source at the core, the CE is expected to be largely convective. In the mechanism proposed by \citet{Tout08} this is expected to drive strong dynamo action giving rise to powerful magnetic fields \citep{Regos95,Tout92}. If sufficiently strong dynamo action occurs in the CE then comparable magnetic fields may be induced in the degenerate core (DC) that then evolves into a WD once the envelope has been removed.  We show here that strong surface fields can result from CE evolution. The strength of such fields is highly dependent on the electrical conductivity of DC, the lifetime of the CE and the variability of the magnetic dynamo.

In section~\ref{ch6.sec.energy} we outline the various sources of energy in the binary/CE system and their relation to the energy requirements of the magnetic dynamo. In section~\ref{ch6.sec.equations} we discuss the governing equations of the system and derive the form of the magnetic field for a general spatially and temporally varying magnetic diffusivity via the static induction equation. Then in section~\ref{ch6.sec.method} we give an overview of the numerical methods we have used to solve the various stages of the problem. In section~\ref{ch6.sec.results} we present our results and we discuss these and conclude in section~\ref{ch6.sec.conclusions}.

\section{CE Evolution and energy constraints}
\label{ch6.sec.energy}

In the absence of detailed hydrodynamic properties, the most favoured models for common envelope evolution are the so--called $\alpha$ \citep{Webbink76,Livio88} and $\gamma$ \citep{Paczynski67,Nelemans00} prescriptions which use a one dimensional parametrization of the transfer of energy and angular momentum respectively between the binary orbit and the envelope. We consider the energy content of a typical envelope compared to the energy necessary to generate the desired magnetic fields.

The primary sources of energy in the CE are the orbital energy of the binary system itself and the gravitational binding energy of the envelope. In the $\alpha$--prescription, energy is transferred from the orbit to the envelope and this leads to the expulsion of the envelope and decay of the orbital separation. This is regarded as the most likely origin of cataclysmic variables \citep{Paczynski76,Meyer79}. Other potential sources are the thermal energy content of the envelope and its rotational energy. However for a virialised cloud, these energies are small compared to the other energy sources. In some systems recombination energy can be similar in magnitude to the binding energy so may become important \citep{Webbink08}.

Let $W,$ $O$ and $M$ be the binding, orbital and magnetic energy content of the binary/CE system respectively.

\begin{eqnarray}
W&=&\eta_{\rm w}\frac{GM_{\rm env}M_{\rm T}}{R_{\rm env}},\\
O&=&\frac{GM_{\rm c}M_{\rm 2}}{a},\\
{\rm and}\\
M&=&\frac{B^2}{2\mu_{\rm 0}}(2\pi a)(\pi r_{\rm I}^2),
\end{eqnarray}

\noindent where $M_{\rm env}$ is the mass of the CE with radius $R_{\rm env}$, $M_{\rm c}$ is the mass of the degenerate core, $M_{\rm 2}$ is the mass of the dense companion, $\eta_{\rm w}$ is a constant of order unity that can be calculated from stellar models. The total mass of the system is $M_{\rm T}=M_{\rm env}+M_{\rm c}+M_{\rm 2}$ and $a$ is the final orbital separation. As we would expect, the total available energy is less than the energy required to produce a field of the desired strength throughout the cloud. However, given that we anticipate that the field is generated by dynamo action, the strongest fields occur where the hydrodynamic motions of the CE are most strongly perturbed. We imagine this region to be a torus of the orbital radius $a$ and cross--sectional radius $r_{\rm I}$ which is a few times $\max(R_{\rm 2},R_{\rm c})$ where $R_{\rm 2}$ is the radius of the dense companion and $R_{\rm c}$ is the radius of the core of the giant. The energy of the cloud is then approximated by

\begin{eqnarray}
W&=&4.6\times10^{38}\,\eta_{\rm{w}}\left(\frac{M_{\rm env}}{M_{\odot}}\right)\left(\frac{M_{\rm T}}{2.6\,M_{\odot}}\right)\left(\frac{R_{\rm env}}{10\,\rm{au}}\right)^{-1}\rm{J},\\
O&=&1.0\times10^{41}\,\left(\frac{M_{\rm c}}{0.6\,M_{\odot}}\right)\left(\frac{M_{\rm 2}}{M_{\odot}}\right)\left(\frac{a}{0.01\,\rm{au}}\right)^{-1}\rm{J},\\
{\rm and}\\
M&=&5.7\times10^{39}\,\left(\frac{B}{10^7\,\rm{G}}\right)^2\left(\frac{a}{0.01\,\rm{au}}\right)\left(\frac{r_{\rm I}}{R_{\odot}}\right)^2\rm{J}.
\label{ch6.eq.me}
\end{eqnarray}

Most of the energy may be derived from the orbital decay of the binary within the CE. The value of $a$ we have taken is for a typical separation with orbital period of $1\,$d and thus represents the final separation of the system. As the orbit decays, the increase in orbital velocity results in stronger local perturbations to the CE and in an increase in dynamo activity that produces stronger magnetic fields provided the CE is still sufficiently dense around the stellar cores. Therefore it is more appropriate to consider the late stage of CE evolution from the point of view of magnetic dynamos.

For the AM Herculis system $M_{\rm{c}}=0.78\,M_{\odot}$ \citep{Gansicke06}, $M_{\rm 2}=0.37\,M_{\odot}$ \citep{Southwell95}, $a=1.1\,R_\odot$ \citep{Kafka05} and $B=1.4\times10^7\,\rm{G}$ \citep{Wickramasinghe85}. Taking $M=O$ gives $r_{\rm I}=2.1\left(B/10^9\,{\rm G}\right)^{-1}\,R_{\odot}$. This represents the limiting radius in which a field of sufficient strength could be generated given the total energy available. This is far larger than the radius of the DC and somewhat larger than the second star and so energetically there is nothing to prevent the necessary field from being generated.

\section{Governing equations for magnetic field evolution}
\label{ch6.sec.equations}

Consider the system evolving according to the static induction equation for a general isotropic magnetic diffusivity, $\eta(\bi{r},t)$, related to the electrical conductivity, $\sigma$ by $\eta=\frac{c^2}{4\pi\sigma}$.  The DC is embedded in an infinite, uniform, vertical, time dependent magnetic field. This is a sensible approximation provided the length scale for variation of the external field defined by the dynamo action in the CE is sufficiently large compared to the radius of the DC. We expect the DC to be spherically symmetric and we further assume the external field to be locally axisymmetric at the surface of the DC. In reality the exact form of the imposed field is uncertain and is likely to support a complex geometry. This is supported by spectropolarimetric observations of WD magnetic field morphologies \citep{Valyavin06}. We consider what effect this might have later on.

Except where stated otherwise, we use spherical polar coordinates $(r,\theta, \phi)$. Outside of the DC ($r>r_{\rm{c}}$) we require

\begin{equation}
\nabla\times\bi{B}=\mathbf{0}
\label{ch6.eq.external}
\end{equation}
\noindent and
\begin{equation}
\bi{B}\to B_z(t)\bi{e}_z \quad {\rm as}\quad r\to\infty.
\end{equation}

\noindent Equation~(\ref{ch6.eq.external}) is satisfied locally around the DC because the magnetic field is a superposition of the curl--free imposed field and the dipole field produced by the DC. In the global field we expect this condition to be broken by the motions of the CE. Inside the DC the magnetic field evolves according to the magneto--hydrodynamic (MHD) induction equation with stationary fluid,

\begin{equation}
\label{ch6.eq.induction}
\frac{d\bi{B}}{dt}=\nabla\times\left(\eta\nabla\times\bi{B}\right).
\end{equation}

\noindent We have ignored the term $\nabla\times\left(\bi{U}\times \bi{B}\right)$, assuming that this term, that results from fluid motions within the DC, is dominated by the diffusive term. If this term were large then we would expect WDs to support dynamo action.  This is not supported by the statistics presented in section \ref{ch6.sec.introduction}. The magnetic field must be continuous at $r=r_{\rm c}$. We proceed by decomposing the magnetic field into poloidal and toroidal parts

\begin{equation}
\bi{B}=\nabla\times\left(T\bi{r}\right)+\nabla\times\left(\nabla\times\left(S\bi{r}\right)\right),
\end{equation}

\noindent where the first term on the right is the toroidal part of the magnetic field and the second term is the poloidal part. In the case of the magnetic field external to the DC, equation~(\ref{ch6.eq.external}) simplifies to

\begin{equation}
{\mathcal L}^2T=0 \quad\rm{and}\quad {\mathcal L }^2(\nabla^2S)=0,
\end{equation}
\noindent where
\begin{equation}
{\mathcal L}^2=-\left\{\frac{1}{\sin\theta}\frac{\partial}{\partial\theta}\left[\sin\theta\frac{\partial}{\partial\theta}\right]+\frac{1}{\sin^2\theta}\frac{\partial^2}{\partial\phi^2}\right\}.
\end{equation}

\noindent Taking $S=\sum_{l=1}^{\infty}S_l(r,t)P_l(\cos\theta)$ (similarly for $T$) and given ${\mathcal L}^2P_l(\cos\theta)=l(l+1)P_l(\cos\theta)$ this implies

\begin{equation}
T=0,\quad  \nabla^2S=0.
\end{equation}

\noindent The condition $\bi{B}\rightarrow B_z\bi{e}_z$ is equivalent to $S\rightarrow\frac{1}{2}B_zr \cos\theta$. The solutions of $\nabla^2S=0$ give $S$ in terms of spherical harmonics. In the far--field the only mode is that corresponding to $l=1$. The other modes decay exponentially without some mechanism to regenerate them. Thus we take $S$ outside of the DC to be of the form

\begin{equation}
S=\left(S_0(t)r^{-2} + \frac{1}{2}B_z(t)r \right)\cos\theta,
\end{equation}

\noindent where $B_z(t)$ is the external field arising from the dynamo activity in the CE. Now consider the field inside the DC. Provided the magnetic diffusivity is spherically symmetric we can again decompose $\bi{B}$ into poloidal and toroidal parts to see that $S$ and $T$ evolve according to

\begin{equation}
\dot{S}=\eta(r,t)\nabla^2S \quad{\rm{and}}\quad \dot{T}=\eta (r,t)\nabla^2T.
\end{equation}

\noindent If we assume that we may write the diffusivity in the self--similar form $\eta(r,t)=\eta_r(r)\eta_t(t)$ then the equation is completely separable. So suppose we write $T(\bi{r},t)=U(t)V(\bi{r})$ then we find that the two functions satisfy

\begin{equation}
\frac{\dot U}{\eta_t U}=\frac{\eta_r\nabla^2 V}{V}=-\lambda^2, \quad \rm{where}\,\lambda\,{\rm is\, a\, complex\, constant}
\end{equation}
\noindent so that
\begin{equation}
\label{ch6.eq.u}
U=\exp\left(-\lambda^2\int\eta_t(t)dt\right)
\end{equation}
\noindent and
\begin{equation}
\nabla^2V+\frac{\lambda^2}{\eta_r}V=0
\label{ch6.eq.helm}.
\end{equation}

In the case of spatially constant diffusivity, equation~(\ref{ch6.eq.helm}) is simply Helmholtz's equation. Choosing solutions which are bounded as $r\rightarrow0$ gives solutions of the form $V\propto j_1(ar)\cos\theta$ where $j_i$ is the $i^{\rm th}$ spherical Bessel function of the first kind and $a=\sqrt{\frac{\lambda^2}{\eta_r}}$. In order to satisfy continuity in the magnetic field at $r=r_{\rm c}$ we must have $j_1(ar_{\rm c})=0$. This gives a real value of $a$ and so $\lambda$ must be real. So with no way to replenish the toroidal field, $T\rightarrow0$ exponentially by equation~(\ref{ch6.eq.u}).

It should be noted that, although we have assumed that all higher order spherical harmonics, azimuthal modes of the toroidal magnetic field, decay exponentially, we have not presented here additional consideration to the relative decay times. Given the vertical form of the imposed external field, the dipole mode is the only one induced by the CE and therefore the only mode we focus on. A similar analysis may be performed for other values of $l$ but these modes are only important if they are produced by the external field.

The equation for $S$ can be treated in a very similar way to $T$. For $r<r_{\rm c}$ we write 

\begin{equation}
\label{ch6.eq.s}
S=\int_{-\infty}^\infty R(r;\gamma)\exp\left(\rm{i} \gamma H\right) \cos\theta \,d\gamma.
\end{equation}

\noindent By taking $H=\int\eta_t(t)dt$ the system is reduced to a Fourier transform. It is possible to use different transform methods with different choices of the parameter $\gamma$. However, because we shall consider an oscillating external field, this is the natural choice. In this form the function is still separable and $R$ solves the equation

\begin{equation}
\nabla^2(R \cos\theta)=\frac{{\rm i} \gamma}{\eta_r} R \cos \theta
\end{equation}
\noindent so that
\begin{equation}
\frac{d^2R}{dr^2}+\frac{2}{r}\frac{dR}{dr}-(\frac{{\rm i}\gamma}{\eta_r}+\frac{2}{r^2})R=0.
\label{ch6.eq.helmholtz}
\end{equation}

\noindent In order to enforce continuity at the boundary we must have $S\,{\rm and}\,\frac{\partial S}{\partial r}$ continuous at $r=r_{\rm c}$ for all $\theta$. These are equivalent to

\begin{equation}
\label{ch6.eq.rearrange1}
\int_{-\infty}^{\infty}\frac{\partial R(r;\gamma)}{\partial r}\bigg|_{r=r_{\rm c}}\exp\left({\rm i}\gamma H\right)\,d\gamma=-\frac{2S_0(t)}{r_{\rm c}^3}+\frac{1}{2}B_z(t)\\
\end{equation}
\noindent and
\begin{equation}
\label{ch6.eq.rearrange2}
\int_{-\infty}^{\infty}R(r_{\rm c};\gamma)\exp({\rm i}\gamma H)\,d\gamma=\frac{S_0(t)}{r_{\rm c}^2}+\frac{1}{2}B_z(t)r_{\rm c}.
\end{equation}

\noindent By taking $\frac{1}{3}(($\ref{ch6.eq.rearrange2}$)-r_{\rm c}.($\ref{ch6.eq.rearrange1}$))$ and $\frac{4}{3}(($\ref{ch6.eq.rearrange2}$)/r_{\rm c}+\frac{1}{2}($\ref{ch6.eq.rearrange1}$))$ these become

\begin{equation}
\frac{1}{3}\int_{-\infty}^{\infty}\left(R(r_{\rm c};\gamma)-\frac{\partial R(r;\gamma)}{\partial r}\bigg|_{r=r_{\rm c}}.r_{\rm c}\right) \exp({\rm i}\gamma H)\,d\gamma=\frac{S_0(H)}{r_{\rm c}^2}
\label{ch6.eq.dipole}
\end{equation}
\noindent and
\begin{equation}
\frac{4}{3}\int_{-\infty}^{\infty}\left(\frac{R(r_{\rm c};\gamma)}{r_{\rm c}}+\frac{1}{2}\frac{\partial R(r;\gamma)}{\partial r}\bigg|_{r=r_{\rm c}}\right) \exp({\rm i}\gamma H)\,d\gamma=B_z(H)
\label{ch6.eq.boundary}.
\end{equation}

\noindent We take $B_z(H)$ as given and so focus for the moment on the second of these equations. This allows us to solve for $S$ and then we can derive the magnetic field within the DC. So if the transform of $B_z(H)$ is $\hat{B}_z(\gamma)$ then we may rewrite the previous equation as

\begin{equation}
\label{ch6.eq.boundary2}
\hat{B}_z(\gamma)=\frac{4}{3}\left(\frac{R(r_{\rm c};\gamma)}{r_{\rm c}}+\frac{1}{2}\frac{\partial R(r;\gamma)}{\partial r}\bigg|_{r=r_{\rm c}}\right).
\end{equation}

\noindent So given the condition $R\to 0$ as $r\to 0$ and this boundary condition we can now fully determine $R$. This in turn fully solves the internal (and external) dipole field of the DC. In the case $T=0$, $S=Q(t)R(r)\cos\theta$, the internal magnetic field (for a particular $\gamma$) is given by

\begin{equation}
\label{ch6.eq.field}
\bi{B}=Q(t)\left(2\frac{R}{r}\cos\theta\,\bi{e}_r-\left(\frac{R}{r}+\frac{\partial R}{\partial r}\right)\sin\theta\,\bi{e}_{\theta}\right).
\end{equation}

\noindent In theory we may solve this system exactly for any imposed field. However, owing to the complexity of the external field and its Fourier transform under the change of variable $t\to H(t)$, numerical computations become extremely difficult. Thus, for the most part, we focus on single modes given by some specific $\gamma$. This is ultimately justified by the slow variation of $\eta$ during the lifetime of the CE.

In the case where the induced magnetic field is not entirely dipolar, the final term in equation~(\ref{ch6.eq.helmholtz}) is modified by some factor for each $l$. In the case of uniform diffusivity this gives spherical Bessel functions of varying order. The forms of these functions in this parameter regime are actually very similar to the first order function considered in section~\ref{ch6.sec.results} so we do not expect including the higher modes to affect the qualitative results significantly for the overall strength and radial variation of the internal field. In addition, given the uncertainties in the geometry of the magnetic field of the CE, inclusion of higher order harmonics is unlikely to give any greater insight.

\section{Numerical Methods}
\label{ch6.sec.method}

We calculate the spatial form of the magnetic field in the presence of an oscillating external field with a sixth--order adaptive step Runge--Kutta algorithm \citep{Press92} applied to equation~(\ref{ch6.eq.field}). The equation was integrated from the centre out to the surface with boundary conditions $R(0)=0$ and a small arbitrary value of $R'(0)$. The solution grows by several thousand orders of magnitude between $r=0$ and $r=r_{\rm c}$. To cope with this variation, we include a subroutine to rescale the solution whenever $R(r)$ or its derivatives exceeds some maximum value, typically $10^{30}$. Once a solution has been found the boundary condition at $r=r_{\rm c}$, given by equation~(\ref{ch6.eq.boundary}), is matched by scaling the entire function. Such arbitrary rescaling is valid because of the linear nature of the governing ODE and the zero boundary condition at $r=0$. In order to calculate the diffusivity field we employed the approximation of \citet{Wendell87} for the electrical conductivity. In addition we needed to calculate the opacity for the given temperature and density. This was done with the data tables and subroutines of the {\sc stars} stellar evolution code \citep{Eggleton73,Pols95}.  In all simulations we have used a CO~WD composition for the DC.

The time evolution of the magnetic field following the dispersal of the external field was calculated with a first order Euler finite step method. The code was run multiple times with a variety of step sizes and spatial resolutions with no significant variation in the solution between runs. The induction equation was simplified by the relations

\begin{equation}
\tilde{R}(r)=R(r)+\frac{r}{2}\frac{\partial R}{\partial r}
\end{equation}
\noindent and
\begin{equation}
R(r)=\frac{1}{r^2}\int^r_02r\tilde{R}(r)dr. \label{ch6.eq.q1}
\end{equation}

\noindent This removes the spherical geometry of the induction equation, reducing it to the form

\begin{equation}
\frac{\partial \tilde{R}}{\partial t}=\eta\frac{\partial^2\tilde{R}}{\partial r^2},
\end{equation}

\noindent which has a zero boundary condition at $r=r_{\rm c}$ in the absence of an external field and is ultimately easier to work with. The function $R$ and subsequently $\bi{B}$ are then recovered from $\tilde{R}$ with the relation in equation~(\ref{ch6.eq.q1}) numerically evaluated with the trapezoidal rule.  The routine does not take into account DC cooling. This is unlikely to affect the evolution of the field during the CE phase and shortly after when the field is evolving rapidly. In the late time evolution however when the field has finished its radial redistribution and is decaying purely exponentially, \citet{Wendell87} showed that the cooling significantly increases the decay timescale and the magnetic field becomes essentially frozen into the dwarf. Therefore we may take the final value for the field strength shortly after it has reached a state of exponential decay. This is typically around $10^7\,$yr after the dispersion of the CE.

\section{Results}
\label{ch6.sec.results}

The magnetic field produced by the CE is heavily dependent on several key factors. Although the magnetic field of the DC and the CE is continuous at the surface, the total magnetic flux able to penetrate the DC depends on the lifetime of the CE. Once the CE has dispersed, the field continues to migrate inwards and the surface field strength decays. The degree to which this happens depends on how much of the field is able to penetrate the DC while the surface field is still maintained. This requires a suitable treatment of the radial conductivity profile of the DC. As the density increases towards the centre of the DC the conductivity rises rapidly, inhibiting further diffusion of the field. This results in the confinement of the field to around the outer $10\,\%$ of the DC by radius.

The structure and orientation of the magnetic field of the CE is also of critical importance. In a convectively driven magnetic dynamo we might expect the orientation of the field to change rapidly. The frequency of the changes has important consequences for the field. Given the geometry of the system driving the fluid motions which in turn give rise to the magnetic dynamo we might also anticipate a preferred direction to the orientation of the field. A field that is maintained in a single direction produces a WD field several orders in magnitude stronger than one in which the orientation varies rapidly and with random orientation.

\subsection{Consequences of CE lifetime}

The lifetime of the CE has a complicated relationship to the original orbital separation of the binary system, the size of the envelope and the properties of the degenerate core and its companion. This is to be expected because of the wide range of magnetic field strengths in MWDs. Here we consider how quickly a field may be built up in the DC as a result of an applied uniform constant vertical magnetic field outside. This mechanism produces stronger magnetic fields than one with varying orientation but it is the best case scenario and places an upper bound on how much field might be retained.  While the MHD properties of CEs are not understood, it is difficult to say to what degree there may be a preferred orientation for the magnetic field so the true physical system may resemble anything in between a uni--directional field and one that reorients itself randomly in any direction.

\begin{figure}
\centering
\includegraphics[width=0.99\textwidth]{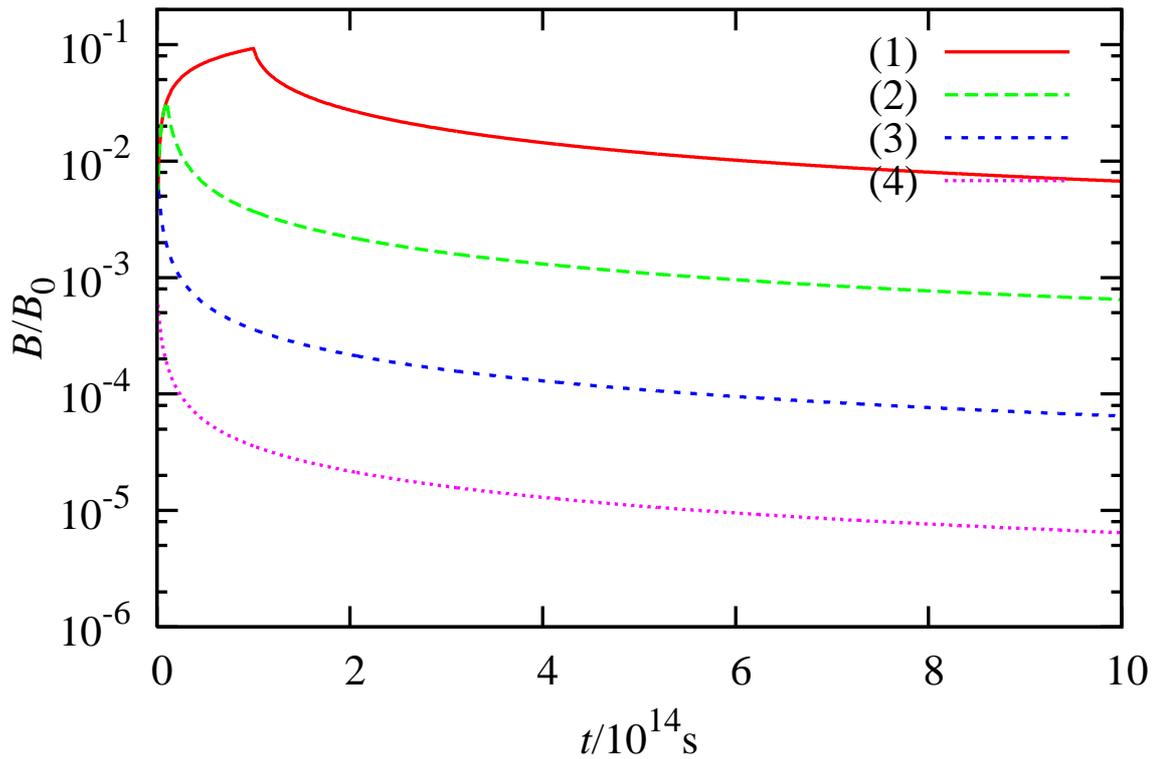}
\caption[Radial magnetic field strength at $0.99$ of the stellar radius for a white dwarf embedded in a uniform vertical field]{The radial magnetic field strength at $r=0.99\,r_{\rm c}$ that results from a constant, uniform, vertical magnetic field applied to a DC for (1) $10^{11}\,{\rm s}$, (2) $10^{12}\,{\rm s}$, (3) $10^{13}\,{\rm s}$ and (4) $10^{14}\,{\rm s}$. Note the peak of the field strength at the end of the phase where the external field is applied. After the removal of the external field there is a period of rapid decay lasting around $2\times 10^{14}\,{\rm s}$ after which the field decays much more slowly on a timescale of around $10^{15}\,\rm{s}$.  The strength of the field is roughly proportional to the lifetime of the external field.}
\label{ch6.fig.applied}
\end{figure}

\begin{figure}
\centering
\includegraphics[width=0.99\textwidth]{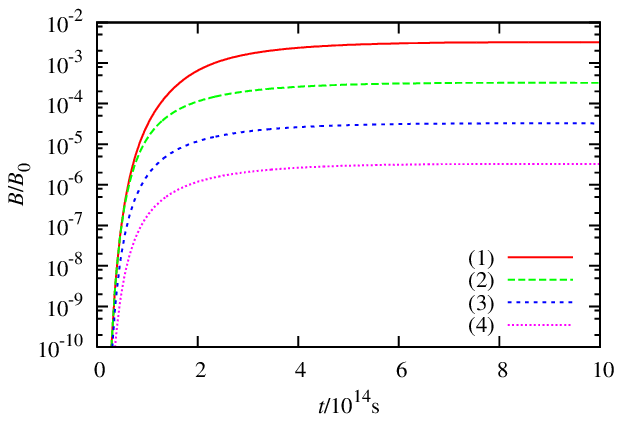}
\caption[Radial magnetic field strength at $0.9$ of the stellar radius for a white dwarf embedded in a uniform vertical field]{The radial magnetic field strength at $r=0.9\,r_{\rm c}$ that results from a constant, uniform, vertical magnetic field applied to a DC for (1) $10^{11}\,{\rm s}$, (2) $10^{12}\,{\rm s}$, (3) $10^{13}\,{\rm s}$ and (4) $10^{14}\,{\rm s}$. The magnetic field does not exhibit the same peaks as at $r=0.99\,r_{\rm c}$ because the saturation of the field takes significantly longer than closer to the surface. The field here continues to grow even after the rapid decay phase of the field closer to the surface as it continues to diffuse inwards. We also see that the strength of the field is roughly proportional to the lifetime of the external field.}
\label{ch6.fig.applied2}
\end{figure}

\begin{figure}
\centering
\includegraphics[width=0.99\textwidth]{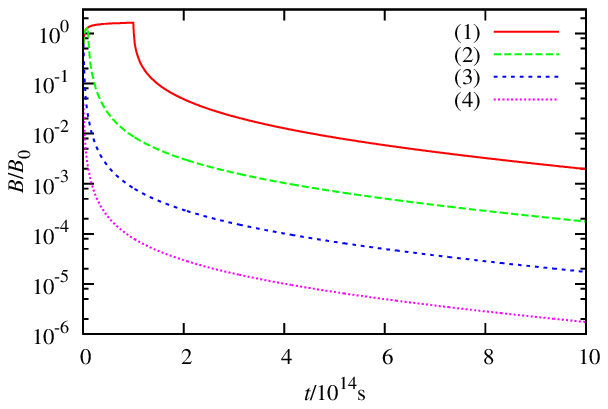}
\caption[Meridional magnetic field strength at $0.99$ of the stellar radius for a white dwarf embedded in a uniform vertical magnetic field]{The meridional magnetic field strength at $r=0.99\,r_{\rm c}$ that results from a constant, uniform, vertical magnetic field applied to a DC for (1) $10^{11}\,{\rm s}$, (2) $10^{12}\,{\rm s}$, (3) $10^{13}\,{\rm s}$ and (4) $10^{14}\,{\rm s}$. Note the peak of the field strength at the end of the phase where the external field is applied.  After the removal of the external field there is a period of rapid decay lasting around $2\times 10^{14}\,{\rm s}$ after which the field decays much more slowly on a timescale of around $10^{15}\,{\rm s}$.  The strength of the meridional field is several orders of magnitude larger than the radial field.  The strength of the field is roughly proportional to the lifetime of the external field.}
\label{ch6.fig.applied3}
\end{figure}

\begin{figure}
\centering
\includegraphics[width=0.99\textwidth]{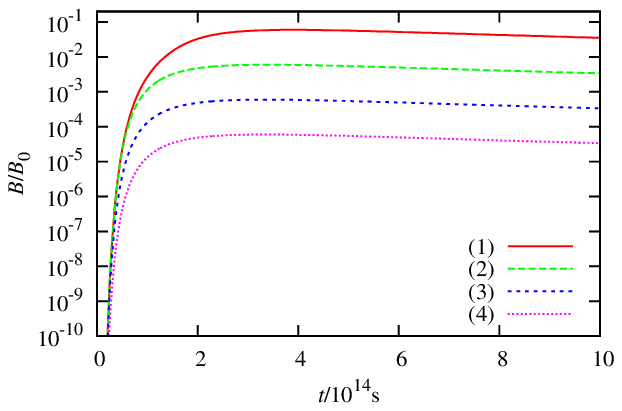}
\caption[Meridional magnetic field strength at $0.9$ of the stellar radius for a white dwarf embedded in a uniform vertical magnetic field]{The meridional magnetic field strength at $r=0.9\,r_{\rm c}$ that results from a constant, uniform, vertical magnetic field applied to a DC for (1) $10^{11}\,{\rm s}$, (2) $10^{12}\,{\rm s}$, (3) $10^{13}\,{\rm s}$ and (4) $10^{14}\,{\rm s}$. The magnetic field does not exhibit the same peaks as at $r=0.99\,r_{\rm c}$ because the saturation of the field takes significantly longer than closer to the surface. The field here continues to grow even after the rapid decay phase of the field closer to the surface as it continues to diffuse inwards. We also see that the strength of the field is roughly proportional to the lifetime of the external field.}
\label{ch6.fig.applied4}
\end{figure}

We applied a constant external field of $\bi{B}_0$ to a DC of radius $0.01\,\rm{R_{\odot}}$, mass $0.6\,\rm{M_{\odot}}$ with diffusivity profile determined by a polytropic index of $3/2$ and temperature $10^5\,{\rm K}$. The equations for electrical conductivity from \citet{Wendell87} are not strongly temperature dependent in the highly degenerate density regime. As such, the diffusivity profile for WDs of temperatures varying between $10^3\,$K and $10^7\,$K only show noticeable differences at the surface. This has no significant effect on the evolution of the DC field in our model. The magnetic field of the WD scales linearly with the applied field. We tested the cases where the field was applied for $10^{11}\,{\rm s},10^{12}\,{\rm s},10^{13}\,{\rm s}$ and $10^{14}\,{\rm s}$. The radial and meridional fields at $r=0.99r_{\rm c}$ and $r=0.9\,r_{\rm c}$ are shown in Figs.~\ref{ch6.fig.applied} to~\ref{ch6.fig.applied4}. For $r=0.99\,r_{\rm c}$ we note the two distinct phases of the solution. First there is the initial growth phase in the presence of the external field. Once the external field is removed the field strength peaks and begins to decay. The strength of the field that remains after the removal of the external field increases in proportion to the lifetime of the external field. Following the removal of the external field, the surface field of the exposed WD decays rapidly for around $10^{14}\,{\rm s}$ before continuing to decay exponentially with a characteristic timescale of around $10^{15}\,{\rm s}$. Whilst this rate of decay is too rapid to explain the existence of long--lived WD magnetic fields, our simulations have not included WD cooling. The results of \citet{Wendell87} indicate that cooling causes the diffusivity to decrease and significantly extends the decay time scale for the magnetic field. This effectively freezes the field.  If we suppose the magnetic field is frozen after $2\times10^{14}\,{\rm s}$ when the field has relaxed to its post CE state then the residual magnetic field strength produced by an external field of strength $B_{\rm ext}$ with lifetime $t_{\rm CE}$, taking into account both radial and meridional components, is approximately

\begin{equation}
B_{\rm res}=B_{\rm ext}\frac{5\,t_{\rm CE}\times10^{-16}}{\,{\rm s}}.
\label{ch6.eq.res}
\end{equation}

From the results at $r=0.9\,r_{\rm c}$ we see that the field continues to diffuse inwards after the removal of the external field until it reaches saturation point. This is because the diffusivity decreases towards the interior of the WD. Once the field reaches a certain point, the diffusion becomes so slow that it is effectively halted. This is extremely important if we wish to build strong surface fields because it prevents any further redistribution of magnetic energy. If we look at the magnetic field further into the star we see that the field doesn't penetrate any deeper than $r=0.5\,r_{\rm c}$ after $10^{15}\,{\rm s}$ and the field inside $r=0.9\,r_{\rm c}$ is extremely weak compared to the surface field.

\subsection{Effect of randomly varying the magnetic field orientation}

If the direction of the applied field is not constant, as we might expect from a dynamo driven field, then typically the final field strength is reduced by some factor based on the degree of variation. For a spatially constant conductivity with oscillating boundary conditions we may solve the induction equation analytically. This gives us some insight into the mechanisms preventing the build up of strong fields in the case of a rapidly varying external field. We also simulated the change in orientation numerically by taking the field generated by applying a magnetic field for a short time in a single orientation and then taking the sum of the same field rotated at random angles at each time step up until the dispersion of the CE.

\subsubsection{Uniform diffusivity DC}
\label{ch6.sec.qparameter}

In the case where $\eta_r$ is a constant we are able to solve for $R$ analytically as in section \ref{ch6.sec.equations}. In this case, $\eta(t)=\eta_r\eta_t(t)$ and equation~(\ref{ch6.eq.helmholtz}) becomes the Helmholtz equation and we seek solutions in the form of spherical Bessel functions. We find that $R\propto j_1\left(\left(\rm{i}\gamma\right)^{\frac{1}{2}}r\right)$. Then from equations~(\ref{ch6.eq.boundary2}) and (\ref{ch6.eq.s}) we find

\begin{equation}
S=\frac{3}{4}\int_{-\infty}^{\infty}\hat{B}_z(\gamma)\left(\frac{j_1\left(\left(\rm{i}\gamma\right)^{\frac{1}{2}}r\right)}{j_1\left(\left(\rm{i}\gamma\right)^\frac{1}{2}r_c\right)+\frac{1}{2}\left(\rm{i}\gamma\right)^\frac{1}{2}j_1'\left(\left(\rm{i}\gamma\right)^\frac{1}{2}r_c\right)}\right)e^{\rm{i} \gamma H}\,d\gamma \,\cos\theta.
\end{equation}

\noindent Consider the strength of the field generated at the surface of the DC by each $\gamma$--mode. We define $Q(\gamma)$ by

\begin{equation}
Q(\gamma)=Re\left[\frac{j_1\left(\left(\rm{i}\gamma\right)^{\frac{1}{2}}r_c\right)}{j_1\left(\left(\rm{i}\gamma\right)^\frac{1}{2}r_c\right)+\frac{1}{2}\left(\rm{i}\gamma\right)^\frac{1}{2}j_1'\left(\left(\rm{i}\gamma\right)^\frac{1}{2}r_c\right)}\right].
\end{equation}

\begin{figure}
\centering
\includegraphics[width=0.99\textwidth]{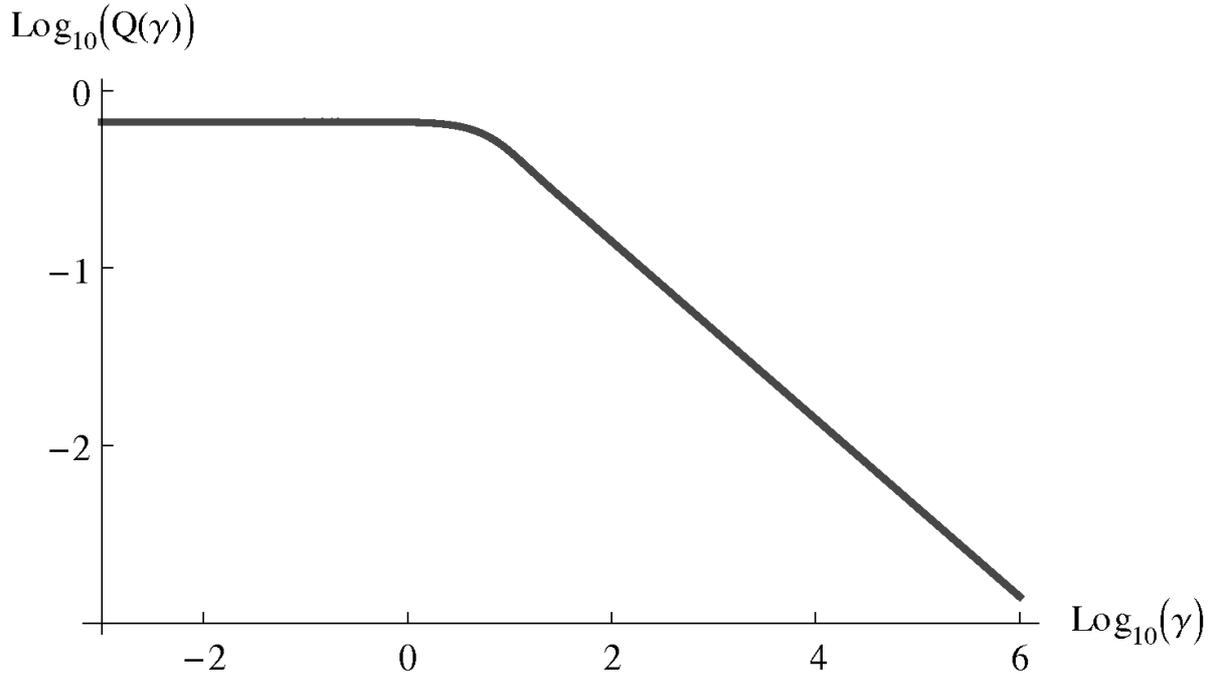}
\caption{Behaviour of the transfer efficiency factor, $Q(\gamma)$ as described in section~\ref{ch6.sec.qparameter}.}
\label{ch6.fig.q}
\end{figure}

\noindent We shall refer to $Q(\gamma)$ as the transfer efficiency because it represents roughly the radial magnetic flux across the surface of the DC relative to the imposed field. Fig.~\ref{ch6.fig.q} shows how $Q(\gamma)$ varies. We see that for $\gamma>1$, $Q(\gamma)$ falls off approximately as $\gamma^{-\frac{1}{2}}$. So the higher $\gamma$ modes of the system are more effectively suppressed but the fall off is slow. This means that the radial magnetic field at the surface of the star for very large $\gamma$ approaches $0$. In this case the boundary conditions are matched by the dipolar field which cancels out the imposed field at the surface of the DC (c.f. equation~(\ref{ch6.eq.rearrange2})).

If we consider a system with constant diffusivity $\eta$, then $H=\eta t$. If we then take $B_z(t)\propto \cos(\alpha t)$ it is easy to show that the transfer efficiency is $Q(\frac{\alpha}{\eta})$. We interpret this by recognising that, if the rate of oscillation is too high compared to the diffusivity each time the external field switches, most of the field generated by the previous oscillation is cancelled out. If the diffusivity is high enough, the oscillations are less effectively cancelled and a residual field is able to build up.

\begin{figure}
\centering
\includegraphics[width=0.99\textwidth]{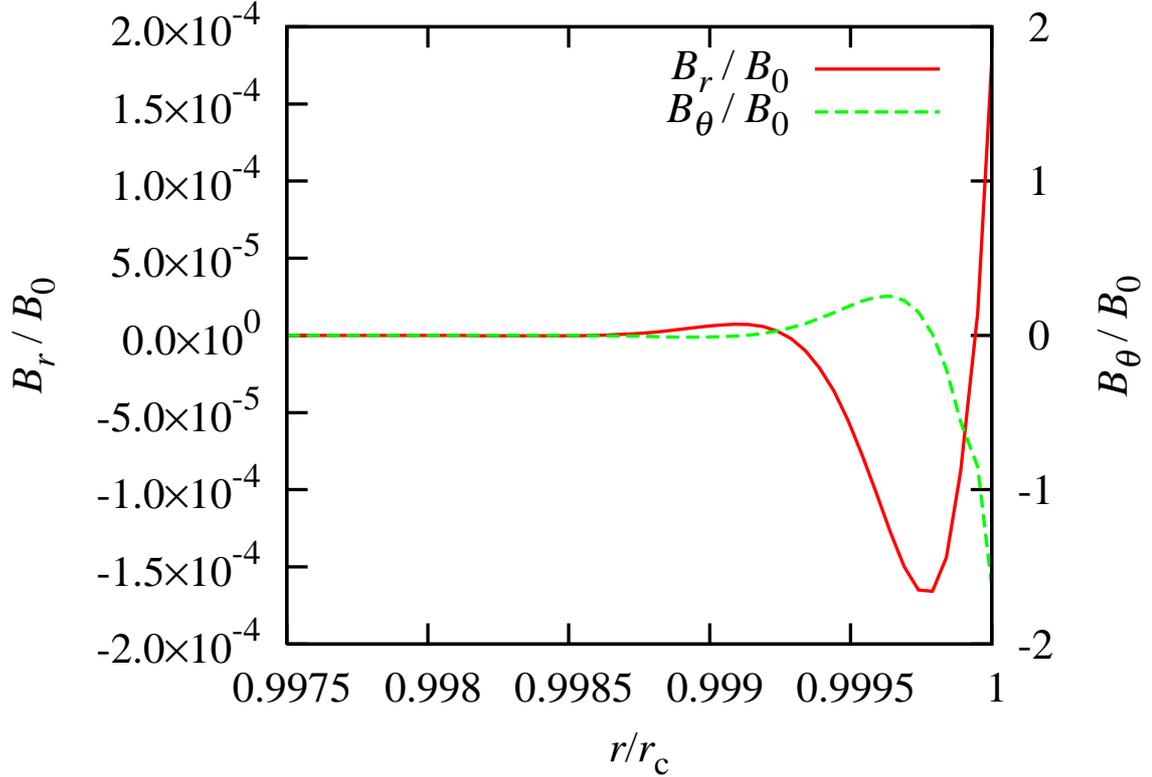}
\caption[Form of the induced field within the DC with spatially uniform diffusivity and varying external field]{Form of the induced field within the DC with a spatially uniform magnetic diffusivity $\eta=221{\rm cm}^2\,{\rm s}^{-1}$ with variation frequency $\alpha=2\times10^{-8}\,{\rm s}^{-1}$ for an imposed field of $B_0$. The radial field is $B_{\rm r}$ and the toroidal field is $B_{\theta}$. We see from the horizontal axis that the field is confined to the outer most regions of the DC. This is because the rapid variation of the external field causes any field that is generated at the surface to smooth out to zero as it diffuses inwards. We also see that the meridional field is several orders of magnitude stronger than the radial field.}
\label{ch6.fig.field}
\end{figure}

Now let us consider the radial form of the induced field. From our discussion above we expect that the field is functionally similar to ${\rm Re}[j_1({\rm i}{\alpha}/{\eta})^{1/2}r]$. Fig.~\ref{ch6.fig.field} shows the form of the field that results from the solution of equations~(\ref{ch6.eq.dipole}),~(\ref{ch6.eq.boundary}) and~(\ref{ch6.eq.field}). We have used the parameters given above and an imposed field strength of $B_0$.  So, although the transfer efficiency is very low and the radial field is suppressed, the strong $R$ gradient produces an internal field parallel to the surface which is comparable in magnitude to the imposed field but decays extremely rapidly with depth. The thickness of this layer behaves asymptotically as $(\frac{\alpha}{\eta})^{-1/2}$. We note that this is the same behaviour as the efficiency factor $Q(\gamma)$. This is reasonable because the efficiency of transfer of magnetic energy from the external field to the DC should scale roughly in proportion to how far the field can penetrate, at least for shallow layers.

Because we are proposing a model in which the magnetic field of a common envelope surrounding a binary system induces the magnetic field of the DC we choose a value of $\alpha$ to reflect this. First consider a far--field varying on a time scale similar to the orbital period of the binary system. We determine this time scale for a $0.6\,M_{\odot}$ DC with a typical red giant companion of mass $6\,M_{\odot}$ and radius $400\,R_{\odot}$ which has filled its Roche Lobe. This gives us an orbital time scale of $t_{\rm orbit}\approx 3.5\times 10^8\,{\rm s}$ and so we choose $\alpha=\frac{2\pi}{t_{\rm orbit}}\approx 2\times 10^{-8}\,{\rm s}^{-1}$. This gives a transfer efficiency of $Q(\frac{\alpha}{\eta})\approx 2\times 10^{-4}$. Alternatively as we suggested earlier, the magnetic dynamo is likely to be stronger when the orbit of the binary has decayed. If we apply the same analysis with $t_{\rm{orbit}}=1\,\rm{day}$ then $Q\approx4\times10^{-6}$. In either case, the field is still confined to a very thin boundary layer and is almost entirely meridional.

\subsubsection{Post CE evolution of the magnetic field}

Following the initial generation of the magnetic field within the DC we foresee two possible cases for the subsequent evolution. Either the CE is dispersed on a time scale shorter than the variation of magnetic diffusivity or the diffusivity varies by a significant factor over the lifetime of the CE. In the absence of a magnetic field, the lifetime of the CE phase is estimated to be on the order of $10^3\,$yr \citep{Taam78}.  The time scale for orbital decay as a result of a dynamo driven wind is estimated to be a few Myr \citep{Regos95}.  If energy from orbital decay is transformed to magnetic energy rather than unbinding the envelope then the lifetime of the CE phase is prolonged though from the energy considerations of section \ref{ch6.sec.energy} the lifetime is likely to lie much closer to the lower bound. Depending on the degree to which this happens the lifetime of the dynamo may lie anywhere between these two bounds. Typically the diffusivity varies on a time scale of $t_\eta\approx 10^{16}\,{\rm s}\approx3\times 10^{10}\,{\rm yr}$, much longer than the probable lifetime of the CE.  However our understanding of CE evolution is sufficiently poor that these estimates may bear little resemblance to reality.

The decay of an unsupported magnetic field occurs on a time scale $t_{B}\approx\beta\frac{r_{\rm c}^2}{\eta}$ where $\beta$ is a constant related to the field geometry. For an unsupported field $\beta=\pi^2$. In the case where $\eta=221\,{\rm cm}^2\,{\rm s}^{-1}$, which is a good approximation for the surface layers of a DC, and $r_{\rm c}=0.01\,R_{\odot}$ we get $t_{B}\approx 2\beta\times10^{15}\,\rm{s}$.  \citet{Wendell87} showed that the magnetic field of a WD without an imposed field has a decay time which is much longer than the evolutionary age of the WD over its whole lifetime and so the magnetic field is essentially frozen into the WD. We observe that, in our setup where the internal field is supported by a rapidly oscillating external field, the strong confinement of the field to the outermost layers of the DC produces strong gradients which lead to rapid magnetic diffusion. Once the field is removed, the system rapidly relaxes to the solution given by zero external field. The field then decays on an ohmic time scale of $t_{B}$ with $\beta=\pi^2$ \citep{Proctor94}.

\begin{figure}
\centering
\includegraphics[width=0.99\textwidth]{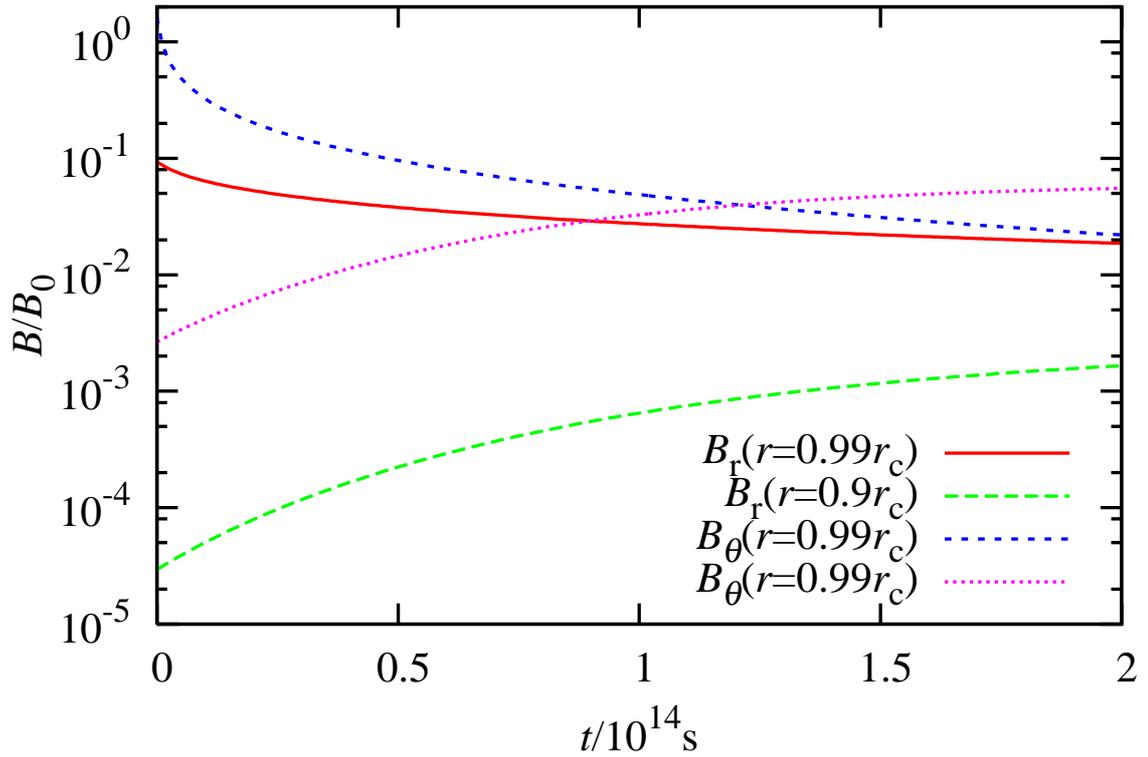}
\caption[Decay of the magnetic field of a WD upon removal of a constant external magnetic field]{Decay of the magnetic field of a WD upon removal of a constant external magnetic field applied for $10^{14}\,{\rm s}$. The plot shows the evolution of the evolution of the radial field, $B_{\rm r}$, and the toroidal field, $B_{\theta}$ at $\frac{r}{r_{\rm c}}=0.99$ and $\frac{r}{r_{\rm c}}=0.9$. Although the surface meridional field is initially much stronger than the radial field it decays much faster until the two are roughly equal $2\times10^{14}\,\rm{s}$ after the external field is removed. We also see that, even after the rapid initial decay of the surface field seems to have finished, the field $0.9\,r_0$ is still growing as it diffuses inwards.}
\label{ch6.fig.decay}
\end{figure}

Fig.~\ref{ch6.fig.decay} shows the evolution of the field produced by a DC with diffusivity profile determined from a polytropic structure. The field strengths of the radial and meridional fields are shown at $r/r_{\rm c}=0.99$ and $0.9$. We see that, once the external field vanishes, the field spreads slowly inwards. The field was calculated below $r/r_{\rm c}=0.9$ but even by the end of the simulation, no field had reached as deep as $r/r_{\rm c}=0.5$. The surface magnetic field decays by about a factor of $10^2$ as it relaxes owing to the redistribution of the magnetic energy over the radius. This occurs on a time scale of around $2\times10^{13}\,{\rm s}$. After the initial rapid decay, the surface field continues to decay exponentially on a much longer timescale of around $10^{14}\,{\rm s}$. This is faster than predicted analytically but as Fig.~\ref{ch6.fig.decay} shows the field at $r/r_{\rm c}=0.9$ is still growing so the system has not fully relaxed. It is likely that the decay timescale grows as the field moves towards spatial equilibrium. WD cooling which has not been included here would also prevent additional decay.

We have neglected secondary effects that are produced as a result of the finite time the magnetic dynamo of the CE takes to decay. We expect though, given that the lifetime of the CE is estimated to be significantly shorter than the relaxation time for the field, the magnetic dynamo is likely to decay on an even shorter time scale. Thus the effect on the solution should be negligible.

\subsubsection{Numerically evolved field with random orientation}

\begin{table}
\begin{center}
\begin{tabular}{cccc}
\hline
$t{\rm{dyn}}/\,\rm{s}$&$B_{\rm{R}}/\rm{G}$&$B_{\rm{M}}/\rm{G}$&$B_{\rm{T}}/\rm{G}$\\
\hline
\multicolumn{4}{c}{No preferred orientation}\\
\hline
$10^6$&$1.9\times10^{-6}$&$-1.8\times10^{-6}$&$1.3\times10^{-5}$\\
$10^7$&$-3.3\times10^{-6}$&$4.0\times10^{-7}$&$-4.1\times10^{-6}$\\
$10^8$&$-3.6\times10^{-6}$&$-2.6\times10^{-6}$&$-1.6\times10^{-6}$\\
$10^9$&$4.2\times10^{-5}$&$-1.6\times10^{-5}$&$3.9\times10^{-6}$\\
\hline
\multicolumn{4}{c}{Half--plane restricted orientation}\\
\hline
$10^6$&$2.3\times10^{-3}$&$-4.2\times10^{-6}$&$-1.1\times10^{-5}$\\
$10^7$&$2.9\times10^{-3}$&$-2.5\times10^{-6}$&$3.8\times10^{-6}$\\
$10^8$&$1.9\times10^{-3}$&$-9.2\times10^{-6}$&$8.7\times10^{-6}$\\
$10^9$&$2.0\times10^{-3}$&$1.6\times10^{-5}$&$1.4\times10^{-5}$\\
\hline
\end{tabular}
\caption[Residual surface magnetic field strengths generated by a varying external field]{Residual surface magnetic field strengths generated by an external field of unit strength which reorients at intervals of $t_{\rm dyn}$. $B_{\rm R}$ is the radial field, $B_{\rm M}$ is the meridional field and $B_{\rm T}$ is the toroidal field. The external field lasts for a total of $10^{13}\,{\rm s}$ and the residual field is taken at $2\times10^{14}\,{\rm s}$}
\label{ch6.tab.orient}
\end{center}
\end{table}

In order to simulate a randomly varying external field in a more three--dimensional sense we considered a supposition of multiple copies of a base field rotated into random orientations applied in sequence. The base field was generated by applying a constant vertical field for a range of times much shorter than the lifetime of the CE. Such a superposition is valid because the induction equation is linear.  So if the field generated by applying a constant vertical magnetic field for time $\delta t$ is $\tilde{\bi{B}}(t)$ then the total field is

\begin{equation}
\bi{B}(t)=\sum_{k=1}^N\bi{A}\left(\theta_k,\phi_k\right)\tilde{\bi{B}}\left(t-k\delta t\right),
\end{equation}

\noindent where $\theta_k$ and $\phi_k$ are randomly generated angles that describe a uniform distribution of spatial orientations and $N\delta t$ is the duration of the magnetic dynamo. The rotation matrix $\bi{A}(\theta,\phi)$ for a vector in Cartesian coordinates is

\begin{equation}
A(\theta,\phi)=\begin{pmatrix}
\cos\theta & \sin\phi\sin\theta & \cos\phi\sin\theta \\
0 & \cos\phi & -\sin\phi \\
-\sin\theta & \sin\phi\cos\theta & \cos\phi\cos\theta
\end{pmatrix}.
\end{equation}

\noindent The results are shown in Table~\ref{ch6.tab.orient}. We find that, in the case of a purely random orientation, the residual field is smaller than the constant external field case by a factor of around $1,000$. This is to be expected because, on average, each field orientation is opposed by one pointing in the opposite direction. The fluctuations arise from the random nature of the field orientation and the different times at which each field is applied. It seems certain that the magnetic field strength that would be needed to generate observed magnetic fields would be extremely large in this case.

\subsubsection{The effect of introducing a preferred direction to the external field}

Owing to the nature of the proposed magnetic dynamo it seems reasonable to suggest that there may be a preferred direction to the magnetic field. If the field is generated by perturbations in the CE owing to the orbital motion of the DC and its companion, the magnetic field lines may show a tendency to align with the orbit. This contrasts with the dynamo mechanism in accretion discs which is a result of the magneto--rotational instability producing eddies that have no preferred direction. Therefore we restrict the magnetic field vector to lie within a hemisphere. The results are also included in Table~\ref{ch6.tab.orient}. We find that the residual radial field can now reach strengths close to those produced by the constant external field. This does not apply to the meridional field or the toroidal field because different orientations may still oppose each other and give typically weak field strengths.

If we now vary the CE lifetimes (Table~\ref{ch6.tab.orient2}) we see that in the case of purely random orientations, the overall field strength is largely unaffected by changes to it. In the case where the field has preferred alignment, the residual radial field strength increases roughly linearly with CE lifetime as in the case of constant external field. It is also possible to produce fields with similar field strength to the uni--directional case. This suggests that while dynamo action may result in fluctuating magnetic field orientations with time, the residual field strength can be maintained provided that the dynamo has a preferred direction on average.

\begin{table}
\begin{center}
\begin{tabular}{cccc}
\hline
$t_{\rm CE}/\rm{s}$&$B_{\rm R}/B_0$&$B_{\rm M}/B_0$&$B_{\rm T}/B_0$\\
\hline
\multicolumn{4}{c}{No preferred orientation}\\
\hline
$10^{11}$&$4.7\times10^{-6}$&$1.8\times10^{-6}$&$4.0\times10^{-7}$\\
$10^{12}$&$3.2\times10^{-6}$&$-4.8\times10^{-6}$&$-8.5\times10^{-7}$\\
$10^{13}$&$4.2\times10^{-5}$&$-1.6\times10^{-5}$&$3.9\times10^{-6}$\\
$10^{14}$&$-6.7\times10^{-5}$&$-3.3\times10^{-5}$&$-4.9\times10^{-5}$\\
\hline
\multicolumn{4}{c}{Half--plane restricted orientation}\\
\hline
$10^{11}$&$1.9\times10^{-5}$&$3.0\times10^{-7}$&$2.8\times10^{-6}$\\
$10^{12}$&$2.0\times10^{-4}$&$1.1\times10^{-6}$&$1.5\times10^{-5}$\\
$10^{13}$&$2.0\times10^{-3}$&$-2.3\times10^{-5}$&$1.3\times10^{-5}$\\
$10^{14}$&$3.1\times10^{-2}$&$-1.2\times10^{-4}$&$9.5\times10^{-5}$\\
\hline
\end{tabular}
\caption[Residual surface field strengths generated by an external field of strength $B_0$ which reorients at intervals of $10^9\,{\rm s}$]{Residual surface magnetic field strengths generated by an external field of strength $B_0$ which reorients at intervals of $10^9\,{\rm s}$. $B_{\rm R}$ is the radial field, $B_{\rm M}$ is the meridional field and $B_{\rm T}$ is the toroidal field. The external field lasts for $t_{\rm CE}$ and the residual field is taken at $2\times10^{14}\,{\rm s}$}
\label{ch6.tab.orient2}
\end{center}
\end{table}

\section{Conclusions}
\label{ch6.sec.conclusions}

The lack of magnetic WDs in wide binary systems suggests that the origin of their strong fields is a product of some feature of binary interaction. Indeed, the argument that the fields originate through flux conservation during the collapse of Ap/Bp stars does not explain their high frequency in interacting binaries. By building on the proposal that the magnetic field is generated by dynamo action within a common convective envelope we have shown that strong fields may be transferred to the DC.  These fields can then be preserved upon dissipation of the envelope. 

Following the dissipation of the CE, we have found that the system rapidly relaxes on a timescale of about $7\,{\rm Myr}$ to the state we would anticipate given no external field. Although there is no significant dissipation of magnetic energy during this time, the redistribution of magnetic flux towards the interior of the WD results in significant decay of the surface field by around a factor of at least $10-100$. Further decay is prevented by the increasing electrical conductivity towards the interior of the WD. This prevents field diffusing beyond around $r=0.9r_{\rm c}$. Once the field has relaxed to its new configuration it continues to decay but on a timescale much longer than the time scale for cooling of the WD so we expect any further loss of field strength to be minimal.

The maximal rate of transfer of field energy from the CE to the DC occurs when the field is kept fixed. In this case the residual strength of the field following dissipation of the CE is linearly related to the lifetime of the strong field in the CE. If we combine equations~(\ref{ch6.eq.me}) and (\ref{ch6.eq.res}) we find

\begin{equation}
M=5.7\times10^{35}\left(\frac{B_{\rm res}}{10^7\,{\rm G}}\right)^2\left(\frac{a}{0.01\,\rm{au}}\right)\left(\frac{r_{\rm I}}{0.01R_{\odot}}\right)^2\left(\frac{t_{\rm CE}}{2\times10^{15}\,{\rm s}}\right)^{-2}\,{\rm J},
\end{equation}

\noindent where $M$ is the total energy of the CE magnetic field as in equation~(\ref{ch6.eq.me}). If we assume that all of the energy released through orbital decay is transformed into magnetic energy then this requires an envelope lifetime of $1.1\times10^{12}\,$s or $3.6\times10^4\,$yr to produce a $10^7\,$G magnetic field. This lifetime is extended if there is variation in the direction of the field produced by the dynamo or there are energy sinks other than the magnetic field. Because the required envelope lifetime scales linearly with $B_{\rm res}$, weaker fields could be produced on a much shorter time scale. Energetically there is nothing to prevent formation of a $10^9\,$G MWD. As shown, the final field strength scales proportionally with the CE lifetime. If the dynamo is confined to a smaller region or the CE lifetime is extended then this strength of field could be produced. Scenarios this extreme seem unlikely and this is complemented by the rarity of MWDs with such strong magnetic fields. We do not rule out the possibility that the strongest fields may be produced by the merger of a white dwarf and the core of a giant star which have both been strongly magnetized during CE phases.

We have also shown that the production of sufficiently strong fields is almost certainly dependent on some preferred orientation of the magnetic dynamo. In the case where there is no preferred direction the DC magnetic field is confined to a layer of thickness about $(\frac{\alpha}{\eta})^{-1/2}$, where $\alpha$ is the frequency for field variation and $\eta$ is the magnetic diffusivity, and is almost entirely meridional. Despite the generation of strong surface fields during the CE phase in this case, the total magnetic energy of the DC field is constrained by the depth of the magnetic layer. Consequently once the CE disperses and the field of the exposed WD begins to diffuse inwards, the majority of the field strength is lost. This result is confirmed by the numerical simulations. Typically a field strength of $10^{-5}$ relative to the average magnetic field strength of the dynamo is the largest possible residual field. This may be sufficient to produce WDs with weaker fields but the formation of the strongest WD magnetic fields would require far more energy than the system could provide. In the case where the field has some preferred orientation, the proportion of field retained is increased up to a few percent depending on the lifetime of the CE dynamo.

Whilst the arguments we have presented here suggest that the origin of the fields of MWDs may be dynamo action in CEs of closely interacting binaries, there are still many uncertainties. Most of these are the result of insufficient understanding of the formation and evolution of CEs. The field structures we have used to construct our models are simple but complex field geometries are observed in WDs such as Feige~7 and KPD 0253+5052. If CE dynamos are responsible for the origins of MWDs, they should support complex geometries. Further progress depends on a better understanding of the physical processes and energy transfer within the CE.

\begin{savequote}[60mm]
I left the ending ambiguous, because that is the way life is. (Bernardo Bertolucci)
\end{savequote}

\chapter{Conclusions}
\label{ch7}

Throughout this work we have considered a number of different aspects of rotation and magnetic fields in relation to stellar evolution. In many respects, these two phenomena are very closely linked. Magnetic fields show a number of instabilities which cause them to decay rapidly compared to stellar lifetimes. In order for strong magnetic fields to endure in the long term they must either be sustained by some sort of contemporary dynamo or the decay time scale must be extended, such as in degenerate stars. We have considered both the effects of a simple $\alpha$--$\Omega$ dynamo on the evolution of massive main-sequence stars and the formation of magnetic fields in white dwarfs during the common-envelope phase of evolution.
The study of magnetic fields in stars is complicated because of our lack of knowledge about the effect of rotation on stellar evolution. Whilst a great deal of work has gone into modifying the equations for stellar structure to incorporate the effects of rotation, there is much uncertainty regarding the properties of rotationally-driven turbulence. We have examined a number of popular models in order to understand the different predictions of each model and how those differences become evident in observations of populations of rotating stars.

\section{Populations of rotating stars}

In chapter \ref{ch2} we introduced the stellar evolution code, {\sc rose}. Adapted from the Cambridge stellar evolution code {\sc stars}, {\sc rose} is able to simulate rotating stars with a variety of common physical models. The code can include the effects of magnetic field evolution with a simple $\alpha$--$\Omega$ dynamo and a model for magnetic braking. We have also included the capability to evolve the angular momentum distribution in convective zones under a range of different assumptions. All of these factors lead to a great degree of variability in the predictions of simulations of rotating stars. It is important to be aware of these variations and the limitations of any particular model before making predictions involving rotating stars. By adapting the population synthesis code, {\sc starmaker} \citep{Brott11}, for use with our stellar evolution calculations we have been able to predict how these different populations would appear as observations and how those predictions relate to the data from the VLT-FLAMES survey of massive stars \citep{Dufton06,Hunter09}.

In chapter~\ref{ch3} we looked at the specific differences between a number of models for rotation, primarily those of \citet{Talon97} and \citet{Heger00}. Under the assumptions of \citet{Talon97}, main-sequence stars appear more luminous for stars less massive than $10$\Msun\ than the model based on \citet{Heger00}. Stars more massive than $10$\Msun\ appear hotter and more luminous according to \citet{Talon97} than \citet{Heger00}. This difference is also reflected when we consider the effect on the changes of surface composition in rotating stars. We find that very massive stars ($M\approx 60$\Msun) show similar degrees of nitrogen and helium-3 enrichment under the assumptions of both physical models. However, when we look at less massive stars ($M\approx 10$\Msun) we find that the amount of nitrogen enrichment is much lower in stars simulated with the model of \citet{Heger00} than those that use the model of \citet{Talon97}. This supports the idea that higher luminosities arise because of stronger mixing. Stronger mixing leads to more hydrogen being transported between the radiative envelope and the convective core and causes additional nuclear burning within the core. The amount of nitrogen enhancement shows only slight mass dependence when simulations are performed with the model of \citet{Talon97}. On the contrary, the model of \citet{Heger00} exhibits strong mass-dependency. 

We have also found that there is a significant difference between the metallicity-dependence of the two different physical models. We performed simulations for stars at very low metallicity ($Z=0.001$). In these stars we found that the relative mixing between \citet{Heger00} at high and low metallicities is much greater than with \citet{Talon97}. At very low metallicity, very massive stars simulated using the model of \citet{Heger00} are more enriched than those that use \citet{Talon97}. The difference between the two models at lower masses is much less pronounced than at higher metallicities. Unfortunately, there is little prospect of observing this tendency because our ability to measure the changes in the surface chemical abundance is currently limited to our own Galaxy and local galaxies such as the Large and Small Magellanic Clouds. We note that the tendencies are similar for the enrichment of helium-3 but there are differences. Whilst we have examined these differences briefly in this dissertation the study of how different elements are affected by rotational turbulence is an important area of research for understanding the properties of rotational mixing. Some work on this topic has been undertaken by \citet{Frischknecht10} but there is still significant work to be done. 

Also of interest is the way in which angular momentum is transported within convective zones. All current models of which we are aware have convective zones that rotate as solid bodies. This assumption works well but is not necessarily consistent with mixing-length theory. We simulated models using a number of different diffusion coefficients in convective zones but under the assumption of uniform specific angular momentum. There is little difference between the models with different diffusion coefficients because the diffusion is sufficiently strong to keep the convective zone well mixed for a wide range range of coefficients around that predicted by mixing-length theory. However, there is a significant difference between those models that assume solid body rotation in convective zones compared to those with uniform specific angular momentum. For a given surface velocity, stars with uniform specific angular momentum in convective zones have significantly higher total angular momentum than those with cores rotating as solid bodies. This leads to more extended cores owing to the lower effective gravity. More importantly, the change in behaviour of the angular momentum distribution leads to a strong shear layer at the convective boundary. This leads to more mixing between the core and the envelope and hence more chemical enrichment at the surface. However, whilst the internal behaviour is quite different, the effects on the observable properties are fairly small and could easily be masked if the free parameters in the radiative angular momentum diffusion coefficients were modified slightly.

In chapter~\ref{ch4} we considered how the results of chapter~\ref{ch3} would be reflected in observed stellar populations. The Hunter diagram \citep{Hunter09} has become the favoured diagnostic tool for analysing observed populations as it highlights the correlation between the rotation rate and surface nitrogen enrichment. The Hunter diagram does not distinguish a star's age or its mass, for which alternative measures are needed. The effective surface gravity may be used in conjunction with the surface nitrogen enrichment to show how rotational mixing varies with stellar mass. It is particularly useful for single-aged populations of stars because this minimises the degeneracy in the relationship between surface gravity and stellar mass. There is always be some degeneracy between these two variables in rotating populations owing to the effect of centrifugal forces but this effect is small. When we plot nitrogen enrichment against effective surface gravity for synthetic populations, where all the stars have the same age, we find that the populations are confined to a very specific region. Stars with surface gravity below a certain limit have already evolved into giants. Stars with surface gravity above another limit are less massive and have not had sufficient time to become enriched. The effect persists for populations generated with ongoing star formation although the relationship is not as clear. Populations evolved with the model of \citet{Talon97} predict that surface nitrogen is enriched over a narrower range of surface gravities than populations generated with the model of \citet{Heger00}. Populations that use the model of \citet{Talon97} have many more moderately enriched stars than those that use the model of \citet{Heger00} but the maximal enrichment is higher in these populations.

This trend continues for lower metallicities. In chapter~\ref{ch3} we considered how the models vary at very low metallicities ($Z=0.001$). In chapter~\ref{ch4} we considered populations with initial composition of the Large Magellanic Cloud, where the metallicity is not as low. At this metallicity the qualitative behaviour is similar to that at Galactic metallicity and not the very low metallicity case of chapter~\ref{ch3}. When we simulate populations at this metallicity, those that use the model of \citet{Talon97} produce a very confined band of enriched stars. Populations evolved with the model of \citet{Heger00} predict a much greater spread in the population and a greater maximal enrichment. As at Galactic metallicity, this is largely because of the differences in the mass dependencies of each model.

When we introduce selection effects into the simulated populations we find that it becomes far more difficult to distinguish between the different models. Both the models of \citet{Talon97} and \citet{Heger00} give reasonably good fits to the data from the VLT-FLAMES survey of massive stars except that the predicted maximum mixing is higher than observed in each case. This is not unexpected given the difficulty in measuring the nitrogen enrichment and surface rotation rates in rapidly rotating, highly-enriched stars. Even though the populations behave very differently when a wide range of stellar masses are included, the populations with observational effects included are very similar. This is because the mass range of the data from the VLT-FLAMES survey is rather narrow. Only stars with masses between around $10$\Msun\ and $20$\Msun\ remain. Under these conditions, the very different mass dependencies of each model cannot be resolved.

\section{Populations of magnetic, rotating stars}

Introducing magnetic fields into the evolution of rotating stars significantly affects their evolution through additional chemical mixing and magnetic braking. We use a simple radiative dynamo based on the work of \citet{Tayler73} and \citet{Spruit99} and a magnetic braking model based on that of \citet{ud-Doula02}. This has significantly improved on previous efforts to replicate observed populations of rotating, magnetic stars. The key difference with the new model is that the poloidal and toroidal components of the magnetic field are evolved as independent variables via a pair of advection-diffusion equations coupled to the angular momentum evolution equation.

When we examine the relation between mass and surface field strength in this new model we find that two distinct behaviours arise. In stars more massive than around $15$\Msun\ the hydrodynamic turbulence dominates the transport of angular momentum and a dynamo cannot be sustained. In less massive stars the magneto-rotational turbulence dominates and a stable dynamo is established. This leads to two distinct populations of stars, the former composed of very massive stars that do not support a strong field and evolve as discussed in chapters \ref{ch3} and \ref{ch4}. In this population there is a strong positive correlation between surface nitrogen enrichment and rotation rate. The second population supports magnetic fields and its stars are spun down rapidly by the action of magnetic braking. These stars appear in the Hunter diagram \citep{Hunter09} as highly enriched, slowly-rotating stars. It has been suggested that in the past that this population arises owing to the action of magnetic fields but this is the first time that the two distinct types of behaviour have arisen from a single model for the evolution of rotating magnetic stars.

\section{Highly magnetic white dwarfs}

In chapter \ref{ch6} we looked at a different population of magnetic stars, that of white dwarfs with very strong magnetic fields. Owing to the significantly higher frequency of magnetic white dwarfs in close interacting binaries, there is strong evidence that these fields are a feature of binary star evolution. We have considered how the field might arise through dynamo action during a common envelope stage of evolution. The final state of the field that can be generated in the white dwarf is strongly dependent on the form of the field generated in the common envelope. A static field in the common envelope could produce a magnetic field of order $10^7$\,G in around $10^4$\,yr. If the field of the common envelope varies with time this timescale is extended. The lifetime of the external field generated by the common envelope is directly proportional the strength of the residual field strength of the white dwarf so weaker fields may be generated much faster. If the field external to the white dwarf varies rapidly with time then the depth of the residual field in the white dwarf is greatly reduced and so when the common envelope dissipates, the total magnetic energy transferred to the white dwarf is significantly less than if the field was static. As the surface magnetic energy diffuses inwards, the surface magnetic field strength of the white dwarf goes through an initial stage of rapid decay and so, in this case, it is much harder to produce strong residual fields.

Energetically, there is nothing to prevent the generation of stronger fields. The limiting factors for the strength of the white dwarf's magnetic field are the life time of the common envelope and the strength of the dynamo-generated field in the common envelope which may be increased if the field is confined to a smaller region in the envelope. It is possible that the strongest white dwarfs may be produced by merging two cores orbiting in a common envelope which have both been highly magnetised by the field in the envelope.

\section{Future work}

Whilst we have made significant progress in the simulation of the properties of rotation and magnetic fields in stars, there are still many questions that remain open. For main-sequence stars, can we find observational tests that can better distinguish the difference between different models of stellar rotation? In particular can we devise tests that match with available data in different mass ranges? For instance, what are the observational differences in the abundance of other elements for different stellar masses? Also, if we can then identify those models that better reflect the observational data, can we bring those models together into a common stellar evolution and population synthesis code? This would allow us to use more advanced statistical techniques to further refine the model and our knowledge of stellar rotation.

We have also shown that magnetic fields play a critical role in the evolution of stars. If slowly-rotating, highly-enriched stars are produced from the magnetic evolution of stars less massive than $15$\Msun\ then any model that ignores this effect cannot properly estimate the properties of stellar populations that bridge this mass range. The model we have presented for the magnetic evolution of rotating stars is at a very preliminary stage and we expect significant refinements to be made in the future. Of particular interest is how the mass limit for a stable dynamo varies with the free parameters defined in the dynamo model. Because this cut-off is currently close to the lower end of the observable mass range in the observations of the VLT-FLAMES survey of massive stars, even a small change to this limit could have a significant effect on the populations. Analyses such as these could better refine the model by matching the predicted populations to the proportion of expected slowly-rotating, highly-enriched stars in the VLT-FLAMES survey and also the proportions of magnetic stars observed as part of the MiMeS collaboration.

There are still many open questions about the evolution of common envelopes even before the introduction of complicating factors such as magnetic fields. We have presented a simple model of magnetic field generation in white dwarfs embedded in a common envelope. However, observed highly magnetic white dwarfs often have complex field geometries so further work on the structure and evolution of magnetic fields generated in common envelopes is essential for further progress in this area.

\appendix     
\begin{savequote}[60mm]
Percy: I intend to discover, this very afternoon, the secret of alchemy. The hidden art of turning base things into gold. 

Blackadder: I see. And the fact that this secret has eluded the most intelligent people since the dawn of time doesn't dampen your spirits? 

Percy: Oh, no. I like a challenge! 

(Blackadder II, Money)
\end{savequote}

\chapter[Stellar structure derivations]{Derivations for the effects of rotation on stellar structure}
\label{ap1}

\section{The structure parameter, $f_P$}
\label{ap1.fp}

We follow the derivation of \citet{Maeder09} for the structure parameter $f_P$. We define the potential, $\Psi$ as

\begin{equation}
\Psi=\Phi-\frac{1}{2}\Omega^2r^2\sin^2\theta,
\end{equation}

\noindent where the gravitational potential is defined by $\diffb{\Phi}{r}=Gm_r/r^2$, $r$ is the radius, $m_r$ is the mass inside radius $r$ and $\Omega$ is the angular velocity. In polar coordinates, the components of $\nabla\Psi$ are

\begin{equation}
\diffb{\Psi}{r}=\diffb{\Phi}{r}-\Omega^2r\sin^2\theta-r^2\sin^2\theta\Omega\diffb{\Omega}{r}
\end{equation}
\noindent and 
\begin{equation}
\frac{1}{r}\diffb{\Psi}{\theta}=\frac{1}{r}\diffb{\Phi}{\theta}-\Omega^2r\sin\theta\cos\theta-\Omega^2r\sin\theta\cos\theta - r^2\sin^2\theta\Omega\frac{1}{r}\diffb{\Omega}{\theta}.
\end{equation}

\noindent The effective gravity is $\bi{g}_{\rm eff}=(-g_{{\rm eff},r},g_{{\rm eff},\theta},0)$, where the components are given by

\begin{equation}
g_{{\rm eff},r}=\diffb{\Phi}{r}-\Omega^2r\sin^2\theta
\end{equation}

\noindent and

\begin{equation}
g_{{\rm eff},\theta}=\Omega^2r\sin\theta\cos\theta.
\end{equation}

\noindent Comparing the terms of $\bi{g}_{\rm eff}$ and $\nabla\Psi$ we see that

\begin{equation}
\bi{g}_{\rm eff}=-\nabla\Psi-r^2\sin^2\theta\Omega\nabla\Omega.
\end{equation}

\noindent As in section~\ref{ch2.sec.rotstructure} we associate a radius $r_P$ to the isobar for pressure $P$ such that the volume inside the isobar, $V_P$, is

\begin{equation}
V_P\equiv\frac{4\pi}{3}r_P^3.
\end{equation}

\noindent We define the mean value of any quantity $\xi$ over an isobar as

\begin{equation}
<\xi>\equiv\frac{1}{S_P}\oint_{\Psi={\rm const}}\xi\, {\rm d}S_P,
\end{equation}

\noindent where $S_P$ is the total surface area of the isobar and d$S_P$ is an element of the surface. Unless the star is rotating as a solid body or the angular velocity has cylindrical symmetry, the effective gravity cannot be described by the gradient of a potential field. Instead we define $\upsilon=|{\rm d}\Omega/{\rm d}\Psi|$ such that $\nabla\Omega=\upsilon\nabla\Psi$. We can do this on the assumption that $\Omega$ is constant along isobars. This tells us that $\nabla\Omega$ is parallel to $\nabla P$ and from the Navier-Stokes equation we further conclude they are also parallel to $\nabla \Psi$. The average distance between two neighbouring isobars is ${\rm d}n\approx{\rm d}r_P$ and

\begin{equation}
\label{ap1.eq.geff}
g_{\rm eff}=|\bi{g}_{\rm eff}|=(1-r^2\sin^2\theta\Omega\upsilon)\difft{\Psi}{n}.
\end{equation}

\noindent The equation for hydrostatic equilibrium (equation~(\ref{ch2.eq.hydrostatic}) for a non-rotating star) similarly becomes

\begin{equation}
\label{ap1.eq.pressure}
\difft{P}{n}=-\rho(1-r^2\sin^2\theta\Omega\upsilon)\difft{\Psi}{n}.
\end{equation}

\noindent We wish to describe the pressure gradient relative to the independent variable $m_P$, the mass inside radius $r_P$. This gives

\begin{align}
{\rm d}m_P=\int_{\Psi={\rm const}}\rho{\rm d}n\,{\rm d}S_P&={\rm d}\Psi\int_{\Psi=const}\rho\difft{n}{\Psi}\,{\rm d}S_P\nonumber\\
&={\rm d}\Psi\int_{\Psi={\rm const}}\rho\frac{1-r^2\sin^2\theta\Omega\upsilon}{g_{\rm eff}}\,{\rm d}S_P.
\end{align}

\noindent Integrating this equation gives

\begin{equation}
\label{ap1.eq.mpsi}
\difft{\Psi}{m_P}=\frac{1}{\rho(1-r^2\sin^2\theta\Omega\upsilon)<g_{\rm eff}^{-1}>S_P}.
\end{equation}

\noindent When combined with equation~(\ref{ap1.eq.pressure}) this gives

\begin{equation}
\label{ap1.eq.fp}
\difft{P}{m_P}=\frac{-1}{<g_{\rm eff}^{-1}>S_P}.
\end{equation}

\noindent We define $f_P$ as 

\begin{equation}
f_P=\frac{4\pi r_P^4}{G m_p S_P<g_{\rm eff}^{-1}>}
\end{equation}

\noindent so that the equation for hydrostatic equilibrium is
 
\begin{equation}
\difft{P}{m_P}=-\frac{Gm_P}{4\pi r_P^4}f_P.
\end{equation}

\noindent This has the same form as the equation for hydrostatic equilibrium for non-rotating stars except for the factor $f_P$.

\section{The structure parameter, $f_T$}
\label{ap1.ft}

To derive the equation for $f_T$ we use many of the identities used in appendix~\ref{ap1.fp}. The radiative flux, $F$, is given by

\begin{equation}
F=-\frac{4acT^4}{3\kappa\rho}\difft{T}{n}=-\frac{4acT^4}{3\kappa\rho}\difft{T}{m_P}\difft{m_P}{\Psi}\difft{\Psi}{n}.
\end{equation}

\noindent We substitute the relations from equations~(\ref{ap1.eq.mpsi}) and~(\ref{ap1.eq.geff}) to find that

\begin{equation}
F=-\frac{4acT^3}{3\kappa\rho}\difft{T}{m_P}\rho<g_{\rm eff}^{-1}>S_P<g_{\rm eff}>.
\end{equation}

\noindent The luminosity, $L_P$ is then

\begin{equation}
\label{ap1.eq.lpft}
L_P=-\frac{4ac}{3}<g_{\rm eff}^{-1}>S^2_P<\frac{T^3g_{\rm eff}}{\kappa}\difft{T}{m_P}>\approx -\frac{4ac\overline{T}^3S_P^2}{3\overline{\kappa}}\difft{\overline{T}}{m_P}<g_{\rm eff}><g_{\rm eff}^{-1}>,
\end{equation}

\noindent where an over bar denotes the average of a quantity over an isobar. By comparing equations~(\ref{ap1.eq.lpft}) and~(\ref{ch2.eq.nablar}) we define

\begin{equation}
\nabla_{r,P}=\difft{\ln P}{\ln T_P}=\frac{f_T}{f_P}\frac{3\kappa PL_P}{16\pi a c G m_P\overline{T}^4},
\end{equation}

\noindent where

\begin{equation}
f_T=\left(\frac{4\pi r_P^2}{S_P}\right)^2\frac{1}{<g_{\rm eff}><g_{\rm eff}^{-1}>}
\end{equation}

\noindent and we have used equation~(\ref{ap1.eq.fp}). By repeating the derivation of the equation for radiative equilibrium in radiative zones from section~\ref{ch2.sec.rotstructure} we get

\begin{align}
\difft{\ln T}{\ln P}&=-\nabla_{r,P}\difft{\ln P}{m_p}\nonumber \\
&=-\frac{G m_P}{4\pi r_P^4}f_P\nabla_P,
\end{align}

\noindent where

\begin{equation}
\nabla_P=\min\left(\nabla_{\rm ad},\nabla_{r,P}\right)
\end{equation}

\noindent and we have assumed the adiabatic gradient has remained roughly unchanged.

\section{The Von Zeipel Theorem}
\label{ap1.vonzeipel}

The energy flux, $\bi{F}$, through a star rotating as a solid body with angular velocity, $\Omega$, is

\begin{align}
\bi{F}(\Omega,\theta)&=-\chi\nabla T(\Omega,\theta) \nonumber \\
&=\chi\difft{T}{P}\nabla P(\Omega,\theta) \nonumber \\
&=-\rho\chi\difft{T}{P}\bi{g}_{\rm eff}(\Omega,\theta), \label{ap1.eq.vz1}\\
\end{align}

\noindent where $\chi=4acT^3/3\kappa\rho$. The first identity comes from the equation for hydrostatic equilibrium. The second identity is possible because the star is rotating as a solid body and so the equipotentials and isobars coincide. The total luminosity, $L$, on an equipotential surface, $S_P$ is

\begin{equation}
L=\int_{S_P}\bi{F}(\Omega,\theta)\cdot \hat{\bi{n}}\,{\rm d}S_P=\rho\chi\difft{T}{P}\int_{S_P}\nabla\Psi(\Omega,\theta)\cdot \hat{\bi{n}}\,{\rm d}S_P,
\end{equation}

\noindent where $\hat{\bi{n}}$ is a unit vector normal to the surface, ${\rm d}S_P$ is a surface element of $S_P$ and $\Psi$ is the total force potential. We know from Poisson's equation that $\nabla^2\Psi=4\pi G\rho-2\Omega^2$ and so

\begin{equation}
\label{ap1.eq.vz2}
L=\rho\chi\difft{T}{P}\int_{V_P}\nabla^2\Psi\,{\rm d}V_P=\rho\chi\difft{T}{P}\int_{V_P}4\pi G\rho - 2\Omega^2 \,{\rm d}V_P
\end{equation}

\noindent where $S_P$ encloses $V_P$. By combining equations~(\ref{ap1.eq.vz1}) and~(\ref{ap1.eq.vz2}) we get

\begin{equation}
\rho\chi\difft{T}{P}=\frac{L}{4\pi G m_r\left(1-\frac{\Omega^2}{2\pi G\overline{\rho}_m}\right)},
\end{equation}

\noindent where $\overline{\rho}_m$ is the average density contained within mass, $m$. If we take this relation at the stellar surface we get

\begin{equation}
\bi{F}(\Omega,\theta)=-\frac{L}{4\pi G M^*}\bi{g}_{\rm eff},
\end{equation}

\noindent where

\begin{equation}
M^*=M\left(1-\frac{\Omega^2}{2\pi G\overline{\rho}_M}\right)
\end{equation}

\noindent and $M$ is the stellar mass. The effective temperature is given by $L=\sigma T_{\rm eff}^4$ where $\sigma$ is the Stefan–-Boltzmann constant. We conclude that

\begin{equation}
T_{\rm eff}=\left(\frac{L}{4\pi \sigma G M^*}\right)g^{1/4}_{\rm eff}.
\end{equation}

\noindent We conclude that $T_{\rm eff}\propto g_{\rm eff}(\Omega,\theta)^{1/4}$. Because the effective gravity is lower at the  equator than at the poles, owing to the centrifugal force, the surface temperature of a rotating star is higher at the poles than at the equator.

\backmatter
\bibliographystyle{plainnat}
\bibliography{thesis}

\end{document}